\pdfoutput=1
\documentclass[a4paper,openany,titlepage,twoside]{report}
\usepackage{a4wide}
\usepackage{graphicx}
\usepackage[T1]{fontenc}
\usepackage[utf8]{inputenc}
\usepackage[UKenglish]{babel}
\usepackage{siunitx}
\usepackage[dvipsnames]{xcolor}
\usepackage{amsmath,amssymb}
\usepackage{mathtools}
\usepackage{dsfont}
\usepackage{upgreek}
\usepackage{tikz}
\usepackage{ifthen}
\usepackage{calculator}
\usepackage{tikz-feynman}
\usepackage{braket}
\usepackage{placeins}
\usepackage{bm}
\usepackage[boxruled,lined,commentsnumbered]{algorithm2e}
\usepackage{xfrac}
\usepackage[final]{changes}
\usepackage{subfigure}
\usepackage{setspace}
\usepackage{datetime}
\usepackage{csquotes}
\usepackage{appendix}
\usepackage[backend=bibtex,style=ieee,citestyle=numeric-comp]{biblatex}
\usepackage[pdftitle={The Hubbard Model on the Honeycomb Lattice with Hybrid Monte Carlo},
pdfauthor={Johann Ostmeyer}]{hyperref}
\usepackage{cleveref}

\graphicspath{{figures/},
	{ising-hmc-paper/pics/},
	{hankel-paper/plots/},
	{hubbard-critical-coupling/data/},
	{hubbard-critical-coupling/figures/},
	{hubbard-observables/data/},
	{hubbard-observables/figure/}
}

\newdateformat{monthyeardate}{%
	\monthname[\THEMONTH] \THEYEAR}
\newdateformat{daymonthyeardate}{%
\THEDAY th\ \monthname[\THEMONTH] \THEYEAR}

\DeclareMathOperator{\im}{i}
\DeclareMathOperator{\ord}{\mathcal{O}}
\DeclareMathOperator{\acosh}{acosh}

\DeclareMathOperator{\real}{Re}
\DeclareMathOperator{\rt}{Rt}
\DeclareMathOperator{\tr}{Tr}
\DeclareMathOperator{\ti}{Ti}

\DeclareMathOperator{\thetaord}{\Theta}

\newcommand{\littleo}[1]{\ensuremath{o\left(#1\right)}}
\newcommand{\bigtheta}[1]{\ensuremath{\thetaord\left(#1\right)}}

\newcommand{\up}{\ensuremath{\uparrow}}
\newcommand{\down}{\ensuremath{\downarrow}}

\newcommand{\Appref}[1]{Appendix~\ref{sec:#1}}

\newcommand{\effm}{effective mass}

\newcommand{\spc}{single-particle correlator}

\newcommand{\efford}[1]{\ensuremath{\ord_{\text{e}}(#1)}}
\newcommand{\erwartung}[1]{\ensuremath{\left\langle#1\right\rangle}}

\newcommand{\pdagger}{{\phantom{\dagger}}}
\newcommand{\kappadelta}{\ensuremath{\tilde{\kappa}}}
\newcommand{\phidelta}{\ensuremath{\tilde{\phi}}}
\newcommand{\mdelta}{\ensuremath{\tilde{m}_s}}

\newcommand{\meff}{\ensuremath{m_\text{eff}}\xspace}
\newcommand{\Z}{\ensuremath{\mathcal{Z}}}
\newcommand{\D}{\ensuremath{\mathcal{D} }}
\newcommand{\nt}{\ensuremath{N_t}\xspace}
\newcommand{\delsqu}[2]{\ensuremath{\frac{\partial^2 #1}{\partial #2 ^2}}}
\newcommand{\aaverage}[1]{\average{\average{#1}}}
\newcommand{\disconnected}[2]{\ensuremath{\average{#1 #2} - \average{#1}\average{#2}}}

\newcommand{\matr}[2]{\left(\begin{array}{#1}#2\end{array}\right)}
\newcommand{\adjoint}{\ensuremath{{}^{\dagger}}}
\newcommand{\conjugate}{\ensuremath{{}^{*}}}
\newcommand{\operator}{\ensuremath{\mathcal{O}}}

\newcommand{\SF}{\ensuremath{S_{F}}\xspace}
\newcommand{\SAF}{\ensuremath{S_{AF}}\xspace}

\newcommand{\qm}{\ensuremath{q_{-}}\xspace}

\newcommand{\grapheneColor}[2]{
	\foreach \x in {0,...,#1}
	{
		\MODULO{\x}{2}{\xmod} 
		\foreach \y in {0,...,#2}
		{
			\tikzset{yshift={\y*1.732cm+Mod(\x,2)*0.866cm}, xshift={\x*1.5cm}}
			\begin{scope}
				\draw[line width=2pt] (0,0)--(1,0);
				\ifthenelse{\xmod=0 \OR \y<#2 \AND \x>0}{\draw[line width=2pt] (0,0)--(-0.5, 0.866);}{}
				\ifthenelse{\xmod>0 \OR \y>0 \AND \x>0}{\draw[line width=2pt] (0,0)--(-0.5, -0.866);}{}
				\ifthenelse{\x=1 \AND \y=0}{\node (A1) at (0,0) {}; \node (A2) at (1.5, 0.866) {}; \node (A3) at (1.5, -0.866) {};}{}
				
				\fill[red] (0,0) circle (4pt);
				
				\fill[blue] (1,0) circle (4pt);
				\ifthenelse{\xmod=0 \OR \y<#2 \AND \x>0}{\fill[blue] (-0.5, 0.866) circle (4pt);}{}
				\ifthenelse{\y > 0 \OR \xmod > 0 \AND \x>0}{\fill[blue] (-0.5, -0.866) circle (4pt);}{}
			\end{scope}
}}}

\newcommand{\grapheneColorShift}[4]{
	\foreach \x in {0,...,#1}
	{
		\MODULO{\x}{2}{\xmod} 
		\foreach \y in {0,...,#2}
		{
			\tikzset{yshift={\y*1.732cm+Mod(\x,2)*0.866cm+#4}, xshift={\x*1.5cm+#3}}
			\begin{scope}
				\draw[line width=2pt] (0,0)--(1,0);
				\ifthenelse{\xmod=0 \OR \y<#2 \AND \x>0}{\draw[line width=2pt] (0,0)--(-0.5, 0.866);}{}
				\ifthenelse{\xmod>0 \OR \y>0 \AND \x>0}{\draw[line width=2pt] (0,0)--(-0.5, -0.866);}{}
				\ifthenelse{\x=1 \AND \y=0}{\node (A1) at (0,0) {}; \node (A2) at (1.5, 0.866) {}; \node (A3) at (1.5, -0.866) {};}{}
				
				\fill[red] (0,0) circle (4pt);
				
				\fill[blue] (1,0) circle (4pt);
				\ifthenelse{\xmod=0 \OR \y<#2 \AND \x>0}{\fill[blue] (-0.5, 0.866) circle (4pt);}{}
				\ifthenelse{\y > 0 \OR \xmod > 0 \AND \x>0}{\fill[blue] (-0.5, -0.866) circle (4pt);}{}
			\end{scope}
}}} \let\oldeqref\eqref
\usepackage{xspace}
\usepackage{bbm}

\newcommand{\Secref}[1]{Section~\ref{sec:#1}}

\newcommand{\figref}[1]{Fig.~\ref{fig:#1}\xspace}
\newcommand{\Figref}[1]{Figure~\ref{fig:#1}\xspace}
\renewcommand{\eqref}[1]{(\ref{eq:#1})\xspace}

\renewcommand{\Ref}[1]{Ref.~\cite{#1}}
\newcommand{\Refs}[1]{Refs.~\cite{#1}}

\newcommand{\includeepslatex}[1]{\resizebox{0.443\textwidth}{!}{{\Large\input{pics/#1}}}}

\newcommand{\goesto}{\ensuremath{\rightarrow}}
\newcommand{\infinity}{\infty}

\newcommand{\order}[1]{\ensuremath{\mathcal{O}\left(#1\right)}\xspace}

\newcommand{\Reals}{\mathbb{R}\xspace}

\DeclareMathOperator{\sech}{sech}

\newcommand{\oneover}[1]{\ensuremath{\frac{1}{#1}}}                             %
\newcommand{\inverse}{\ensuremath{^{-1}}}                                       %
\newcommand{\half}{\ensuremath{\frac{1}{2}} }                                   %

\newcommand{\average}[1]{\ensuremath{\left\langle #1 \right\rangle}\xspace}

\newcommand{\transpose}{\ensuremath{{}^{\top}}}

\newcommand{\eto}[1]{\ensuremath{\mathrm{e}^{#1}}}
\newcommand{\md}{\ensuremath{\mathrm{d}}}

\newcommand{\del}[2]{\ensuremath{\frac{\partial #1}{\partial#2}}}

\newcommand{\tK}{\ensuremath{\tilde{K}}}

\let\builtinLaTeX\LaTeX
\def\LaTeX{\builtinLaTeX\xspace}
 
\makeatletter%
\begin{document}	
	\pagenumbering{roman}
	\pdfbookmark{Titlepage}{cover}
	\onehalfspacing
	\begin{titlepage}
		\begin{center}
			\Huge \textbf{The Hubbard Model\\ on the Honeycomb Lattice\\ with Hybrid Monte Carlo} \\
			\vspace*{\baselineskip}
			
			\Huge Johann Ostmeyer\\
			\Huge
			\vspace*{\baselineskip}

			\LARGE Dissertation in Physics\\
			 designed in the\\
			 Helmholtz Institute for Radiation and Nuclear Physics\\
			 submitted to the\\
			 Faculty of Mathematics and Natural Sciences\\
			 of the\\
			 University of Bonn\\
			
			\Huge \vspace*{\baselineskip}
			\LARGE Bonn, June 2021
		\end{center}
	\end{titlepage}

	\vspace*{\fill}
	
	Published 2021\\
	
	Date of the oral exam: November 4th 2021
	
	\begin{enumerate}
		\item Reviewer: Prof. Carsten Urbach
		\item Reviewer: Prof. Thomas Luu
		\item Reviewer: Prof. Simon Stellmer
		\item Reviewer: Prof. Frank Neese
		\item Reviewer: Prof. Uwe-Jens Wiese
	\end{enumerate}

	Dissertation in Physik zur Erlangung des Doktorgrades (Dr.\ rer.\ nat.)
	angefertigt im Helmholtz-Institut für Strahlen- und Kernphysik
	mit Genehmigung der Mathematisch-Naturwissenschaftlichen Fakultät
	der Rheinischen Friedrich-Wilhelms-Universität Bonn.

\clearpage{}%
\pdfbookmark{Publications}{publications}
\chapter*{Publications}
The publications~\cite{ising,generalisedProny,semimetalmott,more_observables} are part of this thesis. Furthermore~\cite{my_tensor_networks,beer_mat,leveragingML,qsimulatR,hadron,my_beer_tapping} have been published during my PhD studies.
\nocite{acceleratingHMC}

\printbibliography[heading=none,keyword=own]\clearpage{}%
\clearpage{}%
\pdfbookmark{Abstract}{abstract}
\chapter*{Abstract}

We take advantage of recent improvements in the grand canonical Hybrid Monte Carlo (HMC) algorithm, to perform
a precision study of the single-particle gap in the hexagonal Hubbard model, with on-site electron-electron interactions. After
carefully controlled analyses of the Trotter error, the thermodynamic limit, and finite-size scaling with inverse temperature, we find a
critical coupling of $U_c/\kappa=\num{3.835(14)}$ and the critical exponent $\nu=\num{1.181(43)}$
for the semimetal-antiferromagnetic Mott insulator quantum phase transition in the hexagonal Hubbard Model.
Based on these results, we provide a unified, comprehensive treatment of all operators that contribute to the anti-ferromagnetic, ferromagnetic, and 
charge-density-wave structure factors and order parameters of the hexagonal Hubbard Model. We expect our findings
to improve the consistency of Monte Carlo determinations of critical exponents. We perform a data collapse analysis and determine the critical exponent $\upbeta=\num{0.898(37)}$. We consider our findings in view
of the $SU(2)$ Gross-Neveu, or chiral Heisenberg, universality class. We also discuss the computational scaling of the HMC algorithm. Our methods are applicable to a wide range of lattice theories 
of strongly correlated electrons.

The Ising model, a simple statistical model for ferromagnetism, is one such theory.
There are analytic solutions for low dimensions and very efficient Monte Carlo methods, such as cluster algorithms, for simulating this model in special cases.
However most approaches do not generalise to arbitrary lattices and couplings.
We present a formalism that allows one to apply HMC simulations to the Ising model, demonstrating how a system with discrete degrees of freedom can be simulated with continuous variables.
Because of the flexibility of HMC, our formalism is easily generalizable to arbitrary modifications of the model, creating a route to leverage advanced algorithms such as shift preconditioners and multi-level methods, developed in conjunction with HMC.

We discuss the relation of a variety of different
methods to determine energy levels
in lattice field theory simulations: the generalised eigenvalue, the Prony,
the generalised pencil of function and the Gardner 
methods. All three former methods can be
understood as special cases of a generalised eigenvalue problem.
We show analytically that the leading corrections to an energy $E_l$
in all three methods due to unresolved states decay asymptotically
exponentially like $\exp(-(E_{n}-E_l)t)$. Using synthetic data we
show that these corrections behave as expected also in practice. We
propose a novel combination of the generalised eigenvalue and the
Prony method, denoted as GEVM/PGEVM, which helps to increase the energy
gap $E_{n}-E_l$. We illustrate its usage and performance using
lattice QCD examples. The Gardner method on the other hand
is found less applicable to realistic noisy data.\clearpage{}%

	\pdfbookmark{Contents}{contents}
	\tableofcontents
	
	\cleardoublepage
	\pagenumbering{arabic}
	\allowdisplaybreaks[1]

\clearpage{}%
\chapter{Introduction}

How much life time did you waste waiting for your computer? Wouldn't it be loverly\footnote{\textit{My Fair Lady} (1956) by Alan Jay Lerner and Frederick Loewe} to have a computer two orders of magnitude faster than anything we have today? How about making this computer much more energy efficient for good measure?

There is reason to believe this hope might not remain science fiction. Nowadays computers are based on silicon transistors and their maximum clock has not improved beyond several $\si{\giga\hertz}$ for the last decade with the world record~\cite{fastest_silicon_computer} of approximately $\SI{8.4}{\giga\hertz}$ dating back to 2011. Roughly at the same time a new type of transistors based on graphene has been introduced and shown to reach clock rates of up to $\SI{100}{\giga\hertz}$~\cite{graphene_transistor}. Since then graphene transistors have been continuously improved~\cite{transistor_review2013} with a first computer purely based on this technology built in 2013 and a first computer executing a `Hello world' program demonstrated in 2019~\cite{first_nanotube_computer,Hills2019ModernMB}.
Clearly, graphene represents a highly interesting material and its experimental investigations have been honoured by the Nobel Prize in Physics in 2010.
The theoretical understanding of graphene is therefore imperative.
In the following we will explain what graphene is, which properties make it an ideal candidate for highly efficient computers and how this work contributes to a better understanding of these properties.\\

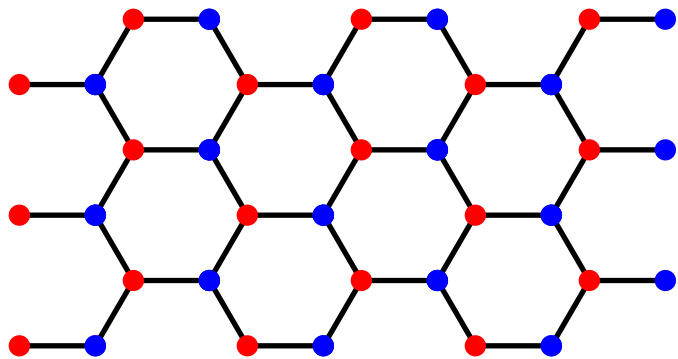
\begin{figure}[ht]
	\centering
	\begin{tikzpicture}
		\grapheneColor{5}{2}
	\end{tikzpicture}
	\caption[Honeycomb lattice of graphene.]{Honeycomb lattice of graphene. The red and the blue points form the two triangular sublattices respectively.}
	\label{fig:honeycomb_lattice}
\end{figure}

Graphene is the only known material consisting of a single atomic layer~\cite{Novoselov,GeimNovoselovReview}. Carbon atoms form a honeycomb lattice consisting out of two triangular Bravais sublattices with each site's nearest neighbours belonging to the opposite sublattice as shown in \cref{fig:honeycomb_lattice}. This means that the lattice can be coloured using two alternating colours. Graphene and derived carbon nanostructures like nanotubes and fullerenes have unique physical properties including unrivalled mechanical strength~\cite{graphene_strength} and extraordinary electromagnetic properties~\cite{CastroNeto2009,Saito1998,overview_graphene,Kotov2012}. We are going to investigate the latter properties throughout this work.

In order to do so, we employ the so-called Hubbard model~\cite{Hubbard:1963} which describes electronic interactions in a simple way. It is assumed that the carbon atoms composing graphene have fixed lattice positions and moreover most of the six electrons per atom are tightly bound to the atoms. On average only one electron per site is allowed to move and thus contribute to the electromagnetic properties of the material. These electrons are confined to the lattice points at any given time, but they can instantly hop from one lattice point to a nearest neighbour. The Pauli principle forbids two or more electrons of the same spin simultaneously at a site. Hence, exactly zero, one or two electrons (of opposite spin) can be at the same lattice point simultaneously. In addition, an on-site interaction $U$ models the repulsive force of the identically charged particles.

There are various extensions to the Hubbard model in this minimalistic form and we are going to comment on some of them later. For now we stick to this form as it will be used in the main part of this thesis. Let us add that we use a particle-hole basis, that is we count the present spin-up particles and the absent spin-down particles, therefore our Hamiltonian reads
\begin{equation}
	H=-\kappa \sum_{\erwartung{x,y}}\left(p^\dagger_{x}p^\pdagger_{y}+h^\dagger_{x}h^\pdagger_{y}\right)+\frac{U}{2}\sum_{x}\rho_x\rho_x\,,\qquad \rho_x = p^\dagger_x p_x -h^\dagger_x h_x\,,\label{eqn:article_hole_hamiltonian}
\end{equation}
where $\erwartung{x,y}$ denotes nearest neighbour tuples, $p$ and $h$ are fermionic particle and hole annihilation operators, $\kappa$ is the hopping amplitude and $\rho_x$ is the charge operator.

\begin{figure}[ht]
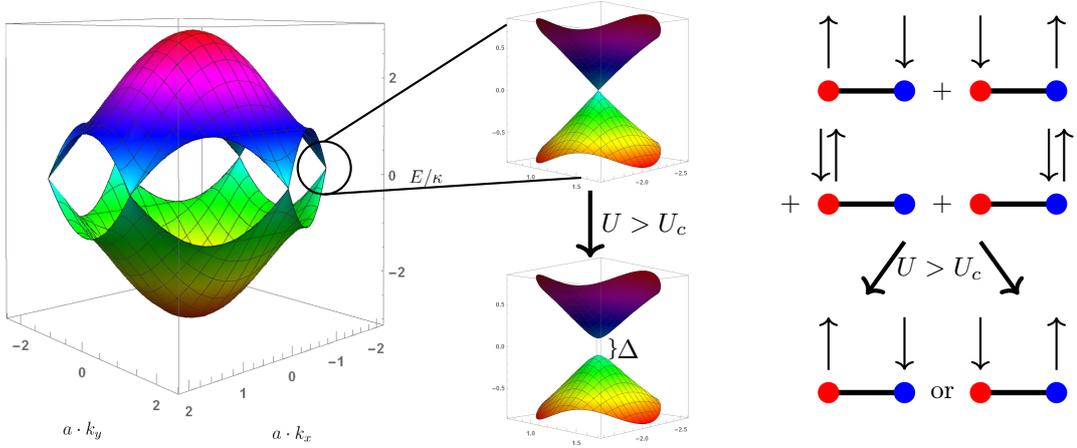

\centering
\hfill
\begin{tikzpicture}
	\node[anchor=south west,inner sep=0] (band) at (0,-1) {\includegraphics[height=0.25\textheight]{band_structure.png}};
	\node[anchor=south west,inner sep=0] (cone) at (6.4,2.4) {\includegraphics[height=0.11\textheight]{dirac_cone.png}};
	\node[anchor=south west,inner sep=0] (open) at (6.4,-1.) {\includegraphics[height=0.11\textheight]{dirac_gap.png}};
	\node at (8,0.3) {\footnotesize$\}$\Large};
	\node at (8.2,0.3) {$\Delta$};
	\begin{scope}[x={(cone.south east)},y={(cone.north west)}]
		\draw[->,ultra thick] (cone) -- (open) node[midway,right] {$U>U_c$};
	\end{scope}
	\color{black}
	\begin{scope}
		\draw[thick] (4.2,2.7) circle (0.35);
		\draw[thick] (4.2,3.05) -- (6.6,4.6);
		\draw[thick] (4.2,2.35) -- (7.58,2.57);
	\end{scope}
\end{tikzpicture}
\hfill
\raisebox{0.1\height}{
	\begin{tikzpicture}
		\grapheneColor{0}{0}
		\draw[thick,->] (0,.3)--(0,1);
		\draw[thick,->] (1,1)--(1,.3);
	\node at (1.5,0) {$+$};		
		\grapheneColorShift{0}{0}{2cm}{0}
		\draw[thick,->] (2,1)--(2,.3);
		\draw[thick,->] (3,.3)--(3,1);
		
		\node at (-.5,-1.5) {$+$};		
		\grapheneColorShift{0}{0}{0}{-1.5cm}
		\draw[thick,->] (-0.1,-.5)--(-0.1,-1.2);
		\draw[thick,->] (.1,-1.2)--(.1,-.5);
		
		\node at (1.5,-1.5) {$+$};		
		\grapheneColorShift{0}{0}{2cm}{-1.5cm}
		\draw[thick,->] (2.9,-.5)--(2.9,-1.2);
		\draw[thick,->] (3.1,-1.2)--(3.1,-.5);
		
		\draw[->,ultra thick] (1.,-2) -- (.5,-2.7) node[midway,right] {$U>U_c$};
		\draw[->,ultra thick] (2.,-2) -- (2.5,-2.7);
		
		\grapheneColorShift{0}{0}{0}{-4cm}
		\draw[thick,->] (0,-3.7)--(0,-3);
		\draw[thick,->] (1,-3)--(1,-3.7);
		\node at (1.5,-4) {or};		
		\grapheneColorShift{0}{0}{2cm}{-4cm}
		\draw[thick,->] (2,-3)--(2,-3.7);
		\draw[thick,->] (3,-3.7)--(3,-3);
	\end{tikzpicture}
\hfill
} 	\caption{Left: The two energy bands (in multiples of the hopping $\kappa$) of the non-interacting Hubbard model as a function of the momentum $k$ normalised by the lattice spacing $a$. Center: Inset showing the Dirac cones. A band gap $\Delta$ separating the bands opens in the phase transition, once a critical coupling $U_c$ is surpassed. The bottom figure is only a qualitative visualisation, not the exact result. Right: The sublattice symmetry is broken at the same critical coupling and the disordered state (a superposition of all possibilities) transitions to an antiferromagnetic order. We show in \Cref{chap:afm_character} that the transitions happen simultaneously.}
	\label{fig:opening_gap}
\end{figure}

There are special cases in which the Hubbard model on the honeycomb lattice can be solved exactly. For instance the tight binding limit~\cite{Bloch1929,tight_binding} with $U=0$ has an analytic solution that features two energy bands touching at the so called Dirac points with a linear (relativistic) dispersion relation~\cite{band_theory} as depicted on the left in \cref{fig:opening_gap}. Furthermore the density of states goes to zero at exactly this point. These two properties define a semimetal and they are in surprisingly good agreement with experimental measurements of graphene which is found to be a good electric conductor.
In contrast to the hopping strength $\kappa\approx \SI{2.7}{\electronvolt}$ well determined experimentally for graphene~\cite{CastroNeto2009,Saito1998}, the coupling $U$ is not known from experiment.
Moreover the general Hubbard model with $U\neq 0$ has neither analytic nor perturbative solutions~\cite{Giuliani2009,PhysRevB.92.045111} and exact numerical solutions become unfeasible for physically interesting numbers of lattice sites because the dimension of the Fock space grows exponentially in size. This necessitates approximate solutions like the stochastic algorithm we introduce below.

By now it is well known that the Hubbard model on the honeycomb lattice undergoes a zero-temperature quantum phase transition at some critical coupling $U_c$~\cite{Meng2010,Assaad:2013xua,Wang:2014cbw,Otsuka:2015iba}. For $U<U_c$ the system is in a conducting semi-metallic state, while above this critical coupling a band gap opens (visualised in the central column of \cref{fig:opening_gap}), so it becomes a Mott insulator.
This is important because it allows one to switch between a conducting and an insulating state which is precisely what transistors do. The changes in $U$ required for this switching can in practice be induced by external electrical field or by mechanical stress. In contrast to silicon transistors, no electrons have to be moved physically in order to perform the switching, thus graphene transistors respond much faster and require less energy.
Experimentally, the value of $U$ in graphene can be confined to the region $U<U_c$ without Mott gap~\cite{Mott_1937,overview_graphene,Kotov2012}, the value of $U_c$ however cannot be measured. $U_c$ therefore has to be determined by theoretical or numerical investigations of the Hubbard model as we do in this work.

\begin{figure}[ht]
	\raisebox{0.06\height}{
		\resizebox{0.5\textwidth}{!}{{\Large
\begingroup
  \inputencoding{latin1}%
  \makeatletter
  \providecommand\color[2][]{%
    \GenericError{(gnuplot) \space\space\space\@spaces}{%
      Package color not loaded in conjunction with
      terminal option `colourtext'%
    }{See the gnuplot documentation for explanation.%
    }{Either use 'blacktext' in gnuplot or load the package
      color.sty in LaTeX.}%
    \renewcommand\color[2][]{}%
  }%
  \providecommand\includegraphics[2][]{%
    \GenericError{(gnuplot) \space\space\space\@spaces}{%
      Package graphicx or graphics not loaded%
    }{See the gnuplot documentation for explanation.%
    }{The gnuplot epslatex terminal needs graphicx.sty or graphics.sty.}%
    \renewcommand\includegraphics[2][]{}%
  }%
  \providecommand\rotatebox[2]{#2}%
  \@ifundefined{ifGPcolor}{%
    \newif\ifGPcolor
    \GPcolortrue
  }{}%
  \@ifundefined{ifGPblacktext}{%
    \newif\ifGPblacktext
    \GPblacktexttrue
  }{}%
  \let\gplgaddtomacro\g@addto@macro
  \gdef\gplbacktext{}%
  \gdef\gplfronttext{}%
  \makeatother
  \ifGPblacktext
    \def\colorrgb#1{}%
    \def\colorgray#1{}%
  \else
    \ifGPcolor
      \def\colorrgb#1{\color[rgb]{#1}}%
      \def\colorgray#1{\color[gray]{#1}}%
      \expandafter\def\csname LTw\endcsname{\color{white}}%
      \expandafter\def\csname LTb\endcsname{\color{black}}%
      \expandafter\def\csname LTa\endcsname{\color{black}}%
      \expandafter\def\csname LT0\endcsname{\color[rgb]{1,0,0}}%
      \expandafter\def\csname LT1\endcsname{\color[rgb]{0,1,0}}%
      \expandafter\def\csname LT2\endcsname{\color[rgb]{0,0,1}}%
      \expandafter\def\csname LT3\endcsname{\color[rgb]{1,0,1}}%
      \expandafter\def\csname LT4\endcsname{\color[rgb]{0,1,1}}%
      \expandafter\def\csname LT5\endcsname{\color[rgb]{1,1,0}}%
      \expandafter\def\csname LT6\endcsname{\color[rgb]{0,0,0}}%
      \expandafter\def\csname LT7\endcsname{\color[rgb]{1,0.3,0}}%
      \expandafter\def\csname LT8\endcsname{\color[rgb]{0.5,0.5,0.5}}%
    \else
      \def\colorrgb#1{\color{black}}%
      \def\colorgray#1{\color[gray]{#1}}%
      \expandafter\def\csname LTw\endcsname{\color{white}}%
      \expandafter\def\csname LTb\endcsname{\color{black}}%
      \expandafter\def\csname LTa\endcsname{\color{black}}%
      \expandafter\def\csname LT0\endcsname{\color{black}}%
      \expandafter\def\csname LT1\endcsname{\color{black}}%
      \expandafter\def\csname LT2\endcsname{\color{black}}%
      \expandafter\def\csname LT3\endcsname{\color{black}}%
      \expandafter\def\csname LT4\endcsname{\color{black}}%
      \expandafter\def\csname LT5\endcsname{\color{black}}%
      \expandafter\def\csname LT6\endcsname{\color{black}}%
      \expandafter\def\csname LT7\endcsname{\color{black}}%
      \expandafter\def\csname LT8\endcsname{\color{black}}%
    \fi
  \fi
    \setlength{\unitlength}{0.0500bp}%
    \ifx\gptboxheight\undefined%
      \newlength{\gptboxheight}%
      \newlength{\gptboxwidth}%
      \newsavebox{\gptboxtext}%
    \fi%
    \setlength{\fboxrule}{0.5pt}%
    \setlength{\fboxsep}{1pt}%
\begin{picture}(7200.00,5040.00)%
    \gplgaddtomacro\gplbacktext{%
      \csname LTb\endcsname%
      \put(814,921){\makebox(0,0)[r]{\strut{}$0$}}%
      \csname LTb\endcsname%
      \put(814,1354){\makebox(0,0)[r]{\strut{}$0.1$}}%
      \csname LTb\endcsname%
      \put(814,1787){\makebox(0,0)[r]{\strut{}$0.2$}}%
      \csname LTb\endcsname%
      \put(814,2220){\makebox(0,0)[r]{\strut{}$0.3$}}%
      \csname LTb\endcsname%
      \put(814,2653){\makebox(0,0)[r]{\strut{}$0.4$}}%
      \csname LTb\endcsname%
      \put(814,3086){\makebox(0,0)[r]{\strut{}$0.5$}}%
      \csname LTb\endcsname%
      \put(814,3520){\makebox(0,0)[r]{\strut{}$0.6$}}%
      \csname LTb\endcsname%
      \put(814,3953){\makebox(0,0)[r]{\strut{}$0.7$}}%
      \csname LTb\endcsname%
      \put(814,4386){\makebox(0,0)[r]{\strut{}$0.8$}}%
      \csname LTb\endcsname%
      \put(814,4819){\makebox(0,0)[r]{\strut{}$0.9$}}%
      \csname LTb\endcsname%
      \put(1246,484){\makebox(0,0){\strut{}$1$}}%
      \csname LTb\endcsname%
      \put(1997,484){\makebox(0,0){\strut{}$1.5$}}%
      \csname LTb\endcsname%
      \put(2748,484){\makebox(0,0){\strut{}$2$}}%
      \csname LTb\endcsname%
      \put(3499,484){\makebox(0,0){\strut{}$2.5$}}%
      \csname LTb\endcsname%
      \put(4250,484){\makebox(0,0){\strut{}$3$}}%
      \csname LTb\endcsname%
      \put(5001,484){\makebox(0,0){\strut{}$3.5$}}%
      \csname LTb\endcsname%
      \put(5752,484){\makebox(0,0){\strut{}$4$}}%
      \csname LTb\endcsname%
      \put(6503,484){\makebox(0,0){\strut{}$4.5$}}%
    }%
    \gplgaddtomacro\gplfronttext{%
      \csname LTb\endcsname%
      \put(198,2761){\rotatebox{-270}{\makebox(0,0){\strut{}$\Delta_0$}}}%
      \put(3874,154){\makebox(0,0){\strut{}$U$}}%
      \csname LTb\endcsname%
      \put(1870,4646){\makebox(0,0)[r]{\strut{}$\beta = 3$}}%
      \csname LTb\endcsname%
      \put(1870,4426){\makebox(0,0)[r]{\strut{}$\beta = 4$}}%
      \csname LTb\endcsname%
      \put(1870,4206){\makebox(0,0)[r]{\strut{}$\beta = 6$}}%
      \csname LTb\endcsname%
      \put(1870,3986){\makebox(0,0)[r]{\strut{}$\beta = 8$}}%
      \csname LTb\endcsname%
      \put(1870,3766){\makebox(0,0)[r]{\strut{}$\beta = 10$}}%
      \csname LTb\endcsname%
      \put(1870,3546){\makebox(0,0)[r]{\strut{}$\beta = 12$}}%
      \csname LTb\endcsname%
      \put(1870,3326){\makebox(0,0)[r]{\strut{}$\beta = \infty$}}%
    }%
    \gplbacktext
    \put(0,0){\includegraphics{gap-pdf}}%
    \gplfronttext
  \end{picture}%
\endgroup
}}}
	\resizebox{0.505\textwidth}{!}{{%
\begingroup
  \inputencoding{latin1}%
  \makeatletter
  \providecommand\color[2][]{%
    \GenericError{(gnuplot) \space\space\space\@spaces}{%
      Package color not loaded in conjunction with
      terminal option `colourtext'%
    }{See the gnuplot documentation for explanation.%
    }{Either use 'blacktext' in gnuplot or load the package
      color.sty in LaTeX.}%
    \renewcommand\color[2][]{}%
  }%
  \providecommand\includegraphics[2][]{%
    \GenericError{(gnuplot) \space\space\space\@spaces}{%
      Package graphicx or graphics not loaded%
    }{See the gnuplot documentation for explanation.%
    }{The gnuplot epslatex terminal needs graphicx.sty or graphics.sty.}%
    \renewcommand\includegraphics[2][]{}%
  }%
  \providecommand\rotatebox[2]{#2}%
  \@ifundefined{ifGPcolor}{%
    \newif\ifGPcolor
    \GPcolortrue
  }{}%
  \@ifundefined{ifGPblacktext}{%
    \newif\ifGPblacktext
    \GPblacktexttrue
  }{}%
  \let\gplgaddtomacro\g@addto@macro
  \gdef\gplbacktext{}%
  \gdef\gplfronttext{}%
  \makeatother
  \ifGPblacktext
    \def\colorrgb#1{}%
    \def\colorgray#1{}%
  \else
    \ifGPcolor
      \def\colorrgb#1{\color[rgb]{#1}}%
      \def\colorgray#1{\color[gray]{#1}}%
      \expandafter\def\csname LTw\endcsname{\color{white}}%
      \expandafter\def\csname LTb\endcsname{\color{black}}%
      \expandafter\def\csname LTa\endcsname{\color{black}}%
      \expandafter\def\csname LT0\endcsname{\color[rgb]{1,0,0}}%
      \expandafter\def\csname LT1\endcsname{\color[rgb]{0,1,0}}%
      \expandafter\def\csname LT2\endcsname{\color[rgb]{0,0,1}}%
      \expandafter\def\csname LT3\endcsname{\color[rgb]{1,0,1}}%
      \expandafter\def\csname LT4\endcsname{\color[rgb]{0,1,1}}%
      \expandafter\def\csname LT5\endcsname{\color[rgb]{1,1,0}}%
      \expandafter\def\csname LT6\endcsname{\color[rgb]{0,0,0}}%
      \expandafter\def\csname LT7\endcsname{\color[rgb]{1,0.3,0}}%
      \expandafter\def\csname LT8\endcsname{\color[rgb]{0.5,0.5,0.5}}%
    \else
      \def\colorrgb#1{\color{black}}%
      \def\colorgray#1{\color[gray]{#1}}%
      \expandafter\def\csname LTw\endcsname{\color{white}}%
      \expandafter\def\csname LTb\endcsname{\color{black}}%
      \expandafter\def\csname LTa\endcsname{\color{black}}%
      \expandafter\def\csname LT0\endcsname{\color{black}}%
      \expandafter\def\csname LT1\endcsname{\color{black}}%
      \expandafter\def\csname LT2\endcsname{\color{black}}%
      \expandafter\def\csname LT3\endcsname{\color{black}}%
      \expandafter\def\csname LT4\endcsname{\color{black}}%
      \expandafter\def\csname LT5\endcsname{\color{black}}%
      \expandafter\def\csname LT6\endcsname{\color{black}}%
      \expandafter\def\csname LT7\endcsname{\color{black}}%
      \expandafter\def\csname LT8\endcsname{\color{black}}%
    \fi
  \fi
    \setlength{\unitlength}{0.0500bp}%
    \ifx\gptboxheight\undefined%
      \newlength{\gptboxheight}%
      \newlength{\gptboxwidth}%
      \newsavebox{\gptboxtext}%
    \fi%
    \setlength{\fboxrule}{0.5pt}%
    \setlength{\fboxsep}{1pt}%
\begin{picture}(5040.00,3772.00)%
    \gplgaddtomacro\gplbacktext{%
      \csname LTb\endcsname%
      \put(814,704){\makebox(0,0)[r]{\strut{}$0$}}%
      \csname LTb\endcsname%
      \put(814,1111){\makebox(0,0)[r]{\strut{}$0.1$}}%
      \csname LTb\endcsname%
      \put(814,1517){\makebox(0,0)[r]{\strut{}$0.2$}}%
      \csname LTb\endcsname%
      \put(814,1924){\makebox(0,0)[r]{\strut{}$0.3$}}%
      \csname LTb\endcsname%
      \put(814,2331){\makebox(0,0)[r]{\strut{}$0.4$}}%
      \csname LTb\endcsname%
      \put(814,2738){\makebox(0,0)[r]{\strut{}$0.5$}}%
      \csname LTb\endcsname%
      \put(814,3144){\makebox(0,0)[r]{\strut{}$0.6$}}%
      \csname LTb\endcsname%
      \put(814,3551){\makebox(0,0)[r]{\strut{}$0.7$}}%
      \csname LTb\endcsname%
      \put(3274,484){\makebox(0,0){\strut{}$U_c$}}%
      \csname LTb\endcsname%
      \put(3411,484){\makebox(0,0){\strut{}}}%
      \csname LTb\endcsname%
      \put(946,484){\makebox(0,0){\strut{}$1$}}%
      \csname LTb\endcsname%
      \put(1357,484){\makebox(0,0){\strut{}$1.5$}}%
      \csname LTb\endcsname%
      \put(1768,484){\makebox(0,0){\strut{}$2$}}%
      \csname LTb\endcsname%
      \put(2178,484){\makebox(0,0){\strut{}$2.5$}}%
      \csname LTb\endcsname%
      \put(2589,484){\makebox(0,0){\strut{}$3$}}%
      \csname LTb\endcsname%
      \put(3000,484){\makebox(0,0){\strut{}$3.5$}}%
      \csname LTb\endcsname%
      \put(3821,484){\makebox(0,0){\strut{}$4.5$}}%
      \csname LTb\endcsname%
      \put(4232,484){\makebox(0,0){\strut{}$5$}}%
      \csname LTb\endcsname%
      \put(4643,484){\makebox(0,0){\strut{}$5.5$}}%
    }%
    \gplgaddtomacro\gplfronttext{%
      \csname LTb\endcsname%
      \put(209,2127){\rotatebox{-270}{\makebox(0,0){\strut{}$m_s$}}}%
      \put(2794,154){\makebox(0,0){\strut{}$U$}}%
    }%
    \gplbacktext
    \put(0,0){\includegraphics{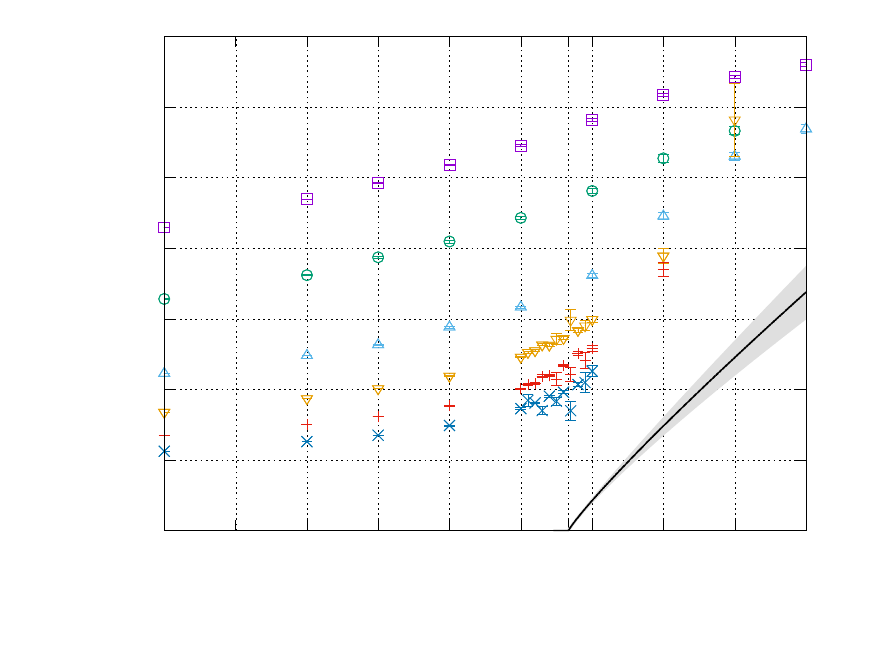}}%
    \gplfronttext
  \end{picture}%
\endgroup
}}
	\caption{All quantities in units of $\kappa$ and after the thermodynamic and continuum limit extrapolations. $\beta$ is the inverse temperature.
		The single-particle gap $\Delta_0(U,\beta)$ (left) and the AFMI order parameter (staggered magnetization) $m_s$ (right).  
		We also show $\Delta_0(U,\beta = \infty)$ and $m_s(U,\beta = \infty)$ as solid black lines with error band (see \Secref{zero temperature extrapolation}). The legend from the left plot applies to both.}
	\label{fig:both_order_parameters}
\end{figure}

It has also been established for some time that an antiferromagnetic (AFM) order is formed in the insulating state (see fig.~\ref{fig:opening_gap}, right) and it has been conjectured that both, insulating and AFM, transitions happen simultaneously. In this thesis we show unambiguously that this indeed is the case. \Cref{fig:both_order_parameters} shows order parameters of both transitions, the single particle gap $\Delta_0$ and the staggered magnetisation $m_s$ which measures the difference between the two sublattices' magnetisations.
In the zero-temperature limit they obtain non-zero values at precisely the same critical coupling $U_c/\kappa=\num{3.835(14)}$. Hence in total we observe a semimetal-antiferromagnetic Mott insulator (SM-AFMI) transition~\cite{more_observables}. We also present a high precision analysis of $U_c$ and the critical exponents $\nu$ and $\upbeta$~\cite{semimetalmott,more_observables} of the phase transition. All these results are presented in \Cref{chap:phase_transition,chap:afm_character} of this thesis. In particular \cref{tab:crit_values} provides an overview of the values of $U_c$, $\nu$ and $\upbeta$ found in the literature to date.

The arguably most prominent extension to the Hubbard model is the addition of long range interactions and all these considerations would significantly loose importance if long range interactions were required in order to describe graphene realistically. It has been found however, at least for a Coulomb potential, that such a change does not crucially influence the physics~\cite{Buividovich:2018crq}. Though the critical parameters and, more generally, the universality class of the phase transition change, its SM-AFMI nature remains.

The transition to AFM order features spontaneous symmetry breaking (SSB) which means that the system has to choose one option from a set of equivalent possibilities. In other words, the ground state solution has a lower symmetry than the original problem. This is illustrated on the bottom right of \cref{fig:opening_gap}.
SSB is ubiquitous in nature. We do not even have to go to the quantum world to find examples of SSB. For instance, recently in~\cite{beer_mat} we explained the flight of rotating discs like beer mats and, in particular, why they always end up with backspin.
Not only is this insight indispensable for any pub visit, fascinating\footnote{We made it to the front pages of several media channels, among them \href{https://www.thetimes.co.uk/article/they-flipped-the-science-behind-throwing-beer-mats-mdwwg3clv}{\textit{The Times, London} (June 23 2021)}.} physicists and the layman alike.
It also features SSB of different kinds.
For backspin to be the preferred direction, a twofold SSB is required. First of all gravity breaks the $SO(3)$ rotational symmetry to an $SO(2)$ residual symmetry confining the stable rotation axes. In addition, more subtly, the existence of air maximally breaks the $\mathbb{R}^3$ Galilean (or Lorentzian) invariance under boosts defining a direction of flight and thus allowing to distinguish back- and topspin.

The SM-AFMI phase transition is of second order and falls into the $SU(2)$ Gross-Neveu ``chiral Heisenberg'' universality class~\cite{gross_neveu_orig,heisenberg_gross_neveu}. This means that the model undergoes an SSB from the original $\mathbb{Z}_2\times SU(2)_\text{sp}\times U(1)_\text{ch}\times U(1)_\chi$ symmetry group down to a remaining $U(1)_\text{sp}\times U(1)_\text{ch}\times U(1)_\chi$. Here the $\mathbb{Z}_2$ symmetry comes from the discrete reflection or sublattice symmetry of the honeycomb lattice, $SU(2)_\text{sp}$ and $U(1)_\text{sp}$ are the respective unbroken and residual spin rotation symmetries, $U(1)_\text{ch}$ is related to charge conservation and $U(1)_\chi$ is the so called chiral symmetry stemming from translational invariance in real space or, interpreted in momentum space, from a duplication of the $U(1)_\text{ch}$ due to the independence of both Dirac cones.
The Gross-Neveu model is itself worth studying as it is an important tool in particle physics and especially its $SU(2)$ (chiral Heisenberg) version is not well understood yet. Therefore it is of broad interest to develop algorithms for efficient simulations of the Hubbard model and through it the Gross-Neveu model.

Numerous approaches have been utilised to solve the Hubbard model. A particularly broad overview of numerical algorithms can be found in~\cite{Hubbard_review_numerical} (though the review does not deal with the honeycomb lattice) while~\cite{Hubbard_review} provides a very recent and well readable (though not very detailed) overview.
The list of algorithms includes functional renormalisation group~\cite{frg_review} and tensor network~\cite{my_tensor_networks,iPEPS_hubbard_2016} techniques, just to name a few.
However, the majority of algorithms dealing with the Hubbard model, including this work, belong to the class of quantum Monte Carlo (QMC) simulations. Stochastic simulations arise naturally from the probabilistic nature of quantum mechanics and they have proven to be very successful. We further subdivide the QMC algorithms into local and global update methods.
Historically, first simulations successfully predicting the SM-AFMI phase transition relied on local update algorithms~\cite{Meng2010,Sorella:SciRep} whereas global update methods recently bridged the gap between numerics and experiment by simulating lattices of physical size~\cite{semimetalmott,more_observables,ulybyshev2021bridging}.

In this work we use the hybrid Monte Carlo\footnote{also called Hamiltonian Monte Carlo} (HMC) algorithm~\cite{Duane:1987de}, a Markov-chain Monte Carlo (MCMC) method with global updates on continuous fields.
Brower, Rebbi and Schaich (BRS) originally proposed to use the HMC algorithm for simulations of graphene~\cite{Brower:2012zd}. Their formalism stands in stark contrast to the widespread local Blankenbecler-Sugar-Scalapino (BSS)~\cite{Blankenbecler:1981jt} algorithm.
The main advantage of the HMC over local update schemes like the BSS algorithm is its superior scaling with volume $\order{V^{5/4}}$~\cite{Creutz:1988wv} whereas most alternative schemes scale as volume cubed $\order{V^3}$.
In practice BSS usually outperforms BRS on small systems where it is less noisy, but the HMC (i.e.\ BRS) gains the upper hand on large lattices which are essential for approaching the thermodynamic limit.
In addition, the HMC has been heavily optimised, in particular in lattice quantum chromodynamics (QCD)~\cite{OMELYAN2003272,fgmres,hasenbusch,urbach,Clark:2016rdz}, and we utilised many of these improvements for our condensed matter simulations~\cite{acceleratingHMC}.
Notable optimisations of the HMC that are not compatible with our ansatz have been developed in~\cite{ULYBYSHEV2019118,Buividovich:2018crq}.

By the time this work started, HMC simulations of the Hubbard model had been well established~\cite{Brower:2012zd,Smith:2014tha,Luu:2015gpl,acceleratingHMC,Wynen:2018ryx}, so that a reliable and efficient implementation could be used to extract the variety of physical results presented in \Cref{chap:phase_transition,chap:afm_character}. Our optimised methods allowed for the largest lattices simulated to date (20,808 lattice sites) enabling us to perform the first thorough analysis and elimination of all finite size and discretisation effects.

In \Cref{chap:ising} the HMC algorithm is discussed in more detail and applied to the Ising model, a discrete statistical spin system~\cite{Ising:1925,onsager_2d_solution}. We will not discuss physical properties of the Ising model in this work. Such properties can be found in numerous books and reviews, for example in Ref.~\cite{friedli_velenik_2017,gallavotti2013statistical,baierlein_1999,history_ising_model}. Instead we use it as a show case for important concepts of numerical simulations, in particular the HMC. Nevertheless this discussion is not only interesting for pedagogical reasons. In \cref{sec:ising_formalism} we introduce a transformation that allows one to extend the applicability of the HMC to arbitrary Ising-like models, although the HMC originally has been designed for the application to continuous systems only. 

A notable limitation of the HMC algorithm (and any other stochastic method) is posed by the fermionic sign problem\footnote{also called complex phase problem} which is caused by the terms usually defining the probability density obtaining values of varying sign or phase. The sign problem is one of the central problems of computational physics, not only in condensed matter.
It appears in the Hubbard model for instance through a non-zero chemical potential or non-bipartite lattice structure.
Recently, we have made substantial progress alleviating the sign problem with the help of machine learning~\cite{leveragingML} and tensor networks~\cite{my_tensor_networks} and we continue our research on these topics.
In this thesis however we only consider the Hubbard model at half filling, that is without chemical potential, and on the bipartite honeycomb lattice. Thus no sign problem emerges and we can exploit the advantages of the HMC at its best.

As we pointed out earlier, our HMC simulations profit greatly from developments in lattice QCD. But the benefits do not end there. 
We also face another challenge well known from lattice QCD, namely that energy levels have to be extracted from correlation functions. Though some exotic algorithms like the Gardner method~\cite{gardner_original} use a completely different ansatz, most approaches to this challenge like the generalised eigenvalue method (GEVM)~\cite{Michael:1982gb,Luscher:1990ck,Blossier:2009kd} or Prony's method~\cite{Prony:1795,Fleming:2004hs,Beane:2009kya} ultimately reduce the correlator to a set of effective masses. These effective masses are again time dependent functions approximating the lowest few energy levels and ideally featuring clear plateaus at the corresponding energies.

Plateau fits seem to be very easy at first glance. The simplest possible function, a constant, is fitted to some scaling region of data that presumably features only statistical fluctuations and no systematic tendency. In reality it turns out however that it can be extremely difficult to find the best fitting region or, in particularly nasty cases, decide that no plateau can be identified. The crucial problem is to find a balance between statistical and systematic errors. The first is minimised by maximizing the length of the fit range while the latter is very difficult to determine and increases with every point outside the scaling region added to the fit. A variety of different approaches to solve this problem has been developed in the recent past. Thus far none have proven to be the ultimate solution. To name a few examples, to date neural networks could not be trained to reliably identify plateau regions~\cite{Till:thesis} and the usage of MCMC sampling of different regions based on Bayesian statistics seemed very promising up to the point of taking correlations of the data into account~\cite{Fischer:thesis}.

Our novel method presented in \Cref{chap:prony} now combines the GEVM and Prony's method (PGEVM) with the main result being sharper plateaus that start much earlier in time. This significantly simplifies the identification of the scaling region and thus substantially improves the systematic error. \Cref{fig:eta} in \cref{sec:qcd_examples} illustrates this advantage particularly well.

To put it in a nutshell, we developed and implemented a powerful numerical simulation and analysis framework that allowed us to thoroughly characterise the SM-AFMI transition of the Hubbard model on the honeycomb lattice which underlies the usage of graphene as a transistor. Our highly optimised HMC algorithm allows for simulations of physical size systems and can easily be augmented to different geometries or modifications of the Hubbard model.

This cumulative thesis is organised as follows: In \Cref{chap:ising} (based on~\cite{ising}) we present a method to apply the HMC algorithm to the Ising model. Next, in \Cref{chap:prony} (based on~\cite{generalisedProny}) we develop the PGEVM. \Cref{chap:phase_transition,chap:afm_character} (based on~\cite{semimetalmott} and~\cite{more_observables}) both deal with the quantum phase transition of the Hubbard model, \Cref{chap:phase_transition} determining the critical coupling $U_c$ and the critical exponent $\nu$, complemented by the determination of the critical exponent $\upbeta$ in \Cref{chap:afm_character}.
\clearpage{}%
\clearpage{}%
\chapter{The Ising Model with Hybrid Monte Carlo}\label{chap:ising}

\begin{center}
	\textit{Based on~\cite{ising} by J.~Ostmeyer, E.~Berkowitz, T.~Luu, M.~Petschlies and F.~Pittler}
\end{center}

This chapter can be seen as introduction to many concepts that will become important later on using the Ising model as a simple example. We explain the Hubbard-Stratonovich transformation, critical slowing down and most importantly hybrid Monte Carlo simulations. Alternatively, the reader might consider the chapter a curiosity featuring a model with discrete degrees of freedom solved by an algorithm developed for continuous fields.

\renewcommand{\includeepslatex}[1]{\resizebox{0.443\textwidth}{!}{{\Large\input{ising-hmc-paper/pics/#1}}}}

\section{Introduction}\label{sec:ising_intro}

The Ising model is a simple model of ferromagnetism and exhibits a phase transition in dimensions $d\ge 2$.
Analytic solutions determining the critical temperature and magnetization are known for $d=1$ and 2 \cite{onsager_2d_solution}, and in large dimensions the model serves as an exemplary test bed for application of mean-field techniques.
It is also a popular starting point for the discussion of the renormalization group and calculation of critical exponents.

In many cases systems that are seemingly disparate can be  mapped into the Ising model with slight modification.
Examples include certain neural networks~\cite{10.1007/978-3-642-61850-5_18,schneidman:2006}, percolation~\cite{1969PSJJS..26...11K,FORTUIN1972536,Saberi_2010}, ice melt ponds in the arctic~\cite{Ma_2019}, financial markets~\cite{Chowdhury1999AGS,KAIZOJI2002441,SORNETTE2006704}, and population segregation in urban areas\cite{doi:10.1080/0022250X.1971.9989794,doi:10.1119/1.2779882}, to name a few.
In short, the applicability of the Ising model goes well beyond its intended goal of describing ferromagnetic behavior.  Furthermore, it serves as an important pedagogical tool---any serious student of statistical/condensed matter physics as well as field theory should be well versed in the Ising model.

The pedagogical utility of the Ising model extends into numerics as well.
Stochastic lattice methods and the Markov-chain Monte-Carlo (MCMC) concept are routinely introduced via application to the Ising model.
Examples range from the simple Metropolis-Hastings algorithm to more advanced cluster routines, such as Swendsen-Wang~\cite{Swendsen:1987ce} and Wolff~\cite{Wolff:1988uh} and the worm algorithm of Prokof'ev and Svistunov~\cite{PhysRevLett.87.160601}.
Because so much is known of the Ising model, it also serves as a standard test bed for novel algorithms.
Machine learning (ML) techniques were recently applied to the Ising model to aid in identification of phase transitions and order parameters~\cite{Wetzel:2017ooo,Cossu:2018pxj,JMLR:v18:17-527,GIANNETTI2019114639,Alexandrou:2019hgt}.

A common feature of the algorithms mentioned above is that they are well suited for systems with discrete internal spaces, which of course includes the Ising model.
For continuous degrees of freedom the hybrid Monte Carlo (HMC) algorithm~\cite{Duane:1987de} is instead the standard workhorse.
Lattice quantum chromodynamics (LQCD) calculations, for example, rely strongly on HMC.
Certain applications in condensed matter physics now also routinely use HMC~\cite{Luu:2015gpl,acceleratingHMC,Wynen:2018ryx}.
Furthermore, algorithms related to preconditioning and multi-level integration have greatly extended the efficacy and utility of HMC.
With the need to sample posterior distributions in so-called big data applications, HMC has become widespread even beyond scientific applications.

It is natural to ask, then, how to apply the numerically-efficient HMC to the broadly-applicable Ising model.
At first glance, the Ising model's discrete variables pose an obstacle for smoothly integrating the Hamiltonian equations of motion to arrive at a new proposal.
However, in Ref.~\cite{pakman2015auxiliaryvariable} a modified version of HMC was introduced where sampling was done over a mixture of binary and continuous distributions and successfully benchmarked to the Ising model in 1D and 2D.  In our work, we describe how to transform the Ising model to a completely continuous space in arbitrary dimensions and with arbitrary couplings between spins (and not just nearest neighbor couplings). \added{Some of these results have already been published in Ref.~\cite{similar_work} without our knowledge and have thus been `rediscovered' by us. Yet, we propose a novel, more efficient approach for the transformation and we perform a thorough analysis of said efficiency and the best choice of the tunable parameter.}

Furthermore, we hope this paper serves a pedagogical function, as a nice platform for introducing both HMC and the Ising model, and a clarifying function, demonstrating how HMC can be leveraged for models with discrete internal spaces.
So, for pedagogical reasons, our implementation of HMC is the simplest `vanilla' version.
As such, it does not compete well, in the numerical sense, with the more advanced cluster algorithms mentioned above.
However, it seems likely that by leveraging the structure of the Ising model one could find a competitive HMC-based algorithm, but we leave such investigations for the future.

This paper is organized as follows.
In \Secref{ising_formalism} we review the Ising model.
We describe how one can transform the Ising model, which resides in a discrete spin space, into a model residing in a continuous space by introducing an auxiliary field and integrating out the spin degrees of freedom.
The numerical stability of such a transformation is not trivial\footnote{Such stability considerations have been egregiously ignored in the past.}, and we describe the conditions for maintaining stability.
With our continuous space defined, we show in \Secref{ising_hmc} how to simulate the system with HMC.
Such a discussion of course includes a cursory description of the HMC algorithm.
In \Secref{ising_results} we show how to calculate observables within this continuous space, since quantities such as magnetization or average energy are originally defined in terms of spin degrees of freedom which are no longer present. We also provide numerical results of key observables, demonstrating proof-of-principle.
We conclude in \Secref{ising_conclusion}.

\section{Formalism}\label{sec:ising_formalism}

The Ising model on a lattice with $N$ sites is described by the Hamiltonian
\begin{align}
    H
    &=
        -   J\sum_{\average{i,j}}s_is_j
        -   \sum_i h_is_i
    \\
    &=
        -   \half J s\transpose K s^{}
        -   h\cdot s
        \label{eq:hamilton_matrix_notation}
\end{align}
where $s_i=\pm1$ are the spins on sites $i = 1,\ldots,N$, $J$ the coupling between neighbouring spins (denoted by $\average{i,j}$),  $h_i$ is the local external magnetic field, and the $\transpose$ superscript denotes the transpose.
We also define the symmetric connectivity matrix $K$ containing the information about the nearest neighbour couplings.
The factor $\half$ on the nearest-neighbor term~\eqref{hamilton_matrix_notation} accounts for the double counting of neighbour pairs that arises from making $K$ symmetric.
If $h$ is constant across all sites we write
\begin{equation}
    \label{eq:constant_h}
    h
    =
    h_0\left(\begin{array}{c}
            1       \\
            1       \\
            \vdots  \\
            1
        \end{array}\right)
        \,.
\end{equation}
We assume a constant coupling $J$ for simplicity in this work.
The same formalism developed here can however be applied for site-dependent couplings as well.
In this case we simply have to replace the matrix $JK$ by the full coupling matrix.

The partition sum over all spin configurations $\left\{ s_i \right\} \equiv \left\{ s_i \,|\, i = 1,\ldots,N \right\}$
\begin{align}
    Z
    =
    \sum_{\left\{s_i\right\}=\pm 1} \eto{-\beta H}
\end{align}
with the inverse temperature $\beta$ is impractical to compute directly for large lattices because the number of terms increases exponentially,  providing the motivation for Monte Carlo methods.
Our goal is to rewrite $Z$ in terms of a continuous variable so that molecular dynamics (MD) becomes applicable.
The usual way to eliminate the discrete degrees of freedom and replace them by continuous ones is via the Hubbard-Stratonovich (HS) transformation.
For a positive definite matrix $A\in\Reals^{N\times N}$ and some vector $v\in\Reals^N$,  the HS relation reads
\begin{align}
    \eto{\half v\transpose A v}
    =
    \frac{1}{\sqrt{\det A\left(2\pi\right)^N}}
    \int\limits_{-\infty}^{\infty}\left[\prod_{i=1}^{N}\md \phi_i\right]\,
        \eto{-\half \phi\transpose A\inverse\phi^{}+v\cdot \phi}
\end{align}
where we integrate over an \emph{auxiliary field} $\phi$.
The argument of the exponent has been linearized in $v$.
In our case the matrix $J'K$ with
\begin{align}
J'\coloneqq\beta J\label{eq:def_J_prime}
\end{align}
takes the place of $A$ in the expression above.
However, $J'K$ is not positive definite in general, nor is $-J'K$.
The eigenvalues $\lambda$ of $K$ are distributed in the interval
\begin{align}
    \lambda
    \in
    \left[-n,\,n\right]\label{eq:eigen_spectrum_K}
\end{align}
where $n$ is the maximal number of nearest neighbours a site can have.
In the thermodynamic limit $N\goesto\infty$ the spectrum becomes continuous and all values in the interval are reached.
Thus the HS transformation is not stable: the Gaussian integral with negative eigenvalues does not converge.

We have to modify the connectivity matrix in such a way that we can apply the HS transformation.
Therefore we introduce a constant shift $C$ to the $K$ matrix,
\begin{align}
    \tilde{K}
    \coloneqq
    K + C\,\mathds{1},
\end{align}
where $C$ has to have the same sign as $J'$, by adding and subtracting the corresponding term in the Hamiltonian.
Now $\tilde{K}$ has the same eigenspectrum as $K$, but shifted by $C\,$.
Thus if we choose $\left|C\right|>n$, $J'\tK$ is positive definite.
We will take such a choice for granted from now on.
For variable coupling the interval~\eqref{eigen_spectrum_K} might have to be adjusted, but the eigenspectrum remains bounded from below, so $C$ can be chosen large enough to make $J'\tK$ positive definite.

Now we can apply the HS transformation to the partition sum
\begin{align}
    Z
    &=
        \sum_{\left\{s_i\right\}=\pm 1}
            \eto{
                    \half\beta Js\transpose \tK s{}
                -   \half \beta J C s^2
                + \beta h\cdot s
                }
    \\
    &=
        \eto{-\half J' C N}
        \sum_{\left\{s_i\right\}=\pm 1}
            \eto{
                    \half J's\transpose \tK s^{}
                +   h'\cdot s
                }
    \label{eq:absorb_beta}\\
    &=
        \eto{-\half J' C N}
        \sum_{\left\{s_i\right\}=\pm 1}
            \frac{1}{\sqrt{\det \tK\left(2\pi J'\right)^N}}
            \int\limits_{-\infty}^{\infty}\left[\prod_{i=1}^{N}\md \phi_i\right]\,
                \eto{
                    -   \oneover{2J'}\phi\transpose\tK\inverse \phi^{}
                    +   \left(h'+\phi\right)\cdot s
                    }
    \label{eq:did_HS_trafo}\\
    &=
        \frac{\eto{-\half J' C N}}{\sqrt{\det \tK\left(2\pi J'\right)^N}}
        \int\limits_{-\infty}^{\infty}\left[\prod_{i=1}^{N}\md \phi_i\right]\,
            \eto{ - \oneover{2J'}\phi\transpose\tK\inverse\phi}
            \left[\prod_{i=1}^N 2\cosh\left(h'_i+\phi_i\right)\right]
    \label{eq:summed_over_si}\\
    &=
        2^N
        \frac{\eto{-\half J' C N}}{\sqrt{\det \tK\left(2\pi J'\right)^N}}
        \int\limits_{-\infty}^{\infty}\left[\prod_{i=1}^{N}\md \phi_i\right]\,
            \eto{
                -   \oneover{2J'}\phi\transpose \tK\inverse \phi^{}
                +   \sum_i\log\cosh\left(h'_i+\phi_i\right)
                }
    \label{eq:cosh_in_exponent}
\end{align}
where we used in \eqref{absorb_beta} that $s_i^2=1$ for all $i$ and defined $h'\coloneqq \beta h$ in analogy with~\eqref{def_J_prime}.
In \eqref{did_HS_trafo} we performed the HS transformation and in \eqref{summed_over_si} we explicitly evaluated the sum over all the now-independent $s_i$, thereby integrating out the spins.
After rewriting the $\cosh$ term in \eqref{cosh_in_exponent} we are left with an effective action that can be used to perform HMC calculations.
However, we do not recommend using this form directly, as it needs a matrix inversion.

Instead, let us perform the substitution
\begin{align}
    \phi
    &=
    \sqrt{J'}\tK\psi-h'
\end{align}
with the functional determinant $\sqrt{J'}^N\det\tK$.
This \added{substitution is going to bring a significant speed up and has not been considered in Ref.~\cite{similar_work}. It} allows us to get rid of the inverse of $\tK$ in the variable part of the partition sum
\begin{align}
    Z
    &=
    \sqrt{\left(\tfrac 2\pi\right)^N\det\tK}\:
    \eto{-\half J' C N}
    \int\limits_{-\infty}^{\infty}\left[\prod_{i=1}^{N}\md \psi_i\right]\,
        \eto{
            -   \half\psi\transpose \tK\psi^{}
            +   \oneover{\sqrt{J'}}\psi\cdot h'
            -   \oneover{2J'}h'\transpose \tK\inverse h'^{}
            +   \sum_i\log\cosh\left(
                    \sqrt{J'}\left(\tK\psi\right)_i
                    \right)
            }\,.
    \label{eq:phi_psi_substitution}
\end{align}
The only left over term involving an inversion remains in the constant $h'\transpose\tK\inverse h'^{}$.
Fortunately this does not need to be calculated during HMC simulations.
We do need it, however, for the calculation of some observables, such as the magnetisation \eqref{formula_for_m}, and for this purpose it can be calculated once without any need for updates. \added{Let us also remark that the inverse $\tK^{-1}$ does not have to be calculated exactly. Instead it suffices to solve the system of linear equations $\tK x = h'$ for $x$ which can be done very efficiently with iterative solvers, such as the conjugate gradient (CG) method~\cite{templates}.}

A further simplification can be achieved when the magnetic field is constant \eqref{constant_h} and every lattice site has the same number of nearest neighbours $n_0$.
Then we find that
\begin{align}
    \tK h'
    =
    \left(n_0+C\right)h'
\end{align}
and thus
\begin{equation}
    h'\transpose \tK\inverse h'^{}
    =
    h'\transpose\oneover{n_0+C}h'^{}
    =
    \frac{N}{n_0+C}h_0'^2\,.\label{eq:hKh_for_constant_h}
\end{equation}

\section{HMC}\label{sec:ising_hmc}

Hybrid Monte Carlo\footnote{Sometimes `Hamiltonian Monte Carlo', especially in settings other than lattice quantum field theory.} (HMC)~\cite{Duane:1987de} requires introducing a fictitious \emph{molecular dynamics time} and conjugate momenta, integrating current field configurations according to Hamiltonian equations of motion to make a Metropolis proposal.
We multiply the partition sum $Z$ \eqref{phi_psi_substitution} by unity, using the Gaussian identity
\begin{equation}
	\label{eq:momentum distribution}
	\frac{1}{(2\pi)^{N/2}} \int_{-\infty}^{+\infty} \left[\prod_{i=1}^N \md p_i\right]\, \eto{-\frac{1}{2}p_i^2} = 1
\end{equation}
where we have one conjugate momentum $p$ for each field variable $\psi$ in $Z$, and we sample configurations of fields and momenta from this combined distribution.
The conceptual advantage of introducing these momenta is that we can evolve the auxiliary fields $\psi$ with the HMC Hamiltonian~$\mathcal{H}$,
\begin{align}
    \mathcal{H}
    =
        \half p^2
    +   \half\psi\transpose\tK\psi^{}
    -   \oneover{\sqrt{J'}}\psi\cdot h'
    -   \sum_i\log\cosh\left(\sqrt{J'}\left(\tK\psi\right)_i\right)
\end{align}
by integrating the equations of motion (EOM)
\begin{align}
    \dot{\psi}
    &=  +\del{\mathcal{H}}{p}
     =  p
    \\
    \dot{p}
    &=  -\del{\mathcal{H}}{\psi}
     =  -\tK\psi + \oneover{\sqrt{J'}} h' + \sqrt{J'}\tK\tanh\left(\sqrt{J'}\tK\psi\right)
\end{align}
where the $\tanh$ is understood element-wise.

Thus one can employ the Hybrid Monte Carlo algorithm to generate an ensemble of field configurations by a Markov chain.
Starting with some initial configuration $\psi$, the momentum $p$ is sampled according to a Gaussian distribution \eqref{momentum distribution}.
The EOM are integrated to update all the field variables at once.
The integration of the differential equations, or the \emph{molecular dynamics}, is performed by a (volume-preserving) symmetric symplectic integrator (we use leap-frog here, but more efficient schemes can be applied \cite{PhysRevE.65.056706,OMELYAN2003272}) to ensure an unbiased update.
The equations of motion are integrated one \emph{molecular dynamics time unit}, which is held fixed for each ensemble, to produce one \emph{trajectory} through the configuration space; the end of the trajectory is proposed as the next step in the Markov chain.
If the molecular dynamics time unit is very short, the new proposal will be very correlated with the current configuration. If the molecular dynamics time unit is too long, it will be very expensive to perform an update.

The proposal is accepted with the Boltzmann probability $\min\left(1,\,\eto{-\Delta \mathcal{H}}\right)$ where the energy splitting $\Delta \mathcal{H}=\mathcal{H}_\text{new}-\mathcal{H}_\text{old}$ is the energy difference between the proposed configuration and the current configuration.
If our integration algorithm were exact, $\Delta \mathcal{H}$ would vanish and we would always accept the new proposal, by conservation of energy.
The Metropolis-Hastings accept/reject step guarantees that we get the correct distribution despite inexact numerical integration.
So, if we integrate with time steps that are too coarse we will reject more often.  Finer integration ensures a greater acceptance rate, all else being equal.

If the proposal is not accepted as the next step of our Markov chain, it is rejected and the previous configuration repeats.
After each accepted or rejected proposal the momenta are refreshed according to the Gaussian distribution \eqref{momentum distribution} and molecular dynamics integration resumes, to produce the next proposal.

If the very first configuration is not a good representative of configurations with large weight, the Markov chain will need to be \emph{thermalized}---driven towards a representative place---by running the algorithm for some number of updates.
Then, production begins.
An ensemble of $N_\text{cf}$ configurations $\left\{\psi^n\right\}$ is drawn from the Markov chain and the estimator of any observable $O(\psi)$
\begin{align}
    \overline{O}
    &=
    \oneover{N_\text{cf}}\sum\limits_{n=1}^{N_\text{cf}}O\!\left(\psi^n\right)
    \\
    &\text{converges to the expectation value} \nonumber
    \\
    \average{O}
    &= \oneover{Z} \sqrt{\left(\tfrac 2\pi\right)^N\det\tK}\:
    \eto{-\half J' C N}
    \int\limits_{-\infty}^{\infty}\left[\prod_{i=1}^{N}\md \psi_i\right]\,
        O(\psi)\;
        \eto{
            -   \half\psi\transpose\tK\psi^{}
            +   \oneover{\sqrt{J'}}\psi\cdot h'
            -   \oneover{2J'}h'\transpose\tK\inverse h'^{}
            +   \sum_i\log\cosh\left(
                    \sqrt{J'}\left(\tK\psi\right)_i
                    \right)
            }
\end{align}
as the ensemble size $N_\text{cf}\goesto\infinity$, with uncertainties on the scale of $N_\text{cf}^{-1/2}$ as long as the configurations are not noticeably correlated---if their \emph{autocorrelation time} (in Markov chain steps) is short enough.

Not much time has been spent on the tuning of $C$ during this work.
We expect that the choice of $C$ can influence the speed of the simulations.
Clearly $\left|C\right|$ must not be chosen too large because in the limit $\left|C\right|\goesto\infty$ the Hamiltonian can be approximated by
\begin{align}
    \oneover{C}\mathcal{H}
    &=
            \half \psi^2
        -   \sqrt{J'}\sum_i\left|\psi_i\right|
        +   \order{C^{-1}}
\end{align}
with the minima
\begin{align}
	\label{eq:large C minima}
    \psi_i
    &=
    \pm\sqrt{J'}.
\end{align}
Any deviation from a minimum is enhanced by the factor of $C$ and is thus frozen out for large $\left|C\right|$.
This reproduces the original discrete Ising model up to normalisation factors.
Plainly the HMC breaks down in this case.
As the limit is approached, the values for the $\psi_i$ become confined to smaller and smaller regions.
The result is that HMC simulations can get stuck in local minima and the time series is no longer \emph{ergodic}---it cannot explore all the states of the Markov chain---which may yield incorrect or biased results.
From now on we use $\left|C\right|=n+\num{e-5}$; we later show the effect of changing $C$ in \Figref{compare_C}.

A large coupling (or low temperature) $J'$ introduces an ergodicity problem as well: as we expect to be in a magnetized phase, all the spins should be aligned and flipping even one spin is energetically disfavored even while flipping them all may again yield a likely configuration.
This case however is less problematic because there are only two regions with a domain wall between them; the region with all $\psi_i>0$ and the region with all $\psi_i<0$.
The ergodicity issue is alleviated by proposing a global sign flip and performing a Metropolis accept/reject step every few trajectories, similar to that proposed in Ref.~\cite{Wynen:2018ryx}.

\section{Results}\label{sec:ising_results}

Let us again assume constant external field with strength $h_0$ \eqref{constant_h}.
Then the expectation value of the average magnetisation and energy per site read
\begin{align}
    \average{m}
    &   =   \oneover{NZ}\del{Z}{h'}
    \\
    &   =   \oneover{N}\average{\oneover{\sqrt{J'}}\sum_i\psi_i
        -   \frac{N}{n_0+C}\frac{h'_0}{J'}}
    \\
    &   =   \frac{\average{\psi}}{\sqrt{J'}}
        - \oneover{n_0+C}\frac{h'_0}{J'}\,,\label{eq:formula_for_m}\\
    \average{\beta\varepsilon}
    &	=	-\frac{\beta}{NZ}\del{Z}{\beta}\label{eq:ee form}\\
    &	=	\half CJ' + \oneover{n_0+C}\frac{{h'_0}^2}{2J'} - \frac{h'_0}{2\sqrt{J'}}\average{\psi} - \frac{\sqrt{J'}}{2N}\average{\left(\tK \psi\right)\cdot\tanh\left(\sqrt{J'}\tK\psi\right)}\label{eq:formula_for_e}
\end{align}
where $\average{\psi} = \average{\psi_i}$ for any site $i$ due to translation invariance.
Any other physical observables can be derived in the same way.
For example, higher-point correlation functions like spin-spin correlators may be derived by functionally differentiating with respect to a site-dependent $h_i$ (without the simplification of constant external field \eqref{hKh_for_constant_h}). 
We stress here that, although $C$ appears in observables (as in the magnetization~\eqref{formula_for_m} and energy density~\eqref{formula_for_e}), the results are independent of $C$---its value only influences the convergence rate.

\begin{figure}[ht]
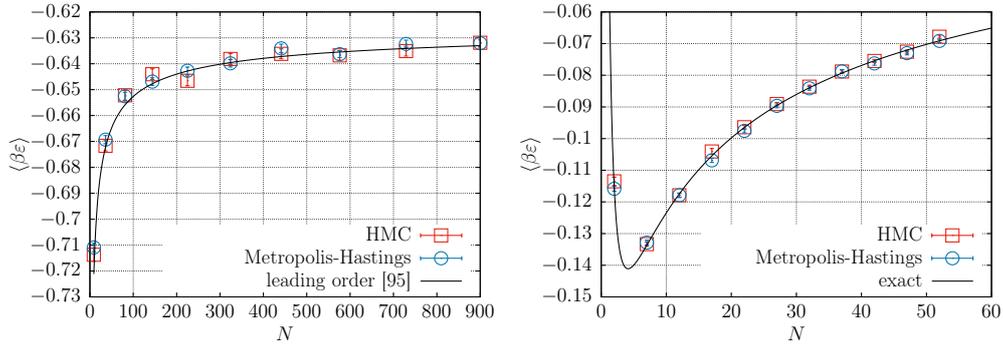

	\centering
	\resizebox{0.443\textwidth}{!}{{\Large\input{pics/energy_2d}}}
	\resizebox{0.443\textwidth}{!}{{\Large\input{pics/energy_alltoall}}}
	\caption{Expectation value of the energy per site for the two dimensional periodic square lattice (left) and the lattice with all-to-all coupling (right) for the HMC and the Metropolis-Hastings algorithms at critical coupling and $h=0$ with lattice sizes $N$.}
	\label{fig:magnetisation}
\end{figure}

In \Figref{magnetisation} we demonstrate that the HMC algorithm\footnote{Our code is publicly available under \url{https://github.com/HISKP-LQCD/ising_hmc}.} indeed produces correct results.
The left panel shows the average energy per site at the critical point~\cite{onsager_2d_solution} of the two-dimensional square lattice with periodic boundary conditions.
We choose to scale the number of integration steps per trajectory with the lattice volume as 
$N_\text{step} = \left\lfloor\log N\right\rfloor$, which empirically leads to acceptance rates between 70\% and 80\% for a broad range of lattice sizes and dimensions.
The results from the HMC simulations are compared to the results obtained via the local Metropolis-Hastings algorithm with the same number $N_\text{cf}$ of sweeps (a sweep consists of $N$ spin flip proposals).
In addition we show the leading order analytic results \cite{ising_2d_exact} $\average{\beta\varepsilon}\approx-\log\left(1+\sqrt{2}\right)\left(\frac{\sqrt{2}}{2}+\frac{1}{3\sqrt{N}}\right)+\order{N^{-1}}$.
We not only find that the results are compatible, but also that the errors of both stochastic methods are comparable.  The right panel shows the average energy per site in the case where the coupling is no longer nearest neighbor, but the extreme opposite with all-to-all couplings.  The Hamiltonian we use in this case is, up to an overall constant, the ``infinite-range'' Ising model~\cite{negele1988quantum}.  This model has analytic solutions for physical observables as a function of the number of lattice sites $N$ which we show for the case of the average energy (black line).  We provide a description of this model, as a well as a derivation of the exact solution for the average energy, in \ref{sect:infiniteRangeIsing}.  Our numerical results agree very well with the exact result.

\begin{figure}[ht]
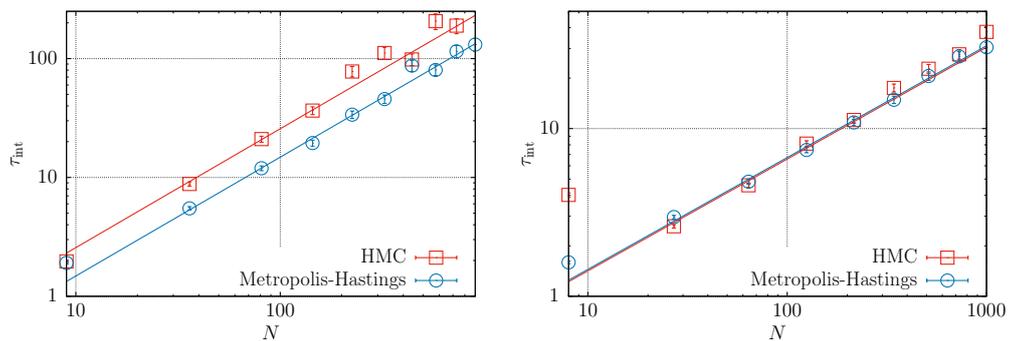

	\centering
	\resizebox{0.443\textwidth}{!}{{\Large\input{pics/autocorrelation}}}
	\resizebox{0.443\textwidth}{!}{{\Large\input{pics/autocorrelation_3d}}}
	\caption{Integrated autocorrelation time of $|m|$ for the HMC and the Metropolis-Hastings algorithms at critical coupling and $h=0$ for the $d=2$ (left) and $d=3$ (right) dimensional periodic square lattice with size $N$. The lines are fits of the form $\tau_\text{int}=\alpha N^\frac 2d$ for $N>10$.}
	\label{fig:tau_int}
\end{figure}

Since it is not the aim of this work to present physical results, but rather to introduce an alternative formulation for simulating the Ising model and generalizations thereof,
we do not compute other observables explicitly, nor do we investigate their dependence on other parameters.
On the other hand it is not sufficient that the algorithm in principle produces correct results---we must also investigate its efficiency.
A good measure for the efficiency is the severity of \emph{critical slowing down}---that the integrated autocorrelation time\footnote{$\tau_\text{int}$ and its error have been calculated according to the scheme proposed in Ref~\cite{monte_carlo_errors}.} $\tau_\text{int}$ diverges at the critical point as some power $\gamma$ of the system size $\tau_\text{int}\propto N^\gamma$.
One could expect that, being a global update algorithm, the HMC does not suffer as much from critical slowing down as Metropolis-Hastings.
\Figref{tau_int} however shows that both algorithms have dynamic exponent $z \equiv d\gamma \approx 2$ in $d=2$ and $d=3$ dimensions (see Ref.~\cite{Pelissetto:2000ek} and references within for a discussion of the critical coupling and exponents in $d=3$).
Still one has to keep in mind that a Metropolis-Hastings sweep takes less time than an HMC trajectory and the HMC trajectories become logarithmically longer as $N$ grows.
In our implementation we find the proportionality
	\begin{align}
		T_\text{HMC}\approx 4N_\text{step}T_\text{MH}
	\end{align}
where $T_\text{HMC}$ is the time required for one HMC trajectory and $T_\text{MH}$ the time required for one Metropolis-Hastings sweep.

\begin{figure}[ht]
		\centering
		\resizebox{0.443\textwidth}{!}{{\Large\input{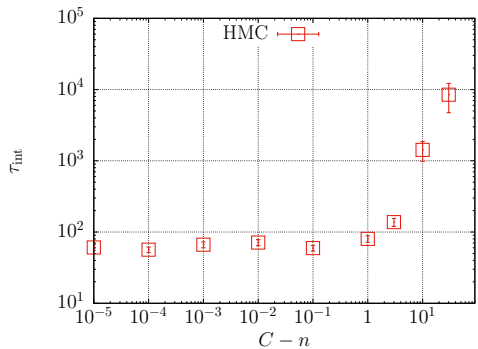}}}
		\caption{Integrated autocorrelation time of $|m|$ for the HMC algorithm at critical coupling and $h=0$ for the $d=2$ dimensional periodic square lattice with size $N=15^2$ against the shift $C$ reduced by the number of nearest neighbours $n$.}
		\label{fig:compare_C}
	\end{figure}

Last but not least let us study the impact of the shift parameter $C$ on the efficiency of the algorithm by means of the 
  autocorrelation for the absolute magnetization $|m|$.
As explained earlier, when $C$ becomes very large the potential becomes very steep around the local minima \eqref{large C minima}.
When this localization becomes important we expect the autocorrelation to increase with $C$, as transitions from one local minimum to another become less likely.
This behaviour can be seen in \Figref{compare_C}.
We find that the autocorrelation is constant within errors below some critical value, in this case $C_\text{crit}\approx n+1$, and increases rapidly for larger $C$.
So, as long as the potential is not too deep HMC can explore the whole configuration space.
A very large $C$ causes wells from which it is difficult to escape, while $C$s just large enough to ensure stability yield very flat, smooth potentials.
We see in \figref{compare_C} that as long as the shift is small $|C|-n \ll 1$ its specific value is irrelevant and does not need to be tuned.
	
\section{Conclusion}\label{sec:ising_conclusion}

In this paper we showed how to apply the HMC algorithm to the Ising model, successfully applying an algorithm that uses only continuous state variables to a system with discrete degrees of freedom.
We find that the HMC algorithm generalises the Ising model very well to arbitrary geometries without much effort.
It has been presented here in the most simple form.
In this simple form the HMC is an extremely inefficient algorithm if applied to the Ising model.
Although more flexible than the most efficient methods, such as cluster algorithms, it loses as compared even to the Metropolis-Hastings algorithm.
The coefficient by which the Metropolis-Hastings algorithm surpasses the HMC decreases with dimension, so that HMC might be preferable in case of an extremely high number of nearest neighbours---in the case of less local coupling, for example.

Moreover, for physical systems that suffer from sign problems, one may hope to leverage complex Langevin, Lefschetz thimble, or other contour-optimizing methods (for a dramatically incomplete set of examples, consider, respectively, \Ref{Aarts:2010gr,Fodor:2015doa}, \Refs{Cristoforetti:2012su,Fujii:2013sra,Alexandru:2017oyw}, and \Refs{Alexandru:2017czx,Mori:2017pne,Kashiwa:2018vxr} and references therein).
The formulation in terms of continuous variables presented here is well-suited for these methods, while the methods that deal directly with the original discrete variables such as the Metropolis-Hastings, cluster, and worm algorithms, for example, are non-starters.
In that sense, our exact reformulation and HMC method can be seen as the first step towards solving otherwise-intractable problems.

The HMC algorithm could be optimised by more efficient integrators and different choices of $C$, just to name the most obvious possibilities.
Many more methods have been developed to improve HMC performance and it is expected that some of them could also speed up the Ising model.

\section*{Acknowledgements}

We thank Paulo Bedaque, Matthias Fischer, Michael Kajan, Ulf Mei{\ss}ner, Marcel Nitsch, Carsten Urbach and Jan-Lukas Wynen for their constructive criticisms.  This work was done in part through financial support from the Deutsche Forschungsgemeinschaft (Sino-German CRC 110).
E.B. is supported by the U.S. Department of Energy under Contract No. DE-FG02-93ER-40762. 
\begin{subappendices}
\section{The ``infinite-range Ising model''}\label{sect:infiniteRangeIsing}

For the calculations shown on the right panel of Figure~\ref{fig:magnetisation} we used the following Hamiltonian,
\begin{align}
H({\bm s})&=-\frac{1}{2}\frac{J}{N}\sum_{i\ne j} s_is_j-h\sum_is_i\\
&=\frac{1}{2}J-\frac{1}{2}\frac{J}{N}\sum_{i, j} s_is_j-h\sum_is_i\ ,
\end{align}
where in the second line we used the fact that $s_i^2=1\ \forall i$ and there is no restriction in the sum over spin couplings.  With the exception of the \emph{self-energy} term $\frac{1}{2}J$ in the second line above,  the remaining terms  constitute the ``infinite-range Ising model''\cite{negele1988quantum}.  From now on we assume that $J>0$, but a similar calculation can be done for $J<0$.  

The partition function for this Hamiltonian can be exactly determined.  Applying the HS transformation as described in the equations leading up to the partition sum \eqref{cosh_in_exponent} one obtains
\begin{equation}
Z=e^{\frac{1}{2}\beta J}\int_{-\infty}^{\infty} \frac{d \phi}{\sqrt{2 \pi \beta \hat{J}}} e^{-\frac{\phi^{2}}{2 \beta \hat{J}}}[2 \cosh (\beta h \pm \phi)]^{N}\ ,
\end{equation}
where we define $\hat J=J/N$.  Expanding the $\cosh$ terms into exponentials allows one to formally integrate over the HS fields, obtaining
\begin{equation}\label{eq:all to all partition function}
Z=e^{\frac{1}{2}\beta J}\sum_{n=0}^{N}\begin{pmatrix}
N \\
n
\end{pmatrix} 
f(\beta \hat{J}, \beta h, N-2 n)\ ,
\end{equation}
and
\begin{equation}
f(\beta \hat{J}, \beta h, x) \equiv e^{\frac{1}{2} \beta \hat{J} x^{2}+\beta h x}\ .
\end{equation}
Using the definition of the energy density \eqref{ee form} with the partition sum \eqref{all to all partition function} gives our analytic expression for the internal energy,
\begin{equation}
\langle\beta \varepsilon\rangle=\frac{1}{2}\beta\hat J-\frac{\beta}{N Z}
\sum_{n=0}^{N}
\begin{pmatrix}
N \\
n
\end{pmatrix}\left[\frac{1}{2}\hat{J}(N-2n)^2+h(N-2n)\right]f(\beta \hat{J}, \beta h, N-2 n)\ .
\end{equation}
Note the relative sign difference between the terms on the right hand side above.  For sufficiently small $N$ the self-energy term wins out, otherwise the second term dominates.

An analogous calculation for the magnetisation with an extrapolation of $N\goesto \infty$ leads to the critical point $\beta J=1$. We used this value in Figure~\ref{fig:magnetisation}.

 \end{subappendices}

\let\eqref\oldeqref
\clearpage{}%
\clearpage{}%
\chapter[On the generalised eigenvalue method]{On the generalised eigenvalue method and its relation to Prony and generalised pencil of function methods}\label{chap:prony}

\begin{center}
	\textit{Based on~\cite{generalisedProny} by M.~Fischer, B.~Kostrzewa, J.~Ostmeyer, K.~Ottnad, M.~Ueding and C.~Urbach}
\end{center}

The Ising model, as discussed in the previous chapter, makes a good introduction to Monte Carlo simulations for many reasons. One such reason is that many results are readily obtained without any sophisticated analysis. The energy for instance can be calculated by a single sum over local contributions. In contrast, lattice field theories like QCD and our formulation of the Hubbard model (see \Cref{chap:phase_transition,chap:afm_character}) usually do not allow for such a simplistic treatment. Instead, the canonical way to proceed in these cases is the calculation of non-local correlators. These correlators in principle encode the full energy spectrum. In practice however noise and limited data allow only for the extraction of the ground state energy\footnote{Strictly speaking, the ground state or vacuum energy cannot be calculated in this setting. Rather we refer to the smallest energy difference from the vacuum as the lowest energy level here.} and possibly several low-lying excited state energies. In this chapter we present a method specifically designed to extract the largest possible amount of information from a given set of correlators.

\newcommand{\fig}[1]{Figure~\ref{#1}}
\newcommand{\tab}[1]{Table~\ref{#1}}
\newcommand{\eq}[1]{Eq.~(\ref{#1})}

\renewcommand{\l}{\left}
\renewcommand{\r}{\right}
\newcommand{\gev}{\,\mathrm{GeV}}
\newcommand{\mev}{\,\mathrm{MeV}}
\newcommand{\fm}{\,\mathrm{fm}}
\renewcommand{\order}[1]{\ensuremath{\mathcal{O}\l({#1}\r)}}
\newcommand{\phys}{\ensuremath{\mathrm{phys}}}
\newcommand{\syserr}[1]{\ensuremath{(#1)_\mathrm{sys}}}
\newcommand{\staterr}[1]{\ensuremath{(#1)_\mathrm{stat}}}

\renewcommand{\eto}[1]{\ensuremath{\mathrm{e}^{#1}}}
\renewcommand{\md}{\ensuremath{\mathrm{d}}}
\newcommand{\ordnung}[1]{\ensuremath{\ord\left(#1\right)}}
\newcommand{\gm}{Gardner method}
\newcommand{\Meff}{\ensuremath{M_\mathrm{eff}}}
\newcommand{\tMeff}{\ensuremath{\tilde{M}_\mathrm{eff}}}

\section{Introduction}

In lattice field theories one is often confronted with the task to
extract energy levels from noisy Monte Carlo data for Euclidean
correlation functions, which have the theoretical form
\begin{equation}
  \label{eq:corr}
  C(t)\ =\ \sum_{k=0}^\infty\ c_k\, e^{-E_k t}
\end{equation}
with real and distinct energy levels $E_{k+1}>E_k$ and real coefficients
$c_k$. It is well known that this task represents an ill-posed problem
because the exponential functions do not form an orthogonal system of
functions.

Still, as long as one is only interested in the ground state $E_0$
and the statistical accuracy is high enough to be able to work at
large enough values of $t$, the task can be accomplished by making use
of the fact that
\begin{equation}
  \label{eq:limit}
  \lim_{t\to\infty} C(t)\ \approx\ c_0 e^{-E_0 t}\,,
\end{equation}
with corrections exponentially suppressed with increasing $t$ due to ground 
state dominance. However, in lattice quantum chromodynamics, the
non-perturbative approach to quantum chromodynamics (QCD), the signal 
to noise ratio for $C(t)$ deteriorates exponentially with increasing
$t$~\cite{Lepage:1989}. Moreover, at large Euclidean times there 
can be so-called thermal pollutions (see e.g.\ Ref.~\cite{Feng:2010es})
to the correlation functions, which, if not accounted for, render the
data at large $t$ useless. And, once one is interested in excited
energy levels  $E_k\,,\ k>0$, alternatives to the ground state
dominance principle need to be found.

The latter problem can be tackled applying the so-called generalised
eigenvalue method (GEVM) -- originally proposed in
Ref.~\cite{Michael:1982gb} and further developed in
Ref.~\cite{Luscher:1990ck}. It is by now well established in lattice
QCD applications and allows one to estimate ground and excited states
for the price that a correlator matrix needs to be computed instead of
a single correlation function. Moreover, the systematics of this
method are well understood~\cite{Luscher:1990ck,Blossier:2009kd}. 

An alternative method, originally proposed by de
Prony~\cite{Prony:1795}, represents an algebraic method to determine
in principle all the energy levels from a single correlation
function. However, it is well known that the Prony method can become
unstable in the presence of noise. The Prony method was first used for
lattice QCD in Refs.~\cite{Fleming:2004hs,Beane:2009kya}. For more
recent references see
Refs.~\cite{Fleming:2006zz,Berkowitz:2017smo,Cushman:2019hfh} and also
Appendix~\ref{sec:gardner}. For an application of the Prony method
in real time dynamics with Tensor networks see Ref.~\cite{Banuls:2019qrq}.

In this paper we discuss the relation among generalised eigenvalue,
Prony and generalised pencil of function (GPOF) methods and trace them all
back to a generalised eigenvalue problem. This allows us to derive the
systematic effects due to so-called excited state contributions for
the Prony and GPOF methods using perturbation theory invented for the
GEVM~\cite{Blossier:2009kd}. In addition, we propose a combination of
the GEVM and the Prony method, the latter of which we also formulate
as a generalised eigenvalue method and denote it as Prony GEVM (PGEVM).
The combination we propose is to apply first the GEVM to a correlator
matrix and extract the so-called principal correlators, which are
again of the form \eq{eq:corr}. Then we apply the PGEVM to the principal
correlators and extract the energy levels. In essence: the GEVM is
used to separate the contributing exponentials in distinct principal
correlators  with reduced pollutions compared to the original
correlators. Then the PGEVM is applied only to obtain the ground state
in each principal correlator, the case where it works best.

By means of synthetic data we verify that the PGEVM works as expected
and that the systematic corrections are of the expected form. 
Moreover, we demonstrate that with the combination GEVM/PGEVM example
data from lattice QCD simulations can be analysed: we study the pion
first, where we are in the situation that the ground state can be
determined with other methods with high confidence. Thereafter we also
look at the $\eta$-meson and $I=1, \pi=\pi$ scattering, both of which
require the usage of the GEVM in the first place, but where also noise
is significant. 

The paper is organised as follows: in the next section we introduce the GEVM 
and PGEVM and discuss the systematic errors of PGEVM. After briefly explaining 
possible numerical implementations, we present example applications using both 
synthetic data and data obtained from lattice QCD simulations. In the end we 
discuss the advantages and disadvantages of our new method, also giving an 
insight into when it is most useful.

\section{Methods}
\label{sec:method}

Maybe the most straightforward approach to analysing the correlation
function \eq{eq:corr} for the ground state energy $E_0$ is to use the
so-called effective mass defined as
\begin{equation}
  \label{eq:effmass}
  M_\mathrm{eff}(t_0, \delta t)\ =\ -\frac{1}{\delta t}\log\left(
  \frac{C(t_0+\delta t)}{C(t_0)}\right)\,.
\end{equation}
In the limit of large $t_0$ and fixed $\delta t$, $M_\mathrm{eff}$
converges to $E_0$. The correction due to the first excited state
$E_1$ is readily computed:
\begin{equation}
  \begin{split}
    M_\mathrm{eff}(t_0, \delta t)\ \approx&\ E_0 + \frac{c_1}{c_0}
    e^{-(E_1-E_0)t_0}\\
    &\times \left( 1 - e^{-(E_1-E_0)\delta t}\right)\,\frac{1}{\delta t}\,.
  \end{split}
\end{equation}
It is exponentially suppressed in $t_0$ and the energy difference between
first excited and ground state. It is also clear from this formula
that taking the limit $\delta t\to\infty$ while keeping $t_0$ fixed
leads to a worse convergence behaviour than keeping $\delta t$ fixed
and changing $t_0$. In this section we will 
discuss how both of the two above equations generalise.

\subsection{The generalised eigenvalue method (GEVM)}

We first introduce the GEVM. The method is important for being able to
determine ground and excited energy levels in a given
channel. Moreover, it helps to reduce excited state contaminations to
low lying energy levels. 

Using the notation of Ref.~\cite{Blossier:2009kd}, one considers
correlator matrices of the form
\begin{equation}
  \label{eq:corrM}
  C_{ij}(t)\ =\ \langle \hat O_i^{}(t')\ \hat O_j^\dagger(t'+t)\rangle\ =\ \sum_{k=0}^\infty e^{-E_k t} \psi_{ki}^*\psi_{kj}\,,
\end{equation}
with energy levels $E_k > 0$ and $E_{k+1} > E_k$ for all values of
$k$. The $\psi_{ki}=\langle 0|\hat O_i|k\rangle$ are matrix elements of
$n$ suitably chosen operators $\hat O_i$ with $i=0, ..., n-1$. Then, the
eigenvalues or so-called principal 
correlators $\lambda(t, t_0)$ of the generalised eigenvalue problem (GEVP)
\begin{equation}
  C(t)\, v_k(t, t_0)\ =\ \lambda^0_k(t, t_0)\, C(t_0)\, v_k(t, t_0)\,,
  \label{eq:GEVP}
\end{equation}
can be shown to read
\begin{equation}
  \label{eq:lambdaGEVP}
  \lambda^0_k(t, t_0)\ =\ e^{-E_k(t - t_0)}
\end{equation}
for $t_0$ fixed and $t\to\infty$.
Clearly, the correlator matrix $C(t)$ will for every practical
application always be square but finite with dimension $n$. This will
induce corrections to \eq{eq:lambdaGEVP}. The corresponding 
corrections were derived in Ref.~\cite{Luscher:1990ck,Blossier:2009kd}
and read to leading order 
\begin{equation}
  \label{eq:t0corr}
  \lambda_k(t, t_0) = b_k \lambda^0_k(1+\mathcal{O}(e^{-\Delta E_k t}))
\end{equation}
with $b_k>0$ and
\begin{equation}
  \Delta E_k\ =\ \min_{l\neq k}|E_l -E_k|\,.
\end{equation}
Most notably, the principal correlators $\lambda_k(t_0, t)$ are at
fixed $t_0$ again a sum of exponentials. As was shown in
Ref.~\cite{Blossier:2009kd}, for $t_0>t/2$ the leading corrections are
different compared to \eq{eq:t0corr}, namely of order
\begin{equation}
  \label{eq:t0corr2}
  \exp[-(E_n - E_k)t]\,.
\end{equation}

\subsection{The Prony method}

For the original Prony method~\cite{Prony:1795}, we restrict ourselves
first to a finite 
number $n$ of exponentials in an Euclidean correlation function $C^0$
\begin{equation}
  \label{eq:C0}
  C^0(t)\ =\ \sum_{k=0}^{n-1} c_k\, e^{-E_k t}\,.
\end{equation}
The $c_k$ are real, but not necessarily positive constants and $t$ is
integer--valued.  Thus, we focus on one matrix element of the correlator
matrix \eq{eq:corrM} from above or other correlators with the
appropriate form. We assume now $E_k\neq 0$ for all $k\in\{0,
\ldots, n-1\}$ and that all  the $E_k$ are distinct. Moreover, we
assume the order $E_{k+1} > E_k$ for all $k$.
Then, Prony's method is a generalisation of the effective mass
\eq{eq:effmass} in the form of a matrix equation
\begin{equation}
  \label{eq:prony}
  H\cdot x = 0\,,
\end{equation}
with an $n\times(n+1)$ Hankel matrix $H$ 
\[
  H=
  \begin{pmatrix}
    C^0(t) & C^0(t+1) & \ldots & C^0(t+n) \\
    C^0(t+1) & C^0(t+2) & \ldots & C^0(t+n+1) \\
    \vdots & \vdots & \ddots & \vdots \\
    C^0(t+n-1) & C^0(t+n) & \ldots & C^0(t+2n-1) \\
  \end{pmatrix}
\]
and a coefficient vector $x = (x_0, \ldots, x_{n-1}, 1)$ of length
$n+1$. After solving for $x$, the exponentials are obtained from $x$
by the roots of  
\[
x_0 + x_1\left(e^{-E_l}\right) + x_2 \left(e^{-E_l}\right)^2 + \ldots
+ \left(e^{-E_l}\right)^n = 0\,. 
\]
For a further generalisation see Ref.~\cite{Beane:2009kya} and
references therein.

\subsection{The Prony GEVM (PGEVM)}

Next we formulate
Prony's method \eq{eq:prony} as a generalised eigenvalue
problem (see also Ref.~\cite{sauer:2013}). Let $H^0(t)$ be a $n\times
n$ Hankel matrix for $i,j=0, 1, 2, \ldots, n-1$ defined by 
\begin{equation}
  \begin{split}
    H^0_{ij}(t)\ &=\ C^0(t + i\Delta + j\Delta)\\
    &=\ \sum_{k=0}^{n-1} e^{-E_kt}\, e^{-E_ki\Delta}\, e^{-E_kj\Delta}
    c_k\,,\\ 
  \end{split}
\end{equation}
with integer $\Delta>0$.
$H^0(t)$ is symmetric, but not necessarily positive definite.
We are going to show that the energies $E_0, \ldots, E_{n-1}$ can be
determined from the generalised eigenvalue problem
\begin{equation}
  \label{eq:hankelgevp}
  H^0(t)\, v_l\ =\ \Lambda^0_l(t, \tau_0)\, H^0(\tau_0)\, v_l\,.
\end{equation}
The following is completely analogous to the corresponding proof of
the GEVM in Ref.~\cite{Blossier:2009kd}. 
Define a square matrix
\begin{equation}
  \chi_{ki}\ =\ e^{-E_k i\Delta}\,.
\end{equation}
and re-write $H^0(t)$ as
\[
H^0_{ij}(t)\ =\ \sum_{k=0}^{n-1} c_k e^{-E_k t}\chi_{ki}\chi_{kj}\,.
\]
Note that $\chi$ is a square Vandermonde matrix
\[
\chi\ =\
\begin{pmatrix}
  1 & e^{-E_0\Delta} & e^{-2E_0\Delta} & \ldots & e^{-(n-1)E_{0}\Delta}\\
  1 & e^{-E_1\Delta} & e^{-2E_1\Delta} & \ldots & e^{-(n-1)E_{1}\Delta}\\
  \vdots & \vdots & \vdots & \ldots & \vdots \\
  1 & e^{-E_{n-1}\Delta} & e^{-2E_{n-1}\Delta} & \ldots & e^{-(n-1)E_{n-1}\Delta}\\
\end{pmatrix}
\]
with all coefficients distinct and, thus, invertible.
Now, like in Ref.~\cite{Blossier:2009kd}, introduce the dual vectors
$u_k$ with 
\[
(u_k, \chi_l)\ =\ \sum_{i=0}^{n-1} (u_{k}^*)_i \chi_{li}\ =\ \delta_{kl}
\]
for $k,l \in\{0, \ldots, n-1\}$. With these we can write
\begin{equation}
  \begin{split}
    H^0(t)\, u_l\ &=\ \sum_{k=0}^{n-1} c_k e^{-E_k t}\chi_k \chi_k^* u_l\ =\
    c_l e^{-E_l t} \chi_l\\
    &=\ e^{-E_l(t-\tau_0)}\ c_l e^{-E_l \tau_0} \chi_l\\
    &=\ e^{-E_l(t-\tau_0)} H^0(\tau_0)\, u_l\\
  \end{split}
\end{equation}
Thus, the GEVP \eq{eq:hankelgevp} is solved by
\begin{equation}
  \label{eq:hankelsol}
  \Lambda^0_k(t, \tau_0)\ =\ e^{-E_k(t-\tau_0)}\,,\quad
  v_k\ \propto\ u_k\,.
\end{equation}
Moreover, much like in the case of the GEVM
we get the orthogonality 
\begin{equation}
  \label{eq:ortho}
  (u_l,\, H^0(t) u_k)\ =\ c_l e^{-E_l t}\delta_{lk}\,,\quad k,l\in\{0, \ldots,
  n-1\}
\end{equation}
for all $t$-values, because $H^0(t)\, u_k\propto \chi_k$.

\subsubsection{Global PGEVM}

In practice, there are two distinct ways to solve the
GEVP \eq{eq:hankelgevp}: one can fix $\tau_0$ and determine
$\Lambda^0_k(\tau_0, t)$ 
as a function of $t$. In this case the solution \eq{eq:hankelsol}
indicates that for each $k$ the eigenvalues decay exponentially in
time. On the other hand, one can fix $\delta t = t-\tau_0$ and determine
$\Lambda^0_k(\tau_0, \delta t)$ as a function of $\tau_0$. In this
case the solution \eq{eq:hankelsol} reads
\[
\Lambda^0_k (\tau_0, \delta t)= e^{-E_k\,\delta t}\ =\ \mathrm{const}\,,
\]
because $\delta t$ is fixed.

The latter approach allows to formulate a global PGEVM. Observing that
the matrices $\chi$ do not depend on $\tau_0$, one can reformulate the
GEVP~\eq{eq:hankelgevp} as follows
\begin{equation}
  \label{eq:globalhankel}
  \sum_{\tau_0} H^0(\tau_0+\delta t)\, v_l\ =\ \Lambda^0_l(\delta t)\,
  \sum_{\tau_0} H^0(\tau_0)\, v_l\,, 
\end{equation}
since $\Lambda^0_k$ does not depend on $\tau_0$. However, this works
only as long as there are only $n$ states contributing and all these
$n$ states are resolved by the PGEVM, as will become
clear below. If this is not the case, pollutions and resolved states
will change roles at some intermediate $\tau_0$-value.

\subsubsection{Effects of Additional States}

Next, we ask the question what corrections to the above result we
expect if there are more than $n$ states contributing, i.e.\@
a correction term
\begin{equation}
  C^1(t)\ =\ \sum_{k=n}^{\infty} c_k\, e^{-E_k t}
\end{equation}
to the correlator and a corresponding correction to the Hankel matrix
\[
\epsilon H_{ij}^1(t)\ =\ \sum_{k=n}^\infty C^1(t + i + j)\,.
\]
(We have set $\Delta=1$ for simplicity.)
We assume that we work at large enough $t$ such that these corrections
can be considered as a small perturbation. Then it turns out that the
results of Refs.~\cite{Luscher:1990ck,Blossier:2009kd} apply directly
to the PGEVM and all systematics are identical (\eq{eq:t0corr} or
\eq{eq:t0corr2}). 

However, there is one key difference between GEVM and PGEVM. The GEVM
with periodic boundary conditions is not able to distinguish the
forward and backward propagating terms in
\[
c\left(e^{-E t} \pm e^{-E(T-t)}\right)\,,
\]
as long as they come with the same amplitude. In fact, the eigenvalue
$\lambda^0$ will in this case also be a $\cosh$ or
$\sinh$~\cite{Irges:2006hg}. In contrast, the PGEVM can distinguish
these two terms. As a consequence, the backward propagating part needs
to be treated as a perturbation like excited states and $\Lambda^0$ is
no longer expected to have a $\cosh$ or $\sinh$ functional form in the
presence of periodic boundary conditions.

This might seem to be a disadvantage at first sight. However, we will
see that this does not necessarily need to be the case.

Concerning the size of corrections there are two regimes to
consider~\cite{Blossier:2009kd}: when $\tau_0$ is fixed at small or
moderately large values and $\Lambda$ is studied as a function of
$t\to\infty$ the corrections of the form~\eq{eq:t0corr}
apply~\cite{Luscher:1990ck}.  When, on the other hand, $\tau_0$
is fixed but $\tau_0\geq t/2$ is chosen and the effective
masses~\eq{eq:effmass} of the eigenvalues are studied, corrections
are reduced to $\mathcal{O}(e^{-\Delta E_{n,l}t})$ with
$\Delta E_{m,n} = E_m - E_n$~\cite{Blossier:2009kd}.

$\tau_0\geq t/2$ is certainly fulfilled if we fix $\delta t$ to some
(small) value. However, for this case $\Lambda^0(t, \tau_0)$ is
expected to be independent of both, $t$ and $\tau_0$ when ground state
dominance is reached and $\Meff$ is, thus, not applicable. Therefore,
we define alternative effective masses 
\begin{equation}
  \label{eq:Eeff}
  \tilde{M}_{\mathrm{eff},l}(\delta t,\tau_0)\ =\ -\frac{\log\left(\Lambda_l(\delta t,
  \tau_0)\right)}{\delta t}
\end{equation}
and apply the framework from Ref.~\cite{Blossier:2009kd} to determine
deviations of $\tilde{M}_{\mathrm{eff},l}$ from the true $E_l$. The authors of
Ref.~\cite{Blossier:2009kd} define $\epsilon = e^{-(E_{n}-E_{n-1})\tau_0}$
and expand
\begin{equation}
  \Lambda_l = \Lambda^0_l + \epsilon \Lambda^1_l + \epsilon^2\Lambda^2_l
  + \ldots\,,
\end{equation}
where we denote the eigenvalues of the full problem as $\Lambda(t, \tau_0)$.
Already from here it is clear that in the situation with $\delta t$
fixed and $\tau_0\to\infty$ the expansion parameter $\epsilon$ becomes
arbitrarily small. Simultaneously with $\tau_0$ also $t\to\infty$. The
first order correction (which is dominant for $\tau_0\geq t/2$) to
$\Lambda_l$ reads 
\begin{equation}
  \begin{split}
    \Lambda_l(\delta t, \tau_0) = e^{-E_l\delta t} 
    +& \frac{c_n}{c_l} e^{-(\Delta E_{n,l})\tau_0}\,\\
    \times &\left(e^{-\Delta
        E_{n,l}\delta t}-1\right) c_{l, n}\\
  \end{split}
\end{equation}
with the definition of $\Delta E_{m,n}$ from above and constant
coefficients 
\[
c_{l, n}\ =\ (v_l^0, \chi_{n})(\chi_{n}, v_l^0)\,.
\] 
These corrections are decaying exponentially in $\tau_0$ with a
decay rate determined by $\Delta E_{n,l}$ as expected from
Ref.~\cite{Blossier:2009kd}. For the effective energies we find 
\begin{equation}
  \label{eq:Eeffcorr}
  \begin{split}
    \tilde{M}_{\mathrm{eff},l}(\delta t,\tau_0)\ \approx \ E_l 
    &+\frac{c_n}{c_l}
    e^{-(\Delta E_{n,l})\tau_0}\,\\
    &\times \left(e^{E_l\delta
      t}-e^{-E_{n}\delta t} \right)
    \frac{c_{l, n}}{\delta t}\,,\\
  \end{split}
\end{equation}
likewise with corrections decaying exponentially in $\tau_0$, again
with a rate set by $\Delta E_{n,l}$.

\subsection{Combining GEVM and PGEVM}

There is one straightforward way to combine GEVM and PGEVM: we noted
already above that the principal correlators of the GEVM are again a
sum of exponentials, and, hence, the PGEVM can be applied to them.
This means a sequential application of first the
GEVM with a correlator matrix of size $n_0$ to determine principal
correlators $\lambda_k$ and then of the PGEVM with size $n_1$ and the
$\lambda_k$'s as input, which we denote as GEVM/PGEVM. This
combination allows us to work with two 
relatively small matrices, which might help to stabilise the method
numerically. Moreover, the PGEVM is 
applied only for the respective ground states in the principal
correlators and only relatively small values of $n_1$ are needed. 

An additional advantage lies in the fact that $\lambda_k$ is a sum of
exponentials with only positive coefficients, because it represents a
correlation function with identical operators at source and sink. As a
consequence, the Hankel matrix $H^0$ is positive definite. 

\subsection{Generalised Pencil of Function (GPOF)}

For certain cases, the PGEVM can actually be understood as a special
case of the generalised pencil-of-function (GPOF) method, see
Refs.~\cite{Aubin:2010jc,Aubin:2011zz,Schiel:2015kwa,Ottnad:2017mzd}
and references therein. Making use of the time evolution operator, we
can define a new operator
\begin{equation}
  \hat O_{\Delta t}(t')\ \equiv\ \hat O(t'+ \Delta t) = \exp(H
  \Delta t)\ \hat O(t')\ \exp(-H \Delta t)\,.
\end{equation}
This allows us to write
\begin{equation}
  \langle \hat O_i^{}(t')\ \hat O_j^\dagger(t'+t + \Delta t)\rangle\ =\
  \langle \hat O_i^{}(t')\ \hat O_{\Delta t,j}^\dagger(t'+t)\rangle\,,
\end{equation}
which is the same as $C_{ij}(t + \Delta t)$. Using $i=j$ and the
operators $O_i$, $O_{\Delta t, i}$, $O_{2\Delta t, i}, \dots$ one
defines the PGEVM based on a single correlation function.
Note, however, that the PGEVM is more general as it is also
applicable to sum of exponentials not stemming from a two-point
function.

The generalisation is now straightforward by combining $\hat O_i$ and 
$\hat O_{m\Delta t, i}$ for $i=0, \ldots, n_0-1$ and $m=0, \ldots,
n_1-1$. These operators define a Hankel matrix $\mathcal{H}^0$ with
size  $n_1$ of correlator matrices of size $n_0$ as follows
($\Delta=1$ for simplicity) 
\begin{equation}
  \mathcal{H}^0_{\alpha\beta}\ =\ \sum_{k=0}^{n'-1} e^{-E_k t}
  \eta_{k\alpha}\eta_{k\beta}^*\,, 
\end{equation}
with
\begin{equation}
  (\eta_{k})_{i n_0 + j}\ =\ e^{-E_k i}\psi_{kj}\,,
\end{equation}
for $j=0, \ldots,n_0-1$ and $i=0, \ldots, n_1-1$.
Then $n'=n_0\cdot n_1$ is the number of energies that can be resolved.
$\mathcal{H}$ is hermitian, positive definite and the
same derivation as the one from the previous subsection leads to the
GEVP
\[
\mathcal{H}^0(t)\, v_k\ =\ \Lambda_k(t, \tau_0)\, \mathcal{H}^0(\tau_0)\, v_k
\]
with solutions
\[
\Lambda^0_k(t, \tau_0) = e^{-E_k(t - \tau_0)}\,.
\]
In this case the matrix $\mathcal{H}^0$ is positive definite, but
potentially large, which might lead to numerical instabilities. This
can be alleviated by using only for a limited subset of
operators $\hat O_i$ their shifted versions $\hat O_{m\Delta t,i}$,
preferably for those $\hat O_i$ contributing the least noise.

Note that with $\Delta t=0$ GPOF defines the
GEVM. Therefore, GPOF is a generalisation of both, GEVM and
PGEVM. We also remark that without noise, GEVM/PGEVM with $n_0$ and
$n_1$ is exactly equivalent to GPOF with $n'=n_0\cdot n_1$. However, with
noise this is no longer the case.

\section{Numerical Implementation}

In case the Hankel matrix $H^0$ is positive definite,
one can compute the Cholesky decomposition $C(t_0)=L\cdot L^T$. Then
one solves the ordinary eigenvalue problem
\[
L^{-1}\, C(t)\, L^{-T}\, w_k = \lambda_k w_k
\]
with $w_k= L^T v_k$.

If this is not the case, the numerical solution of the PGEVM can proceed along
two lines. The first is to compute the inverse of $H^0(\tau_0)$ for
instance using a QR-decomposition and then solve the ordinary
eigenvalue problem for the matrix $A=H^0(\tau_0)^{-1}H^0(t)$. Alternatively, one may take
advantage of the symmetry of both $H^0(t)$ and $H^0(\tau_0)$. One
diagonalises both $H^0(t)$ and $H^0(\tau_0)$ with diagonal eigenvalue
matrices $\Lambda_t$ and $\Lambda_{\tau_0}$ and orthogonal eigenvector matrices
$U_t$ and $U_{\tau_0}$. Then, the eigenvectors of the generalised problem
are given by the matrix
\[
U\ =\ U_{\tau_0}\, \Lambda_{\tau_0}^{-1/2}\, U_t
\]
and the generalised eigenvalues read
\[
\Lambda\ =\ U^T\, H^0(t)\, U\,.
\]
Note that $U$ is in contrast to $U_t$ and $U_{\tau_0}$ not orthogonal.

\subsection{Algorithms for sorting GEVP states}
\label{sec:sorting}

Solving the generalized eigenvalue problem in Eq.~(\ref{eq:GEVP}) for
an $n\times n$ correlation function matrix $C(t)$ (or Hankel matrix
$H$) with $t>t_0$, results in an a priori unsorted set $\l\{s_k(t)
|k\in[0,...,n-1]\r\}$ of states $s_k(t)=(\lambda_k(t,t_0),
v_k(t,t_0))$ on each timeslice $t$ defined by an eigenvalue
$\lambda_k(t,t_0)$ and an eigenvector $v_k(t,t_0)$. In the
following discussion we assume that the initial order of states is
always fixed on the very first timeslice $t_0+1$ by sorting the states
by eigenvalues, i.e.\ choosing the label $n$ by requiring
$\lambda_0(t_0+1,t_0) > \lambda_1(t_0+1,t_0) > ... >
\lambda_{n-1}(t_0+1,t_0)$, s.t.\ the vector of states reads
$(s_{0}(t_0+1), ..., s_{n-1}(t_0+1) )$. \par  

After defining the initial ordering of states, there are many different possibilities to sort the remaining states for $t>t_0$. In general, this requires a prescription that for any unsorted vector of states $(s_{(k=0)}(t), ..., s_{(k=n-1)}(t) )$ yields a re-ordering $s_{\epsilon(k)}(t)$ of its elements. The permutation $\epsilon(k)$ may depend on some set of reference states $(s_{0}(\tilde{t}), ... , s_{n-1}(\tilde{t}))$ at time $\tilde{t}$ which we assume to be in the desired order. However, for the algorithms discussed here, such explicit dependence on a previously determined ordering at a reference time $\tilde{t}$ is only required for eigenvector-based sorting algorithms. Moreover, $\tilde{t}$ does not necessarily have to equal $t_0+1$. In fact, the algorithms discussed below are in practice often more stable for choosing e.g.\ the previous timeslice $t-1$ to determine the order of states at $t$ while moving through the available set of timeslices in increasing order. \par

\subsubsection{Sorting by eigenvalues}
This is arguably the most basic way of sorting states; it simply consists of repeating the ordering by eigenvalues that is done at $t_0$ for all other values of $t$, i.e.\ one chooses $\epsilon(k)$ independent of any reference state and ignoring any information encoded in the eigenvectors, s.t.\@
\begin{equation}
 \lambda_0(t,t_0) > \lambda_1(t,t_0) > ... > \lambda_{n-1}(t,t_0) \,.
 \label{eq:eval_sort}
\end{equation}
The obvious advantage of this method is that it is computationally fast and trivial to implement. However, it is not stable under noise which can lead to a rather large bias and errors in the large-$t$ tail of the correlator due to incorrect tracking of states. This is an issue for systems with a strong exponential signal-to-noise problem (e.g.\ the $\eta$,$\eta'$-system) as well as for large system sizes $n$. Moreover, the algorithm fails by design to correctly track crossing states, which causes a flipping of states at least in an unsupervised setup and tends to give large point errors around their crossing point in $t$. \par

\subsubsection{Simple sorting by eigenvectors}
Sorting algorithms relying on eigenvectors instead of eigenvalues generally make use of orthogonality properties. A simple method is based on computing the scalar product
\begin{equation}
 c_{kl} = \langle v_l(\tilde{t}), v_k(t)\rangle \,,
 \label{eq:evec_simple_sort}
\end{equation}
where $v_l(\tilde{t})$ denote eigenvectors of some (sorted) reference states $s_l(\tilde{t})$ at $\tilde{t}<t$ and $v_k(t)$ belongs to a state $s_k(t)$ that is part of the set which is to be sorted. For all values of $k$ one assigns $k\rightarrow \epsilon(k)$, s.t.\@ $\l| c_{kl} \r| \stackrel{!}{=} \mathrm{max}$. If the resulting map $\epsilon(k)$ is a permutation the state indexing at $t$ is assigned according to $s_{k}(t) \rightarrow s_{\epsilon(k)}(t)$. Otherwise sorting by eigenvalues is used as a fallback. \par

This method has some advantages over eigenvalue-based sorting methods: It can in principle track crossing states and flipping or mixing of states in the presence of noise are less likely to occur. The latter is especially an issue for resampling (e.g.\ bootstrap or jackknife), i.e.\ if state assignment fails only on a subset of samples for some value(s) of $t$, leading to large point errors and potentially introducing a bias. On the downside, the resulting order of states from this method is in general not unambiguous for systems with $n>2$ and the algorithm is not even guaranteed to yield a valid permutation $\epsilon(k)$ for such systems in the presence of noise, hence requiring a fallback.\par

\subsubsection{Exact sorting by eigenvectors}
\label{subec:exact_sorting}
Any of the shortcomings of the aforementioned methods are readily avoided by an approach that uses volume elements instead of scalar products. This allows to obtain an unambiguous state assignment based on (globally) maximized orthogonality. The idea is to consider the set of all possible permutations $\{\epsilon(k)\}$ for a given $n\times n$ problem and compute 
\begin{equation}
  \label{eq:evec_exact_sort}
  \begin{split}
    c_{\epsilon} =\prod_k & \l| \mathrm{det}\l(
    v_{\epsilon(0)}(\tilde{t}), \ldots , v_{\epsilon(k-1)}(\tilde{t})\right.\right.
    ,\\ &\left.\left.v_k(t), v_{\epsilon(k+1)}(\tilde{t}) , \ldots , v_{\epsilon(n-1)}(\tilde{t})\r) \r| \,,\\
  \end{split}
\end{equation}
for each $\epsilon$. This can be understood as assigning a score for how well each individual vector $v_k(t)$ fits into the set of vectors at the reference timeslice $\tilde{t}$ at a chosen position $\epsilon(k)$ and computing a global score for the current permutation $\epsilon$ by taking the product of the individual scores for all vectors $v_k(t)$. The final permutation is then chosen s.t.\ $c_{\epsilon}\stackrel{!}{=}\mathrm{max}$. \par

Unlike the method using the scalar product, this method is guaranteed
to always give a unique solution, which is optimal in the sense that
it tests all possible permutations and picks the global
optimum. Therefore, the algorithm is most stable under noise and well
suited for systems with crossing states. Empirically, this results in
e.g.\ the smallest bootstrap bias at larger values of $t$ compared to
any other method described here. A minor drawback of the approach is
that it is numerically more expensive due to the required evaluations
of (products of) volume elements instead of simple scalar
products. However, this becomes only an issue for large system sizes
and a large number of bootstrap (jackknife) samples.

We remark that another, numerically cheaper method leading to a definite
state assignment can be obtained by replacing the score function in
Eq.~\eqref{eq:evec_exact_sort} by
\begin{equation}
  c_{\epsilon} = \prod_k \l| (v_k(\tilde{t}) \cdot v_{\epsilon(k)}(t)) \r| \,.
\end{equation}
For a $2\times2$ problem both methods give identical results. However,
for the general $n\times n$ case with $n>2$ they are no longer equivalent. 
This is because the method based on Eq.~(\ref{eq:evec_exact_sort}) uses more
information, i.e.\ the individual score for each $\epsilon(k)$ entering
the product on the r.h.s.\ is computed against $(n-1)$ vectors on the
reference time-slice instead of just a single vector as it is the case
for the method based on the scalar product.

\subsubsection{Sorting by minimal distance}
\label{subsec:mindistsort}
While the methods discussed above work all fine for the standard case
where the GEVP is solved with fixed time $t_0$ (or $\tau_0$) and
$\delta t$ is varied, the situation is different for $\tMeff$ with
$\delta t$ fixed: there are $t$-values for which it is numerically not
easy to separate wanted states from pollutions, because they are of
very similar size in the elements of the sum of exponentials entering
at these specific $t$-values. However, when looking at the bootstrap
histogram of all eigenvalues, there is usually a quite clear peak at
the expected energy value for all $t$-values with not too much noise.

Therefore, we implemented an alternative sorting for this situation
which goes by specifying a target value $\xi$. Then we chose among all
eigenvalues for a bootstrap replicate the one which is closest to
$\xi$. The error is computed from half of the $16$\% to $84$\%
quantile distance of the bootstrap distribution and the central value
as the mean of $16$\% and $84$\% quantiles. For the central value one
could also use the median, however, we made the above choice to have
symmetric errors.

This procedure is much less susceptible to large outliers in the
bootstrap distribution, which appear because of the problem discussed
at the beginning of this sub-section.

For the numerical experiments shown below we found little to no
difference in between sorting by \emph{eigenvalues} and
any of the sorting by \emph{vectors}. Thus, we will work with sorting by
\emph{eigenvalues} for all cases where we study $\Lambda_l(t, \tau_0)$ with
$\tau_0$ fixed. On the other hand, specifying a target value $\xi$
and sort by \emph{minimal distance}
turns out to be very useful for the case $\Lambda_l(\delta t, \tau_0)$
with $\delta t$ fixed. As it works much more reliably than the other
two approaches, we use this sorting by \emph{minimal distance} for the
$\delta t$ fixed case throughout this paper.

The methods used in this paper are fully implemented in a R package
called \emph{hadron}~\cite{hadron}, which is freely available
software. 

\section{Numerical Experiments}
\label{sec:experiments}

\begin{figure*}[th]
  \centering
  \subfigure{\includegraphics[width=.4\textwidth,page=1]{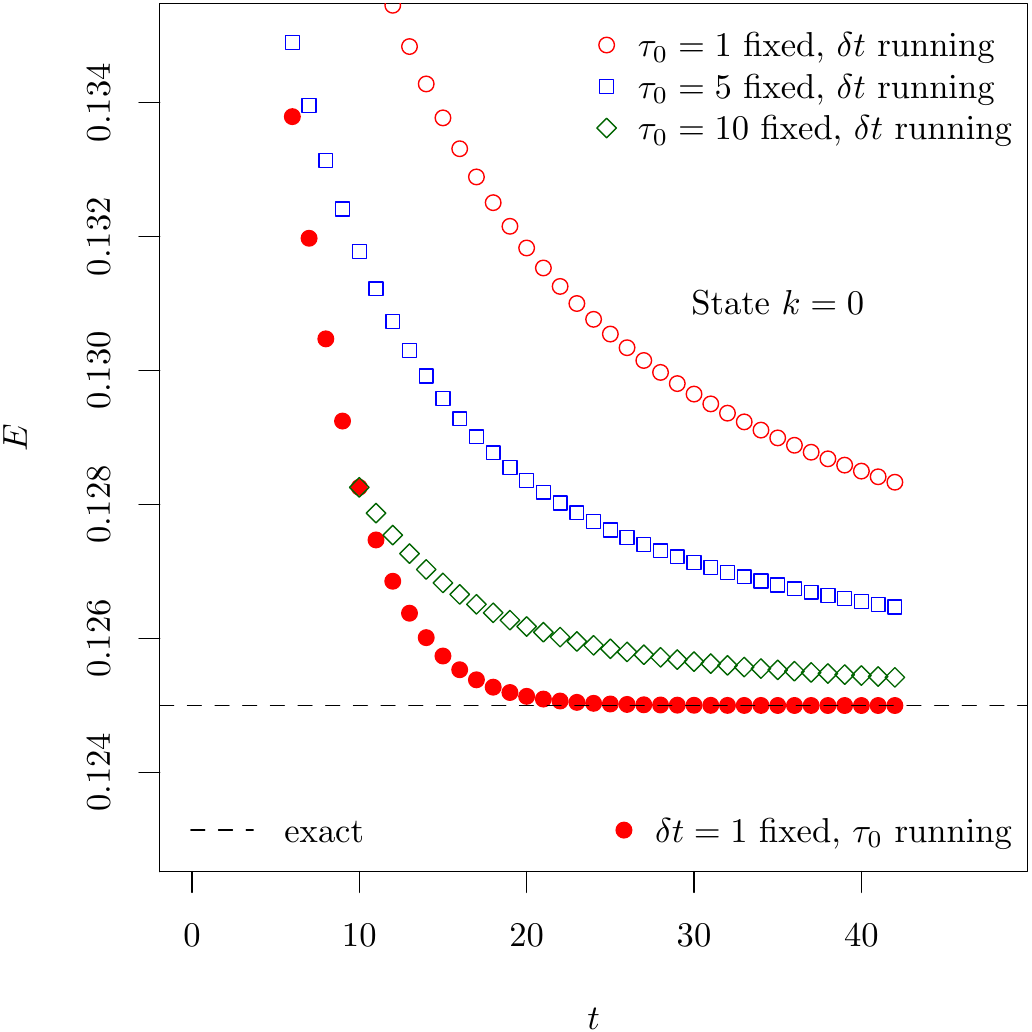}}\qquad
  \subfigure{\includegraphics[width=.4\textwidth,page=2]{experiments}}
  \caption{Effective energies from the PGEVM with $n=2$ applied
  to synthetic data containing three states with $\Delta=1$. Open symbols correspond
  to $\tMeff$ \eq{eq:Eeff} of the Prony principal
  correlator with $\tau_0$
  fixed, while filled symbols are for $\delta t$ fixed. In the
  left panel we show the ground state with $k=0$, in the right one the
  first excited state, both for different choices of $\tau_0$.}
  \label{fig:synthetic}
\end{figure*}

In this section we first apply the PGEVM to synthetic data. With
this we investigate whether additional states not accounted for by the
size of the Prony GEVP lead to the expected distortions in the
principal correlators and effective masses. At this stage the energy
levels and amplitudes are not necessarily chosen realistically, because
we would first like to understand the systematics.

In a next step we apply the combination of GEVM and PGEVM to
correlator matrices from lattice QCD simulations. After applying the
framework to the pion, we have chosen two realistic examples, the
$\eta$-meson and the $\rho$-meson. 

\subsection{Synthetic Data}

As a first test we apply the PGEVM alone to synthetic data. We
generate a correlator
\begin{equation}
C_s(t)\ =\ \sum_{k=0}^{2} c_k\, e^{-E_k t}
\end{equation}
containing three states with $E_k=(0.125, 0.3, 0.5), k=0,1,2$ and
$t\in\{0, \ldots, 48\}$. The amplitudes $c_k$ have been chosen all equal to
$1$. 

\begin{figure}
  \centering
  \includegraphics[width=.4\textwidth,page=3]{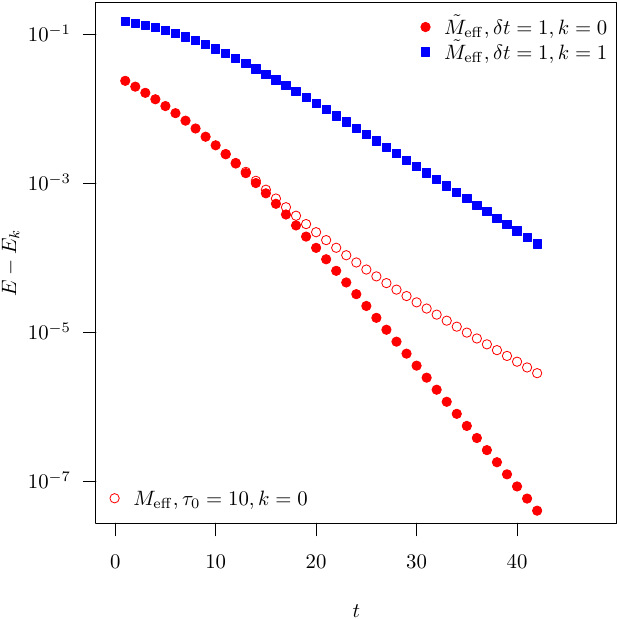}
  \caption{Effective $E$ minus the exact ground
    state energy $E_0=0.125$ for $k=0$ and $E_1=0.3$ for $k=1$ as a
    function of $t$ on a logarithmic scale for $\Delta=1$. Filled symbols correspond
    to $\tMeff$ \eq{eq:Eeff} with $\delta t=1$ fixed, open symbols to
    $\Meff$ \eq{eq:effmass} for $\tau_0=10$ fixed.}
  \label{fig:synthetic4}
\end{figure}

We apply the PGEVM to this correlator $C_s$ with
$n=2$. This allows us to resolve only two states and we would like to see
how much the third state affects the two extracted states. The result
is plotted in \fig{fig:synthetic}. We plot $\tMeff$ of \eq{eq:Eeff} as
a function of $t$, filled symbols correspond to $\delta t=1$
fixed. Open symbols correspond to $\tau_0$ fixed with values
$\tau_0=1, 5$ and $\tau_0=10$. In the left panel we show the ground
state $k=0$, in the right one the second state $k=1$ resolved by the
PGEVM.  The solid lines represent the input values for $E_0$ and
$E_1$, respectively.  

One observes that the third state not resolved by the PGEVM leads
to pollutions at small values of $t$. These pollutions are clearly
larger for the case of fixed $\tau_0$, as expected from our discussion
in section~\ref{sec:method}. The relative size of the pollutions is
much larger in the second state with $k=1$ than in the state with
$k=0$, which is also in line with the expected pollution.

\begin{figure}
  \centering
  \includegraphics[width=.4\textwidth,page=4]{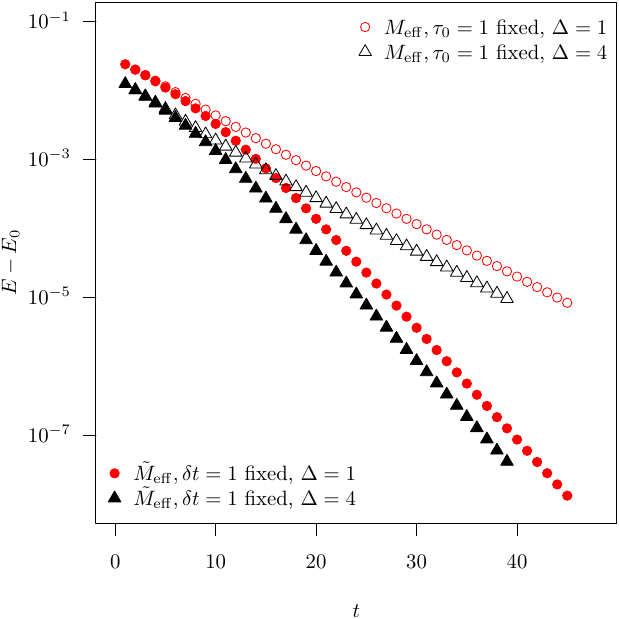}
  \caption{$E-E_0$ for $k=0$ on a logarithmic scale as a function of
    $t$ for different values of $\Delta$. Filled symbols correspond to
    $\tMeff$ with $\delta t=1$ fixed, open symbols to $\Meff$ for $\tau_0=1$ fixed.}
  \label{fig:synthetic5}
\end{figure}

We remark in passing that the not shown values for
$\Meff$ of \eq{eq:effmass} of 
the eigenvalue $\Lambda_k(t, \tau_0)$ at fixed $\tau_0$ are almost
indistinguishable on the scale of \fig{fig:synthetic} from $\tMeff$
with $\delta t$ fixed. For the tiny differences and the influence of
$\tau_0$ thereon see Figures~\ref{fig:synthetic4} and~\ref{fig:synthetic5}.

In \eq{eq:Eeffcorr} we have discussed
that we expect corrections in $\tMeff$ and $\Meff$ to decay
exponentially in $t = \delta t + \tau_0$. We can test this by
subtracting the exactly known energy $E_k$ from the PGEVM
results. Therefore, we plot in \fig{fig:synthetic4} effective masses
minus the exact $E_k$ values as a function of $t$. Filled symbols
correspond to $\tMeff$ with $\delta t=1$ and open symbols (only $k=0$)
to $\Meff$ with $\tau_0=10$. The asymptotically exponential
convergence in $t$ is nicely visible for both effective mass
definitions and also for $k=0$ and $k=1$. For $\tMeff$ the
decay rate is to a good approximation $E_2-E_0$ for $k=0$ and $E_2-E_1$ for
$k=1$, respectively, as expected from \eq{eq:Eeffcorr}. For $\Meff$
the asymptotic logarithmic decay rate is approximately $E_1-E_0$ and,
thus, worse as expected from \eq{eq:t0corr}.

So far we have worked solely with $\Delta=1$. In
\fig{fig:synthetic5} we investigate the dependence of $\tMeff$ and
$\Meff$ on $\Delta$: we plot $E-E_0$ on a logarithmic scale as a
function of $t$ for $\Delta=1$ and $\Delta=4$. While $\Delta$ has no
influence on the convergence rate, it reduces the amplitude of the
pollution for both $\tMeff$ and $\Meff$ by shifting the data points to the 
left.  The reason is that a larger $\Delta$ allows to reach larger times in the 
Hankel matrices at the same $t$. A smaller $\Delta$ on the other hand allows to 
go to larger $t$, thus the advantage of increased $\Delta$ is negligible.

\begin{figure}
  \centering
  \includegraphics[width=.4\textwidth,page=5]{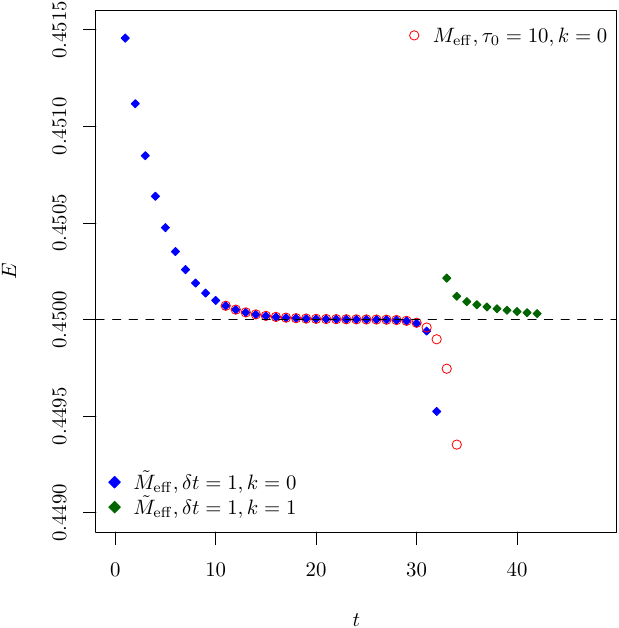}
  \caption{Effective energies as a function of $t$
    for synthetic data including a backpropagating state for the
    $k=0$ ground state obtained by the PGEVM with $\Delta=1$. Filled symbols
    correspond to $\tMeff$ with $\delta t=1$ fixed and open symbols to
    $\Meff$ with $\tau_0=10$ fixed.}
  \label{fig:synthetic2}
\end{figure}

In order to see the effect of so-called back-propagating states,
we next investigate a correlator 
\begin{equation}
C_s(t)\ =\ \sum_{k=0}^{2} c_k\, \left(e^{-E_k t}+ \delta_{k0}e^{-E_k
(T-t)}\right) 
\end{equation}
with a back-propagating contribution to the ground state
$E_0$ only (see also Ref.~\cite{Bailas:2018car}). Energies are chosen as
$E_i=(0.45,0.6,0.8)$ and the 
amplitudes are $c_i=(1, 0.1, 0.01)$ with $T=96$. The
result for the ground state effective energy determined from the PGEVM
principal correlator is shown in \fig{fig:synthetic2}. We show $\Meff$
from the principal correlator for $\tau_0=10$ fixed as open red symbols. The
filled symbols correspond to $\tMeff$ for $k=0$ and $k=1$ with $\delta
t=1$ fixed. Both is again for $\Delta=1$.

One observes a downward bending of the two $k=0$ effective masses
starting around $t=28$. The difference between $\tau_0$ fixed and
$\delta t$ fixed is only visible in the $t$-range where the bending
becomes significant. Obviously, in this region the contribution of the
forward and backward propagating states becomes comparable in size,
while the state with $k=2$ becomes negligible. Interestingly, for
$\delta t=1$ fixed the state of interest is then contained in the $k=1$
state while the $k=0$ states drop to the state with energy $-E_0$ (not
visible in the figure). The (not fully shown) state with $k=1$
decays from a value $0.65$ towards $0.6$ in the time range from 0 to about
30, after which it abruptly drops to a value of about $0.45$, which
is visible in the figure.

It becomes clear that there is an intermediate region in $t$, in
this case from $t=28$ to $t=38$, where the different contributions to
the correlator cannot be clearly distinguished by the PGEVM using
$\tMeff$. Around $t=28$ contributions by the $k=2$ state have become
negligible, while the backward propagating state becomes important. At
this point the state with $k=1$ becomes the pollution and the PGEVM
resolves forward and backward propagating states. This transition will
also be visible for the lattice QCD examples discussed next.

\subsection{Lattice QCD Examples}\label{sec:qcd_examples}

As a first lattice QCD example 
we start with the charged pion, which gives rise to
one of the cleanest signals in any correlation function extracted from
lattice QCD simulations. In particular, the signal to noise ratio is
independent of $t$. From now on quantities are given in units of the
lattice spacing $a$, i.e.\ $aE$, $aM$, $t/a$, \ldots are dimensionless
real numbers. However, for simplicity we set $a=1$.

The example we consider is the B55.32 ensemble generated with
$N_f=2+1+1$ dynamical quark flavours by ETMC~\cite{Baron:2010bv} at a
pion mass of about $350\ \mathrm{MeV}$. For details on the ensemble we
refer to Ref.~\cite{Baron:2010bv}. The correlation functions for the
pion have been computed with the so-called one-end-trick and spin
dilution, see Ref.~\cite{Boucaud:2008xu} on $4996$ gauge
configurations. The time extent is $T=64$ lattice points, the spatial
one $L=T/2$.

\begin{figure}[th]
  \centering
  \includegraphics[width=.4\textwidth,page=1]{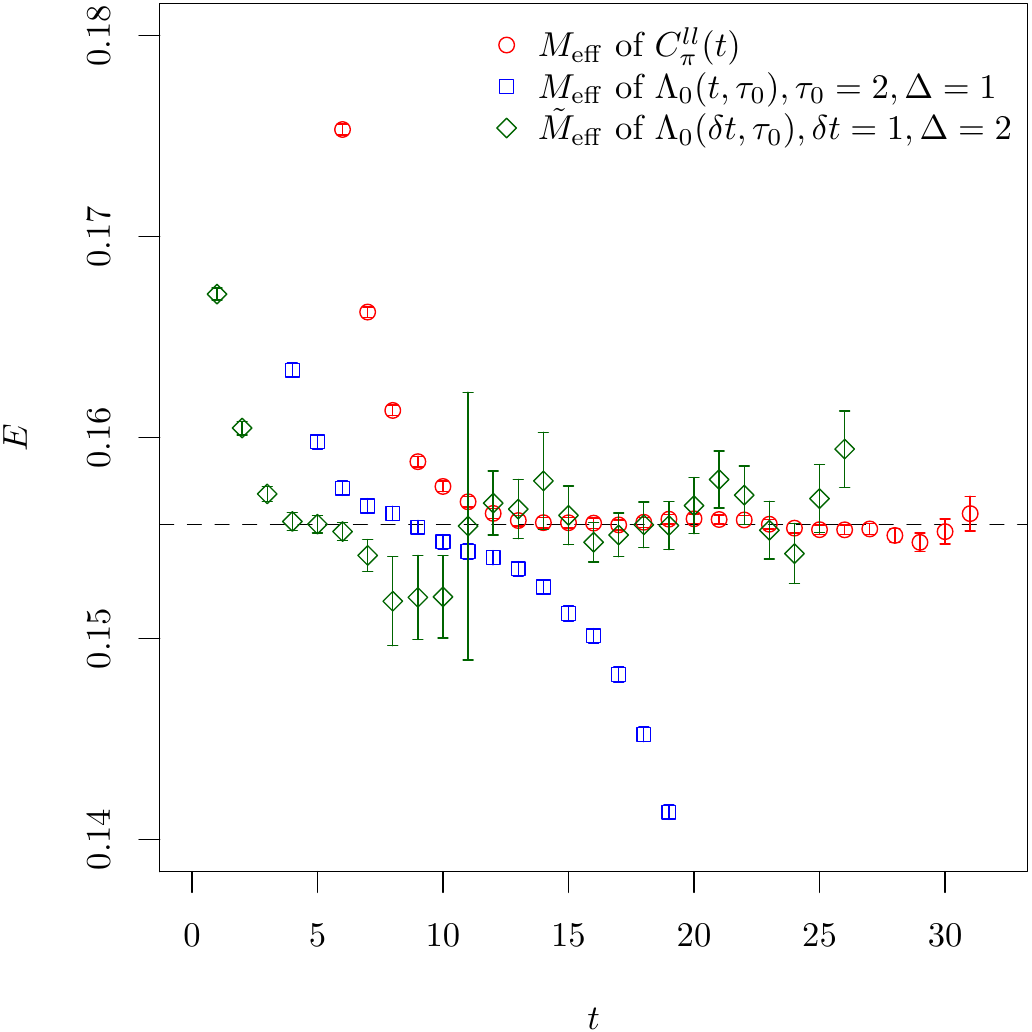}
  \caption{Effective energies $aE$ as a function of $t$
    computed from the local-local two-point 
    pion correlation function on ensemble B55.32. The red circles
    represent the $\cosh$ effective masses \eq{eq:cosheffmass} computed from the
    single twopoint correlator directly. The blue squares are the effective masses
    $\Meff$ computed from the PGEVM principal correlator with
    $\tau_0=2$ and $n=2$
    fixed. The green diamonds represent $\tMeff$
    computed from the PGEVM principal correlator with $\delta t=1$ and $n=2$
    fixed. The dashed line represents the mean value of a fit with a
    two parameter $\cosh$ model to the
    original correlator.}
  \label{fig:pion1}
\end{figure}

\subsubsection{Pion}

We look at the single pion two-point correlation function
$C_\pi^{ll}(t)$ computed with local sink and local source using the
standard operator $\bar u\, i\gamma_5 d$ projected to zero
momentum. Since the pion is relatively light, 
the backpropagating state due to periodic boundary conditions is
important. For this reason, we compute the cosh effective
mass from the ratio
\begin{equation}
  \label{eq:cosheffmass}
  \frac{C_\pi^{ll}(t+1)}{C_\pi^{ll}(t)} = \frac{e^{-E_\pi (t+1)} +
    e^{-E_\pi(T-(t+1))}}{e^{-E_\pi t} + e^{-E_\pi(T-t)}} 
\end{equation}
by solving numerically for $E_\pi$. The corresponding result is shown
as red circles in \fig{fig:pion1} as a function of $t$. The
effective masses $\Meff$ computed from the PGEVM principal
correlator with $\tau_0=2$, $n=2$ and $\Delta=1$ fixed are shown as
blue squares. One observes that excited states are reduced but the
pollution by the backward propagating state ruins the plateau. As
green diamonds we show the $\tMeff$ for the principal correlator with
$\delta t=1$, $n=2$ and $\Delta=2$ fixed. Here, we used a target
value $\xi=0.16$ -- chosen by eye -- to identify the
appropriate state during 
resampling, see section~\ref{sec:sorting}. The plateau
starts as early as $t=5$, there is an intermediate region where
forward and backward propagating states contribute similarly, and
there is a region for large $t$, where again the ground state is
identified. The apparent jump in the data at $t=11$ is related to
coupling to a different state than on previous timeslices and is
accompanied by a large error because the sorting of states is
performed for each bootstrap sample. Coupling to a different state is
allowed for the method with fixed $\delta t$ as the $\tau_0$ of the
GEVP changes for every timeslice. In fact, this feature is a key
difference to the methods with fixed $\tau_0$ for which the set of
states is unambiguously determined by the initial choice of $\tau_0$,
see the discussion in section~\ref{subsec:mindistsort}.

\begin{figure}[th]
  \centering
  \includegraphics[width=.4\textwidth,page=2]{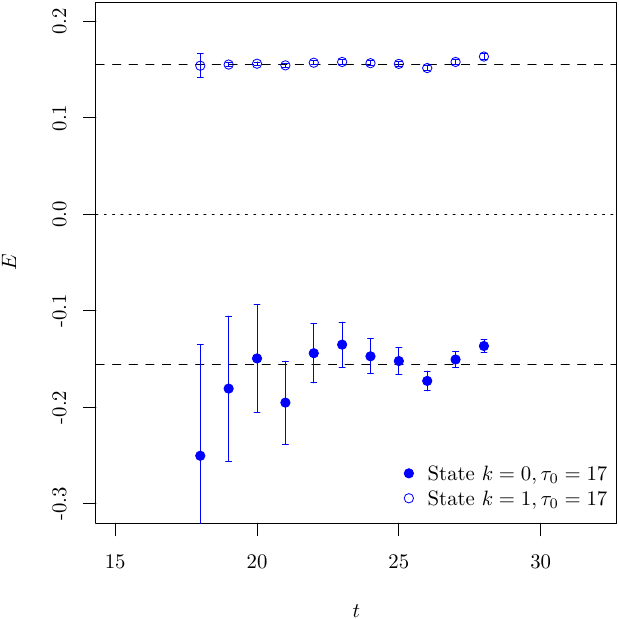}
  \caption{Effective energies $aE$ as a function of $t$
    computed from the local-local twopoint 
    pion correlation function on ensemble B55.32. We show effective
    masses $\Meff$ of the PGEVM principal correlators with $k=0$ and $k=1$ and
    $\tau_0=17$ and $n=2$ fixed.} 
  \label{fig:pion2}
\end{figure}

Once all the excited states have become negligible, the PGEVM can also
resolve both forward and backward propagating states (see also
Ref.~\cite{Schiel:2015kwa}). For the example 
at hand this is shown in \fig{fig:pion2} with $\tau_0=17$ and
$n=2$ fixed. For this to work it is important to chose $\tau_0$ large
enough, such that excited states have decayed
sufficiently. Interestingly, the noise is mainly projected into the
state with negative energy. 

\begin{figure}[th]
  \centering
  \includegraphics[width=.4\textwidth,page=3]{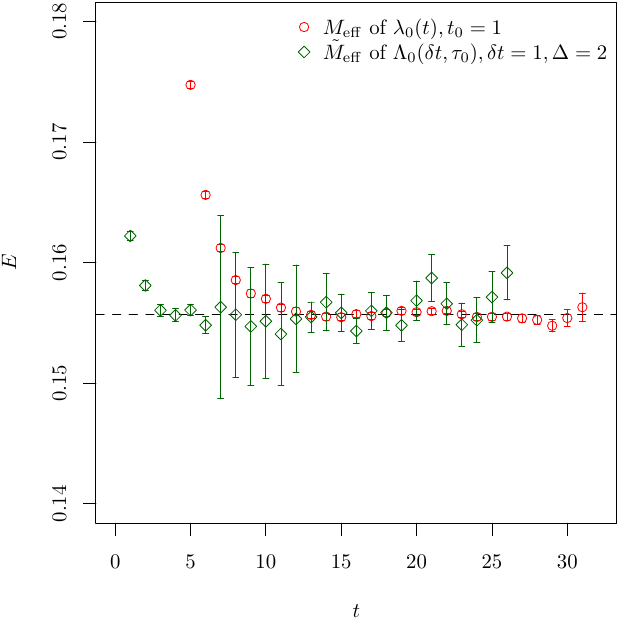}
  \caption{Like \fig{fig:pion1}, but starting with a GEVM
    principal correlator.} 
  \label{fig:pion3}
\end{figure}

In \fig{fig:pion3} we visualise the improvement realised by
combining GEVM with PGEVM. Starting with a $2\times 2$ correlator
matrix built from local and fuzzed operators, we determine the GEVM
principal correlator $\lambda_0(t)$ using $t_0=1$. The $\cosh$ effective
mass of $\lambda_0$ is shown as red circles in
\fig{fig:pion3}. In green we show $\tMeff$
of the PGEVM principal correlator $\Lambda_0$ obtained
with $\delta t= 1$, $n_1=2$ and $\Delta =2$ fixed.

\begin{figure}[th]
  \centering
  \includegraphics[width=.4\textwidth,page=4]{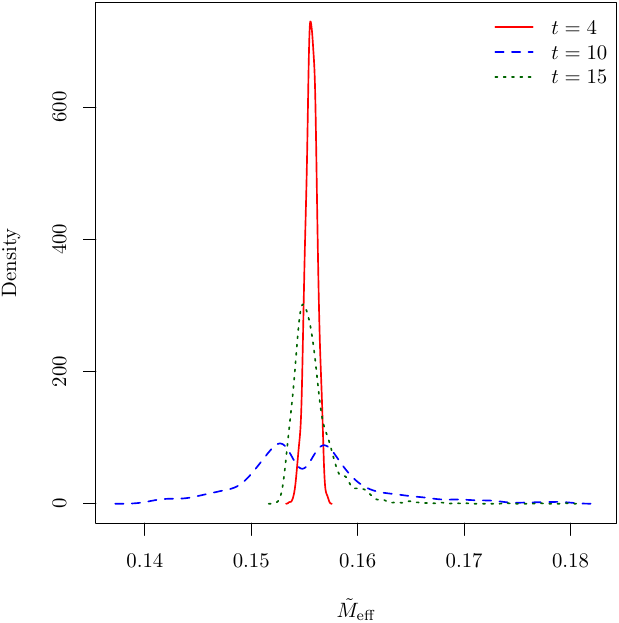}
  \caption{Density of bootstrap replicates for $\tMeff$ at different
    $t$-values for the data of \fig{fig:pion3}}
  \label{fig:pion4}
\end{figure}

Compared to \fig{fig:pion3}, the plateau in $\tMeff$
starts as early as $t=3$. However, in particular at larger
$t$-values the noise is also increased compared to the PGEVM directly
applied to the original correlator. It should be clear that the pion
is not the target system for an analysis combining GEVM and PGEVM,
because its energy levels can be extracted without much systematic
uncertainty directly from the original correlator. However, it serves
as a useful benchmark system, where one can also easily check for
correctness. 

In \fig{fig:pion4} we plot the (interpolated) bootstrap sample
densities of $\tMeff$ for three $t$-values: $t=4$, $t=10$ and
$t=15$. They correspond to the green diamonds in \fig{fig:pion3}. One
observes that at $t=4$ the distribution is approximately
Gaussian. At $t=15$ the situation is similar, just that the
distribution is a bit skew towards larger $\tMeff$-values. In the
intermediate region with $t=10$ there is a two peak structure visible,
which is responsible for the large error. It is explained -- see above
-- by the inability of the method with $\delta t=1$ to distinguish the
different exponentials contributing to $\lambda_0$. 

\begin{table}
  \centering
  \begin{tabular*}{1.\linewidth}{@{\extracolsep{\fill}}lrrrrr}
    \hline
    & $t_1$ & $t_2$ & $\Delta$ & $M_\pi$ & $\chi^2_\mathrm{red}$\\
    \hline
    $M_\mathrm{eff}$ of $C_\pi^{ll}$ &15 & 30 & - & $0.15567(12)$ & $1.17$\\
    $\tMeff$ of PGEVM & 4 & 20 & 1 & $0.15539(25)$ & $1.00$\\
    $\tMeff$ of GEVM/PGEVM & 3 & 20 & 2 & $0.15569(25)$ & $0.66$\\
    \hline
  \end{tabular*}
  \caption{Results for $M_\pi$ of fits to various pion effective
    energies, see red circles and green diamonds of \fig{fig:pion1}
    for the first two rows and green diamonds of \fig{fig:pion2} for
    the third row. The fit ranges are $[t_1, t_2]$.}
  \label{tab:pionres}
\end{table}

In \tab{tab:pionres} we have compiled fit results obtained for the
pion: the first row corresponds to a fit to the effective mass of the
correlator $C_\pi^{ll}$ in the fit range indicated by $t_1, t_2$,
which was chosen by eye. The
second row represents the fit to $\tMeff$ with $\delta t=1$
fixed obtained with PGEVM on $C_\pi^{ll}$ directly (green diamonds in
\fig{fig:pion1}). The last row is the same, but for the combination of
GEVM/PGEVM (green diamonds in \fig{fig:pion3}).
The agreement is very good, even though the PGEVM and GEVM/PGEVM
errors are larger than the ones obtained from the correlator directly.
In the last column we give the value of the correlated
$\chi^2_\mathrm{red}=\chi^2/\mathrm{dof}$, where one observes that
the fits are roughly comparable in terms of fit quality.

\subsubsection{$\eta$-meson}

\begin{figure}[th]
  \centering
  \includegraphics[width=.4\textwidth,page=1]{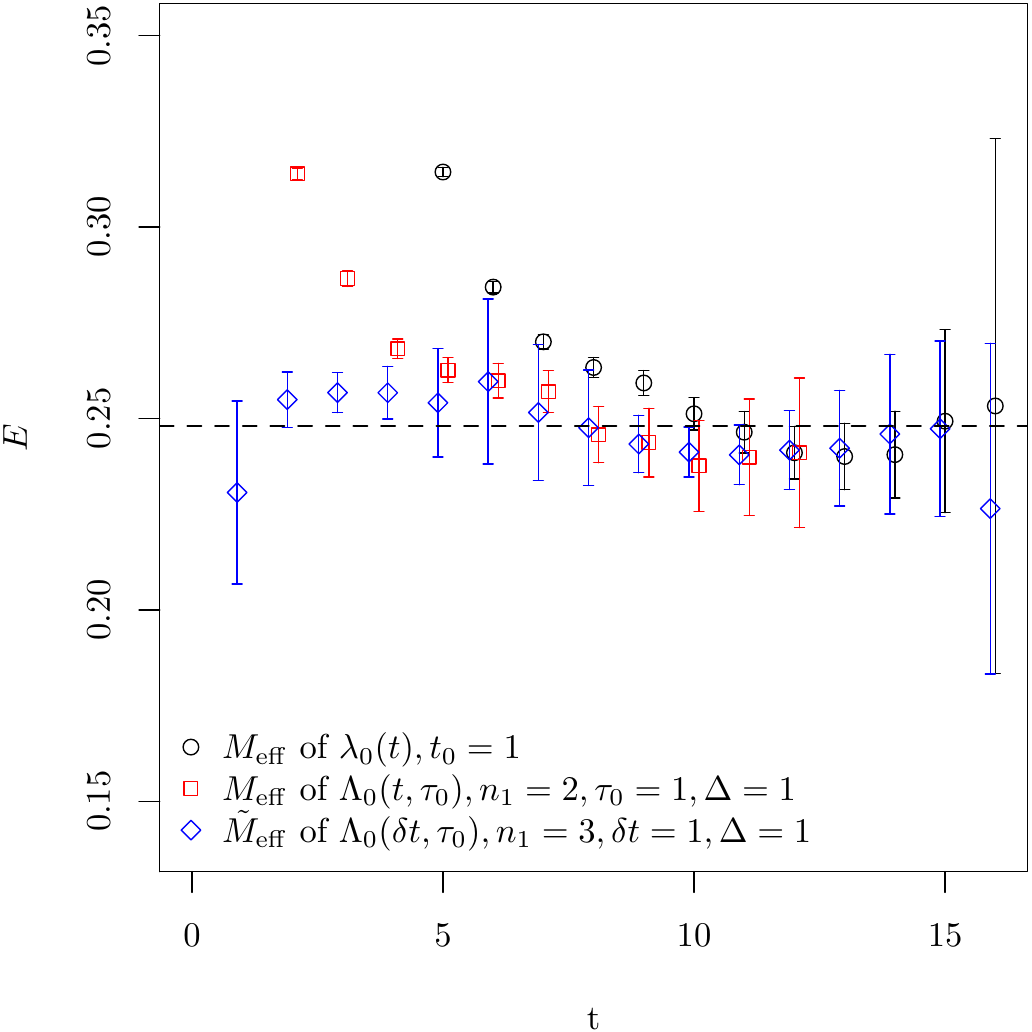}
  \caption{Effective energies for the $\eta$-meson as a function of $t$ for
  the GEVM principal correlator $\lambda_0$ and from
  the GEVM/PGEVM principal correlator $\Lambda_0$ with $n_1=2$ and
  $n_1=3$, respectively. The data is for the B55.32 ETMC ensemble. The
  dashed horizontal line represents the result quoted in
  Ref.~\cite{Ottnad:2017bjt}.} 
  \label{fig:eta}
\end{figure}

As a next example we study the $\eta/\eta^\prime$ system,
where due to mixing of flavour singlet and octet states the
GEVM cannot be avoided in the first place. In addition, due to
large contributions by fermionic disconnected diagrams the correlators
are noisy making the extraction of energy levels at late Euclidean
times difficult. The $\eta/\eta^\prime$ analysis on the B55.32
ensemble was first carried out
in Refs.~\cite{Ottnad:2012fv,Michael:2013gka,Ottnad:2017bjt} using a
powerful method to subtract excited states we can compare to. However,
this excited state subtraction method is based on some (well founded)
assumptions. 

The starting point is a $3\times 3$ correlator matrix $C_{ij}^\eta(t)$
with light, strange and charm flavour singlet operators and local
operators only. We apply the GEVM with $t_0=1$ and extract the first
principal correlator $\lambda_0(t)$ corresponding to the $\eta$-state,
which is then input to the PGEVM. 

In \fig{fig:eta} we show the effective mass of the $\eta$-meson for
this GEVM principal correlator $\lambda_0(t)$ as black circles. 
In addition we show as red squares the effective masses of $\Lambda_0$
obtained from the PGEVM applied to this principal correlator with
$n_1=2$, $\tau_0=1$ and $\Delta=1$. The blue diamonds represent $\tMeff$
of $\Lambda_0$ obtained with $n_1=3$, $\delta t=1$ and $\Delta =1$
fixed. The dashed horizontal line indicates the results obtained 
using excited state subtraction~\cite{Ottnad:2017bjt}. For better
legibility we show the effective masses for each of the three cases
only up to a certain $t_\mathrm{max}$ after which errors become too
large. Moreover, the two PGEVM results are slightly displaced horizontally.
Note that $\tMeff$ with $n_1=2$ is in between $\tMeff$
with $n_1=3$ and $\Meff$ with $n_1=2$. We did not attemt a
comparison here, but wanted to show the potential of the method.

One observes two things: excited state pollutions are
significantly reduced by the application of the PGEVM to
the GEVM principal correlator $\lambda_0$. However, also noise
increases. But, since in the effective masses of $\lambda_0$ there are
only $5$ points which can be interpreted as a plateau, the usage of
PGEVM can increase the confidence in the analysis. 

In the corresponding $\eta^\prime$ principal correlator the noise is
too large to be able to identify a plateau for any of the cases
studied for the $\eta$. 

\begin{table}
  \centering
  \begin{tabular*}{1.\linewidth}{@{\extracolsep{\fill}}lrrrrr}
    \hline
    & $t_1$ & $t_2$ & $\Delta$ & $M_\eta$ & $\chi^2_\mathrm{red}$\\
    \hline
    $M_\mathrm{eff}$ of $\lambda_0$ & 10 & 16 & - & $0.2467(29)$ & $0.40$\\
    $M_\mathrm{eff}$ of $\Lambda_0$, $t_0=1$ & 7 & 14 & 1 &
    $0.2425(38)$ & $0.10$\\
    $\tilde{M}_\mathrm{eff}$ of $\Lambda_0$, $\delta t=1$ & 1 & 15 & 1
    & $0.2504(36)$ & $0.29$\\
    \hline
    Ref.~\cite{Ottnad:2017bjt} & - & - & - & $0.2481(08)$\\
    \hline
  \end{tabular*}
  \caption{Results of fits to effective $\eta$ energies, see
    \fig{fig:eta}. The fitrange is given by $[t_1, t_2]$.}
  \label{tab:etares}
\end{table}

In \tab{tab:etares} we present fit results to the different
$\eta$ effective masses from \fig{fig:eta}. The agreement among the
different definitions, but also with the literature value is
reasonable within errors.

\subsubsection{$I=1, \pi-\pi$-scattering} 

\begin{figure*}[th]
  \centering
  \subfigure{\includegraphics[width=.4\textwidth,page=2]{rho}}\qquad
  \subfigure{\includegraphics[width=.4\textwidth,page=1]{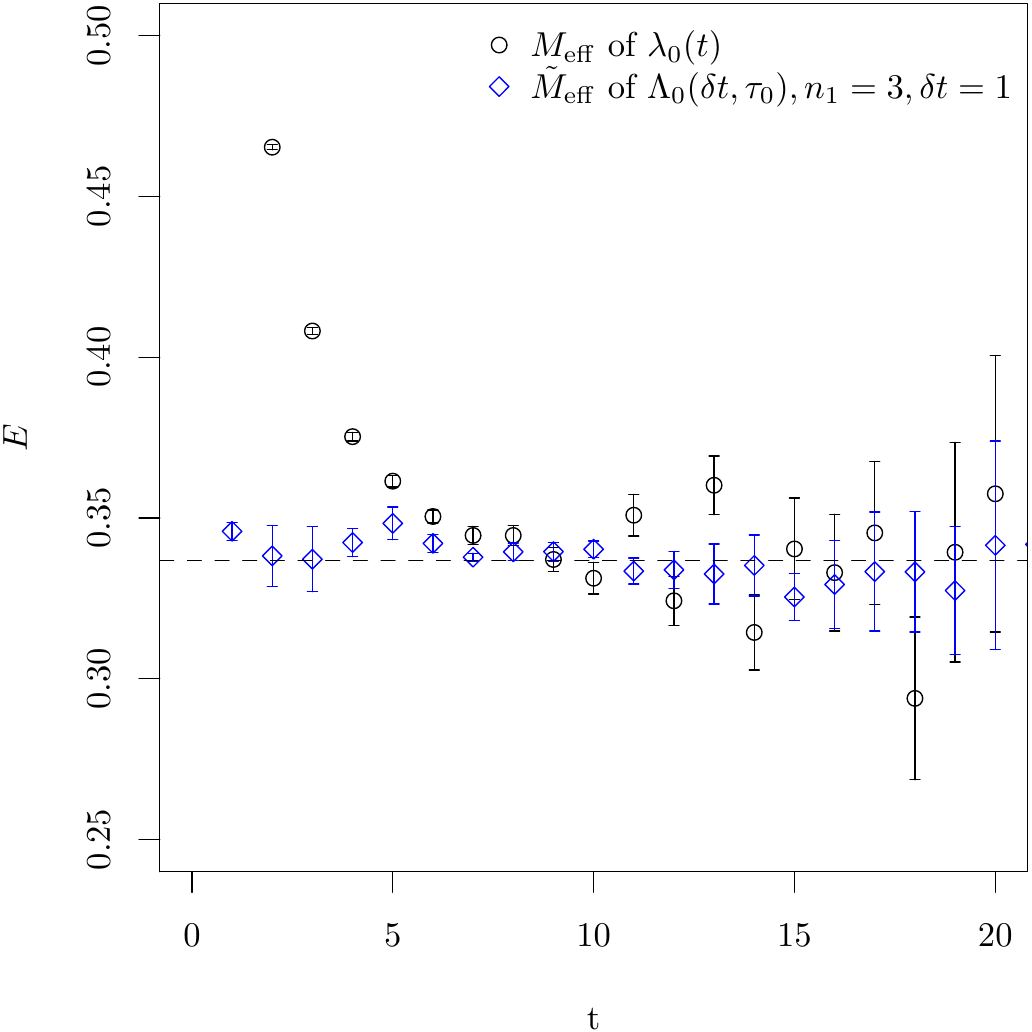}}
  \caption{Effective energies for $I=1, \pi-\pi$-scattering as a function of
    $t$. Left: $A_1$ irrep with total momentum square equal to $1$
    in lattice units. Right: $T_{1u}$ irrep with total zero
    momentum. In both cases the ground state energy level is
    shown. The dashed horizontal lines represent the fit results to
    $\Meff$ of $\lambda_0$, see Tables~\ref{tab:rhores2} and
    \ref{tab:rhores1}.} 
  \label{fig:rho}
\end{figure*}

Finally, we investigate correlator matrices for the $I=1,
\pi-\pi$-scattering. The 
corresponding correlator matrices were determined as part of a Lüscher
analysis including moving frames and all relevant lattice irreducible
representations (irreps). A detailed discussion of the framework and the theory can be
found in Ref.~\cite{Werner:2019hxc}. Here we use the
$N_f=2$ flavour ensemble cA2.30.48 generated by
ETMC~\cite{Abdel-Rehim:2015pwa,Liu:2016cba}, to which we apply
in Ref.~\cite{Fischer:2020fvl} the
same methodology as discussed in Ref.~\cite{Werner:2019hxc}.

The first example corresponds to the ground state in the $A_1$
irreducible representation with total squared momentum equal to $1$ in
units of $4\pi^2/L^2$, for which the results are shown in the left
panel of \fig{fig:rho}. In this case the effective mass computed from
the GEVM principal correlator $\lambda_0$ shows a reasonable plateau
(black circles). The red squares show $\Meff$ of
$\Lambda_0$ with $n_1=2$, $\tau_0=1$ and $\Delta=2$ fixed. Even though
the plateau starts at earlier times, noise is increasing
quickly. Actually, we no longer display the energies from $t>17$ due to
too large error bars for better legibility. When using $\tMeff$ with
$n_1=3$, $\delta t=1$ and $\Delta=1$, a plateau can be
identified from $t=1$ on and with a very reasonable signal to noise
ratio. 

\begin{table}
  \centering
  \begin{tabular*}{1.\linewidth}{@{\extracolsep{\fill}}lrrrr}
    \hline
    & $t_1$ & $t_2$ & $aW_\rho$ & $\chi^2_\mathrm{red}$\\
    \hline
    $M_\mathrm{eff}$ of $\lambda_0$ & 9 & 20 & $0.28411(26)$ & $0.89$\\
    $M_\mathrm{eff}$ of $\Lambda_0$ with $t_0=3$ & 3 & 15 &
    $0.28446(50)$ & $1.16$\\ 
    $\tilde{M}_\mathrm{eff}$ of $\Lambda_0$ with $\delta t=1$ & 1 & 20
    & $0.2838(10)$ & $0.95$\\ 
    \hline
  \end{tabular*}
  \caption{Results of fits to effective energy levels
    for $I=1, \pi-\pi$-scattering for the $A_1$ 
    irrep, see left panel of \fig{fig:rho}.}
  \label{tab:rhores2}
\end{table}

Fit results to the effective masses for the $A_1$ irrep are compiled
in \tab{tab:rhores2}. Here one notices that, despite the visually much
longer plateau range, the error on the fitted mass is significantly
larger for $\tMeff$ than for the other two methods. The overall
agreement is very good, though.

The same can be observed in the right panel of \fig{fig:rho} for the
$T_{1u}$ irrep. However,
this time it is not straightforward to identify a plateau in $\Meff$
of $\lambda_0$ shown as black circles. Using $\tMeff$ instead
with $n_1=3$, $\delta t=1$ and $\Delta = 1$ fixed improves
significantly over the traditional effective masses.

\begin{table}
  \centering
  \begin{tabular*}{1.\linewidth}{@{\extracolsep{\fill}}lrrrr}
    \hline
    & $t_1$ & $t_2$ & $aW_\rho$ & $\chi^2_\mathrm{red}$\\
    \hline
    $M_\mathrm{eff}$ of $\lambda_0$ & 9 & 20 & $0.33680(67)$  & $2.03$\\
    $\tilde{M}_\mathrm{eff}$ of $\Lambda_0$ with $\delta t=1$ & 2 & 20
    & $0.3377(16)$ & $0.86$\\
    \hline
  \end{tabular*}
  \caption{Results of fits to effective energy levels for $I=1,
    \pi-\pi$-scattering for the $T_{1u}$ 
    irrep, see right panel of \fig{fig:rho}.}
  \label{tab:rhores1}
\end{table}

Fit results for the $T_{1u}$ irrep are compiled in
\tab{tab:rhores1}. The conclusion is similar to the one from the $A_1$
irrep. 

\subsection{GPOF versus GEVM/PGEVM}
As discussed in section~\ref{sec:method}, the sequential application
of GEVM/PGEVM is equivalent to GPOF when applied to noise-free
data. For data with noise, however, we experienced an advantage of
GEVM/PGEVM over GPOF, which becomes larger with increasing matrix
size.

\begin{figure}[th]
  \centering
  \includegraphics[width=.4\textwidth,page=1]{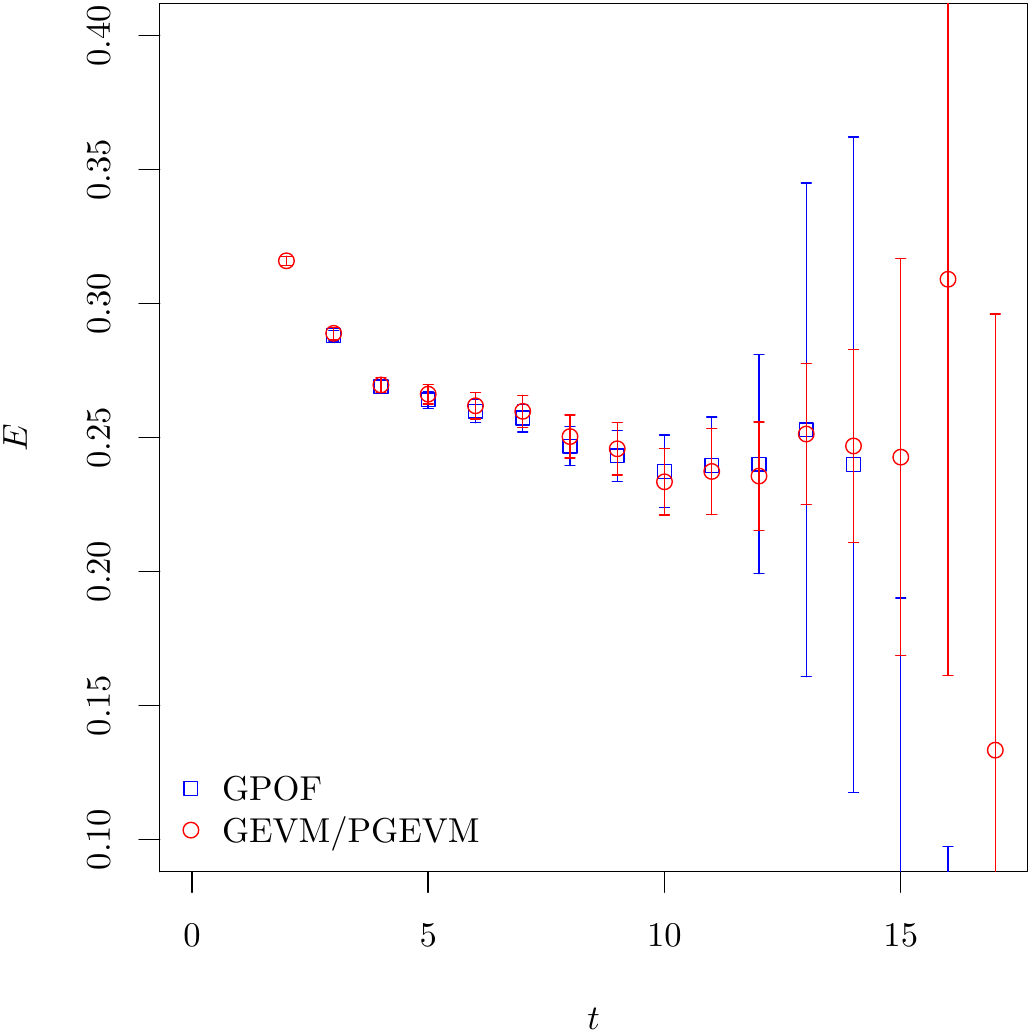}
  \caption{$\Meff$ of GPOF and GEVM/PGEVM compared for
      the $\eta$ state, see text for details.}
  \label{fig:GPOFvsPGEVM}
\end{figure}

While a systematic comparison is beyond the scope of this article,
we show in \fig{fig:GPOFvsPGEVM} a comparison of GPOF versus
GEVM/PGEVM for the case of the $\eta$-meson. We compare $\Meff$ for
GEVM/PGEVM with $n_0=3$ and $n_1=2$ and GPOF with $n'=6$. For small
$t$-values the agreement is very good. From $t=12$ onwards, however,
the errors of GPOF are significantly larger than the ones from
GEVM/PGEVM and from $t=15$ on the mean values start to differ
significantly.

When doubling the matrix size, the difference becomes more
pronounced: while the GEVM/PGEVM effective mass stays almost
unchanged compared to \fig{fig:GPOFvsPGEVM}, GPOF shows large statistical
uncertainties and fluctuating mean values from small $t$-values on.

\section{Discussion}

In this paper we have first discussed the relation among the
generalised eigenvalue, the Prony and the generalised pencil of
function methods: they are all special cases of a generalised
eigenvalue method. This fact allows one to discuss systematic effects
stemming from finite matrix sizes used to resolve the infinite tower
of states. The results previously derived for the generalised
eigenvalue method~\cite{Luscher:1990ck,Blossier:2009kd} can be
transferred and generalised to the other methods. In particular,
pollutions due to unresolved states decay exponentially in time.

At the beginning of the previous section we have demonstrated with
synthetic data that the PGEVM works as expected. In
particular, we could confirm that pollutions due to unresolved
excited states vanish exponentially in $t$. This exponential
convergence to the wanted state is faster if $\tMeff$ \eq{eq:Eeff} with $\delta t$ fixed is
used, as expected from the perturbative description. Increasing the
footprint of the Hankel matrix by increasing the parameter $\Delta$
helps in reducing the amplitude of the polluting terms. 

Still using synthetic data, we have shown that backward propagating
states affect PGEVM effective energies at large times. But, PGEVM
makes it also possible to distinguish forward from backward
propagating states.

As a first example for data with noise we have looked at the
pion. There are three important conclusions to be drawn here: first,
the PGEVM can also resolve forward and backward propagating states in
the presence of noise. Second, $\tMeff$ computed for fixed
$\delta t$ is advantageous compared to $\Meff$ at fixed
$t_0$, because in this case strong effects from the backward
propagating pion can be avoided.
And finally, combining GEVM and PGEVM sequentially leads to a
reduction of excited state contributions.

The next two QCD examples are for the $\eta$ meson and the $\rho$
meson where one must rely on the variational method. Moreover, the
signal to noise ratio decays exponentially such that excited state
reduction is imperative.

For the case of the $\eta$ meson the combined GEVM/PGEVM leads
to increased confidence in the
extracted energy levels. For the 
$I=1, \pi-\pi$-scattering a strong improvement is visible. The latter is likely due
to the large input correlator matrix to the GEVM. This leads to a
large gap relevant for the corrections due to excited states and,
therefore, to small excited states in the PGEVM principal correlator.

Interestingly, for the $\rho$-meson example studied here also the
signal to noise ratio in the PGEVM principal correlator at fixed
$\delta t$ is competitive if not favourable compared to the effective
mass of the GEVM principal correlator.

We have also introduced sorting by minimal distance, requiring an
input parameter value $\xi$. It is important to have in mind that the
choice of $\xi$ will influence the result and potentially introduce a
bias. Moreover, this sorting is, like sorting by value, susceptible to
misidentifications during resampling procedures.

Last but not least let us emphasise that the novel method presented here is not 
always advantageous and many other methods have been developed for the analysis 
of multi-exponential signals, each with their own strengths and weaknesses. We 
are especially referring to the recent developments of techniques based on the 
use of ordinary differential equations~\cite{ODE_methods} and the Gardner 
method~\cite{gardner_original}, for the latter see
appendix~\ref{sec:gardner}. Both methods are in principle capable of
extracting the full energy spectrum. However, the Gardner method 
becomes unreliable in the case of insufficient data and precision,
while we have not tested the ODE method here. But the results in
Ref.~\cite{ODE_methods} look promising.

\section{Summary}

In this paper we have clarified the relation among different methods
for the extraction of energy levels in lattice QCD available in the
literature. We have proposed and tested a new combination of
generalised eigenvalue and Prony method (GEVM/PGEVM), which helps to reduce excited state contaminations.

We have first discussed the systematic effects in the Prony GEVM stemming
from states not resolved by the method. They decay exponentially fast
in time with $\exp(-\Delta E_{n, l} t_0)$ with $\Delta E_{n, l}
=E_{n} -E_l$ the difference of the first not resolved energy level
$E_{n}$ and the level of interest $E_l$. Using synthetic data we
have shown that this is indeed the leading correction.

Next we have applied the method to a pion system and discussed its
ability to also determine backward propagating states, given high
enough statistical accuracy, see also
Ref.~\cite{Schiel:2015kwa}. Together with the results from the 
synthetic data we could also conclude that working at fixed
$\delta t$ is clearly advantageous compared to working at fixed $t_0$,
at least for data with little noise.

Finally, looking at lattice QCD examples for the $\eta$-meson and the
$\rho$-meson, we find that excited state contaminations can be reduced
significantly by using the combined GEVM/PGEVM. While it is not
clear whether also the statistical precision can be improved,
GEVM/PGEVM can significantly improve the confidence in the extraction
of energy levels, because plateaus start early enough in Euclidean
time. This is very much in line with the findings for the Prony method
in the version applied by the NPLQCD
collaboration~\cite{Beane:2009kya}.

The GEVM/PGEVM works particularly well, if in the first step
the GEVM removes as many intermediate states as possible and,
thus, the gap $\Delta E_{n, l}$ becomes as large as possible in the
PGEVM with moderately small $n$. The latter is important to
avoid numerical instabilities in the PGEVM.

~\\%
\section*{Acknowledgements}
The authors gratefully acknowledge the Gauss Centre for Supercomputing
e.V. (www.gauss-centre.eu) for funding this project by providing
computing time on the GCS Supercomputer JUQUEEN~\cite{juqueen} and the
John von Neumann Institute for Computing (NIC) for computing time
provided on the supercomputers JURECA~\cite{jureca} and JU\-WELS~\cite{juwels} at Jülich
Supercomputing Centre (JSC).
This project was funded in part by the DFG as a project in the
Sino-German CRC110.
The open source software packages tmLQCD~\cite{Jansen:2009xp,Abdel-Rehim:2013wba,Deuzeman:2013xaa}, 
Lemon~\cite{Deuzeman:2011wz}, 
QUDA~\cite{Clark:2009wm,Babich:2011np,Clark:2016rdz} and R~\cite{R:2019} have 
been used.

\begin{subappendices}
\section{The Gardner method}
\label{sec:gardner}

	The Gardner method is a tool for the analysis of multicomponent exponential 
	decays. It completely avoids fits and uses Fourier transformations instead. 
	This global approach makes it extremely powerful, but also unstable. In 
	this section we discuss why we do not find the \gm\ applicable to 
	correlator analysis of lattice theories.

	\subsection{The algorithm}
	The most general form of a multicomponent exponential decaying function 
	$f(t)$ is
	\begin{align}
	f(t) &= \int_0^\infty g(\lambda)\eto{-\lambda 
		t}\,\md\lambda\label{eq:general_mult_exp_fun}
	\end{align}
	with some integrable function $g(\lambda)$ and $t$ bound from below, WLOG 
	$t\ge 0$. In the common discrete case we get
	\begin{align}
	g(\lambda) &= \sum_{i=0}^{\infty}A_i\delta(\lambda-E_i)
	\end{align}
	where the $A_i\in \mathbb R$ are the amplitudes, the $E_i$ are the decay 
	constants, often identified with energy levels, and $\delta$ denotes the 
	Dirac-Delta distribution. Gardner et al.~\cite{gardner_original} proposed 
	to multiply equation~\eqref{eq:general_mult_exp_fun} by $t=\exp(x)$ and 
	substitute $\lambda=\exp(-y)$ in order to obtain the convolution
	\begin{align}
	\eto{x}f\left(\eto{x}\right) &= 
		\int_{-\infty}^{\infty}g\left(\eto{-y}\right)\exp\left(-\eto{x-y}\right)\eto{x-y}\,\md 
		y\,.
	\end{align}
	This equation can now easily be solved for $g(\lambda)$ using Fourier 
	transformations. We define
	\begin{align}
	F(\mu) &\coloneqq 
		\frac{1}{\sqrt{2\pi}}\int_{-\infty}^{\infty}\eto{x}f\left(\eto{x}\right)\eto{\im 
		\mu x}\,\md x\,,\label{eq:ft_data_times_t}\\
	K(\mu) &\coloneqq 
		\frac{1}{\sqrt{2\pi}}\int_{-\infty}^{\infty}\exp\left(-\eto{x}\right)\eto{x}\eto{\im 
		\mu x}\,\md x\label{eq:ft_exp_e_e}\\
	&= \frac{1}{\sqrt{2\pi}}\Gamma(1+\im \mu)
	\end{align}
	and obtain
	\begin{align}
	g(\eto{-y}) &= 
		\frac{1}{2\pi}\int_{-\infty}^\infty\frac{F(\mu)}{K(\mu)}\eto{-\im 
		y\mu}\,\md \mu\,.\label{eq:ft_back_f_by_k}
	\end{align}
	The Fourier transformation in equation~\eqref{eq:ft_exp_e_e} has been 
	solved analytically, yielding the complex Gamma function $\Gamma$.

	The peaks of $g(\eto{-y})$ indicate the values of the $E_i$ by their 
	positions and the normalised amplitudes $A_i E_i$ by their heights. The 
	normalisation is due to the substitution 
	$g(\lambda)\mapsto\eto{-y}g(\eto{-y})$.

	\subsection{Numerical Precision}
	The Fourier integrals~\eqref{eq:ft_data_times_t} 
	and~\eqref{eq:ft_back_f_by_k} have to be solved numerically. We used the 
	extremely efficient algorithms \textit{double exponential 
	formulas}~\cite{dbl_exp_trafo} for low frequencies $\le 2\pi$ and 
	\textit{double exponential transformation for Fourier-type 
	integrals}~\cite{dbl_exp_osci} for high frequencies $\ge 2\pi$.

	These techniques allow to achieve machine precision of floating point 
	double precision arithmetics with $\lesssim 100$ function evaluations. This 
	however can only work as long as the result of the integral has the same 
	order of magnitude as the maximum of the integrated function. It turns out that 
	this is not the case for the given integrals. $F(\mu)$ decays exponentially 
	in $\ordnung{\exp(-\frac{\pi}{2}|\mu|)}$ (at the same rate as $K(\mu)$) if 
	$f(t)$ follows equation~\eqref{eq:general_mult_exp_fun}. Thus, as $|\mu|$ 
	grows, the sum of values $\eto{x}f\left(\eto{x}\right)\in\ordnung{1}$ 
	approaches zero more and more, loosing significant digits. To avoid this 
	effect one would have to employ higher precision arithmetic.

	With double precision arithmetics the values of $F(\mu)$ become completely 
	unreliable in the region $|\mu|\gtrsim 20$ where $F(\mu)$ approaches 
	machine precision. In practice we find that only $F(|\mu| \lesssim 10)$ is 
	precise enough to be trusted.
	\begin{figure}[htb]
		\centering
		\includegraphics[width=0.4\textwidth, page=2]{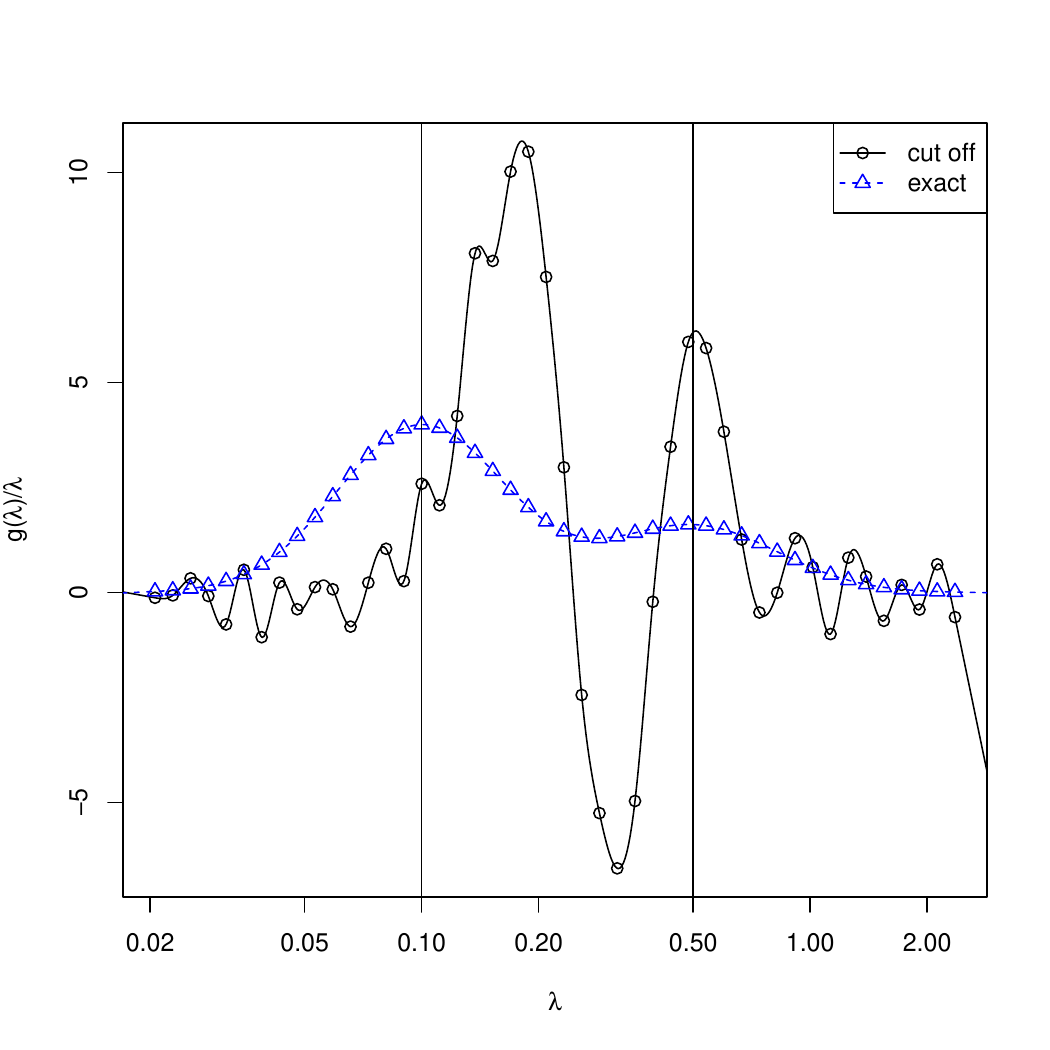}
		\caption{\gm\ applied to $f(t)=\eto{-\num{0.1}t}+2\eto{-\num{0.5}t}$. 
		Lines are cubic splines between the calculated points and guides to the 
		eye only. The black circles are obtained via a cubic spline to the 
		discrete set $\left\{\log(f(t))\,|\;t\in\{0,\dots,20\}\right\}$ with 
		$f(t>20)=0$. The blue triangles are obtained using the exact functional 
		form.}
		\label{fig:gardner_trafo_cut}
	\end{figure}

	\subsection{Limited data}
	In the case relevant for this work the data is limited to a noisy time 
	series $f(t)+\nu(t)$, $t\in\{0,\dots,n\}$, where $\nu(t)$ is an error. Thus 
	we have to deal with three difficulties, namely a discrete set, a finite 
	range and noise. Additional problems are the aforementioned limitation in 
	precision for high frequencies and possible small gaps between decay 
	constants $E_i$ that cannot be resolved. Ref.~\cite{gardner_review} 
	summarises a large number of improvements to the \gm\ and we are going to 
	mention the relevant ones explicitly below.

	\paragraph{Limited precision of $F(\mu)$}
	at high frequencies leads to a divergence of $\frac{F(\mu)}{K(\mu)}$ and 
	thus to a divergent integral in equation~\eqref{eq:ft_back_f_by_k}. If one 
	does not have or want to spend the resources for arbitrary precision 
	arithmetics, one is therefore forced to dampen the integrand 
	in~\eqref{eq:ft_back_f_by_k}. Gardner et al.~\cite{gardner_original} 
	originally proposed to simply introduce a cut off to the integral. It turns 
	out that this cut off leads to sinc-like oscillations of $g(\eto{-y})$, 
	i.e.\@ a high number of slowly decaying spurious peaks. These oscillations 
	can be removed by introducing a convergence factor of the form 
	$\exp(-\frac{\mu^2}{2w^2})$ instead of the cut off~\cite{gardner_gauss}. 
	The effective convolution of the exact result $g(\eto{-y})$ with a Gaussian 
	only smoothes $g(\eto{-y})$ but does not introduce oscillations. We chose 
	$w=2$ for our test runs. This choice does not always yield optimal results, 
	but it is very stable.

        \begin{figure}[htb]
	  \centering
	  \includegraphics[width=0.4\textwidth, page=4]{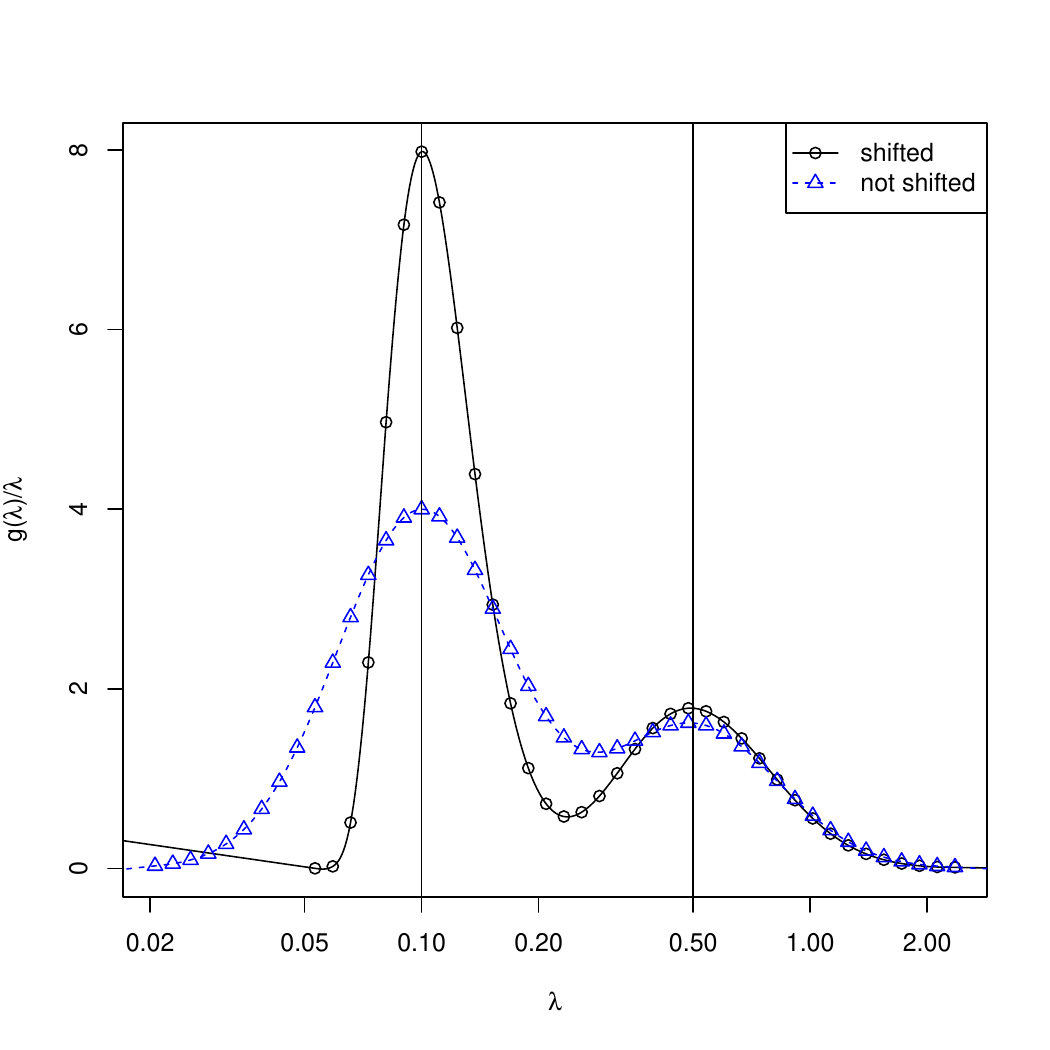}
	  \caption{\gm\ applied to $f(t)=\eto{-\num{0.1}t}+2\eto{-\num{0.5}t}$. 
	    Lines are cubic splines between the calculated points and guides to the 
	    eye only. The black circles are obtained via a cubic spline to the 
	    discrete set 
	    $\left\{\log(f(t)\eto{\num{.05}t})\,|\;t\in\{0,\dots,20\}\right\}$ with 
	    linear extrapolation. The blue triangles are obtained using the exact 
	    functional form.}
	  \label{fig:gardner_trafo_shift}
          
	\end{figure}

	\paragraph{Discrete data}
	is probably easiest to compensate. The exponential of a cubic spline of 
	$\log(f(t))$ yields a very precise interpolation of the data. Typically for 
	test functions the relative error is less than $10^{-4}$. Usually this is 
	far below noise level.

        \begin{figure}[htb]
	  \centering
	  \includegraphics[width=0.4\textwidth, page=3]{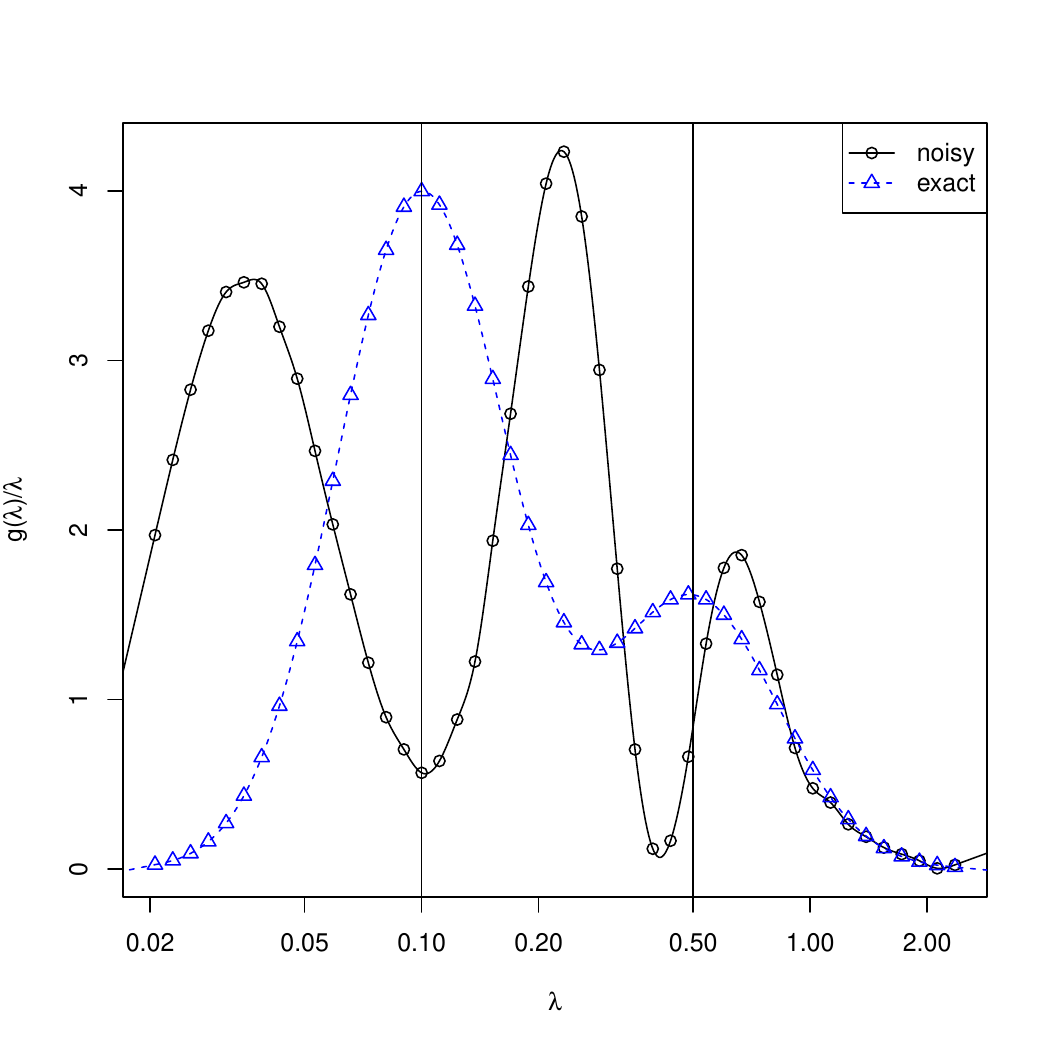}
	  \caption{\gm\ applied to $f(t)=\eto{-\num{0.1}t}+2\eto{-\num{0.5}t}$. 
	    Lines are cubic splines between the calculated points and guides to the 
	    eye only. The black circles are obtained via a cubic spline to the 
	    discrete set 
	    $\left\{\log(f(t))\,|\;t\in\{0,\dots,19\}\right\}\cup\left\{\log\left(\frac{f(19)+f(20)}{2}\right)\right\}$ 
	    with linear extrapolation. The blue triangles are obtained using the 
	    exact functional form.}
	  \label{fig:gardner_trafo_noise}
	\end{figure}

	\paragraph{Finite time range}
	is a much more severe problem. The exponential tail of $f(t)$ for 
	$t\rightarrow\infty$ carries a lot of information, especially about the 
	lowest decay modes. Thus extrapolation of the data essentially fixes the 
	ground state energy which we are usually most interested in. An 
	extrapolation of some kind is necessary, as a cut off completely obscures 
	the result (see \fig{fig:gardner_trafo_cut}). For a proper 
	extrapolation one would need to know at least the smallest $E_i$ in 
	advance, removing the necessity to apply the \gm\ in the first place. In 
	our test runs we used a linear extrapolation of the splines to the 
	log-data.

	Provencher~\cite{gardner_damped} proposes to multiply the complete time 
	series by a damping term of the form $t^\alpha\eto{-\beta t}$ with 
	$\alpha,\beta>0$ instead of $t$. This leads to a suppression of the region 
	beyond the data range, but it also moves the peaks of $g(\eto{-y})$ closer 
	together, thus decreasing the resolution. Still, Pro\-ven\-cher does not remove 
	the necessity of an extrapolation completely. In addition the method 
	introduces two parameters that have to be tuned.

	Let us remark here that, given a reliable extrapolation or very long 
	measurement, the inverse of Provencher's method can be used to improve 
	resolution: choose $\beta$ with $\min(E_i) < \beta < 0$ and so separate the lowest lying 
	peak from the others. We show the advantage of such a shift of the decay 
	constants in \fig{fig:gardner_trafo_shift}.

	\paragraph{Noisy data}
	is not a significant problem by itself, as long as the magnitude is known. 
	Fluctuations can be captured by the bootstrap or other error propagating 
	methods. Severe problems arise if noise is combined with the aforementioned 
	finite range. Then extrapolations based on the last few points (e.g.\@ with 
	the spline method) become very unreliable. We show this effect in 
	\fig{fig:gardner_trafo_noise} where we slightly increased the value 
	of the very last data point.

	\subsection{Applicability in practice}
\begin{figure}[htb]
		\centering
		\includegraphics[width=0.4\textwidth, page=5]{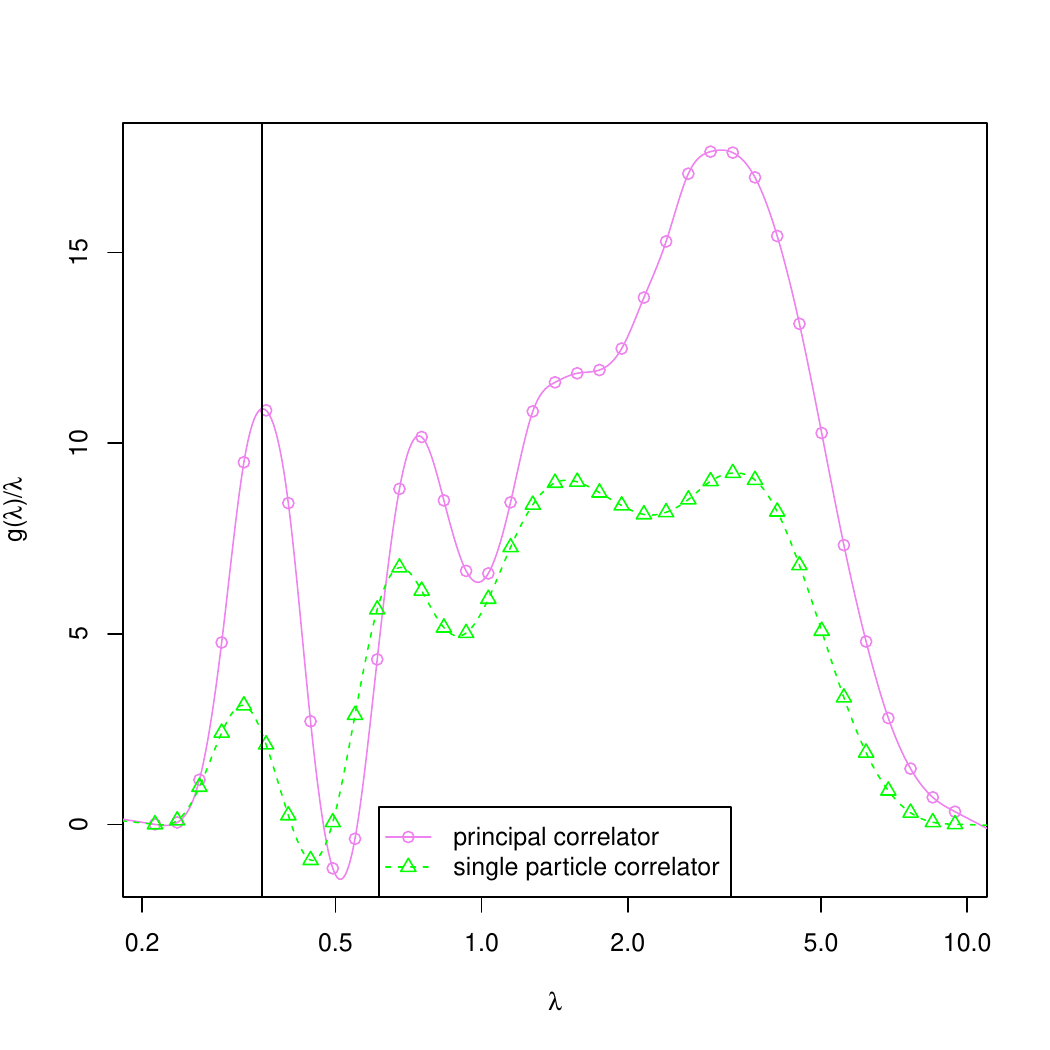}
		\caption{\gm\ with cubic spline inter- and extrapolation and 
		$\beta=\num{-.2}$ applied to the principal correlator obtained from a 
		GEVP and the single particle correlator of a pion. Lines are cubic 
		splines between the calculated points and guides to the eye only. The 
		vertical line at $\lambda=\num{.353}$ shows the ground state obtained 
		from the principal correlator via to state $\chi^2$-fit.}
		\label{fig:gardner_trafo_data}
	\end{figure}

	We applied the method to some data obtained from lattice QCD simulations. 
	With some fine tuning of $\beta$ and a sensible truncation of the data (we 
	removed points below noise level and regions not falling monotonously) one 
	can obtain very good results. Note especially the high resolution of the 
	ground state in \fig{fig:gardner_trafo_data}, but the relevant 
	exited states can be resolved as well.

	Nevertheless we have to conclude that the \gm\ is not broadly applicable to 
	real data commonly obtained from lattice simulations. One reason is that it 
	requires fine tuning of several parameters to obtain good results. The main 
	problem however is the absence of a reliable extrapolation of noisy data 
	from the limited time range. The algorithm does not fail gracefully, i.e.\@ 
	there is no obvious check whether the result for $g(\eto{-y})$ is correct or 
	not. Thus even though the \gm\ can yield very precise results, one cannot 
	automatise it and rely on the correctness of the output.

	As a last remark we would like to add that the \gm\ is also orders of 
	magnitude costlier in terms of computing resources than simpler methods 
	like $\chi^2$-fits.
 \end{subappendices}\clearpage{}%
\clearpage{}%
\chapter[Semimetal-Mott Insulator Quantum Phase Transition]{The Semimetal-Mott Insulator Quantum Phase Transition of the Hubbard Model on the Honeycomb Lattice}\label{chap:phase_transition}

\begin{center}
	\textit{Based on~\cite{semimetalmott} by J.~Ostmeyer, E.~Berkowitz, S.~Krieg, T.~A.~L\"{a}hde, T.~Luu and C.~Urbach}
\end{center}

\newcommand{\critU}{\ensuremath{\num{3.834(14)}}}
\newcommand{\critMu}{\ensuremath{\num{0.844(31)}}}
\newcommand{\critNu}{\ensuremath{\num{1.185(43)}}} %
\newcommand{\critZeta}{\ensuremath{\num{0.924(17)}}}
\newcommand{\critExp}{\ensuremath{\num{1.095(37)}}}

\newcommand{\critUf}{\ensuremath{\num{3.877(46)}}}
\newcommand{\critUg}{\ensuremath{\num{3.876(79)}}}

\newcommand{\DeltaInf}{\ensuremath{\num{0.057(11)}}}
\newcommand{\coeffU}{\ensuremath{\num{0.479(93)}}}

Now that the simulation and analysis techniques are understood, we can investigate the central topic of this thesis, namely the Hubbard model on the honeycomb lattice and its quantum phase transition.
In this chapter we focus on the electrical properties of the Hubbard model. They are determined via the single particle gap, which is a measure for conductivity. It has been determined using plateau fits, though the complete machinery developed in \Cref{chap:prony} could not be utilised for two reasons. As we are only interested in the ground state here,  we calculated only one correlator deeming a GEVP obsolete. We also encounter a very significant back propagating part and therefore have to use a $\cosh$-ansatz instead of a purely exponential one. Nevertheless the thorough analysis of plateau fits culminating in~\cite{generalisedProny} (\Cref{chap:prony}) had many useful by-products. One of them (not included in the original publication~\cite{semimetalmott}) is a possible bias introduced through direct plateau fits as described in \cref{sec:cosh_bias}.

In this chapter we determine the critical coupling $U_c$ and the critical exponent $\nu$. Our preliminary results for the critical exponent $\beta$ on the other hand have been found inaccurate and corrected in \Cref{chap:afm_character}. \Cref{tab:crit_values} summarising all the results has been updated accordingly.

\section{Introduction 
\label{sec:intro}}

Monte Carlo (MC) simulations of strongly correlated electrons in carbon nano-materials~\cite{GeimNovoselovReview,CastroNeto2009,Kotov2012} is an 
emerging topic in both the condensed matter~\cite{Khveshchenko2004323,heisenberg_gross_neveu,PhysRevB.86.115447} and
nuclear physics communities~\cite{Drut:2008rg,Hands:2008id}. The basis of such studies is the Hubbard model, 
a Hamiltonian approach which reduces, at weak electron-electron coupling, to the tight-binding description of atomic orbitals in a lattice 
of carbon ions~\cite{Saito1998,Wehling2011,Tang2015}. The properties of the Hubbard model on a honeycomb lattice are thought to resemble those of graphene.
MC simulations of the Hubbard model are closely related to 
problems of current interest in atomic and nuclear physics, such as the unitary Fermi gas~\cite{Chen:2003vy,Bulgac:2005pj,Bloch:2008zzb,Drut:2010yn}
and nuclear lattice effective field theory~\cite{Borasoy:2006qn,Lee:2008fa,Lahde:2013uqa,Lahde:2019npb}.

Our objective is to take advantage of this recent algorithmic development and perform, for the first time, a precision calculation of the single-particle
gap $\Delta$ of the hexagonal Hubbard model, in the grand canonical ensemble. Whether such a gap exists or not, is determined by
the relative strength of the on-site electron-electron coupling $U$, and the nearest-neighbor hopping amplitude $\kappa$.
Prior MC work in the canonical ensemble has established the existence of a second-order transition into a gapped, anti-ferromagnetic, Mott insulating (AFMI) phase
at a critical coupling of $U_c/\kappa \simeq 3.8$~\cite{Meng2010,Assaad:2013xua,Wang:2014cbw,Otsuka:2015iba}. The existence of an intermediate spin-liquid (SL)
phase~\cite{Meng2010} at couplings slightly below $U_c$, appears to now be disfavored~\cite{Otsuka:2015iba}.
Long-range interactions in graphene~\cite{Son:2007ja,Smith:2014tha} are thought to frustrate the AFMI transition, as these
favor charge-density wave (CDW) symmetry breaking~\cite{Buividovich:2016tgo,Buividovich:2018yar} over AFMI. The critical exponents of the AFMI transition should 
fall into the $SU(2)$ Gross-Neveu (GN), or chiral Heisenberg, universality class~\cite{gross_neveu_orig,heisenberg_gross_neveu}.

The observability of the AFMI transition in graphene is of interest for fundamental as well as applied physics.
While $\kappa$ is well constrained from density functional theory (DFT) and experiment~\cite{CastroNeto2009}, the on-site coupling $U$ is more difficult to 
determine theoretically for graphene~\cite{Tang2015}, although the physical value of $U/\kappa$ is commonly believed to be insufficient to trigger the AFMI 
phase in samples of suspended graphene or with application of biaxial strain~\cite{Wehling2011,Tang2015}. The AFMI phase may be more easily observed in the presence
of external magnetic fields~\cite{Gorbar:2002iw,Herbut:2008ui}, and the Fermi velocity at the Dirac point may still be strongly renormalized due to interaction effects~\cite{Tang2015}.
The reduced dimensionality of fullerenes and carbon nanotubes may increase the importance of electron-electron interactions in such systems~\cite{Luu:2015gpl}.

Let us summarize the layout of our paper. Our lattice fermion operator and Hybrid Monte Carlo (HMC) algorithm is introduced in \Secref{formalism}.
We describe in \Secref{results} how correlation functions and $\Delta$ are computed from MC simulations in the grand canonical ensemble. We also give details on our
extrapolation in the temporal lattice spacing (or Trotter error) $\delta$ and system (lattice) size $L$, and provide results for $\Delta$ as a function of $U/\kappa$ and inverse temperature $\beta$.
In \Secref{analysis}, we analyze these results using finite-size scaling in $\beta$, and provide our best estimate $U_c/\kappa=\critU$ for the critical coupling at which $\Delta$ becomes
non-zero. We also provide a preliminary estimate of the critical exponent of the AFMI order parameter, under the assumption that
the opening of the gap coincides with the AFMI transition. In \Secref{conclusion}, we compare our results with other studies of the Hubbard model and the chiral Heisenberg universality class, 
and discuss possible extensions of our work to carbon nanotubes, fullerenes and topological insulators.

\section{Formalism 
\label{sec:formalism}}

The Hubbard model is a theory of interacting fermions, that can hop between nearest-neighbor sites. We focus on the two-dimensional honeycomb lattice, 
which is bipartite in terms of $A$~sites and $B$~sites. The Hamiltonian is given by
\begin{equation}
H := - \sum_{xy} 
(a\adjoint_x h_{xy}^{} a_y^{} + b\adjoint_x h_{xy}^{} b_y^{})
+ \frac{1}{2} \sum_{xy} \rho_x^{} V_{xy}^{} \rho_y^{},
\label{Ham_def}
\end{equation}
where we have applied a particle-hole transformation to a theory of spin \up\ and spin \down\ electrons. Here, 
$a^\dag$ and $a$ are creation and annihilation operators for particles (spin-up electrons), and $b^\dag$ and $b$ are similarly
for holes (spin-down electrons). As usual, the signs of the $b$ operators have been switched for $B$~sites.
The matrix $h_{xy} := \kappa \delta_{\langle x,y \rangle}$ describes nearest-neighbor hopping, while $V_{xy}$ is the potential between particles on different sites, 
and $\rho_x := b\adjoint_x b_x - a\adjoint_x a_x$ is the charge operator. In this work, we study the Hubbard model with on-site interactions only, such that $V_{xy} = U\delta_{xy}$;
the ratio $U/\kappa$ determines whether we are in a strongly or weakly coupled regime.

Hamiltonian theories such as~(\ref{Ham_def}) have for a long time been studied with lattice MC methods~\cite{Paiva2005,Beyl:2017kwp}, as this allows for 
a fully \textit{ab initio} stochastic evaluation of the thermal trace, or Grassmann path integral. There is a large freedom of choice in the construction of lattice MC algorithms, 
including the discretization of the theory, the choice of Hubbard-Stratonovich (or auxiliary field) transformation, and the algorithm used to update the auxiliary field variables.
This freedom can be exploited to optimize the algorithm with respect to a particular computational feature. These pertain to the scaling of the computational effort with system (lattice) 
size $L$, inverse temperature $\beta$, number of time slices $N_t=\beta/\delta$, interaction strength $U/\kappa$, and electron number density (for simulations away from half filling).
Hamiltonian theories are often simulated with an exponential (or compact) form of both the kinetic and 
potential energy contributions to the partition function (or Euclidean time projection amplitude), and with random Metropolis updates of the auxiliary fields (which may be either 
discrete or continuous). In condensed matter and atomic physics, such methods are referred to as the Blankenbecler-Sugar-Scalapino (BSS) 
algorithm~\cite{Blankenbecler:1981jt}.

In Lattice Quantum Chromodynamics (QCD), the high dimensionality of the theory and the need to precisely approach the continuum limit have led to the development of specialized algorithms 
which optimize the computational scaling with $L$. These efforts have culminated in the HMC algorithm, which combines elements of the 
Langevin, Molecular Dynamics (MD), and Metropolis algorithms~\cite{Duane:1987de}. The application of HMC to the Hubbard model~(\ref{Ham_def}) has proven to be 
surprisingly difficult, due to problems related to ergodicity, symmetries of the Hamiltonian, and the correct approach to the (temporal) continuum limit.
For a thorough treatment of these from the point of view of HMC, see \Refs{Luu:2015gpl,Wynen:2018ryx,Ostmeyer:thesis}.
In order to realize the expected $\sim V^{5/4}$ computational scaling (where $V = 2L^2$), a suitable conjugate gradient (CG) 
method has to be found for the numerical integration of the MD equations of motion.
The Hasenbusch preconditioner~\cite{hasenbusch} from Lattice QCD has recently been found to work for the Hubbard model as well~\cite{acceleratingHMC}.
The resulting combination of HMC with the Hubbard model is referred to as the Brower-Rebbi-Schaich (BRS) algorithm~\cite{Brower:2011av,Brower:2012zd}, which is closely related to the 
BSS algorithm. The main differences are the linearized kinetic energy (or nearest-neighbor hopping) term, and the purely imaginary auxiliary field, which is updated using HMC moves in the
BRS algorithm.

The BSS algorithm has preferentially been used within the canonical ensemble~\cite{Meng2010,Assaad:2013xua,Otsuka:2015iba}. This entails a
projection Monte Carlo (PMC) calculation, where particle number is conserved and one is restricted to specific many-body Hilbert spaces.
PMC is highly efficient at accessing zero-temperature (or ground state) properties, especially when the number of particles is constant, for
instance the $A$ nucleons in an atomic nucleus. For the Hubbard model on the honeycomb lattice at half-filling, the fully anti-symmetric trial wave function 
encodes a basis of $2L^2$ electrons to be propagated in Euclidean time. Due to this scaling of the number of trial wave functions, PMC and grand canonical 
versions of the BSS algorithm both exhibit $\sim V^3$ scaling (with random, local Metropolis updates). In contrast to PMC, the grand canonical
formalism resides in the full Fock space, and no trial wave function is used. Instead, Boltzmann-weighted thermal expectation values of observables are extracted.
At low temperatures and large Euclidean times, spectral observables are measured relative to the ground state of the Fock space, which is the half-filling state 
(an explicit example is given in \Secref{results}).
With HMC updates, such an algorithm has been found to scale as $\sim V^{5/4}$~\cite{Creutz:1988wv,acceleratingHMC}.
A drawback of the grand canonical ensemble is the explicit inverse temperature $\beta$. Thus, the limit $\beta \to \infty$ is taken by extrapolation, or
more specifically by finite-size scaling. This $\beta$-dependence may be considerable, though observable-dependent. While PMC simulations are 
not free of similar effects (due to contamination from excited state contributions), 
they are typically less severe due to the absence of backwards-propagating states in Euclidean time.

For a number of reasons, HMC updates have proven difficult for the BSS algorithm. First, the exponential form of the fermion operator $M$ causes $\det M$
to factorize into regions of positive and negative sign. Though this does not imply a sign problem at half filling (the action $S \sim \log |\det M|^2$), 
it does introduce boundaries in the energy landscape of the theory, which HMC trajectories in general cannot cross without special and computationally very expensive 
methods~\cite{Fodor:2003bh,Cundy:2005pi}. For $\beta \to \infty$, this fragmentation effect increases dramatically, and causes an ergodicity problem with HMC.
Second, while this problem can be circumvented by a complex-valued auxiliary field, the resulting ``complexified'' HMC algorithm shows poor (roughly cubic) scaling with $V$~\cite{Beyl:2017kwp}.
It is interesting to note how the BRS algorithm avoids this ergodicity problem. Due to the linearization of the hopping term in the fermion operator (with imaginary auxiliary field), 
the boundaries impassable to HMC are reduced in dimension and can be avoided~\cite{Wynen:2018ryx}. Naturally, the BSS and BRS formulations become equivalent in the temporal
continuum limit. Then, the ergodicity problem would eventually be recovered when $\delta \to 0$ (where MC simulations are in any case not practical).
A drawback specific to BRS is that spin symmetry is explicitly broken for $\delta \neq 0$, due to the linearization of the hopping term~\cite{Buividovich:2016tgo}.
Hence, the choice of BSS versus BRS represents a tradeoff between the retention of more symmetries at finite $\delta$, and faster convergence to the 
continuum limit (BSS), or improved ergodicity and computational scaling with $V$ (BRS).

Here, we apply the BRS algorithm with HMC updates to the Hubbard model~(\ref{Ham_def}). This entails the stochastic evaluation of a path integral over a Hubbard-Stratonovich
field $\phi$.
The exact form of the fermion operator $M$ depends on the choice of discretization
for time derivatives. We adopt the conventions of \Ref{Luu:2015gpl}, which used a ``mixed differencing'' operator, with forward differencing in time for $A$ sites, and backward 
for $B$ sites. For reference, we note that this scheme does not introduce a fermion doubling problem. For the non-interacting theory, mixed differencing gives $\ordnung{\delta^2}$ discretization 
errors (per time step). Numerically, mixed differencing has been shown to approach the limit $\delta \to 0$ faster than pure forward or backward differencing when $U > 0$. 
The explicit form of $M$ is
\begin{equation}
\begin{split}
M^{AA}_{(x,t)(y,t')} & = \delta_{xy}^{} \left(\delta_{t+1,t^\prime}^{} - \delta_{t,t^\prime}^{} \exp(-i\phidelta_{x,t}^{}) \right), \\
M^{BB}_{(x,t)(y,t')} & = \delta_{xy}^{} \left(\delta_{t,t^\prime} - \delta_{t-1,t'} \exp(-i\phidelta_{x,t}^{}) \right), \\
M^{AB}_{(x,t)(y,t')} = M^{BA}_{(x,t)(y,t^\prime)} & =- \kappadelta\,\delta_{\erwartung{x,y}} \delta_{t,t^\prime},
\end{split}
\label{eqn_ferm_op}
\end{equation}
where the dependence on $A$ and $B$ sites has been written out. All quantities multiplied by $\delta$ have been denoted by a tilde. While the hopping term
in Eq.~(\ref{eqn_ferm_op}) has been linearized, the auxiliary field $\tilde\phi$ enters through exponential ``gauge links'' familiar from Lattice QCD. As first discussed 
in Refs.~\cite{Brower:2011av,Brower:2012zd}, such gauge links contain a ``seagull term'', which needs to be correctly handled in order to recover the physical $\delta \to 0$ limit.
This condition is satisfied by Eq.~(\ref{eqn_ferm_op}), and further details are given in \Appref{mixed_bias}. 

\section{The gap 
\label{sec:results}}

\subsection{The \spc}

We shall now describe the procedure of obtaining the single-particle gap $\Delta$ as a function of $U/\kappa$ and $\beta$, from a given ensemble of auxiliary-field 
configurations. We recall that $a$ and $a\adjoint$ are annihilation and creation operators for quasiparticles, and similarly $b$ and $b\adjoint$ for (quasi-)holes.
For instance, by creating and destroying quasiparticles at different locations and times, we obtain the correlator
\begin{equation}
C_{xy}^{}(t)
:= \left\langle a_{x,t}^\pdagger a\adjoint_{y,0} \right\rangle
= \frac{1}{\Z} \int \D\phi\; M[\phi] \inverse_{(x,t),(y,0)} \exp(-S[\phi])
= \left\langle M[\phi] \inverse_{(x,t),(y,0)} \right\rangle,
\end{equation}
as an ensemble average, 
where $S[\phi]$ is the Euclidean action and $\Z$ is the partition function---the integral without $M$.
We have used Wick contraction to replace the operators with the fermion propagator.

We now move to the Heisenberg picture and express the correlator 
as a thermal trace
\begin{equation}
C_{xy}(t) = \left\langle a_{x,t}^\pdagger a\adjoint_{y,0} \right\rangle
= \frac{1}{\Z} \tr\bigg\{a_{x,t}^\pdagger a\adjoint_{y,0} \exp(-\beta H) \bigg\}
= \frac{1}{\Z} \tr\bigg\{\exp(-H(\beta-t)) \, a_x^\pdagger \exp(-Ht) a\adjoint_y \bigg\},
\end{equation}
and by inserting the identity (resolved in the interacting basis), we find the spectral decomposition
\begin{align}
\label{eq:spectral decomposition}
C_{xy}^{}(t) = \left\langle a_{x,t}^\pdagger a\adjoint_{y,0} \right\rangle & =
\frac{1}{\sum_i \exp(-\beta E_i)} 
\sum_{m,n} \exp(-\beta E_m) \exp(-(E_n-E_m)t) 
\, z_{mxn}^{\vphantom{*}} z\conjugate_{myn}, \\
\label{eq:overlap factor}
z_{mxn}^{} & :=
\left\langle m \middle| a_x^{} \middle| n \right\rangle,
\end{align}
where the summation indices $m$ and $n$ denote interacting eigenstates with energies $E_m^{}$ and $E_n^{}$, respectively, and the $z_{ijk}^{}$ are referred to as ``overlap factors''.
At large $\beta$ and in the limit of large Euclidean time, the correlator decays exponentially,
\begin{equation}
\lim_{t\goesto\infty} \lim_{\beta\goesto\infty} 
C_{xy}^{}(t) \simeq \exp(-(E_1-E_0)t) \, z_{0x1}^{\vphantom{*}} z\conjugate_{0y1},
\end{equation}
where the energy $E_1^{}$ is measured relative to the energy $E_0$ of ground state of the Fock space, assuming that the associated 
overlap factors are non-zero. 

To continue it is convenient to Fourier transform the fermion propagator, which is a function of the lattice coordinates $x$ and $y$, to a function of the (lattice) momenta $k$ and $p$\footnote{We note that under the ensemble average the correlators are diagonal in $k$ and $p$ due to translational invariance.},
\begin{equation}
C_{kp}^{}(t)=\frac{1}{L^4}\sum_{x,y}\exp(-ikx-ipy) \, C_{xy}^{}(t)\ .
\end{equation}
Repetition of this procedure yields an ensemble of ``measurements'', the average of which is our MC estimate of $C_{kp}(t)$.
As there are two Dirac points $K$ and $K^\prime$, by symmetry we have\footnote{Because of the underlying sublattices $A$ and $B$, 
there are in principle two independent correlators for each momentum $k$~\cite{Luu:2015gpl}.  However, these two correlators 
are degenerate at each Dirac point $K$ and $K^\prime$; we average them to construct $C_{KK}$ and $C_{K^\prime K^\prime}$, respectively.}
\begin{equation}
C_{KK}^{}(t) = C_{K^\prime K^\prime}^{}(t),
\end{equation}
which holds for the expectation values, but not on a configuration-by-configuration basis. By choosing to Fourier transform $x$ and $y$ to the Dirac momenta $K$ or $K^\prime$, we adjust the overlap factors so that we can take
$E_1^{}$ to refer to the energy at the Dirac point. We define the ``effective mass''
\begin{equation}
\meff := E_1^{}-E_0^{},
\end{equation}
which can be extracted from the correlator according to
\begin{equation}
\label{eq:usual effective mass}
\meff(t) = -\lim_{\beta\goesto\infty} \partial_t \ln C_{KK}^{}(t),
\qquad
\meff = \lim_{t\goesto\infty} \meff(t),
\end{equation}
such that the single-particle gap is given by
\begin{equation}
\Delta = 2 \meff,
\end{equation}
for a specific value of $L$ and $\delta$. We discuss the extrapolations in these variables in Section~\ref{sec:gap_results}.

By symmetry, the two Dirac points are indistinguishable, and the correlator expectation values are symmetric in time around 
$\beta/2$ (or $\nt/2$ in units of $\tau=t/\delta$). We therefore average and fold the correlator according to
\begin{equation}
\label{eq:symmetrized}
C(t) := \frac{1}{4} \big( 
C_{KK}^{}(t) + C_{K^\prime K^\prime}(t) + 
C_{KK}^{}(\beta-t) + C_{K^\prime K^\prime}(\beta-t)\big),
\end{equation}
on a configuration-by-configuration basis, 
which increases our numerical precision without the need for generating additional MC configurations. %
In most cases, thermal effects due to backwards-propagating states cannot be completely neglected, and the isolation of an unambiguous exponential decay is difficult.
Instead of obtaining $\Delta$ from
\begin{align}
\label{eq:const fit}
\meff(\tau)\,\delta = \ln\frac{C(\tau)}{C(\tau+1)},
\end{align}
we use the symmetry of the correlator at the Dirac point about $\tau=\beta/2$ and calculate
\begin{equation}
\label{eq:cosh fit}
\meff(\tau)\, \delta = \cosh^{-1}\left(\frac{C(\tau+1)+C(\tau-1)}{2C(\tau)}\right)\ ,
\end{equation}
where $C(\tau)$ is the folded and symmetrized correlator \eqref{eq:symmetrized}.
This hyperbolic cosine form \eqref{eq:cosh fit} is found to be more accurate than the exponential form \eqref{eq:const fit}, 
as the quasiparticle masses are rather small, and the backward-propagating contributions non-negligible (due to the finite extent in $\beta$).
Once the optimal region from which to extract the effective mass has been found (for details, see \Appref{plateau_estimation}), 
we fit the correlator in this region to the form
\begin{align}
C(\tau) = a \cosh\left(\meff\,\delta\left(\tau-N_t/2 \right)\right),
\end{align}
with $a$ and $\meff$ as fit parameters. In \Figref{plateau_fit}, we show the difference between the hyperbolic cosine fit and a direct fit of $\meff(\tau)$. The former is shown
by a blue line and band (statistical error), and the latter by an orange dashed line. While the results are in general agreement, the direct fit of $\meff(\tau)$ tends to overestimate the
gap for small {\effm}es and high statistical noise levels. The estimated systematic error due to the choice of the fit range (as explained in \Appref{plateau_estimation}) is 
shown by the red dot-dashed line. The statistical errors are obtained by a bootstrap procedure, and the total error has been estimated by adding the statistical and
systematic uncertainties in quadrature. The data analysis described in this and the following sections has mostly been performed in the \texttt{R} language~\cite{R_language}, 
in particular using the \texttt{hadron} package~\cite{hadron}.

\begin{figure}[t]
\centering
\includegraphics[width=.45\textwidth]{{b_8_l_15_t_64_u_3.5_plateau}.pdf}
\includegraphics[width=.45\textwidth]{{b_12_l_6_t_72_u_3.85_plateau}.pdf}
\caption{Examples of effective mass determinations from the {\spc}s, extracted from ensembles of auxiliary field configurations.
The blue line with error band gives $\meff$ with statistical error, obtained from a hyperbolic cosine effective mass \eqref{eq:cosh fit}.
The length of the blue band indicates the fitting region. For comparison, a constant fit to the effective mass \eqref{eq:usual effective mass} is 
shown by the dashed orange line. The dot-dashed red line shows the estimation of the systematic error, as explained in \Appref{plateau_estimation}.
Note that the red and orange lines have been extended outside of the fitting region, for clearer visibility.
Left panel: $\kappa\beta=8$, $L=15$, $N_t=64$, $U/\kappa=\num{3.5}$. Right panel: $\kappa\beta=12$, $L=6$, $N_t=72$, $U/\kappa=\num{3.85}$.
The timeslice index $\tau$ is integer-valued. The effective mass \meff is given in units of $\kappa$.
\label{fig:plateau_fit}}
\end{figure}

\subsection{Lattice artifacts \label{sec:lattice_artefacts}}

Once we have determined the single-particle gap $\Delta$, we still need to consider the limits $\delta \to 0$, $L \to \infty$, and $\beta \to \infty$. We shall discuss the
first two limits here, and return to the issue of finite-size scaling in $\beta$ later.

In general, grand canonical MC simulations of the Hubbard model are expected to show very mild finite-size effects which vanish exponentially with $L$, as found by
\Ref{Wang:2017ohn}. This situation is more favourable than in canonical simulations, where observables typically scale as a power-law in $L^{-1}$~\cite{Wang:2017ohn}.
Let us briefly consider the findings of other recent MC studies.
For the extrapolation in $L$, \Refs{Assaad:2013xua,Meng2010} 
(for $\Delta$ and the squared staggered magnetic moment $m_s^2$), \Ref{Otsuka:2015iba} (for $m_s^2$) and \Ref{Buividovich:2018yar} 
(for the square of the total spin per sublattice), find a power-law dependence of the form $a+bL\inverse+cL^{-2}$.
On the other hand, \Ref{Ulybyshev_2017} found little or no dependence on $L$ for the
conductivity.
A side effect of a polynomial dependence on $L^{-1}$ is that (manifestly positive) extrapolated quantities may become
negative in the limit $L \to \infty$.

The continuous time limit $\delta \to 0$ was taken very carefully in \Ref{Otsuka:2015iba},
by simulations at successively smaller $\delta$ until the numerical results stabilized. With the exponential (or compact)
kinetic energy term used in \Ref{Otsuka:2015iba}, the Trotter error of observables should scale as $\ordnung{\delta^2}$. As shown, for instance in 
\Ref{Wynen:2018ryx}, discretization errors of observables for our linearised kinetic energy operator are in general of $\ordnung\delta$. Even with an
exponential kinetic energy operator, some extrapolation in $\delta$ is usually required~\cite{Wynen:2018ryx}. 
For further details concerning the limit $\delta \to 0$, see \Appref{mixed_bias}.

In \Appref{thermal_gap}, we argue that the residual modification of $\Delta$ due to the finite lattice extent $L$ should be proportional to $L^{-3}$.
This is not expected for all observables, but only for those that satisfy two conditions. First, the observable should not (for single MC configurations)
have errors proportional to $L^{-2}$, which are not cancelled by an ensemble average. This condition is satisfied by the correlation functions at the Dirac points, 
but not in general for (squared) magnetic or other locally defined quantities. For example, $m_s^2$ exhibits $\sim L^{-2}$ fluctuations which are 
positive for every MC configuration. As the average of these positive contributions does not vanish, the error is effectively proportional to $L^{-2}$.
Second, the correlation length $\xi$ has to fulfil $\xi \ll L$, such that the error contribution $\sim\exp(-L/\xi)$ as in \Ref{Wang:2017ohn} remains suppressed.
While we expect $L^{-3}$ scaling to hold far from phase transitions, this may break down 
close to criticality, where $\xi \to \infty$. We find that deviations from inverse cubic scaling are small when
\begin{align}
L \gg \frac{\kappa\beta}{\pi},
\end{align}
where $\pi/\beta$ is the minimum Matsubara frequency dominating the correlation length ($\kappa\beta/\pi\gtrsim\xi$). We employed $\kappa \beta\le 12$ implying the requirement $L\gg 4$. Numerically, we find that for $L \ge 9$ our observed dependence on $L$ is entirely governed by inverse cubic scaling (see Figures~\ref{fig:2d_gap_fit} and~\ref{fig:finite_size_always_cubic}).

\subsection{Extrapolation method}

In this study, we have chosen to perform a simultaneous extrapolation in $\delta$ and $L$, by a two-dimensional chi-square minimization.
We use the extrapolation formula
\begin{align}
\mathfrak{O}(L,N_t^{}) =
\mathfrak{O} + a_1^{} L^{-3} + a_2^{} N_t^{-2},
\label{eqn_white_2d}
\end{align}
where $\mathfrak{O}$ is an observable with expected Trotter error of $\ordnung{\delta^2}$, and $a_1, a_2$ are fit parameters.
This is similar to the procedure of \Ref{White:1989zz} for expectation values of the Hamiltonian.
Before we describe our fitting procedure in detail, let us note some advantages of Eq.~(\ref{eqn_white_2d}) relative to a method where observables 
are extracted at fixed $(U,\beta)$ by first taking the temporal continuum limit $\delta \to 0$, 
\begin{align}
\mathfrak{O}(L,N_t^{}) =
\mathfrak{O}(L) + a_2^{}(L) N_t^{-2},
\label{eqn_white_1d_step1}
\end{align}
where each value of $\mathfrak{O}(L)$ and $a_2^{}(L)$ is obtained from a separate chi-square fit, where $L$ is
held fixed. This is followed by
\begin{align}
\mathfrak{O}(L) =
\mathfrak{O} + a_1^{} L^{-3},
\label{eqn_white_1d_step2}
\end{align}
as the final step. Clearly, a two-dimensional chi-square fit using Eq.~(\ref{eqn_white_2d}) involves a much larger
number of degrees of freedom relative to the number of adjustable parameters. This feature makes it easier to judge the quality
of the fit and to identify outliers, which is especially significant when the fit is to be used for extrapolation. While we have presented arguments for the 
expected scaling of $\Delta$ as a function of $L$ and $\delta$, in general the true functional dependence on these variables is not \textit{a priori} known. Thus, the only unbiased check on the 
extrapolation is the quality of each individual fit. This criterion is less stringent, if the data for each value of $L$ is individually extrapolated 
to $\delta \to 0$.

The uncertainties of the fitted parameters have been calculated via 
parametric bootstrap, which means that the bootstrap samples have been generated by drawing from independent normal distributions, defined by every single 
value of $\Delta(L,N_t^{})$ and its individual error.

\subsection{Results \label{sec:gap_results}}

\begin{figure}[t]
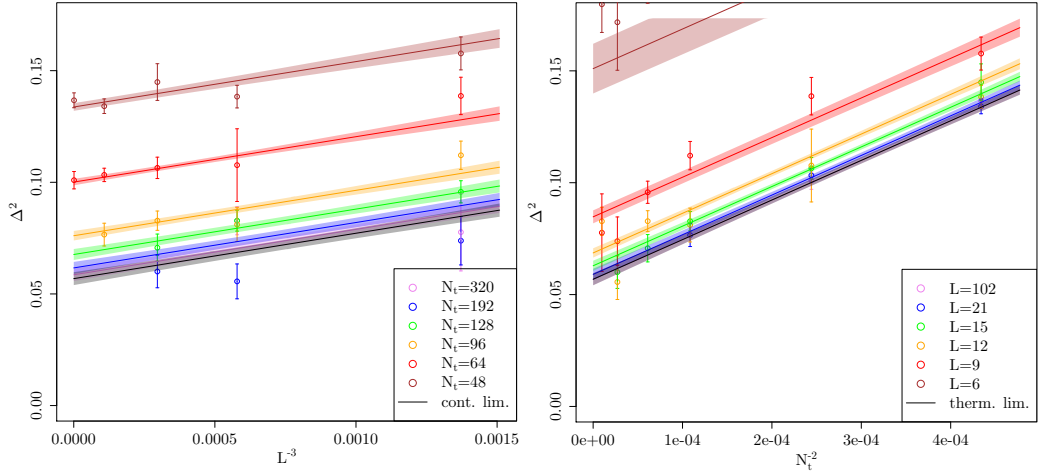

\centering
\raisebox{0.016\height}{\includegraphics[width=.45\textwidth]{{b_8_u_3.5_m_0_gap_l}.pdf}}
\includegraphics[width=.45\textwidth]{{b_8_u_3.5_m_0_gap_t}.pdf}
\caption{Simultaneous two-dimensional fit of $\Delta(N_t,L)$ (in units of $\kappa$) using Eq.~\eqref{eqn_gap_artefacts}, for $\kappa\beta=8$ and $U/\kappa=\num{3.5}$.
Note that Eq.~\eqref{eqn_gap_artefacts} only incorporates effects of $\ordnung{L^{-3}}$ and $\ordnung{N_t^{-2}}$. Data points for $L<9$ have been omitted from the fit, 
but not from the plot. Very small lattices lead to large values of $\Delta$, which are not visible on the scale of the plot.
This fit has $\chi^2/\text{d.o.f.} \simeq \num{1.1}$, corresponding to a p-value of $\simeq \num{0.34}$.
\label{fig:2d_gap_fit}}
\end{figure}

Given the fermion operator described in \Secref{formalism}, we have used Hasenbusch-accelerated HMC~\cite{hasenbusch,acceleratingHMC}
to generate a large number of ensembles at different values of $L$, $U/\kappa$, $\kappa\beta$, and $\kappa\delta=\kappa\beta/ N_t$.
We have used six inverse temperatures $\kappa\beta\in\{3, 4, 6, 8, 10, 12\}$\footnote{
These correspond to a highest temperature of $T\approx\SI{1.04e4}{\kelvin}$ and a lowest temperature of $T\approx\SI{2.6e3}{\kelvin}$. According to \Ref{temperature_phase_trans} 
this range is well below the critical temperature $T_c\approx\SI{1.3e4}{\kelvin}$, above which the sublattice symmetry is restored. %
} and a number of couplings in the range $U/\kappa \in [1.0, 5.5]$, which is expected to bracket the critical coupling $U_c/\kappa$ of the AFMI transition. For each temperature and 
coupling, we scanned a large range in $L$ and $\kappa\delta$, to provide reliable extrapolations to the physical limits. These are
$L \in [3, 102]$ \footnote{
	We generated several very large lattices for a few sets of parameters to check the convergence.
	As we observed a nearly flat dependence of $\Delta$ on $L\gtrsim12$, we did most of the analysis with medium-sized lattices ($L\le 21$) in order to conserve
	computation time.
} and $\kappa\delta$ from~$1/4$ down to~$1/40$. Our values of $L$ were chosen to be 0 (mod 3) so that the
non-interacting dispersion has momentum modes exactly at the Dirac points. In other words, we select the lattice geometry to ensure that $E_1-E_0 = 0$ in the absence of
an interaction-induced (Mott) gap. 

Let us briefly summarize our procedure for computing $\Delta$. We have calculated the \spc\ $C(t)$ 
as an expectation value of the inverse fermion matrix, at both independent Dirac-points $K, K^\prime$, and averaged over them to increase our statistics.
For each set of parameters simulated, those correlators have been fitted with an effective mass in order to extract $\Delta$.
We are then in a position to take the limits $\delta \to 0$ and $L \to \infty$, which we accomplish by a simultaneous fit to the functional form
\begin{align}
\Delta^2(L,N_t^{}) =
\Delta_0^2 + c_0^{} N_t^{-2}
+ c_1^{} L^{-3}\,,
\label{eqn_gap_artefacts}
\end{align}
where the fitted quantities are the (infinite-volume, temporal continuum) gap $\Delta_0$, and the leading corrections proportional to $c_0^{}$ and $c_1^{}$.
Around $\delta = 0$, we find the leading $L$-dependence
\begin{align}
\Delta(L) = \Delta_0^{} + \frac{c_1^{}}{2\Delta_0^{}}L^{-3} + \ordnung{L^{-6}},
\end{align}
where $c_1^{}$ is numerically found to be very small. Also, around $L = \infty$, we have
\begin{align}
\Delta(N_t^{}) = \Delta_0^{} + \frac{c_0^{}}{2\Delta_0^{}} N_t^{-2} + \ordnung{N_t^{-4}}\,.
\end{align}
It should be noted that inclusion of a term of order $N_t^{-1}$ did not improve the quality of the fit. This observation 
can be justified by the expected suppression of the linear term due to our mixed differencing scheme. 
Though the term of $\ordnung\delta$ is not removed analytically, it appears small enough to be numerically unresolvable.

\begin{figure}[t]
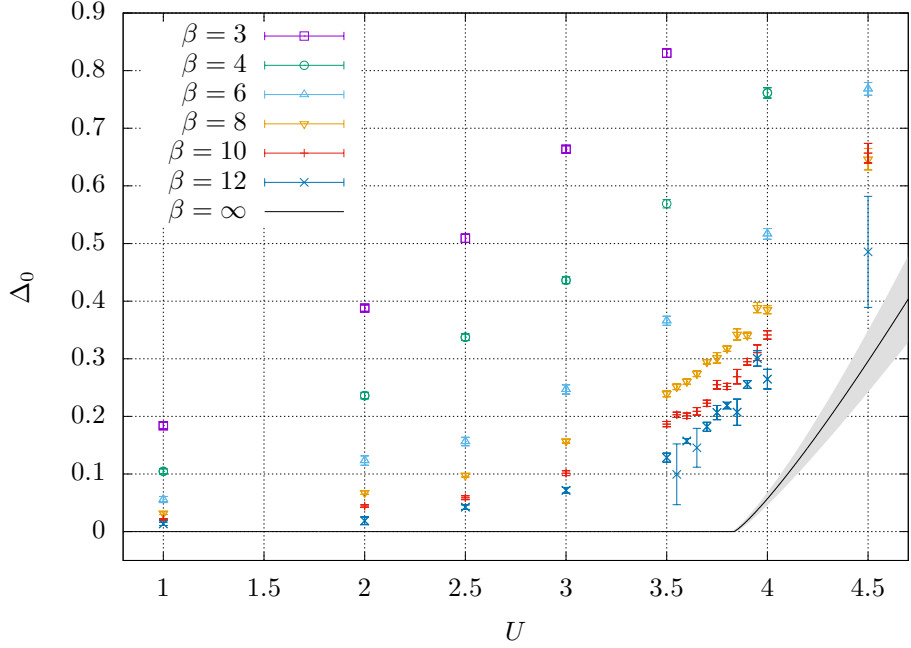

\begingroup
  \inputencoding{latin1}%
  \makeatletter
  \providecommand\color[2][]{%
    \GenericError{(gnuplot) \space\space\space\@spaces}{%
      Package color not loaded in conjunction with
      terminal option `colourtext'%
    }{See the gnuplot documentation for explanation.%
    }{Either use 'blacktext' in gnuplot or load the package
      color.sty in LaTeX.}%
    \renewcommand\color[2][]{}%
  }%
  \providecommand\includegraphics[2][]{%
    \GenericError{(gnuplot) \space\space\space\@spaces}{%
      Package graphicx or graphics not loaded%
    }{See the gnuplot documentation for explanation.%
    }{The gnuplot epslatex terminal needs graphicx.sty or graphics.sty.}%
    \renewcommand\includegraphics[2][]{}%
  }%
  \providecommand\rotatebox[2]{#2}%
  \@ifundefined{ifGPcolor}{%
    \newif\ifGPcolor
    \GPcolortrue
  }{}%
  \@ifundefined{ifGPblacktext}{%
    \newif\ifGPblacktext
    \GPblacktexttrue
  }{}%
  \let\gplgaddtomacro\g@addto@macro
  \gdef\gplbacktext{}%
  \gdef\gplfronttext{}%
  \makeatother
  \ifGPblacktext
    \def\colorrgb#1{}%
    \def\colorgray#1{}%
  \else
    \ifGPcolor
      \def\colorrgb#1{\color[rgb]{#1}}%
      \def\colorgray#1{\color[gray]{#1}}%
      \expandafter\def\csname LTw\endcsname{\color{white}}%
      \expandafter\def\csname LTb\endcsname{\color{black}}%
      \expandafter\def\csname LTa\endcsname{\color{black}}%
      \expandafter\def\csname LT0\endcsname{\color[rgb]{1,0,0}}%
      \expandafter\def\csname LT1\endcsname{\color[rgb]{0,1,0}}%
      \expandafter\def\csname LT2\endcsname{\color[rgb]{0,0,1}}%
      \expandafter\def\csname LT3\endcsname{\color[rgb]{1,0,1}}%
      \expandafter\def\csname LT4\endcsname{\color[rgb]{0,1,1}}%
      \expandafter\def\csname LT5\endcsname{\color[rgb]{1,1,0}}%
      \expandafter\def\csname LT6\endcsname{\color[rgb]{0,0,0}}%
      \expandafter\def\csname LT7\endcsname{\color[rgb]{1,0.3,0}}%
      \expandafter\def\csname LT8\endcsname{\color[rgb]{0.5,0.5,0.5}}%
    \else
      \def\colorrgb#1{\color{black}}%
      \def\colorgray#1{\color[gray]{#1}}%
      \expandafter\def\csname LTw\endcsname{\color{white}}%
      \expandafter\def\csname LTb\endcsname{\color{black}}%
      \expandafter\def\csname LTa\endcsname{\color{black}}%
      \expandafter\def\csname LT0\endcsname{\color{black}}%
      \expandafter\def\csname LT1\endcsname{\color{black}}%
      \expandafter\def\csname LT2\endcsname{\color{black}}%
      \expandafter\def\csname LT3\endcsname{\color{black}}%
      \expandafter\def\csname LT4\endcsname{\color{black}}%
      \expandafter\def\csname LT5\endcsname{\color{black}}%
      \expandafter\def\csname LT6\endcsname{\color{black}}%
      \expandafter\def\csname LT7\endcsname{\color{black}}%
      \expandafter\def\csname LT8\endcsname{\color{black}}%
    \fi
  \fi
    \setlength{\unitlength}{0.0500bp}%
    \ifx\gptboxheight\undefined%
      \newlength{\gptboxheight}%
      \newlength{\gptboxwidth}%
      \newsavebox{\gptboxtext}%
    \fi%
    \setlength{\fboxrule}{0.5pt}%
    \setlength{\fboxsep}{1pt}%
\begin{picture}(7200.00,5040.00)%
    \gplgaddtomacro\gplbacktext{%
      \csname LTb\endcsname%
      \put(814,921){\makebox(0,0)[r]{\strut{}$0$}}%
      \csname LTb\endcsname%
      \put(814,1354){\makebox(0,0)[r]{\strut{}$0.1$}}%
      \csname LTb\endcsname%
      \put(814,1787){\makebox(0,0)[r]{\strut{}$0.2$}}%
      \csname LTb\endcsname%
      \put(814,2220){\makebox(0,0)[r]{\strut{}$0.3$}}%
      \csname LTb\endcsname%
      \put(814,2653){\makebox(0,0)[r]{\strut{}$0.4$}}%
      \csname LTb\endcsname%
      \put(814,3086){\makebox(0,0)[r]{\strut{}$0.5$}}%
      \csname LTb\endcsname%
      \put(814,3520){\makebox(0,0)[r]{\strut{}$0.6$}}%
      \csname LTb\endcsname%
      \put(814,3953){\makebox(0,0)[r]{\strut{}$0.7$}}%
      \csname LTb\endcsname%
      \put(814,4386){\makebox(0,0)[r]{\strut{}$0.8$}}%
      \csname LTb\endcsname%
      \put(814,4819){\makebox(0,0)[r]{\strut{}$0.9$}}%
      \csname LTb\endcsname%
      \put(1246,484){\makebox(0,0){\strut{}$1$}}%
      \csname LTb\endcsname%
      \put(1997,484){\makebox(0,0){\strut{}$1.5$}}%
      \csname LTb\endcsname%
      \put(2748,484){\makebox(0,0){\strut{}$2$}}%
      \csname LTb\endcsname%
      \put(3499,484){\makebox(0,0){\strut{}$2.5$}}%
      \csname LTb\endcsname%
      \put(4250,484){\makebox(0,0){\strut{}$3$}}%
      \csname LTb\endcsname%
      \put(5001,484){\makebox(0,0){\strut{}$3.5$}}%
      \csname LTb\endcsname%
      \put(5752,484){\makebox(0,0){\strut{}$4$}}%
      \csname LTb\endcsname%
      \put(6503,484){\makebox(0,0){\strut{}$4.5$}}%
    }%
    \gplgaddtomacro\gplfronttext{%
      \csname LTb\endcsname%
      \put(198,2761){\rotatebox{-270}{\makebox(0,0){\strut{}$\Delta_0$}}}%
      \put(3874,154){\makebox(0,0){\strut{}$U$}}%
      \csname LTb\endcsname%
      \put(1870,4646){\makebox(0,0)[r]{\strut{}$\beta = 3$}}%
      \csname LTb\endcsname%
      \put(1870,4426){\makebox(0,0)[r]{\strut{}$\beta = 4$}}%
      \csname LTb\endcsname%
      \put(1870,4206){\makebox(0,0)[r]{\strut{}$\beta = 6$}}%
      \csname LTb\endcsname%
      \put(1870,3986){\makebox(0,0)[r]{\strut{}$\beta = 8$}}%
      \csname LTb\endcsname%
      \put(1870,3766){\makebox(0,0)[r]{\strut{}$\beta = 10$}}%
      \csname LTb\endcsname%
      \put(1870,3546){\makebox(0,0)[r]{\strut{}$\beta = 12$}}%
      \csname LTb\endcsname%
      \put(1870,3326){\makebox(0,0)[r]{\strut{}$\beta = \infty$}}%
    }%
    \gplbacktext
    \put(0,0){\includegraphics{gap-pdf}}%
    \gplfronttext
  \end{picture}%
\endgroup
 \caption{The single-particle gap $\Delta_0(U,\beta)$, with all quantities in units of $\kappa$, after the thermodynamic and continuum limit extrapolations.  
We also show $\Delta_0(U,\beta \to \infty)$ as a solid black line with error band (see \Secref{zero temperature extrapolation}). For 
$U< U_c \simeq \critU$ the zero-temperature gap vanishes. 
\label{fig:delta_against_U}}
\end{figure}

An example of such a fit is shown in \Figref{2d_gap_fit}. We find that Eq.~\eqref{eqn_gap_artefacts} describes our data well with only a minimal set of parameters for large $L$ and 
small $\delta$. Lattices with $L\le6$ have been omitted from the fit, as such data does not always lie in the scaling region where the data shows 
$\ordnung{L^{-3}}$ convergence, as explained in \Secref{lattice_artefacts}.
Our results for $\Delta_0$ are shown
in \Figref{delta_against_U} for all values of $U/\kappa$ and $\kappa\beta$, along with an extrapolation (with error band) to zero temperature ($\beta\to\infty$).
For details as to the zero-temperature gap, see \Secref{analysis}.

\section{Analysis \label{sec:analysis}}

We now analyze the single-particle gap $\Delta_0$ as a function of coupling $U-U_c$ and inverse temperature $\beta$, in order to determine the critical coupling
$U_c/\kappa$ of the quantum (AFMI) phase transition, along with some of the associated critical exponents. Our MC results for $\Delta_0$ as a function of $U$ are shown in \Figref{delta_against_U}.
We make use of the standard finite-size scaling (FSS) method~\cite{newmanb99, dutta_aeppli, Shao_2016}, whereby a given observable $\mathfrak{O}$ is described by
\begin{equation}
\mathfrak{O} = (U-U_c)^q \mathcal{F}_{\mathfrak{O}}^{}(L/\xi, \beta/\xi_t^{}), 
\label{eqn:scaling_1}
\end{equation}
in the vicinity of $U_c$, where $q$ is the relevant critical exponent. The scaling function $\mathcal{F}$ accounts for the effects of finite spatial system size $L$ and inverse temperature $\beta$.
The spatial and temporal correlation lengths $\xi$ and $\xi_t$ are
\begin{equation}
\xi \sim (U-U_c)^{-\nu},
\quad
\xi_t^{} \sim (U-U_c)^{-\nu_t^{}} = (U-U_c)^{-z\nu} \sim \xi^z,
\label{eqn:corr}
\end{equation}
such that in the thermodynamic limit we can express the FSS relation as
\begin{equation}
\mathfrak{O} = \beta^{-q/(z\nu)} F_{\mathfrak{O}}^{}(\beta^{1/(z\nu)}(U-U_c)),
\label{eqn:scaling_5}
\end{equation}
which is analogous to \Refs{Assaad:2013xua, Otsuka:2015iba}, apart from the scaling argument here being
$\beta$ instead of $L$, and the correlation length exponent $z\nu$ picked up a factor of the dynamical exponent $z$.

We assume the standard scaling behavior~\cite{Herbut:2009qb} for the single-particle gap 
\begin{equation}
\Delta_0^{} \sim \xi_t^{-1} \sim (U-U_c)^{z\nu},
\end{equation}
as extracted from the asymptotic behavior of the correlator $C(t)$. Therefore, Eq.~\eqref{eqn:scaling_5} gives
\begin{equation}
\Delta_0^{} = \beta^{-1} 
F(\beta^{1/(z\nu)}(U-U_c)),
\label{eqn:scaling_c2}
\end{equation}
as the FSS relation for $\Delta_0$. Our treatment of Eq.~\eqref{eqn:scaling_c2} 
is similar to the data-collapsing procedure of Refs.~\cite{beach2005data,Campostrini:2014lta}.
We define a universal (not explicitly $\beta$-dependent), smooth function $F$ such that
\begin{align}
u & \coloneqq \beta^\mu (U-U_c^{}), \quad \mu:= 1/(z\nu), \\
f & \coloneqq \beta \Delta_0^{},\\
f & \overset{!}{=} F(u),
\label{eq:scaling_for_collapse}
\end{align}
and by adjusting the parameters $\mu$ and $U_c$, we seek to minimize the dependence of the observable $\Delta_0$ on the system size~\cite{19881,Campostrini:2014lta}, in our
case the inverse temperature $\beta$. In practice, these parameters are determined such that all points of the (appropriately scaled) gap $\Delta_0$ lie on a single line in a $u$-$f$ plot.

In this work, we have not computed the AFMI order parameter (staggered magnetization) $m_s$. However, we can make use of the findings of \Ref{Assaad:2013xua} to
provide a first estimate of the associated critical exponent $\beta$ (which we denote by $\tilde\beta$, to avoid confusion with the inverse temperature $\beta$). Specifically, 
\Ref{Assaad:2013xua} found that describing $\Delta_0/U$ according to the FSS relation
\begin{equation}
\Delta_0^{}/U = \beta^{-\tilde\beta/(z\nu)} 
G(\beta^{1/(z\nu)}(U-U_c)),
\label{eqn:scaling_gg}
\end{equation}
produced a scaling function $G$ which was indistinguishable from the true scaling function for $m_s$, in spite of the \textit{ansatz} $m_s \sim \Delta/U$ being a
mean-field result. In our notation, this corresponds to
\begin{align}
g \coloneqq \beta^\zeta \Delta_0^{}/U & \overset{!}{=} G(u), \quad \zeta:= \tilde\beta/(z\nu),
\label{eq:scaling_for_collapse_magn}
\end{align}
similarly to Eq.~\eqref{eq:scaling_for_collapse}.

It should be noted that Refs.~\cite{Toldin:2014sxa, Otsuka:2015iba} also considered sub-leading corrections to the FSS relation for $m_s$.
Here, data points outside of the scaling region have been omitted instead. When taken together with additional fit parameters, 
they do not increase the significance of the fit. On the contrary, they reduce its stability. As a cutoff, we take $\beta U < 8$ for Eq.~\eqref{eq:scaling_for_collapse} and $\beta U < 10$ for 
Eq.~\eqref{eq:scaling_for_collapse_magn}. 
The reason for this limited scaling region can be understood as follows. The gap vanishes at $U=0$ regardless of $\beta$, thus $\Delta_0(u=-\beta^\mu U_c)=0$, and therefore data points with small 
$\beta$ and $U$ no longer collapse onto $F(u)$. Similarly, as we show in Appendix~\ref{sec:thermal_gap}, for $U \ll U_c$ we have $\Delta_0\sim U\beta^{-2}$, which implies $g\sim\beta^{\zeta-2}\approx\beta^{-1}$, and represents a strong deviation from the expected scaling. With decreasing $\beta$, the effect increases and materialises at larger $U$.
We shall revisit the issue of sub-leading corrections in a future MC study of $m_s$.

\begin{figure}[ht]
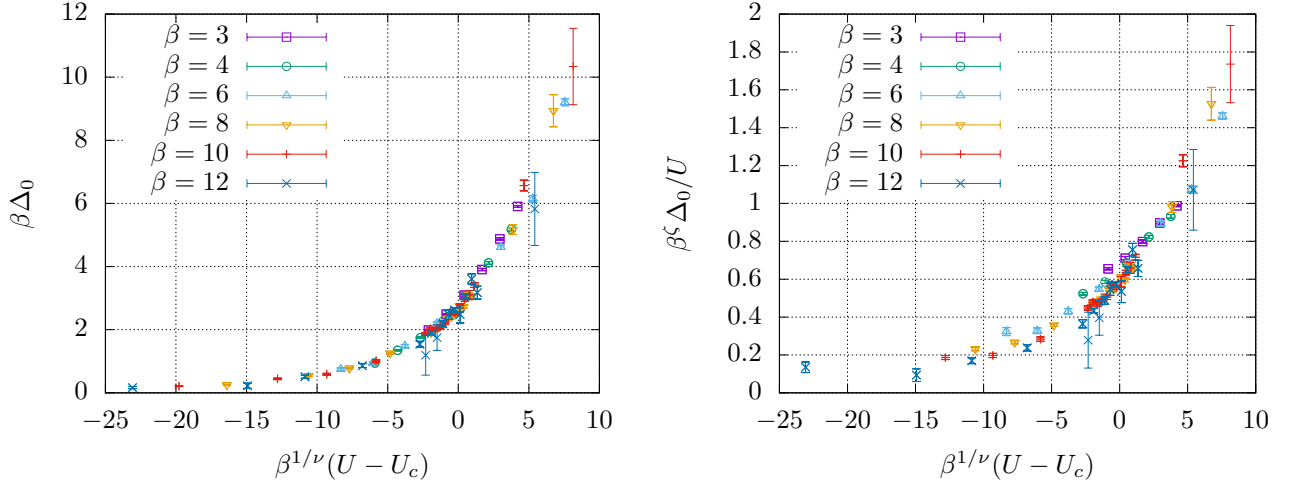

\begingroup
  \inputencoding{latin1}%
  \makeatletter
  \providecommand\color[2][]{%
    \GenericError{(gnuplot) \space\space\space\@spaces}{%
      Package color not loaded in conjunction with
      terminal option `colourtext'%
    }{See the gnuplot documentation for explanation.%
    }{Either use 'blacktext' in gnuplot or load the package
      color.sty in LaTeX.}%
    \renewcommand\color[2][]{}%
  }%
  \providecommand\includegraphics[2][]{%
    \GenericError{(gnuplot) \space\space\space\@spaces}{%
      Package graphicx or graphics not loaded%
    }{See the gnuplot documentation for explanation.%
    }{The gnuplot epslatex terminal needs graphicx.sty or graphics.sty.}%
    \renewcommand\includegraphics[2][]{}%
  }%
  \providecommand\rotatebox[2]{#2}%
  \@ifundefined{ifGPcolor}{%
    \newif\ifGPcolor
    \GPcolortrue
  }{}%
  \@ifundefined{ifGPblacktext}{%
    \newif\ifGPblacktext
    \GPblacktexttrue
  }{}%
  \let\gplgaddtomacro\g@addto@macro
  \gdef\gplbacktext{}%
  \gdef\gplfronttext{}%
  \makeatother
  \ifGPblacktext
    \def\colorrgb#1{}%
    \def\colorgray#1{}%
  \else
    \ifGPcolor
      \def\colorrgb#1{\color[rgb]{#1}}%
      \def\colorgray#1{\color[gray]{#1}}%
      \expandafter\def\csname LTw\endcsname{\color{white}}%
      \expandafter\def\csname LTb\endcsname{\color{black}}%
      \expandafter\def\csname LTa\endcsname{\color{black}}%
      \expandafter\def\csname LT0\endcsname{\color[rgb]{1,0,0}}%
      \expandafter\def\csname LT1\endcsname{\color[rgb]{0,1,0}}%
      \expandafter\def\csname LT2\endcsname{\color[rgb]{0,0,1}}%
      \expandafter\def\csname LT3\endcsname{\color[rgb]{1,0,1}}%
      \expandafter\def\csname LT4\endcsname{\color[rgb]{0,1,1}}%
      \expandafter\def\csname LT5\endcsname{\color[rgb]{1,1,0}}%
      \expandafter\def\csname LT6\endcsname{\color[rgb]{0,0,0}}%
      \expandafter\def\csname LT7\endcsname{\color[rgb]{1,0.3,0}}%
      \expandafter\def\csname LT8\endcsname{\color[rgb]{0.5,0.5,0.5}}%
    \else
      \def\colorrgb#1{\color{black}}%
      \def\colorgray#1{\color[gray]{#1}}%
      \expandafter\def\csname LTw\endcsname{\color{white}}%
      \expandafter\def\csname LTb\endcsname{\color{black}}%
      \expandafter\def\csname LTa\endcsname{\color{black}}%
      \expandafter\def\csname LT0\endcsname{\color{black}}%
      \expandafter\def\csname LT1\endcsname{\color{black}}%
      \expandafter\def\csname LT2\endcsname{\color{black}}%
      \expandafter\def\csname LT3\endcsname{\color{black}}%
      \expandafter\def\csname LT4\endcsname{\color{black}}%
      \expandafter\def\csname LT5\endcsname{\color{black}}%
      \expandafter\def\csname LT6\endcsname{\color{black}}%
      \expandafter\def\csname LT7\endcsname{\color{black}}%
      \expandafter\def\csname LT8\endcsname{\color{black}}%
    \fi
  \fi
    \setlength{\unitlength}{0.0500bp}%
    \ifx\gptboxheight\undefined%
      \newlength{\gptboxheight}%
      \newlength{\gptboxwidth}%
      \newsavebox{\gptboxtext}%
    \fi%
    \setlength{\fboxrule}{0.5pt}%
    \setlength{\fboxsep}{1pt}%
\begin{picture}(4896.00,3772.00)%
    \gplgaddtomacro\gplbacktext{%
      \csname LTb\endcsname%
      \put(682,704){\makebox(0,0)[r]{\strut{}$0$}}%
      \csname LTb\endcsname%
      \put(682,1179){\makebox(0,0)[r]{\strut{}$2$}}%
      \csname LTb\endcsname%
      \put(682,1653){\makebox(0,0)[r]{\strut{}$4$}}%
      \csname LTb\endcsname%
      \put(682,2128){\makebox(0,0)[r]{\strut{}$6$}}%
      \csname LTb\endcsname%
      \put(682,2602){\makebox(0,0)[r]{\strut{}$8$}}%
      \csname LTb\endcsname%
      \put(682,3077){\makebox(0,0)[r]{\strut{}$10$}}%
      \csname LTb\endcsname%
      \put(682,3551){\makebox(0,0)[r]{\strut{}$12$}}%
      \csname LTb\endcsname%
      \put(814,484){\makebox(0,0){\strut{}$-25$}}%
      \csname LTb\endcsname%
      \put(1340,484){\makebox(0,0){\strut{}$-20$}}%
      \csname LTb\endcsname%
      \put(1867,484){\makebox(0,0){\strut{}$-15$}}%
      \csname LTb\endcsname%
      \put(2393,484){\makebox(0,0){\strut{}$-10$}}%
      \csname LTb\endcsname%
      \put(2920,484){\makebox(0,0){\strut{}$-5$}}%
      \csname LTb\endcsname%
      \put(3446,484){\makebox(0,0){\strut{}$0$}}%
      \csname LTb\endcsname%
      \put(3973,484){\makebox(0,0){\strut{}$5$}}%
      \csname LTb\endcsname%
      \put(4499,484){\makebox(0,0){\strut{}$10$}}%
    }%
    \gplgaddtomacro\gplfronttext{%
      \csname LTb\endcsname%
      \put(198,2127){\rotatebox{-270}{\makebox(0,0){\strut{}$\beta\Delta_0$}}}%
      \put(2656,154){\makebox(0,0){\strut{}$\beta^{1/\nu}(U-U_c)$}}%
      \csname LTb\endcsname%
      \put(1738,3378){\makebox(0,0)[r]{\strut{}$\beta = 3$}}%
      \csname LTb\endcsname%
      \put(1738,3158){\makebox(0,0)[r]{\strut{}$\beta = 4$}}%
      \csname LTb\endcsname%
      \put(1738,2938){\makebox(0,0)[r]{\strut{}$\beta = 6$}}%
      \csname LTb\endcsname%
      \put(1738,2718){\makebox(0,0)[r]{\strut{}$\beta = 8$}}%
      \csname LTb\endcsname%
      \put(1738,2498){\makebox(0,0)[r]{\strut{}$\beta = 10$}}%
      \csname LTb\endcsname%
      \put(1738,2278){\makebox(0,0)[r]{\strut{}$\beta = 12$}}%
    }%
    \gplbacktext
    \put(0,0){\includegraphics{gap_collapse0-pdf}}%
    \gplfronttext
  \end{picture}%
\endgroup
\begingroup
  \inputencoding{latin1}%
  \makeatletter
  \providecommand\color[2][]{%
    \GenericError{(gnuplot) \space\space\space\@spaces}{%
      Package color not loaded in conjunction with
      terminal option `colourtext'%
    }{See the gnuplot documentation for explanation.%
    }{Either use 'blacktext' in gnuplot or load the package
      color.sty in LaTeX.}%
    \renewcommand\color[2][]{}%
  }%
  \providecommand\includegraphics[2][]{%
    \GenericError{(gnuplot) \space\space\space\@spaces}{%
      Package graphicx or graphics not loaded%
    }{See the gnuplot documentation for explanation.%
    }{The gnuplot epslatex terminal needs graphicx.sty or graphics.sty.}%
    \renewcommand\includegraphics[2][]{}%
  }%
  \providecommand\rotatebox[2]{#2}%
  \@ifundefined{ifGPcolor}{%
    \newif\ifGPcolor
    \GPcolortrue
  }{}%
  \@ifundefined{ifGPblacktext}{%
    \newif\ifGPblacktext
    \GPblacktexttrue
  }{}%
  \let\gplgaddtomacro\g@addto@macro
  \gdef\gplbacktext{}%
  \gdef\gplfronttext{}%
  \makeatother
  \ifGPblacktext
    \def\colorrgb#1{}%
    \def\colorgray#1{}%
  \else
    \ifGPcolor
      \def\colorrgb#1{\color[rgb]{#1}}%
      \def\colorgray#1{\color[gray]{#1}}%
      \expandafter\def\csname LTw\endcsname{\color{white}}%
      \expandafter\def\csname LTb\endcsname{\color{black}}%
      \expandafter\def\csname LTa\endcsname{\color{black}}%
      \expandafter\def\csname LT0\endcsname{\color[rgb]{1,0,0}}%
      \expandafter\def\csname LT1\endcsname{\color[rgb]{0,1,0}}%
      \expandafter\def\csname LT2\endcsname{\color[rgb]{0,0,1}}%
      \expandafter\def\csname LT3\endcsname{\color[rgb]{1,0,1}}%
      \expandafter\def\csname LT4\endcsname{\color[rgb]{0,1,1}}%
      \expandafter\def\csname LT5\endcsname{\color[rgb]{1,1,0}}%
      \expandafter\def\csname LT6\endcsname{\color[rgb]{0,0,0}}%
      \expandafter\def\csname LT7\endcsname{\color[rgb]{1,0.3,0}}%
      \expandafter\def\csname LT8\endcsname{\color[rgb]{0.5,0.5,0.5}}%
    \else
      \def\colorrgb#1{\color{black}}%
      \def\colorgray#1{\color[gray]{#1}}%
      \expandafter\def\csname LTw\endcsname{\color{white}}%
      \expandafter\def\csname LTb\endcsname{\color{black}}%
      \expandafter\def\csname LTa\endcsname{\color{black}}%
      \expandafter\def\csname LT0\endcsname{\color{black}}%
      \expandafter\def\csname LT1\endcsname{\color{black}}%
      \expandafter\def\csname LT2\endcsname{\color{black}}%
      \expandafter\def\csname LT3\endcsname{\color{black}}%
      \expandafter\def\csname LT4\endcsname{\color{black}}%
      \expandafter\def\csname LT5\endcsname{\color{black}}%
      \expandafter\def\csname LT6\endcsname{\color{black}}%
      \expandafter\def\csname LT7\endcsname{\color{black}}%
      \expandafter\def\csname LT8\endcsname{\color{black}}%
    \fi
  \fi
    \setlength{\unitlength}{0.0500bp}%
    \ifx\gptboxheight\undefined%
      \newlength{\gptboxheight}%
      \newlength{\gptboxwidth}%
      \newsavebox{\gptboxtext}%
    \fi%
    \setlength{\fboxrule}{0.5pt}%
    \setlength{\fboxsep}{1pt}%
\begin{picture}(4896.00,3772.00)%
    \gplgaddtomacro\gplbacktext{%
      \csname LTb\endcsname%
      \put(814,704){\makebox(0,0)[r]{\strut{}$0$}}%
      \csname LTb\endcsname%
      \put(814,989){\makebox(0,0)[r]{\strut{}$0.2$}}%
      \csname LTb\endcsname%
      \put(814,1273){\makebox(0,0)[r]{\strut{}$0.4$}}%
      \csname LTb\endcsname%
      \put(814,1558){\makebox(0,0)[r]{\strut{}$0.6$}}%
      \csname LTb\endcsname%
      \put(814,1843){\makebox(0,0)[r]{\strut{}$0.8$}}%
      \csname LTb\endcsname%
      \put(814,2128){\makebox(0,0)[r]{\strut{}$1$}}%
      \csname LTb\endcsname%
      \put(814,2412){\makebox(0,0)[r]{\strut{}$1.2$}}%
      \csname LTb\endcsname%
      \put(814,2697){\makebox(0,0)[r]{\strut{}$1.4$}}%
      \csname LTb\endcsname%
      \put(814,2982){\makebox(0,0)[r]{\strut{}$1.6$}}%
      \csname LTb\endcsname%
      \put(814,3266){\makebox(0,0)[r]{\strut{}$1.8$}}%
      \csname LTb\endcsname%
      \put(814,3551){\makebox(0,0)[r]{\strut{}$2$}}%
      \csname LTb\endcsname%
      \put(946,484){\makebox(0,0){\strut{}$-25$}}%
      \csname LTb\endcsname%
      \put(1454,484){\makebox(0,0){\strut{}$-20$}}%
      \csname LTb\endcsname%
      \put(1961,484){\makebox(0,0){\strut{}$-15$}}%
      \csname LTb\endcsname%
      \put(2469,484){\makebox(0,0){\strut{}$-10$}}%
      \csname LTb\endcsname%
      \put(2976,484){\makebox(0,0){\strut{}$-5$}}%
      \csname LTb\endcsname%
      \put(3484,484){\makebox(0,0){\strut{}$0$}}%
      \csname LTb\endcsname%
      \put(3991,484){\makebox(0,0){\strut{}$5$}}%
      \csname LTb\endcsname%
      \put(4499,484){\makebox(0,0){\strut{}$10$}}%
    }%
    \gplgaddtomacro\gplfronttext{%
      \csname LTb\endcsname%
      \put(198,2127){\rotatebox{-270}{\makebox(0,0){\strut{}$\beta^\zeta\Delta_0/U$}}}%
      \put(2722,154){\makebox(0,0){\strut{}$\beta^{1/\nu}(U-U_c)$}}%
      \csname LTb\endcsname%
      \put(1870,3378){\makebox(0,0)[r]{\strut{}$\beta = 3$}}%
      \csname LTb\endcsname%
      \put(1870,3158){\makebox(0,0)[r]{\strut{}$\beta = 4$}}%
      \csname LTb\endcsname%
      \put(1870,2938){\makebox(0,0)[r]{\strut{}$\beta = 6$}}%
      \csname LTb\endcsname%
      \put(1870,2718){\makebox(0,0)[r]{\strut{}$\beta = 8$}}%
      \csname LTb\endcsname%
      \put(1870,2498){\makebox(0,0)[r]{\strut{}$\beta = 10$}}%
      \csname LTb\endcsname%
      \put(1870,2278){\makebox(0,0)[r]{\strut{}$\beta = 12$}}%
    }%
    \gplbacktext
    \put(0,0){\includegraphics{gap_collapse1-pdf}}%
    \gplfronttext
  \end{picture}%
\endgroup
 	\caption{$f = \beta\Delta_0$ (left panel) and $g = \beta^\zeta\Delta_0/U$ (right panel) as a function of $u = \beta^\mu(U-U_c)$ according to Eqs.~\eqref{eq:scaling_for_collapse} and~\eqref{eq:scaling_for_collapse_magn}, respectively. All quantities are plotted with the parameters for optimal data collapse, and expressed in units of $\kappa$.}
	\label{fig:fully_collapsed_data}
\end{figure}

We perform the data-collapse analysis by first interpolating the data using a smoothed cubic spline for each
value of $\beta$. Next we calculate the squared differences between the interpolations for each pair of two different temperatures and integrate over these squared differences. Last the sum over all the integrals is minimised. The resulting optimal data collapse plots are shown 
in \Figref{fully_collapsed_data}. The optimization of Eq.~\eqref{eq:scaling_for_collapse} is performed first (see \Figref{fully_collapsed_data}, left panel). This 
yields $\mu=\critMu$, therefore $z\nu=\critNu$, and $U_c/\kappa=\critU$, where the errors have been determined using parametric bootstrap. 
Second, the data collapse is performed for Eq.~\eqref{eq:scaling_for_collapse_magn} (see \Figref{fully_collapsed_data}, right panel) 
which gives $\zeta=\critZeta$, such that $\tilde\beta=\critExp$. While the $u$-$f$ collapse is satisfactory, the $u$-$g$ collapse does not 
materialize for $u \ll 0$. This is not surprising, because our scaling \textit{ansatz} does not account for the thermal gap, which we discuss in \Appref{thermal_gap}. 
Moreover, these findings are consistent with the notion that the AFMI state predicted by mean-field theory is destroyed by quantum fluctuations at weak coupling, at which point
the relation $m_s \sim \Delta/U$ ceases to be valid.
Also, note that in \Ref{Assaad:2013xua} the plot corresponding to our \Figref{fully_collapsed_data} starts at $u=-2$, a region where our $u$-$g$ collapse works 
out as well (for sufficiently large $\beta$).

We briefly describe our method for determining the errors of the fitted quantities and their correlation matrix. We use a mixed error propagation scheme, consisting of a parametric bootstrap part 
as before, and an additional influence due to a direct propagation of the bootstrap samples of $\mu$ and $U_c$. For every $(\mu,U_c)$-sample we generate a new data set, following the 
parametric sampling system. Then, the variation of the composite samples should mirror the total uncertainty of the results.
The correlation matrix\footnote{We give enough digits to yield percent-level matching to our full numerical results when inverting the correlation matrix.}
\begin{align}
	\mathrm{corr}(U_c,\, z\nu,\, \tilde\beta) &= \matr{S[round-mode=places,round-precision = 4]S[round-mode=places,round-precision = 4]S[round-mode=places,round-precision = 4]}{
	 1.00000000 & -0.01987849 & -0.2330682\\
	-0.01987849 &  1.00000000 &  0.8649537\\
	-0.23306820 &  0.86495372 &  1.0000000
		}
\end{align}
clearly shows a strong correlation between $z\nu$ and $\tilde\beta$ (due to the influence of $\mu$ on both of them), whereas $U_c$ is weakly correlated.

\begin{figure}[ht]
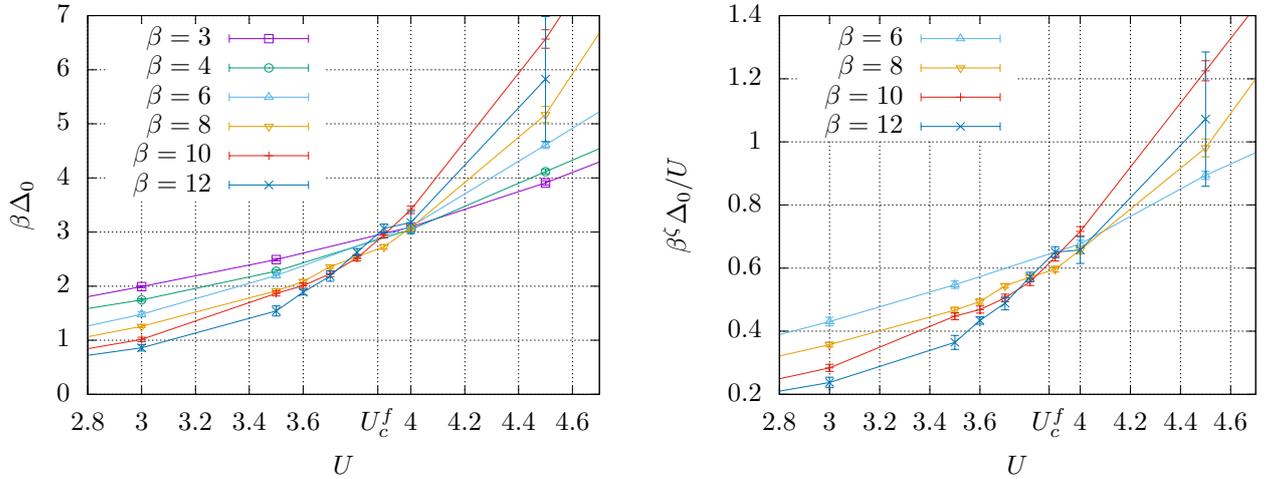

\begingroup
  \inputencoding{latin1}%
  \makeatletter
  \providecommand\color[2][]{%
    \GenericError{(gnuplot) \space\space\space\@spaces}{%
      Package color not loaded in conjunction with
      terminal option `colourtext'%
    }{See the gnuplot documentation for explanation.%
    }{Either use 'blacktext' in gnuplot or load the package
      color.sty in LaTeX.}%
    \renewcommand\color[2][]{}%
  }%
  \providecommand\includegraphics[2][]{%
    \GenericError{(gnuplot) \space\space\space\@spaces}{%
      Package graphicx or graphics not loaded%
    }{See the gnuplot documentation for explanation.%
    }{The gnuplot epslatex terminal needs graphicx.sty or graphics.sty.}%
    \renewcommand\includegraphics[2][]{}%
  }%
  \providecommand\rotatebox[2]{#2}%
  \@ifundefined{ifGPcolor}{%
    \newif\ifGPcolor
    \GPcolortrue
  }{}%
  \@ifundefined{ifGPblacktext}{%
    \newif\ifGPblacktext
    \GPblacktexttrue
  }{}%
  \let\gplgaddtomacro\g@addto@macro
  \gdef\gplbacktext{}%
  \gdef\gplfronttext{}%
  \makeatother
  \ifGPblacktext
    \def\colorrgb#1{}%
    \def\colorgray#1{}%
  \else
    \ifGPcolor
      \def\colorrgb#1{\color[rgb]{#1}}%
      \def\colorgray#1{\color[gray]{#1}}%
      \expandafter\def\csname LTw\endcsname{\color{white}}%
      \expandafter\def\csname LTb\endcsname{\color{black}}%
      \expandafter\def\csname LTa\endcsname{\color{black}}%
      \expandafter\def\csname LT0\endcsname{\color[rgb]{1,0,0}}%
      \expandafter\def\csname LT1\endcsname{\color[rgb]{0,1,0}}%
      \expandafter\def\csname LT2\endcsname{\color[rgb]{0,0,1}}%
      \expandafter\def\csname LT3\endcsname{\color[rgb]{1,0,1}}%
      \expandafter\def\csname LT4\endcsname{\color[rgb]{0,1,1}}%
      \expandafter\def\csname LT5\endcsname{\color[rgb]{1,1,0}}%
      \expandafter\def\csname LT6\endcsname{\color[rgb]{0,0,0}}%
      \expandafter\def\csname LT7\endcsname{\color[rgb]{1,0.3,0}}%
      \expandafter\def\csname LT8\endcsname{\color[rgb]{0.5,0.5,0.5}}%
    \else
      \def\colorrgb#1{\color{black}}%
      \def\colorgray#1{\color[gray]{#1}}%
      \expandafter\def\csname LTw\endcsname{\color{white}}%
      \expandafter\def\csname LTb\endcsname{\color{black}}%
      \expandafter\def\csname LTa\endcsname{\color{black}}%
      \expandafter\def\csname LT0\endcsname{\color{black}}%
      \expandafter\def\csname LT1\endcsname{\color{black}}%
      \expandafter\def\csname LT2\endcsname{\color{black}}%
      \expandafter\def\csname LT3\endcsname{\color{black}}%
      \expandafter\def\csname LT4\endcsname{\color{black}}%
      \expandafter\def\csname LT5\endcsname{\color{black}}%
      \expandafter\def\csname LT6\endcsname{\color{black}}%
      \expandafter\def\csname LT7\endcsname{\color{black}}%
      \expandafter\def\csname LT8\endcsname{\color{black}}%
    \fi
  \fi
    \setlength{\unitlength}{0.0500bp}%
    \ifx\gptboxheight\undefined%
      \newlength{\gptboxheight}%
      \newlength{\gptboxwidth}%
      \newsavebox{\gptboxtext}%
    \fi%
    \setlength{\fboxrule}{0.5pt}%
    \setlength{\fboxsep}{1pt}%
\begin{picture}(4896.00,3772.00)%
    \gplgaddtomacro\gplbacktext{%
      \csname LTb\endcsname%
      \put(550,704){\makebox(0,0)[r]{\strut{}$0$}}%
      \csname LTb\endcsname%
      \put(550,1111){\makebox(0,0)[r]{\strut{}$1$}}%
      \csname LTb\endcsname%
      \put(550,1517){\makebox(0,0)[r]{\strut{}$2$}}%
      \csname LTb\endcsname%
      \put(550,1924){\makebox(0,0)[r]{\strut{}$3$}}%
      \csname LTb\endcsname%
      \put(550,2331){\makebox(0,0)[r]{\strut{}$4$}}%
      \csname LTb\endcsname%
      \put(550,2738){\makebox(0,0)[r]{\strut{}$5$}}%
      \csname LTb\endcsname%
      \put(550,3144){\makebox(0,0)[r]{\strut{}$6$}}%
      \csname LTb\endcsname%
      \put(550,3551){\makebox(0,0)[r]{\strut{}$7$}}%
      \csname LTb\endcsname%
      \put(682,484){\makebox(0,0){\strut{}$2.8$}}%
      \csname LTb\endcsname%
      \put(1084,484){\makebox(0,0){\strut{}$3$}}%
      \csname LTb\endcsname%
      \put(1486,484){\makebox(0,0){\strut{}$3.2$}}%
      \csname LTb\endcsname%
      \put(1887,484){\makebox(0,0){\strut{}$3.4$}}%
      \csname LTb\endcsname%
      \put(2289,484){\makebox(0,0){\strut{}$3.6$}}%
      \csname LTb\endcsname%
      \put(2846,484){\makebox(0,0){\strut{}$U_c^f$}}%
      \csname LTb\endcsname%
      \put(3093,484){\makebox(0,0){\strut{}$4$}}%
      \csname LTb\endcsname%
      \put(3495,484){\makebox(0,0){\strut{}$4.2$}}%
      \csname LTb\endcsname%
      \put(3896,484){\makebox(0,0){\strut{}$4.4$}}%
      \csname LTb\endcsname%
      \put(4298,484){\makebox(0,0){\strut{}$4.6$}}%
    }%
    \gplgaddtomacro\gplfronttext{%
      \csname LTb\endcsname%
      \put(198,2127){\rotatebox{-270}{\makebox(0,0){\strut{}$\beta\Delta_0$}}}%
      \put(2590,154){\makebox(0,0){\strut{}$U$}}%
      \csname LTb\endcsname%
      \put(1606,3378){\makebox(0,0)[r]{\strut{}$\beta = 3$}}%
      \csname LTb\endcsname%
      \put(1606,3158){\makebox(0,0)[r]{\strut{}$\beta = 4$}}%
      \csname LTb\endcsname%
      \put(1606,2938){\makebox(0,0)[r]{\strut{}$\beta = 6$}}%
      \csname LTb\endcsname%
      \put(1606,2718){\makebox(0,0)[r]{\strut{}$\beta = 8$}}%
      \csname LTb\endcsname%
      \put(1606,2498){\makebox(0,0)[r]{\strut{}$\beta = 10$}}%
      \csname LTb\endcsname%
      \put(1606,2278){\makebox(0,0)[r]{\strut{}$\beta = 12$}}%
    }%
    \gplbacktext
    \put(0,0){\includegraphics{gap_crossing-pdf}}%
    \gplfronttext
  \end{picture}%
\endgroup
\begingroup
  \inputencoding{latin1}%
  \makeatletter
  \providecommand\color[2][]{%
    \GenericError{(gnuplot) \space\space\space\@spaces}{%
      Package color not loaded in conjunction with
      terminal option `colourtext'%
    }{See the gnuplot documentation for explanation.%
    }{Either use 'blacktext' in gnuplot or load the package
      color.sty in LaTeX.}%
    \renewcommand\color[2][]{}%
  }%
  \providecommand\includegraphics[2][]{%
    \GenericError{(gnuplot) \space\space\space\@spaces}{%
      Package graphicx or graphics not loaded%
    }{See the gnuplot documentation for explanation.%
    }{The gnuplot epslatex terminal needs graphicx.sty or graphics.sty.}%
    \renewcommand\includegraphics[2][]{}%
  }%
  \providecommand\rotatebox[2]{#2}%
  \@ifundefined{ifGPcolor}{%
    \newif\ifGPcolor
    \GPcolortrue
  }{}%
  \@ifundefined{ifGPblacktext}{%
    \newif\ifGPblacktext
    \GPblacktexttrue
  }{}%
  \let\gplgaddtomacro\g@addto@macro
  \gdef\gplbacktext{}%
  \gdef\gplfronttext{}%
  \makeatother
  \ifGPblacktext
    \def\colorrgb#1{}%
    \def\colorgray#1{}%
  \else
    \ifGPcolor
      \def\colorrgb#1{\color[rgb]{#1}}%
      \def\colorgray#1{\color[gray]{#1}}%
      \expandafter\def\csname LTw\endcsname{\color{white}}%
      \expandafter\def\csname LTb\endcsname{\color{black}}%
      \expandafter\def\csname LTa\endcsname{\color{black}}%
      \expandafter\def\csname LT0\endcsname{\color[rgb]{1,0,0}}%
      \expandafter\def\csname LT1\endcsname{\color[rgb]{0,1,0}}%
      \expandafter\def\csname LT2\endcsname{\color[rgb]{0,0,1}}%
      \expandafter\def\csname LT3\endcsname{\color[rgb]{1,0,1}}%
      \expandafter\def\csname LT4\endcsname{\color[rgb]{0,1,1}}%
      \expandafter\def\csname LT5\endcsname{\color[rgb]{1,1,0}}%
      \expandafter\def\csname LT6\endcsname{\color[rgb]{0,0,0}}%
      \expandafter\def\csname LT7\endcsname{\color[rgb]{1,0.3,0}}%
      \expandafter\def\csname LT8\endcsname{\color[rgb]{0.5,0.5,0.5}}%
    \else
      \def\colorrgb#1{\color{black}}%
      \def\colorgray#1{\color[gray]{#1}}%
      \expandafter\def\csname LTw\endcsname{\color{white}}%
      \expandafter\def\csname LTb\endcsname{\color{black}}%
      \expandafter\def\csname LTa\endcsname{\color{black}}%
      \expandafter\def\csname LT0\endcsname{\color{black}}%
      \expandafter\def\csname LT1\endcsname{\color{black}}%
      \expandafter\def\csname LT2\endcsname{\color{black}}%
      \expandafter\def\csname LT3\endcsname{\color{black}}%
      \expandafter\def\csname LT4\endcsname{\color{black}}%
      \expandafter\def\csname LT5\endcsname{\color{black}}%
      \expandafter\def\csname LT6\endcsname{\color{black}}%
      \expandafter\def\csname LT7\endcsname{\color{black}}%
      \expandafter\def\csname LT8\endcsname{\color{black}}%
    \fi
  \fi
    \setlength{\unitlength}{0.0500bp}%
    \ifx\gptboxheight\undefined%
      \newlength{\gptboxheight}%
      \newlength{\gptboxwidth}%
      \newsavebox{\gptboxtext}%
    \fi%
    \setlength{\fboxrule}{0.5pt}%
    \setlength{\fboxsep}{1pt}%
\begin{picture}(4896.00,3772.00)%
    \gplgaddtomacro\gplbacktext{%
      \csname LTb\endcsname%
      \put(814,704){\makebox(0,0)[r]{\strut{}$0.2$}}%
      \csname LTb\endcsname%
      \put(814,1179){\makebox(0,0)[r]{\strut{}$0.4$}}%
      \csname LTb\endcsname%
      \put(814,1653){\makebox(0,0)[r]{\strut{}$0.6$}}%
      \csname LTb\endcsname%
      \put(814,2128){\makebox(0,0)[r]{\strut{}$0.8$}}%
      \csname LTb\endcsname%
      \put(814,2602){\makebox(0,0)[r]{\strut{}$1$}}%
      \csname LTb\endcsname%
      \put(814,3076){\makebox(0,0)[r]{\strut{}$1.2$}}%
      \csname LTb\endcsname%
      \put(814,3551){\makebox(0,0)[r]{\strut{}$1.4$}}%
      \csname LTb\endcsname%
      \put(946,484){\makebox(0,0){\strut{}$2.8$}}%
      \csname LTb\endcsname%
      \put(1320,484){\makebox(0,0){\strut{}$3$}}%
      \csname LTb\endcsname%
      \put(1694,484){\makebox(0,0){\strut{}$3.2$}}%
      \csname LTb\endcsname%
      \put(2068,484){\makebox(0,0){\strut{}$3.4$}}%
      \csname LTb\endcsname%
      \put(2442,484){\makebox(0,0){\strut{}$3.6$}}%
      \csname LTb\endcsname%
      \put(2958,484){\makebox(0,0){\strut{}$U_c^f$}}%
      \csname LTb\endcsname%
      \put(3190,484){\makebox(0,0){\strut{}$4$}}%
      \csname LTb\endcsname%
      \put(3564,484){\makebox(0,0){\strut{}$4.2$}}%
      \csname LTb\endcsname%
      \put(3938,484){\makebox(0,0){\strut{}$4.4$}}%
      \csname LTb\endcsname%
      \put(4312,484){\makebox(0,0){\strut{}$4.6$}}%
    }%
    \gplgaddtomacro\gplfronttext{%
      \csname LTb\endcsname%
      \put(198,2127){\rotatebox{-270}{\makebox(0,0){\strut{}$\beta^\zeta\Delta_0/U$}}}%
      \put(2722,154){\makebox(0,0){\strut{}$U$}}%
      \csname LTb\endcsname%
      \put(1870,3378){\makebox(0,0)[r]{\strut{}$\beta = 6$}}%
      \csname LTb\endcsname%
      \put(1870,3158){\makebox(0,0)[r]{\strut{}$\beta = 8$}}%
      \csname LTb\endcsname%
      \put(1870,2938){\makebox(0,0)[r]{\strut{}$\beta = 10$}}%
      \csname LTb\endcsname%
      \put(1870,2718){\makebox(0,0)[r]{\strut{}$\beta = 12$}}%
    }%
    \gplbacktext
    \put(0,0){\includegraphics{gap_crossing1-pdf}}%
    \gplfronttext
  \end{picture}%
\endgroup
 	\caption{Single-particle gap $\Delta_0$ scaled by $\beta$ (left panel) and by $\beta^\zeta/U$ (right panel), as a function of $U$. Note that $\Delta_0$ has been extrapolated to infinite volume
	and $\delta \to 0$ (continuum limit). Each line represents a given inverse temperature $\beta$. Here, $\zeta=\critZeta$ as obtained from the data collapse fit.
	The averages of all the crossing points $U_c^f$ and $U_c^g$ have been marked by vertical lines. For $U_c^g$, data with $\beta\le 4$ have been omitted due to the thermal gap discussed 
	in \Appref{thermal_gap}. All quantities are given in units of $\kappa$.
	\label{fig:crossing}}
\end{figure}

From the definitions of the scaling functions $F$ and $G$, one finds that they become independent of $\beta$ at $U = U_c$. Therefore, all lines in a 
$U$-$f$ and $U$-$g$ plot should cross at $U=U_c$. The average of all the crossing points (between pairs of lines with different $\beta$) yields an estimate of $U_c$ as well.
We obtain $U_c^{f}/\kappa=\critUf$ from the crossings in the $U$-$f$ plot (see \Figref{crossing}, left panel) and $U_c^{g}/\kappa=\critUg$ from the $U$-$g$ plot 
(see \Figref{crossing}, right panel). The errors are estimated from the standard deviation of all the crossings $\sigma_c$ and the number of crossings $n_c$ as $\sigma_c/\sqrt{n_c}$.
We find that all three results for $U_c$ are compatible within errors. However, the result from the data collapse analysis is much more precise than the crossing analysis because the data collapse analysis performs a global fit over all the MC data points, not just the points in the immediate vicinity of $U_c$.

\subsection{Zero temperature extrapolation}\label{sec:zero temperature extrapolation}

Let us first consider the scaling properties as a function of $\beta$ for $U = U_c$. As the scaling function reduces to a constant, we have
\begin{align}
\Delta_0^{} \propto \beta^{-1}\,.
\label{eq:scaling_beta}
\end{align}
For $U > U_c$ and $\beta \to \infty$, we have $\mathfrak{O} \sim (U-U_c)^q$ by construction
and therefore
\begin{align}
\Delta_0^{} & = \begin{cases}
0 & U \le U_c \\
c_U^{} (U-U_c)^{z\nu} & U > U_c
\end{cases}\label{eq:scaling_U} %
\end{align}
where $c_U^{}$ %
is a constant of proportionality.

We are now in a position to perform a simple extrapolation of $\Delta_0$ to $\beta \to \infty$, in order to visualize the quantum phase transition.
Let us assume that the scaling in Eq.~\eqref{eq:scaling_beta} holds approximately for $U$ slightly above $U_c$, up to a constant shift $\Delta_0^\infty$, 
the zero-temperature gap. Thus we fit
\begin{align}
\Delta_0^{} & = \Delta_0^\infty + c_\beta^{}\beta^{-1},
\end{align}
at a chosen value of $U/\kappa = 4$. The resulting value of $\Delta_0^\infty/\kappa=\DeltaInf$ then allows us to determine the 
coefficient $c_U = \coeffU$ (in units of $\kappa$) of Eq.~\eqref{eq:scaling_U}. This allows us to plot the 
solid black line with error bands in \Figref{delta_against_U}. Such an extrapolation should be regarded as valid only in the immediate 
vicinity of the phase transition. For $U \gg U_c$ the data seem to approach the mean-field result $\tilde\beta = 1/2$~\cite{Sorella_1992}. Furthermore, 
we note that an inflection point has been observed in~\Ref{Assaad:2013xua} at $U/\kappa\approx 4.1$, though this effect is thought to be an artifact of the extrapolation of the
MC data in the Trotter error $\delta$. The error estimation for the extrapolation $\beta \to \infty$ follows the same scheme as the one for the $u$-$g$ data collapse. 
We obtain the error band in \Figref{delta_against_U} as the area enclosed by the two lines corresponding to the lower bound of $U_c$ and the upper bound 
of $z\nu$ (on the left) and vice versa (on the right). This method conservatively captures any correlation between the different parameters. 

\section{Conclusions}\label{sec:conclusion}

Our work represents the first instance where the grand canonical BRS algorithm has been applied to the hexagonal Hubbard model 
(beyond mere proofs of principle), and we have found highly promising results. We emphasize that previously encountered issues related to the computational scaling and ergodicity 
of the HMC updates have been solved~\cite{Wynen:2018ryx}. 
We have primarily investigated the single-particle gap $\Delta$ (which we assume to be due to the semimetal-AFMI transition) as a function of $U/\kappa$, 
along with a comprehensive analysis of the temporal continuum, thermodynamic and zero-temperature limits. The favorable scaling of the HMC
enabled us to simulate lattices with $L > 100$ and to
perform a highly systematic treatment of all three limits. The latter limit was taken by means of a finite-size scaling analysis, which determines the
critical coupling $U_c/\kappa=\critU$ and the critical exponent $z\nu=\critNu$.
While we have not yet performed a direct MC calculation of the AFMI order parameter $m_s$, our scaling analysis of $\Delta/U$ has enabled an estimate of
the critical exponent $\tilde\beta=\critExp$. Depending on which symmetry is broken, the critical exponents of the hexagonal Hubbard model 
are expected to fall into one of the Gross-Neveu (GN) universality classes~\cite{PhysRevLett.97.146401}. The semimetal-AFMI transition should fall into the GN-Heisenberg 
$SU(2)$ universality class, as $m_s$ is described by a vector with three components. 

The GN-Heisenberg critical exponents have been studied by 
means of PMC simulations of the hexagonal Hubbard model, by the $d = 4-\epsilon$ expansion around the upper 
critical dimension $d$, by large~$N$ calculations, and by functional renormalization group (FRG) methods.
In Table~\ref{tab:crit_values}, we give an up-to-date comparison with our results. Our value for $U_c/\kappa$ is in overall agreement with previous MC simulations. 
For the critical exponents $\nu$ and $\tilde\beta$, the situation is less clear. Our results for $\nu$ (assuming $z = 1$ due to Lorentz invariance~\cite{PhysRevLett.97.146401}) and $\tilde\beta$ agree 
best with the HMC calculation (in the BSS formulation) of Ref.~\cite{Buividovich:2018yar}, followed by the FRG and large~$N$ calculations. 
On the other hand, our critical exponents are systematically larger than most PMC calculations and first-order $4-\epsilon$ expansion results. The agreement appears to be significantly 
improved when the $4-\epsilon$ expansion is taken to higher orders, although the discrepancy between expansions for $\nu$ and $1/\nu$ persists.
\begin{table}
	\caption{Summary of critical couplings $U_c/\kappa$ and critical exponents $\nu$ and $\upbeta$ (called $\tilde\beta$ in the rest of this chapter) obtained by recent MC calculations of 
	various Hubbard models in the Gross-Neveu (GN) Heisenberg universality class, and with other methods for direct calculations of the GN Heisenberg model.
	We include brief comments of special features of each
	calculation. Note the abbreviations HMC (Hybrid Monte Carlo), AF (Auxiliary Field), 
	BSS (Blankenbecler-Sugar-Scalapino) and BRS (Brower-Rebbi-Schaich). These concepts are explained in the main text. 
	Furthermore, we denote FRG (Functional Renormalization Group).
	Our value of $\nu$~($\dag$) is given for $z = 1$~\cite{PhysRevLett.97.146401}.
	Our first estimate of $\upbeta$ from~\cite{semimetalmott} is based on the mean-field result $m_s \sim \Delta/U$~($\ddag$)~\cite{Assaad:2013xua}.
	The asterisk (*) indicates that the $4-\epsilon$ exponents of Ref.~\cite{Herbut:2009vu} were used as input in the
	MC calculation of $U_c$ in Ref.~\cite{Assaad:2013xua}. Also, note the ambiguities~\cite{Otsuka:2015iba} as to the correct number of fermion components in the $4-\epsilon$ expansion
	of Ref.~\cite{Rosenstein:1993zf}.\\
	The table has been updated by our most recent results~\cite{more_observables}, see \cref{chap:afm_character} for details.
	\label{tab:crit_values}}
	\vspace{.2cm}
	\begin{tabular}{l*{3}{S[table-format=1.7]}}
		Method & $U_c/\kappa$ & $\nu$ & $\upbeta$ \\
		\hline
		Grand canonical BRS HMC (\cite{semimetalmott}, present chapter) & 3.834(14) & 1.185(43)$^\dag$ & 1.095(37)$^\ddag$ \\
		Grand canonical BRS HMC (\cite{more_observables}, \cref{chap:afm_character}) & 3.835(14) & 1.181(43)$^\dag$ & 0.898(37) \\
		Grand canonical BSS HMC, complex AF~\cite{Buividovich:2018yar} & 3.90(5) & 1.162 & 1.08(2) \\
		Grand canonical BSS QMC~\cite{Buividovich:2018crq} & 3.94 & 0.93 & 0.75 \\
		Projection BSS QMC~\cite{Otsuka:2015iba} & 3.85(2) & 1.02(1) & 0.76(2) \\
		Projection BSS QMC, $d$-wave pairing field~\cite{Otsuka:2020lhc} & & 1.05(5) & \\
		Projection BSS QMC~\cite{Toldin:2014sxa} & 3.80(1) & 0.84(4) & 0.71(8) \\
		Projection BSS QMC, spin-Hall transition~\cite{Liu:2018sww} & & 0.88(7) & \\
		Projection BSS QMC, pinning field~\cite{Assaad:2013xua} & 3.78 & 0.882* & 0.794* \\
		GN $4-\epsilon$ expansion, 1st order~\cite{Herbut:2009vu, Otsuka:2015iba} & & 0.882* & 0.794* \\
		GN $4-\epsilon$ expansion, 1st order~\cite{Rosenstein:1993zf, Otsuka:2015iba} & & 0.851 & 0.824 \\
		GN $4-\epsilon$ expansion, 2nd order~\cite{Rosenstein:1993zf, Otsuka:2015iba} & & 1.01 & 0.995 \\
		GN $4-\epsilon$ expansion, $\nu$ 2nd order~\cite{Rosenstein:1993zf,heisenberg_gross_neveu} & & 1.08 & 1.06 \\
		GN $4-\epsilon$ expansion, $1/\nu$ 2nd order~\cite{Rosenstein:1993zf,heisenberg_gross_neveu} & & 1.20 & 1.17 \\
		GN $4-\epsilon$ expansion, $\nu$ 4th order~\cite{Zerf:2017zqi} & & 1.2352 & \\
		GN $4-\epsilon$ expansion, $1/\nu$ 4th order~\cite{Zerf:2017zqi} & & 1.5511 & \\
		GN FRG~\cite{heisenberg_gross_neveu} & & 1.31 & 1.32 \\
		GN FRG~\cite{Knorr:2017yze} & & 1.26 & \\
		GN Large $N$~\cite{Gracey:2018qba} & & 1.1823 &
	\end{tabular}
\end{table}

Our results show that the BRS algorithm is now applicable to problems of a realistic size in the field of carbon-based nano-materials.
There are several future directions in which our present work can be developed. For instance, while the AFMI phase may not be directly observable in graphene, 
we note that tentative empirical evidence for such a phase exists in carbon nanotubes~\cite{Deshpande02012009}, along with preliminary theoretical 
evidence from MC simulations presented in Ref.~\cite{Luu:2015gpl}. The MC calculation of the single-particle Mott gap in a (metallic) carbon nanotube is 
expected to be much easier, since the lattice dimension $L$ is determined by the physical nanotube radius used in the experiment (and by the number of unit 
cells in the longitudinal direction of the tube). As electron-electron interaction (or correlation) effects are expected to be more pronounced 
in the (1-dimensional) nanotubes, the treatment of flat graphene as the limiting case of an infinite-radius nanotube would be especially interesting.
Strong correlation effects could be even more pronounced in the (0-dimensional) carbon fullerenes (buckyballs), where we are also faced 
with a fermion sign problem due to the admixture of pentagons into the otherwise-bipartite honeycomb structure~\cite{leveragingML}.
This sign problem has the unusual property of vanishing as the system size becomes
large, as the number of pentagons in a buckyball is fixed by its Euler characteristic to be exactly~$12$, independent of the number of hexagons.
The mild scaling of HMC with system size gives access to very large physical systems ($\sim 10^4$ sites or more), so it may be plausible to put a 
particular experimental system (a nanotube, a graphene patch, or a topological insulator, for instance) into software, for a direct, first-principles 
Hubbard model calculation.

\section*{Acknowledgements}

We thank Jan-Lukas Wynen for helpful discussions on the Hubbard model and software issues. We also thank Michael Kajan for proof reading and for providing a lot of detailed comments.
This work was funded, in part, through financial support from the Deutsche Forschungsgemeinschaft (Sino-German CRC 110 and SFB TRR-55).
E.B. is supported by the U.S. Department of Energy under Contract No. DE-FG02-93ER-40762.
The authors gratefully acknowledge the computing time granted through JARA-HPC on the supercomputer JURECA~\cite{jureca} at Forschungszentrum J\"ulich.
We also gratefully acknowledge time on DEEP~\cite{DEEP}, an experimental modular supercomputer at the J\"ulich Supercomputing Centre.

\begin{subappendices}
\section{Subtleties of the mixed differencing scheme 
\label{sec:mixed_bias}}

As in Ref.~\cite{Luu:2015gpl}, we have used a mixed-differencing scheme for the $A$ and $B$ sublattices, which was first suggested by
Brower \textit{et al.} in Ref.~\cite{Brower:2012zd}. While mixed differencing does not cancel the linear Trotter error completely, it does diminish it significantly,
as shown in Ref.~\cite{Luu:2015gpl}. We shall now discuss some fine points related to the correct continuum limit when mixed differencing is used.

\subsection{Seagull term}

Let us consider the forward differencing used in Ref.~\cite{Luu:2015gpl} for the $AA$ contribution to fermion operator,
\begin{equation}
M^{AA}_{(x,t)(y,t^\prime)} = \delta_{xy}^{} \left\{
-\delta_{t,t^\prime}^{} + \left[\exp(i\phidelta_{x,t}^{})-\mdelta\right] \delta_{t+1,t^\prime}^{} \right\},
\label{ferm_op}
\end{equation}
with the gauge links and an explicit staggered mass $\mdelta$ (not to be confused with the AFMI order parameter) on the time-off-diagonal. We recall
that a tilde means that the corresponding quantity is multiplied by $\delta=\beta/N_t$. An expansion in $\delta$ gives
\begin{align}
M^{AA}_{(x,t)(y,t^\prime)} 
& = \delta_{xy}^{} \left\{
-\delta_{t,t^\prime} + \left[1 + i\phidelta_{x,t}^{} - \frac{\phidelta_{x,t}^2}{2} - \mdelta\right] \delta_{t+1,t^\prime}^{} \right\} + \ordnung{\delta^3}, \\
& = \delta_{xy}^{} \left\{
\delta_{t+1,t^\prime} - \delta_{t,t^\prime} + \left[ i\phidelta_{x,t}^{} - \frac{\phidelta_{x,t}^2}{2} - \mdelta\right] \delta_{t+1,t^\prime}^{} \right\} + \ordnung{\delta^3}, \\
& = \delta_{xy}^{} \left\{
\delta\partial_t^{} +\frac{\delta^2\partial^2_t}{2} + \left[ i\phidelta_{x,t}^{} - \frac{\phidelta_{x,t}^2}{2} - \mdelta\right] 
(\delta_{t,t^\prime}^{} + \delta\partial_t^{})\right\} + \ordnung{\delta^3}, \\
& = \delta_{xy}^{} \left\{
\delta\partial_t^{} + ( i\phidelta_{x,t}^{} - \mdelta^\prime ) \delta_{t,t^\prime} - \delta\mdelta\partial_t^{} \right\} + \ordnung{\delta^3},
\end{align}
where terms proportional to $\delta$ persist in the continuum limit. In the last step we defined
\begin{equation}
\mdelta^\prime := \mdelta + \frac{\phidelta_{x,t}^2}{2} - \frac{\delta^2\partial^2_t}{2} - i\delta\phidelta_{x,t}^{}\partial_t^{},
\label{eqn_effective_staggered_mass}
\end{equation}
as the ``effective'' staggered mass at finite $\delta$.
The same calculation can be performed for the $BB$ contribution of Ref.~\cite{Luu:2015gpl}. This gives
\begin{align}
M^{BB}_{(x,t)(y,t^\prime)}
& = \delta_{xy}^{} \left\{
\delta\partial_t^{} + ( i\phidelta_{x,t}^{} + \mdelta^\prime ) \delta_{t,t^\prime} - \delta\mdelta\partial_t^{} \right\} + \ordnung{\delta^3},
\end{align}
where the effective staggered mass has the opposite sign, as expected.

Let us discuss the behavior of $\mdelta^\prime$ when $m_s \to 0$ and $\delta \to 0$. First, 
$\partial_t^2$ is negative semi-definite, $\phidelta_{x,t}^2$ is positive semi-definite, and $i\phidelta_{x,t}\partial_t$ is indefinite, 
but one typically finds that $\mdelta^\prime \ge \mdelta$. For a vanishing bare staggered mass $\mdelta \to 0$, this creates a non-vanishing 
bias between the sublattices at $\delta \neq 0$, which is due to the mixed differencing scheme. Numerically, we find that this effect prefers $\erwartung{m_A-m_B}>0$.
Second, the ``seagull term'' $\phidelta_{x,t}^2$ is not suppressed in the continuum limit, as first noted in Ref.~\cite{Brower:2012zd}. This happens because,
in the vicinity of continuum limit, the Gaussian part of the action becomes narrow, and 
$\phidelta$ is approximately distributed as $\phidelta \sim \mathcal{N}(0,\sqrt{\delta U})$. Because 
$\phidelta$ scales as $\sqrt{\delta}$, the seagull term is not in $\ordnung{\delta^2}$ but effectively linear in $\delta$, denoted as $\efford{\delta}$.
The seagull term contains important physics, and should be correctly generated by the gauge links.

\subsection{Field redefinition
\label{sec_field_redef}}

Following Brower \textit{et al.} in \Ref{Brower:2012zd}, the seagull term can be absorbed by means of a redefinition of the Hubbard-Stratonovich field,
at the price of generating the so-called ``normal-ordering term'' of Ref.~\cite{Brower:2012zd}, which is of
physical significance. Let us briefly consider how this works in our case. The field redefinition of \Ref{Brower:2012zd} is
\begin{equation}
\phi_x^{} := \varphi_x^{} - \frac{\delta}{2} V_{xy}^{} \varphi_y^{} \psi^*_y \psi_y^{},
\label{phi_shift}
\end{equation}
in terms of the field $\psi$ on which the fermion matrix $M$ acts. For backward differencing of $M$, the full Hamiltonian
includes the terms
\begin{align}
& \frac{1}{2} \sum_t \tilde\phi_{x,t}^{} \tilde V_{xy}^{-1} \tilde \phi_{y,t}^{} +
\sum_{t,t^\prime} \psi^*_{x,t} \bigg[ \delta_{t,t^\prime}^{}
-\exp(-i\tilde\phi_{x,t}^{}) \delta_{t-1,t^\prime}^{} + \tilde m \delta_{t-1,t^\prime}^{} \bigg] \psi_{x,t^\prime}^{}
\nonumber \\
& \qquad =
\ { \frac{1}{2} \int dt \, \phi_x^{} V_{xy}^{-1} \phi_y^{}}
\ { + \int dt \, \psi^*_x \partial_t^{} \psi_x^{}}
\ { + \, i \int dt \, \psi^*_x \phi_x^{} \psi_x^{}}
\ { + \, m \int dt \, \psi^*_x \psi_x^{}}
\ { + \,\frac{\delta}{2} \int dt \, \psi^*_x \phi_x^2 \psi_x^{}}
+ \efford{\delta},
\label{lattice_B}
\end{align}
where the Gaussian term generated by the Hubbard-Stratonovich transformation is included, and $\tilde m$ is left unspecified for the moment.
If we apply the redefinition~\eqref{phi_shift} to the Gaussian term, we find
\begin{align}
\frac{1}{2} \int dt \, \phi_x^{} V_{xy}^{-1} \phi_y^{} & = \frac{1}{2} \int dt \, \varphi_x^{} V_{xy}^{-1} \varphi_y^{}
- \frac{\delta}{2} \int dt \, \varphi_x^{} V_{xz}^{-1} V_{zy}^{} \varphi_y^{} \psi^*_y \psi_y^{} + \efford{\delta},
\nonumber \\
& = \frac{1}{2} \int dt \, \varphi_x^{} V_{xy}^{-1} \varphi_y^{}
- \frac{\delta}{2} \int dt \, \psi^*_x \varphi_x^2 \psi_x^{} + \efford{\delta},
\end{align}
and along the lines of \Ref{Brower:2012zd}, we note that the $\varphi_x^2$ seagull term cancels the $\phi_x^2$ 
seagull term of Eq.~(\ref{lattice_B}) to leading order in $\delta$. As we are performing a path integral over the Hubbard-Stratonovich field,
we need to account for the Jacobian of the field redefinition, which is
\begin{align}
\int \mathcal{D}\tilde\phi =
\int \mathcal{D}\tilde \varphi \, \det \left[\frac{\partial\tilde\phi_{x,t}^{}}{\partial\tilde\varphi_{y,t^\prime}^{}}\right]
& = \int \mathcal{D}\tilde\varphi \, \exp \left[\mathrm{Tr} \log\left(\delta_{xy}^{}\delta_{t,t^\prime}^{} - \frac{\delta}{2}
V_{xy}^{} \psi^*_{y,t} \psi_{y,t}^{} \delta_{t,t^\prime}^{}\right) \right]
\nonumber \\
& \simeq \int \mathcal{D}\varphi \,
\exp\left(- \frac{1}{2} \int dt \,  V_{xx}^{} \psi^*_x \psi_x^{}\right),
\end{align}
where in the last step we used $\log(1+\delta z) = \delta z+\ordnung{\delta^2}$ before taking the continuum limit. 
We conclude that the seagull term in the expansion of the gauge links has the correspondence
\begin{equation}
\frac{\delta}{2} \int dt \, \psi^*_x \phi_x^2 \psi_x^{}
\longleftrightarrow \frac{1}{2} \int dt \, V_{xx}^{} \psi^*_x \psi_x^{},
\label{seagull_norm}
\end{equation}
which is exactly the normal-ordering term proportional to $V_{xx}^{}/2$ of \Ref{Smith:2014tha}.
Hence, as argued in \Ref{Brower:2012zd}, the normal-ordering term should be omitted when gauge links are used, as an equivalent term is dynamically
generated by the gauge links.
This statement is valid when backward differencing is used for both sublattices. In \Appref{alternative derivative}, we discuss
how this argument carries over to the case of forward and mixed differencing.

\subsection{Alternative forward difference 
\label{sec:alternative derivative}}

In case the backward differencing of Eq.~(\ref{lattice_B}) is used for both sublattices as in \Ref{Smith:2014tha}, then simply taking
the usual staggered mass term
\begin{equation}
\tilde m = \mdelta \quad (x \in A),
\qquad
\tilde m = -\mdelta \quad (x \in B),
\end{equation}
suffices to get the correct Hubbard Hamiltonian, as both sublattices receive a dynamically generated normal-ordering term with coefficient $V_{xx}^{}/2$.
However, the mixed-difference lattice action in \Ref{Luu:2015gpl} produces a ``staggered'' normal-ordering term, with $-V_{xx}^{}/2$ for sublattice $A$ and 
$V_{xx}^{}/2$ for sublattice $B$. Hence, with the mixed-difference operator of \Ref{Luu:2015gpl} (forward for sublattice $A$, backward for sublattice $B$), 
we should instead take
\begin{equation}
\tilde m = \tilde V_{00}^{} + \mdelta \quad (x \in A),
\qquad
\tilde m = -\mdelta \quad (x \in B),
\end{equation}
in order to again obtain the physical Hubbard Hamiltonian with normal-ordering and staggered mass terms.
Therefore, in our current work we adopt the alternative forward differencing
\begin{align}
M^{AA}_{(x,t)(y,t^\prime)} & = \delta_{xy}^{} \left\{ \delta_{t+1,t^\prime}^{}
- \left[\exp(-i\phidelta_{x,t}^{}) + \mdelta\right] \delta_{t,t^\prime}^{} \right\},
\label{MAA_modified}
\end{align}
instead of Eq.~\eqref{ferm_op}, which again yields a normal-ordering term $V_{xx}^{}/2$ for sublattice $A$. As in \Ref{Smith:2014tha}, we thus
retain the desirable feature of a completely dynamically generated normal-ordering term. In our actual numerical simulations, we set the bare 
staggered mass $\mdelta = 0$. In our CG solver with Hasenbusch preconditioning, we work with finite $\mdelta$~\cite{hasenbusch}.
The spectrum of the operator~\eqref{MAA_modified} lacks conjugate reciprocity, which causes an ergodicity problem~\cite{Wynen:2018ryx}.
\section{Finding a plateau 
\label{sec:plateau_estimation}}

Here we present an automatized, deterministic method that reliably finds the optimal plateau in a given data set (such as the effective mass $m(\tau)$). Specifically, our method finds 
the region of least slope and fluctuations, and checks whether this region is a genuine plateau without significant drift. If a given time series does not exhibit an acceptable plateau, 
our method returns an explicit error message.

Apart from the time series $m(\tau)$ expected to exhibit a plateau, the algorithm requires two parameters to be chosen in advance. 
The first is the minimal length $\lambda$ a plateau should have. The second is an ``analysis window'' of width $\mu\le\lambda$. 
This controls how many data points are considered in the analysis of local fluctuations. We find that
\begin{align}
\lambda & = \frac{N_t^{}}{6}, \\
\mu & = \log_2^{}(N_t^{}),
\end{align}
are in most cases good choices.

Algorithm~\ref{alg_plateau} describes the procedure in detail. The idea is to find a balance between least statistical and systematic fluctuations. 
Statistical fluctuations decrease with increasing plateau length. This is why we seek to choose the plateau as long as possible, without running into a region 
with large systematic deviations. This property can also be used to our advantage. If we calculate the mean from a given time $\tau_2$ to all the 
previous times, the influence of another point compatible with the mean will decrease with the distance from $\tau_2$. Thus the local fluctuation of the 
running mean decreases, until it reaches a point with significant systematic deviation. This local fluctuation minimum marks the optimal $\tau_1$. The plateau 
then ranges from $\tau_1$ to $\tau_2$. We check that it does not exhibit significant drift, by fitting a linear function and checking if the first order term deviates 
from zero within twice its error. By repeating the analysis for all possible values of $\tau_2$, the globally best plateau can be found, as determined by 
least local fluctuations of the running mean.

	\begin{algorithm}
		\SetKwInOut{Input}{input}\SetKwInOut{Output}{output}
		\Input{$N_t$, $m[0,\dots,\,N_t-1]$, $\lambda$, $\mu$}
		\Output{$\tau_1$, $\tau_2$}
		\BlankLine
		\For{$\tau'=\mu-1,\dots,\,N_t-1$}{
			\For{$\tau=0,\dots,\, \tau'$}{
				$\overline{m}[\tau;\tau'] = \mathrm{mean}\!\left(m[\tau,\dots,\tau']\right)$\;
			}
			\For{$\tau=0,\dots,\, \tau'-\mu+1$}{
				$\sigma[\tau;\tau'] = \mathrm{sd}\!\left(\overline{m}[\tau,\dots,\tau+\mu-1;\tau']\right)$\;
			}
			$\tau_1^*[\tau'] = \underset{\tau\in \left\{0,\dots\,, \tau'-\mu\right\}}{\mathrm{argmin}} \left(\sigma[\tau;\tau']\right)$\;
		}
		$\Lambda_{\phantom{0}} = \left\{\left(\tau,\tau'\right)\,|\:\tau=\tau_1^*[\tau'],\,\tau'-\tau\ge\lambda\right\}$\;
		$\Lambda_0 = \left\{\left(\tau,\tau'\right)\in \Lambda\,|\:m[\tau,\dots,\tau'] \text{ has no significant drift}\right\}$\;
		\uIf{$\Lambda_0 \neq \varnothing$}{
			$\left(\tau_1,\tau_2\right) = \underset{(\tau,\tau')\in \Lambda_0}{\mathrm{argmin}} \left(\sigma[\tau;\tau']\right)$\;
		}\Else{
			No acceptable plateau of requested length found.
		}
		\caption{Finding a fit range for a plateau in a time series. \label{alg_plateau}}
	\end{algorithm}

As every range in the set $\Lambda_0$ from Algorithm~\ref{alg_plateau} is a valid plateau, it allows us to estimate the systematic error due to the choice of plateau. 
We simply repeat the calculation of the relevant observable for all ranges in $\Lambda_0$, and interpret the standard deviation of the resulting set of values as a systematic uncertainty.
\section{Possible bias of direct plateau fits}\label{sec:cosh_bias}

	The correlators obtained by simulations in euclidean space-time decay exponentially and one can assume that after some imaginary time higher energy contributions are negligible. So the ground state dominates and the decay constant, the effective mass, can directly be extracted.

	Let us distinguish two scenarios. In the first case the correlator shows strict exponential decay in the region of relevance. We then write
	\begin{align}
	C_\text{e}(t,\,x)&\propto \eto{-mt}(1+x)
	\end{align}
	where $m$ is the effective mass, $t$ the euclidean time and $x$ some (statistical) noise. It turns out that this ansatz is not universally applicable. If the back-propagating part of the correlator is significant, we have to choose
	\begin{align}
	C_\text{c}(t,\,t_0,\,x)&\propto \cosh\left(m\left(t-t_0\right)\right)(1+x)
	\end{align}
	for the correlator.
	
	\subsection{The exponential case}
	The effective mass can locally be estimated, using e.g.\@ the ``log'' method in the \texttt{hadron} library~\cite{hadron}, by
	\begin{align}
	\hat{m}&=\log\left(\frac{C_\text{e}(t,\,x)}{C_\text{e}(t+1,\,y)}\right)\\
	&=m+\log(1+x)-\log(1+y)\\
	&=m+x-y+\ordnung{x^2}+\ordnung{y^2}\,.
	\end{align}
	This means that, as long as $x$ and $y$ are identically distributed, the estimator is unbiased. Even in the case of different distributions of $x$ and $y$ we obtain the bias
	\begin{align}
	\erwartung{\hat{m}-m}&=\erwartung{x}-\erwartung{y}+\ordnung{\erwartung{x^2}}+\ordnung{\erwartung{y^2}}\\
	&=\ordnung{\sigma_x^2}+\ordnung{\sigma_y^2}
	\end{align}
	with vanishing first order contribution. Plainly correlation between $x$ and $y$ reduces any bias. Here we denote the standard deviation of $x$ and $y$ by $\sigma_x$ and $\sigma_y$ respectively. We also used that the noise itself has to be unbiased (i.e.\@ $\erwartung{x}=\erwartung{y}=0$) because any systematic bias would have been absorbed in the mean of the correlator by this point anyway.

	The only exclusion from this rule would be if $x$ or $y$ had a significant probability to be smaller than $-1$. This would however mean that a significant part of the correlators would be negative, so that the complete analysis would be very much in doubt. In addition in this case one would have to consider higher order terms as the linear approximation of the logarithm would be very inaccurate.

	All in all one can safely assume that this estimator of the effective mass does not suffer from bias in any practically relevant case.
	
	\subsection{The $\cosh$ case}
	There are several approaches to estimate the effective mass locally in the $\cosh$ case. To our knowledge the numerical solution of
	\begin{align}
	\frac{\cosh\left(\hat{m}\left(t-t_0\right)\right)}{\cosh\left(\hat{m}\left(t+1-t_0\right)\right)}&=\frac{C_\text{c}(t,\,t_0,\,x)}{C_\text{c}(t+1,\,t_0,\,y)}
	\end{align}
	yields the most accurate results. The method is called ``solve'' in the \texttt{hadron} library. This equation cannot be investigated analytically for a bias, thus we use the alternative method, called ``acosh'' in \texttt{hadron},
	\begin{align}
	\hat{m}&= \acosh\left(\frac{C_\text{c}(t-1,\,t_0,\,x)+C_\text{c}(t+1,\,t_0,\,x)}{2C_\text{c}(t,\,t_0,\,y)}\right)\\
	&= \acosh\left(\cosh m\,\frac{1+x}{1+y}\right)\\
	&= m + x\coth m - y\coth m + xy\frac{\cosh m}{\sinh^3 m}+\ordnung{x^2}+\ordnung{y^2}\,,\label{eqn_lin_expansion_acosh}
	\end{align}
	where we summarised the noise in the numerator into a single term. One would naively assume that with the same argument as above the first order noise contributions have to vanish in the expectation value. In fact this is the case if $x$ and $y$ are strongly correlated or $\sigma_x$ and $\sigma_y$ are very small. In this context small means that the probability of $\cosh m\,\frac{1+x}{1+y} < 1$ is negligible. Otherwise one has to consider in the calculation of the expectation value that tuples $(x,\,y)$ fulfilling the upper condition are dropped from the analysis. If $x$ and $y$ are distributed according to the probability density functions (pdf) $\rho_\text{X}(x)$ and $\rho_\text{Y}(y)$ respectively, we find that the expectation value of some function $f(x,y)$ is given by
	\begin{align}
	\erwartung{f}&=\frac 1Z \int_{-\infty}^{\infty}\md y\int_{(1+y)/\cosh m-1}^{\infty}\md x\,f(x,y)\rho_\text{X}(x)\rho_\text{Y}(y)\,,\label{eqn_expect_value_cosh}\\
	Z&= \int_{-\infty}^{\infty}\md y\int_{(1+y)/\cosh m-1}^{\infty}\md x\,\rho_\text{X}(x)\rho_\text{Y}(y)\,.\label{eqn_area_of_bias}
	\end{align}
	The lower bound of the inner integral approaches $y$ as $m\rightarrow 0$ independently of the distributions of $x$ and $y$. This is very different from the exponential case where the integration area would only be confined for broadly distributed noise.

	The integral in equation~\eqref{eqn_area_of_bias} is (to our knowledge) not solvable analytically even for the normal distribution, but we can confine its value to
	\begin{align}
	\frac{1}{2} \le Z\le 1
	\end{align}
	because the pdfs are assumed to be normalised and the area integrated over cannot be reduced to less than half.

	We can however solve the integral~\eqref{eqn_expect_value_cosh} for the relevant linear functions if we take $x$ and $y$ to be normally distributed. The expectation values
	\begin{align}
	\erwartung{x}&=\frac{\sigma_x^2 \exp \left(-\frac{( \sech m-1)^2}{2 \left(\sigma_y^2 \left( \sech ^2m\right)+\sigma_x^2\right)}\right)}{\sqrt{2 \pi } \sqrt{\sigma_y^2 \left( \sech ^2m\right)+\sigma_x^2}}\\
	\erwartung{y}&=-\frac{\sigma_y^2 \exp \left(-\frac{2 \left(\sinh ^4\left(\frac m2\right)\right)}{\sigma_x^2 \left( \cosh ^2m\right)+\sigma_y^2}\right)}{\sqrt{2 \pi } \sqrt{\sigma_x^2 \left( \cosh ^2m\right)+\sigma_y^2}}\\
	\erwartung{xy}&=-\frac{\sigma_x^2 \sigma_y^2 ( \sech m-1) ( \sech m) \exp \left(-\frac{( \sech m-1)^2}{2 \left(\sigma_y^2 \left( \sech ^2m\right)+\sigma_x^2\right)}\right)}{\sqrt{2 \pi } \left(\sigma_y^2 \left( \sech ^2m\right)+\sigma_x^2\right)^{3/2}}
	\end{align}
	can be expanded together with the prefactors from equation~\ref{eqn_lin_expansion_acosh} about $m=0$ because this is the interesting limit. In addition we set $\sigma_x=\sigma_y=\sigma$ for simplicity. This yields a first order bias
	\begin{align}
	\hat{m} -m &= \frac{9\sigma}{8\sqrt{\pi}m}+\ordnung{m}+\ordnung{\sigma^2}\,.\label{eqn_predicted_acosh_bias}
	\end{align}
	Thus for $\sigma$ in the order of $m^2$ or larger we get a significant relative bias with positive sign.

	We learn from this estimator that one can only trust the values of the effective mass if all samples produced numeric values and none failed. In this case the linear order bias vanishes. In any other case the effective mass will have a bias.

\section{Thermal gap}\label{sec:thermal_gap}

It is useful to consider the influence of the inverse temperature $\beta$ on the single-particle gap, in order to provide a better understanding
of the scaling of $\Delta$ with $\beta$. Naturally, we are not able to solve the entire problem analytically, so we shall consider small
perturbations in the coupling $U$, and assume that the dispersion relation of graphene is not significantly perturbed by the interaction (which is
expected to be the case when $U$ is small). Let us now compute the expectation value of the number of electrons excited from the ground state.
As we consider exclusively the conduction band, we assume that the particle density follows Fermi-Dirac statistics.  We take the positive-energy part of
\begin{equation}
\omega_k^{} := \kappa\sqrt{\tilde\omega_k^2},
\qquad
\tilde\omega_k^2 = 
3 + 4 \cos(3a k_x^{}/2) \cos(\sqrt{3}a k_y^{}/2) + 2\cos(\sqrt{3} a k_y^{}),
\end{equation}
where $a \simeq 1.42$ angstrom is the nearest-neighbor lattice spacing and we assume that every excited electron contributes an energy $E(U)$ to a ``thermal gap'' $\Delta(\beta)$.
These considerations yield the gap equation
\begin{equation}
\Delta(\beta) = E(U) a^2 f_{\text{BZ}}^{} \int_{k\in\text{BZ}}
\frac{d^2k}{(2\pi)^2} \frac{1}{1+\exp(\beta\omega_k^{})},
\qquad
f_{\text{BZ}}^{} := \frac{3\sqrt{3}}{2},
\label{eqn_therm_gap_inf_L}
\end{equation}
where the factor $f_{\text{BZ}}^{}$ is due to the 
hexagonal geometry of the first Brillouin zone (BZ). It should be emphasized that the thermal gap is not the interaction-driven Mott gap we are studying here, 
even though it is not numerically distinguishable from the latter. A thermal gap can occur even if the conduction and valence bands touch or overlap. 
The physical interpretation of the thermal gap (as explained above) is a measure of the degree of 
excitation above the ground state, based on the number of excited states that are already occupied in thermal equilibrium.

\subsection{Finite temperature}

\begin{figure}[t]
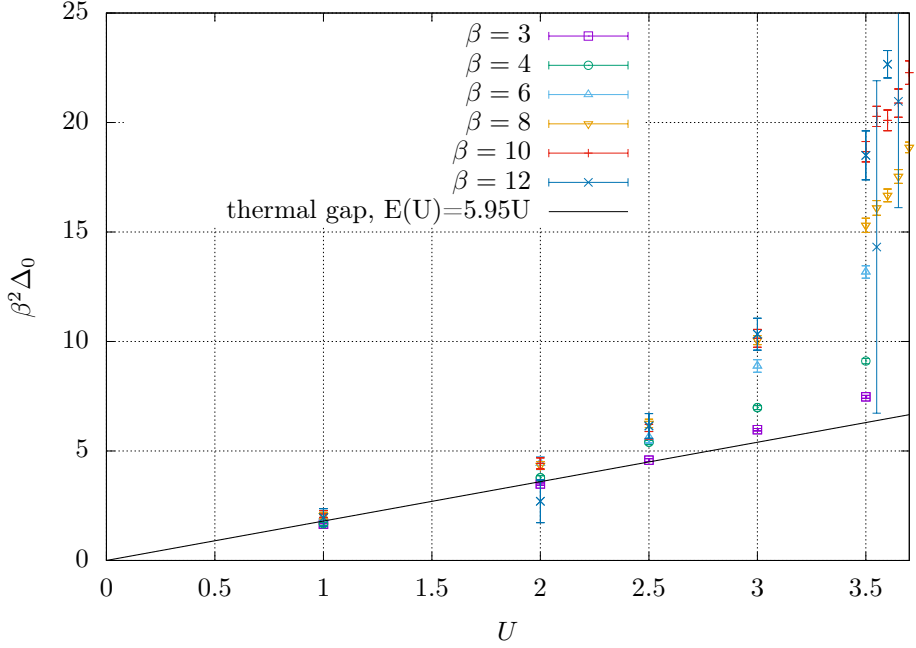

\begingroup
  \inputencoding{latin1}%
  \makeatletter
  \providecommand\color[2][]{%
    \GenericError{(gnuplot) \space\space\space\@spaces}{%
      Package color not loaded in conjunction with
      terminal option `colourtext'%
    }{See the gnuplot documentation for explanation.%
    }{Either use 'blacktext' in gnuplot or load the package
      color.sty in LaTeX.}%
    \renewcommand\color[2][]{}%
  }%
  \providecommand\includegraphics[2][]{%
    \GenericError{(gnuplot) \space\space\space\@spaces}{%
      Package graphicx or graphics not loaded%
    }{See the gnuplot documentation for explanation.%
    }{The gnuplot epslatex terminal needs graphicx.sty or graphics.sty.}%
    \renewcommand\includegraphics[2][]{}%
  }%
  \providecommand\rotatebox[2]{#2}%
  \@ifundefined{ifGPcolor}{%
    \newif\ifGPcolor
    \GPcolortrue
  }{}%
  \@ifundefined{ifGPblacktext}{%
    \newif\ifGPblacktext
    \GPblacktexttrue
  }{}%
  \let\gplgaddtomacro\g@addto@macro
  \gdef\gplbacktext{}%
  \gdef\gplfronttext{}%
  \makeatother
  \ifGPblacktext
    \def\colorrgb#1{}%
    \def\colorgray#1{}%
  \else
    \ifGPcolor
      \def\colorrgb#1{\color[rgb]{#1}}%
      \def\colorgray#1{\color[gray]{#1}}%
      \expandafter\def\csname LTw\endcsname{\color{white}}%
      \expandafter\def\csname LTb\endcsname{\color{black}}%
      \expandafter\def\csname LTa\endcsname{\color{black}}%
      \expandafter\def\csname LT0\endcsname{\color[rgb]{1,0,0}}%
      \expandafter\def\csname LT1\endcsname{\color[rgb]{0,1,0}}%
      \expandafter\def\csname LT2\endcsname{\color[rgb]{0,0,1}}%
      \expandafter\def\csname LT3\endcsname{\color[rgb]{1,0,1}}%
      \expandafter\def\csname LT4\endcsname{\color[rgb]{0,1,1}}%
      \expandafter\def\csname LT5\endcsname{\color[rgb]{1,1,0}}%
      \expandafter\def\csname LT6\endcsname{\color[rgb]{0,0,0}}%
      \expandafter\def\csname LT7\endcsname{\color[rgb]{1,0.3,0}}%
      \expandafter\def\csname LT8\endcsname{\color[rgb]{0.5,0.5,0.5}}%
    \else
      \def\colorrgb#1{\color{black}}%
      \def\colorgray#1{\color[gray]{#1}}%
      \expandafter\def\csname LTw\endcsname{\color{white}}%
      \expandafter\def\csname LTb\endcsname{\color{black}}%
      \expandafter\def\csname LTa\endcsname{\color{black}}%
      \expandafter\def\csname LT0\endcsname{\color{black}}%
      \expandafter\def\csname LT1\endcsname{\color{black}}%
      \expandafter\def\csname LT2\endcsname{\color{black}}%
      \expandafter\def\csname LT3\endcsname{\color{black}}%
      \expandafter\def\csname LT4\endcsname{\color{black}}%
      \expandafter\def\csname LT5\endcsname{\color{black}}%
      \expandafter\def\csname LT6\endcsname{\color{black}}%
      \expandafter\def\csname LT7\endcsname{\color{black}}%
      \expandafter\def\csname LT8\endcsname{\color{black}}%
    \fi
  \fi
    \setlength{\unitlength}{0.0500bp}%
    \ifx\gptboxheight\undefined%
      \newlength{\gptboxheight}%
      \newlength{\gptboxwidth}%
      \newsavebox{\gptboxtext}%
    \fi%
    \setlength{\fboxrule}{0.5pt}%
    \setlength{\fboxsep}{1pt}%
\begin{picture}(7200.00,5040.00)%
    \gplgaddtomacro\gplbacktext{%
      \csname LTb\endcsname%
      \put(682,704){\makebox(0,0)[r]{\strut{}$0$}}%
      \csname LTb\endcsname%
      \put(682,1527){\makebox(0,0)[r]{\strut{}$5$}}%
      \csname LTb\endcsname%
      \put(682,2350){\makebox(0,0)[r]{\strut{}$10$}}%
      \csname LTb\endcsname%
      \put(682,3173){\makebox(0,0)[r]{\strut{}$15$}}%
      \csname LTb\endcsname%
      \put(682,3996){\makebox(0,0)[r]{\strut{}$20$}}%
      \csname LTb\endcsname%
      \put(682,4819){\makebox(0,0)[r]{\strut{}$25$}}%
      \csname LTb\endcsname%
      \put(814,484){\makebox(0,0){\strut{}$0$}}%
      \csname LTb\endcsname%
      \put(1623,484){\makebox(0,0){\strut{}$0.5$}}%
      \csname LTb\endcsname%
      \put(2433,484){\makebox(0,0){\strut{}$1$}}%
      \csname LTb\endcsname%
      \put(3242,484){\makebox(0,0){\strut{}$1.5$}}%
      \csname LTb\endcsname%
      \put(4051,484){\makebox(0,0){\strut{}$2$}}%
      \csname LTb\endcsname%
      \put(4861,484){\makebox(0,0){\strut{}$2.5$}}%
      \csname LTb\endcsname%
      \put(5670,484){\makebox(0,0){\strut{}$3$}}%
      \csname LTb\endcsname%
      \put(6479,484){\makebox(0,0){\strut{}$3.5$}}%
    }%
    \gplgaddtomacro\gplfronttext{%
      \csname LTb\endcsname%
      \put(198,2761){\rotatebox{-270}{\makebox(0,0){\strut{}$\beta^2\Delta_0$}}}%
      \put(3808,154){\makebox(0,0){\strut{}$U$}}%
      \csname LTb\endcsname%
      \put(3982,4646){\makebox(0,0)[r]{\strut{}$\beta = 3$}}%
      \csname LTb\endcsname%
      \put(3982,4426){\makebox(0,0)[r]{\strut{}$\beta = 4$}}%
      \csname LTb\endcsname%
      \put(3982,4206){\makebox(0,0)[r]{\strut{}$\beta = 6$}}%
      \csname LTb\endcsname%
      \put(3982,3986){\makebox(0,0)[r]{\strut{}$\beta = 8$}}%
      \csname LTb\endcsname%
      \put(3982,3766){\makebox(0,0)[r]{\strut{}$\beta = 10$}}%
      \csname LTb\endcsname%
      \put(3982,3546){\makebox(0,0)[r]{\strut{}$\beta = 12$}}%
      \csname LTb\endcsname%
      \put(3982,3326){\makebox(0,0)[r]{\strut{}thermal gap, E(U)=5.95U}}%
    }%
    \gplbacktext
    \put(0,0){\includegraphics{gap_norm-pdf}}%
    \gplfronttext
  \end{picture}%
\endgroup
 	\caption{Illustration of the thermal gap in the weakly coupled regime, as given by Eq.~\eqref{eqn:thermal_gap}. Our MC data for the single-particle gap $\Delta$ from
	\Figref{delta_against_U} is shown multiplied by $\beta^2$. All quantities are expressed in appropriate units of $\kappa$.}
	\label{fig:beta_sq_delta_against_U}
\end{figure}

Let us evaluate Eq.~(\ref{eqn_therm_gap_inf_L}) under the assumption that $\beta$ is large.
Then, the integrand only contributes in the region where $\omega_k^{} \approx 0$, in other words 
near the Dirac points $K$ and $K^\prime$ located at momenta $k_D^{}$.
In the vicinity of a Dirac point, the dispersion relation reduced to the well-known Dirac cone 
\begin{equation}
\omega_k^{} \simeq v_F^{} | k-k_D^{} |,
\qquad
v_F^{} := 3\kappa a/2,
\end{equation}
with Fermi velocity $v_F$. Within this approximation, we may 
sum over the two Dirac points and perform the angular integral, which gives
\begin{align}
\Delta(\beta) & \approx
2 E(U) a^2 f_{\text{BZ}}^{} \int_{k\in\mathbb{R}^2}
\frac{d^2k}{(2\pi)^2} 
\frac{1}{1+\exp(\beta v_F^{} | k | )}
= E(U) a^2 f_{\text{BZ}}^{}
\int_{0}^\infty \frac{dk}{\pi}
\frac{k}{1+\exp(\beta v_F^{} k)}, \\
\label{eqn:thermal_gap}
& = E(U) a^2 f_{\text{BZ}}^{} \times \frac{\pi}{12} (\beta v_F^{})^{-2}
= \frac{\sqrt{3}\pi}{18} \, E(U) (\beta\kappa)^{-2} 
\approx \num{0.3023} \, E(U) (\beta\kappa)^{-2},
\end{align}
where the Fermi-Dirac integral has been evaluated in terms of the polylogarithm function. We expect the
error of this approximation to be exponentially suppressed in $\beta$.

In \Figref{beta_sq_delta_against_U}, we validate Eq.~\eqref{eqn:thermal_gap} using our data for $\Delta$ shown in \Figref{delta_against_U}.
We find that the prediction of quadratic scaling in $\beta$ and the linear approximation in $U$ are quite accurate.
A fit of Eq.~\eqref{eqn:thermal_gap} to our MC data for $\Delta$ in the weak-coupling regime $U/\kappa \le 2$ gives
\begin{align}
E(U) = \num{5.95\pm.15}\,U,
\end{align}
under the (perturbative) assumption $E(U)\propto U$. This fit is shown in \Figref{beta_sq_delta_against_U}, where we plot $\beta^2\Delta$ as a function
of $U$ (in proper units of $\kappa$). In the weakly coupled regime, the MC data for $\beta^2\Delta$ coincide and fall on a straight line. Once the critical coupling $U_c$ is
approached, the points for various $\beta$ separate. As expected, the linear dependence on $U$ persists longest for small $\beta$, as temperature effects 
are dominant over interaction effects. The quadratic scaling with $\beta$ is most accurate for large $\beta$, which is in line with the expected
exponential convergence stated above.

\subsection{Finite lattice size}

Let us also consider the leading correction to the thermal gap due to finite lattice size $L$.
The discretized form of Eq.~(\ref{eqn_therm_gap_inf_L}) is
\begin{equation}
\Delta(L,\beta)=\frac{E(U)}{L^2}\sum_{k\in \text{BZ}}\frac{1}{1+\exp(\beta\omega_k^{})}
\label{eqn_therm_gap_discrete}
\end{equation}
which in general has a very complicated convergence behavior. 
Because of periodic boundary conditions, Eq.~(\ref{eqn_therm_gap_discrete}) is
an effective trapezoidal approximation to Eq.~(\ref{eqn_therm_gap_inf_L}), thus the
convergence is \textit{a priori} expected to scale as $\ordnung{L^{-2}}$.

We shall now obtain a precise leading-order error estimation. Let us discretize the 
first BZ into a regular triangular lattice, with lattice spacing $h\propto L^{-1}$. 
We integrate a function $f(x,y)$ over a single triangle, spanned by the coordinates $(\pm h/2, 0)$ 
and $(0, \sqrt{3}h/2)$,
\begin{equation}
I := \int_{-h/2}^{h/2} dx \int_0^{b(x)} dy \, f(x,y),
\qquad
b(x) := \sqrt{3}h/2-\sqrt{3} | x |,
\end{equation}
and subtract the average of $f(x,y)$ over the corner points multiplied by the area of the triangle,
\begin{equation}
\hat{I} := \frac{\sqrt{3}h^2}{4} \times \frac{1}{3}
\left[ f(-h/2,0)+f(h/2,0)+f(0,\sqrt{3}h/2) \right],
\end{equation}
which gives the (local) error
\begin{equation}
\delta I := 
I-\hat{I}=-\frac{\sqrt{3}}{64} \left[\delsqu{f}{x}(0)+\delsqu{f}{y}(0)\right] h^4 + \ordnung{h^5},
\end{equation}
due to discretization. The global error is obtained by summing over the complete BZ,
\begin{align}
\sum_{k\in \text{BZ}} \delta I(k) & =
-\frac{\sqrt{3}}{64}\sum_{k\in \text{BZ}} \left[\delsqu{f}{x}(k)+\delsqu{f}{y}(k)\right] h^4 + \ordnung{L^2h^5}, \\
&\propto \frac{1}{L^4} \sum_{k\in \text{BZ}} \left[\delsqu{f}{x}(k)+\delsqu{f}{y}(k)\right] + \ordnung{L^{-3}}, \\
&\propto \frac{1}{L^2} \int_{k\in \text{BZ}} d^2k \left[\delsqu{f}{x}(k)+\delsqu{f}{y}(k)\right] + \ordnung{L^{-3}},
\end{align}
which equals
\begin{align}
\sum_{k\in \text{BZ}} \delta I(k) & \propto
\frac{1}{L^2} \oint_{k\in \partial\text{BZ}} \nabla f(k) \cdot d\vec k + \ordnung{L^{-3}},
\label{eqn_integral_1st_bz} \\
& \propto \ordnung{L^{-3}},
\end{align}
where Gauss's theorem has been applied in Eq.~\eqref{eqn_integral_1st_bz}. Hence, 
the projection of $\nabla f$ onto the normal of the BZ is integrated over the boundary of the BZ.
As every momentum-periodic function takes the same values on the opposite edges of the BZ, the result sums up to zero.
Surprisingly, one then finds that the second order error term in $L$ vanishes. For the special case of
$f(k)\propto 1/(1+\exp(\beta\omega_k^{}))$, the gradient in BZ-normal direction vanishes everywhere on the boundary, and the integral in Eq.~\eqref{eqn_integral_1st_bz} 
is trivially zero.

\begin{figure}[t]
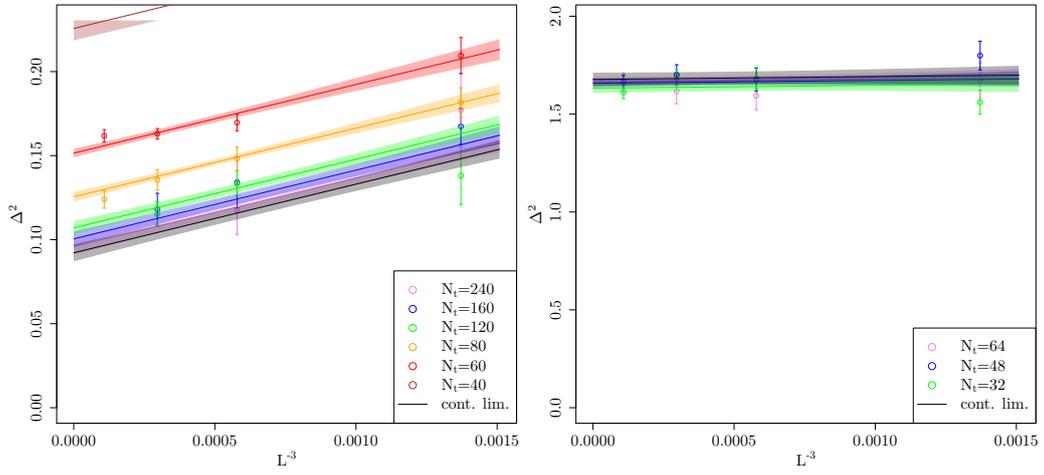

\centering
\includegraphics[width=.45\textwidth]{{b_10_u_3.9_m_0_gap_l}.pdf}
\includegraphics[width=.45\textwidth]{{b_4_u_5_m_0_gap_l}.pdf}
\caption{Simultaneous two-dimensional fit of $\Delta(N_t,L)$ (in units of $\kappa$) using Eq.~\eqref{eqn_gap_artefacts}, 
for $\kappa\beta=10$ and $U/\kappa=\num{3.9}$ (left panel) and $\kappa\beta=4$ and $U/\kappa=\num{5.0}$ (right panel). Only the extrapolations in $L$ are shown.
Data points for $L<9$ have been omitted from the fits, but not from the plots. 
These fits have $\chi^2/\text{d.o.f.} \simeq \num{0.83}$ and p-value of $\simeq \num{0.62}$ (left panel), and
$\chi^2/\text{d.o.f.} \simeq \num{1.1}$ and p-value of $\simeq \num{0.36}$ (right panel).
\label{fig:finite_size_always_cubic}}
\end{figure}

Higher orders in $U$ influencing the thermal gap are not as easy to calculate, but can in principle be dealt with using
diagrammatic techniques in a finite-temperature Matsubara formalism~\cite{bruus_qft}. As an example, we 
know from \Ref{Giuliani2009} that $v_F^{}$ is influenced (at weak coupling) only at $\ordnung{U^2}$.
Let us finally provide some further numerical evidence for the expected cubic finite-size effects in $L$.
In \Figref{2d_gap_fit}, we have already shown that cubic finite-size effects are a good approximation for $U < U_c$, as expected for
states with small correlation lengths. In \Figref{finite_size_always_cubic}, we show that the cubic behavior in $L$ still holds for $U \simeq U_c$
and $U > U_c$.

 \end{subappendices}\clearpage{}%
\clearpage{}%
\chapter[Antiferromagnetic Character of the Phase Transition]{The Antiferromagnetic Character of the Quantum Phase Transition in the Hubbard Model on the Honeycomb Lattice}\label{chap:afm_character}

\begin{center}
	\textit{Based on~\cite{more_observables} by J.~Ostmeyer, E.~Berkowitz, S.~Krieg, T.~A.~L\"{a}hde, T.~Luu and C.~Urbach}
\end{center}

Without further ado, let us resolve any ambiguity left in the previous \Cref{chap:phase_transition} concerning the magnetic nature of the phase transition and determine a reliable value for the critical exponent $\upbeta$.\\

\renewcommand{\Figref}[1]{Figure~\ref{figure:#1}\xspace}

\newcommand{\critUold}{\ensuremath{\num{3.834(14)}}}
\newcommand{\critMuold}{\ensuremath{\num{0.844(31)}}}
\newcommand{\critNuold}{\ensuremath{\num{1.185(43)}}} %
\newcommand{\critZetaold}{\ensuremath{\num{0.924(17)}}}
\newcommand{\critExpold}{\ensuremath{\num{1.095(37)}}}

\newcommand{\DeltaInfold}{\ensuremath{\num{0.057(11)}}}
\newcommand{\coeffUold}{\ensuremath{\num{0.479(93)}}}

\renewcommand{\critU}{\ensuremath{\num{3.835(14)}}}
\renewcommand{\critMu}{\ensuremath{\num{0.847(31)}}}
\renewcommand{\critNu}{\ensuremath{\num{1.181(43)}}} %
\renewcommand{\critZeta}{\ensuremath{\num{0.761(5)}}}
\renewcommand{\critExp}{\ensuremath{\num{0.898(37)}}}

The Fermi-Hubbard model---a prototypical model of electrons hopping between lattice sites~\cite{Hubbard:1963}---has a rich phenomenology of strongly-correlated electrons, requiring nonperturbative 
treatment~\cite{White:1989zz}.
On a honeycomb lattice, where it provides a basis for studying electronic properties of carbon nanosystems like graphene and nanotubes, the Hubbard model is expected to exhibit a second-order quantum phase transition between a (weakly coupled) semi-metallic (SM) state and an antiferromagnetic Mott insulating (AFMI) state as a function of the electron-electron coupling~\cite{Paiva2005}.

It is noteworthy that a fully-controlled \emph{ab initio} characterization of the SM-AFMI transition has not yet appeared; Monte Carlo (MC) calculations have not provided unique, 
generally accepted values for the critical exponents~\cite{Otsuka:2015iba,semimetalmott}. Such discrepancies are primarily attributed to the adverse scaling of MC algorithms with spatial system
size $L$ and inverse temperature $\beta$. 
The resulting systematic error is magnified by an incomplete understanding of the extrapolation of operator expectation values to the thermodynamic and temporal continuum limits.

Lattice Monte Carlo (LMC) simulation of Hamiltonian and Lagrangian theories of interacting fermions is a mature  field of study, which is seeing tremendous progress in the areas of novel computer hardware, algorithms, and theoretical developments.
Efficient Hybrid Monte Carlo (HMC) algorithms \cite{Duane:1987de,hasenbusch} designed for theories with dynamical fermions, coupled with GPU-accelerated supercomputing \cite{Clark:2009wm}, now allow for the direct simulation of systems of the same size as those used in realistic condensed-matter experiments and applications~\cite{ulybyshev2021bridging}.

While much of the focus of LMC efforts continues to be on Lattice QCD, the algorithms and methods so developed are now rapidly finding their place among the large variety of Hamiltonian theories in condensed matter physics.
Recently, preliminary studies of nanotubes \cite{Luu:2015gpl,Berkowitz:2017bsn} have appeared and treatments of the fermion sign problem based on formal developments~\cite{Cristoforetti:2012su,Kanazawa:2014qma} have made promising progress towards first-principles treatments of fullerenes~\cite{leveragingML} and doped systems~\cite{Ulybyshev:2019fte}.
Moreover, our understanding of these algorithms' ergodicity properties~\cite{Wynen:2018ryx} and computational scaling~\cite{Beyl:2017kwp,acceleratingHMC} has also recently been placed on a firm footing.

In this Letter we seek to remove, using Lattice Monte Carlo (LMC) techniques, the systematic uncertainties which affect determinations of the critical exponents of
the SM-AFMI transition in the honeycomb Hubbard model.
We present a unified, comprehensive, and systematically controlled treatment of the anti-ferromagnetic (AFM), ferromagnetic (FM), and charge-density-wave (CDW) order parameters, 
and confirm the AFM nature of the transition from first principles. Building on our determination of the Mott gap~\cite{semimetalmott} we find that the critical coupling $U_c/\kappa=\critU$ and 
the critical exponents---expected to be the exponents of the SU(2) Gross-Neveu, or chiral Heisenberg, universality class~\cite{gross_neveu_orig,heisenberg_gross_neveu}---to be $\nu=\critNu$ and $\upbeta=\critExp$. %

\emph{Method:}
We formulate the grand canonical Hubbard model at half filling in the particle-hole basis and without a bare staggered mass.
Its Hamiltonian reads
\begin{equation}
	H=-\kappa \sum_{\erwartung{x,y}}\left(p^\dagger_{x}p^\pdagger_{y}+h^\dagger_{x}h^\pdagger_{y}\right)+\frac{U}{2}\sum_{x}\rho_x\rho_x\,,\label{eqn_particle_hole_hamiltonian}
\end{equation}
where $p$ and $h$ are fermionic particle and hole annihilation operators, $\kappa$ is the hopping amplitude, $U$ the on-site interaction, and
\begin{equation}
	\rho_x = p^\dagger_x p_x -h^\dagger_x h_x,
	\label{eq:charge_operator}
\end{equation}
is the charge operator.
Using Hasenbusch-accelerated~\cite{hasenbusch,acceleratingHMC} HMC~\cite{Duane:1987de} with the BRS formulation~\cite{Brower:2011av,Brower:2012zd} and a 
mixed time differencing~\cite{Brower:2012zd,Luu:2015gpl,semimetalmott} which has favorable computational scaling and ergodicity properties~\cite{Wynen:2018ryx}, 
we generate ensembles of auxiliary field configurations for different linear spatial extents $L$ with a maximum of $L$=$102$ corresponding to 20,808 lattice sites, interaction strengths $U$ (with fixed hopping $\kappa$), inverse temperatures $\beta = 1/T$, and $N_t$ Trotter steps;\footnote{Throughout this work, we use an upright $\upbeta$ for the critical exponent and a slanted $\beta$ for the inverse temperature.}
See \Ref{semimetalmott} for full details.

We have used these ensembles to compute the Mott gap $\Delta$ as a function of $U$ and $\beta$ to locate a quantum critical point (QCP) at $T = 0$~\cite{semimetalmott} using 
finite-size scaling (FSS)~\cite{newmanb99,dutta_aeppli,Shao_2016}. The single-particle gap $\Delta$ was found to open at $U/\kappa=\critUold$. 
Though this is widely expected to coincide with a semimetal-AFMI transition, we did not characterize the nature of the transition in Ref.~\cite{semimetalmott}.
Instead, we found the correlation length exponent $\nu=\critNuold$ from first principles, and estimated the critical exponent $\upbeta=\critExpold$ for the staggered magnetization $m_s$ under the AFMI assumption and the (mean-field) expectation $m_s \sim \Delta/U$~\cite{Assaad:2013xua}.
In this work we forgo these assumptions and confirm the AFMI character of the transition, finding $\upbeta=\critExp$ in terms of a FSS analysis in the inverse temperature $\beta$.

At half-filling, we compute expectation values of one-point and two-point functions of bilinear local operators, 
the spins 
\begin{equation}
S^i_{x} = \half (p^\dagger_x,\, (-1)^x h_x) \sigma^i (p_x,\, (-1)^x h^\dagger_x)\transpose
\end{equation}
and the charge~\eqref{eq:charge_operator}
where the $\sigma^i$ are Pauli matrices and $(-1)^x$ provides a 
minus sign depending on the triangular sublattice of the honeycomb to which the site $x$ belongs; the sign originates from the particle-hole transformation.
For operators, we consider the uniform magnetization 
\begin{equation}
S^i_{+} = \sum_x S^i_{x},
\end{equation}
where spins are summed coherently, and the staggered magnetization 
\begin{equation}
S^i_{-} = \sum_x (-1)^x S^i_{x},
\end{equation}
which computes the difference between the sublattices.
We expect the extensive one-point functions, the ferromagnetic magnetization $\average{S^i_+}$, antiferromagnetic magnetization $\average{S^i_-}$, the total charge $\average{\rho_+}$, and 
the charge separation $\average{\rho_-}$ (and their respective intensive one-point functions) to vanish by symmetry at half-filling for all $\beta$ and $U$.
However, the two-point functions 
\begin{equation}
S^{ii}_{\pm\pm} = \aaverage{S^i_\pm S^i_\pm},
\end{equation}
and
\begin{equation}
Q^2 = \aaverage{\rho_+ \rho_+}, \quad Q_-^2 = \aaverage{\rho_-\rho_-},
\end{equation}
need not vanish at 
half-filling. The double-bracket notation indicates the connected correlator, 
\begin{equation}
\aaverage{\operator_1 \operator_2} = \disconnected{\operator_1}{\operator_2}.
\end{equation}
These quantities scale quadratically with the spatial volume\footnote{Here, $V$ denotes the number of unit cells---half the number of lattice sites. Thus $V=L^2$ in case of an $L\times L$ lattice.} 
$V$ which must be divided out to compute their respective intensive partners, which we denote with lower case letters.  For example, $s^{ii}_{\pm\pm}=S^{ii}_{\pm\pm}/V^2$ and similarly for $q^2$ and $\qm^2$.
A \emph{finite} non-vanishing $s^{ii}_{++}$ ($s^{ii}_{--}$) indicates ferromagnetic (antiferromagnetic) order, while a non-vanishing $\qm^2$ indicates CDW order.\footnote{Susceptibilities, like the AFM susceptibility $\chi_{\text{AF}}$, can be reconstructed from these observables at half-filling.}

\begin{figure*}[ht]
	\makebox[\linewidth]{
\begingroup
  \inputencoding{latin1}%
  \makeatletter
  \providecommand\color[2][]{%
    \GenericError{(gnuplot) \space\space\space\@spaces}{%
      Package color not loaded in conjunction with
      terminal option `colourtext'%
    }{See the gnuplot documentation for explanation.%
    }{Either use 'blacktext' in gnuplot or load the package
      color.sty in LaTeX.}%
    \renewcommand\color[2][]{}%
  }%
  \providecommand\includegraphics[2][]{%
    \GenericError{(gnuplot) \space\space\space\@spaces}{%
      Package graphicx or graphics not loaded%
    }{See the gnuplot documentation for explanation.%
    }{The gnuplot epslatex terminal needs graphicx.sty or graphics.sty.}%
    \renewcommand\includegraphics[2][]{}%
  }%
  \providecommand\rotatebox[2]{#2}%
  \@ifundefined{ifGPcolor}{%
    \newif\ifGPcolor
    \GPcolortrue
  }{}%
  \@ifundefined{ifGPblacktext}{%
    \newif\ifGPblacktext
    \GPblacktexttrue
  }{}%
  \let\gplgaddtomacro\g@addto@macro
  \gdef\gplbacktext{}%
  \gdef\gplfronttext{}%
  \makeatother
  \ifGPblacktext
    \def\colorrgb#1{}%
    \def\colorgray#1{}%
  \else
    \ifGPcolor
      \def\colorrgb#1{\color[rgb]{#1}}%
      \def\colorgray#1{\color[gray]{#1}}%
      \expandafter\def\csname LTw\endcsname{\color{white}}%
      \expandafter\def\csname LTb\endcsname{\color{black}}%
      \expandafter\def\csname LTa\endcsname{\color{black}}%
      \expandafter\def\csname LT0\endcsname{\color[rgb]{1,0,0}}%
      \expandafter\def\csname LT1\endcsname{\color[rgb]{0,1,0}}%
      \expandafter\def\csname LT2\endcsname{\color[rgb]{0,0,1}}%
      \expandafter\def\csname LT3\endcsname{\color[rgb]{1,0,1}}%
      \expandafter\def\csname LT4\endcsname{\color[rgb]{0,1,1}}%
      \expandafter\def\csname LT5\endcsname{\color[rgb]{1,1,0}}%
      \expandafter\def\csname LT6\endcsname{\color[rgb]{0,0,0}}%
      \expandafter\def\csname LT7\endcsname{\color[rgb]{1,0.3,0}}%
      \expandafter\def\csname LT8\endcsname{\color[rgb]{0.5,0.5,0.5}}%
    \else
      \def\colorrgb#1{\color{black}}%
      \def\colorgray#1{\color[gray]{#1}}%
      \expandafter\def\csname LTw\endcsname{\color{white}}%
      \expandafter\def\csname LTb\endcsname{\color{black}}%
      \expandafter\def\csname LTa\endcsname{\color{black}}%
      \expandafter\def\csname LT0\endcsname{\color{black}}%
      \expandafter\def\csname LT1\endcsname{\color{black}}%
      \expandafter\def\csname LT2\endcsname{\color{black}}%
      \expandafter\def\csname LT3\endcsname{\color{black}}%
      \expandafter\def\csname LT4\endcsname{\color{black}}%
      \expandafter\def\csname LT5\endcsname{\color{black}}%
      \expandafter\def\csname LT6\endcsname{\color{black}}%
      \expandafter\def\csname LT7\endcsname{\color{black}}%
      \expandafter\def\csname LT8\endcsname{\color{black}}%
    \fi
  \fi
    \setlength{\unitlength}{0.0500bp}%
    \ifx\gptboxheight\undefined%
      \newlength{\gptboxheight}%
      \newlength{\gptboxwidth}%
      \newsavebox{\gptboxtext}%
    \fi%
    \setlength{\fboxrule}{0.5pt}%
    \setlength{\fboxsep}{1pt}%
\begin{picture}(5040.00,3772.00)%
    \gplgaddtomacro\gplbacktext{%
      \csname LTb\endcsname%
      \put(814,1142){\makebox(0,0)[r]{\strut{}$1$}}%
      \csname LTb\endcsname%
      \put(814,1690){\makebox(0,0)[r]{\strut{}$1.5$}}%
      \csname LTb\endcsname%
      \put(814,2237){\makebox(0,0)[r]{\strut{}$2$}}%
      \csname LTb\endcsname%
      \put(814,2785){\makebox(0,0)[r]{\strut{}$2.5$}}%
      \csname LTb\endcsname%
      \put(814,3332){\makebox(0,0)[r]{\strut{}$3$}}%
      \csname LTb\endcsname%
      \put(946,484){\makebox(0,0){\strut{}$-25$}}%
      \csname LTb\endcsname%
      \put(1474,484){\makebox(0,0){\strut{}$-20$}}%
      \csname LTb\endcsname%
      \put(2002,484){\makebox(0,0){\strut{}$-15$}}%
      \csname LTb\endcsname%
      \put(2530,484){\makebox(0,0){\strut{}$-10$}}%
      \csname LTb\endcsname%
      \put(3059,484){\makebox(0,0){\strut{}$-5$}}%
      \csname LTb\endcsname%
      \put(3587,484){\makebox(0,0){\strut{}$0$}}%
      \csname LTb\endcsname%
      \put(4115,484){\makebox(0,0){\strut{}$5$}}%
      \csname LTb\endcsname%
      \put(4643,484){\makebox(0,0){\strut{}$10$}}%
    }%
    \gplgaddtomacro\gplfronttext{%
      \csname LTb\endcsname%
      \put(209,2127){\rotatebox{-270}{\makebox(0,0){\strut{}$\beta^{\upbeta/\nu} m_s$}}}%
      \put(2794,154){\makebox(0,0){\strut{}$\beta^{1/\nu}(U-U_c)$}}%
      \csname LTb\endcsname%
      \put(1870,3378){\makebox(0,0)[r]{\strut{}$\beta = 3$}}%
      \csname LTb\endcsname%
      \put(1870,3158){\makebox(0,0)[r]{\strut{}$\beta = 4$}}%
      \csname LTb\endcsname%
      \put(1870,2938){\makebox(0,0)[r]{\strut{}$\beta = 6$}}%
      \csname LTb\endcsname%
      \put(1870,2718){\makebox(0,0)[r]{\strut{}$\beta = 8$}}%
      \csname LTb\endcsname%
      \put(1870,2498){\makebox(0,0)[r]{\strut{}$\beta = 10$}}%
      \csname LTb\endcsname%
      \put(1870,2278){\makebox(0,0)[r]{\strut{}$\beta = 12$}}%
      \csname LTb\endcsname%
      \put(1870,2058){\makebox(0,0)[r]{\strut{}$\beta = \infty$}}%
    }%
    \gplbacktext
    \put(0,0){\includegraphics{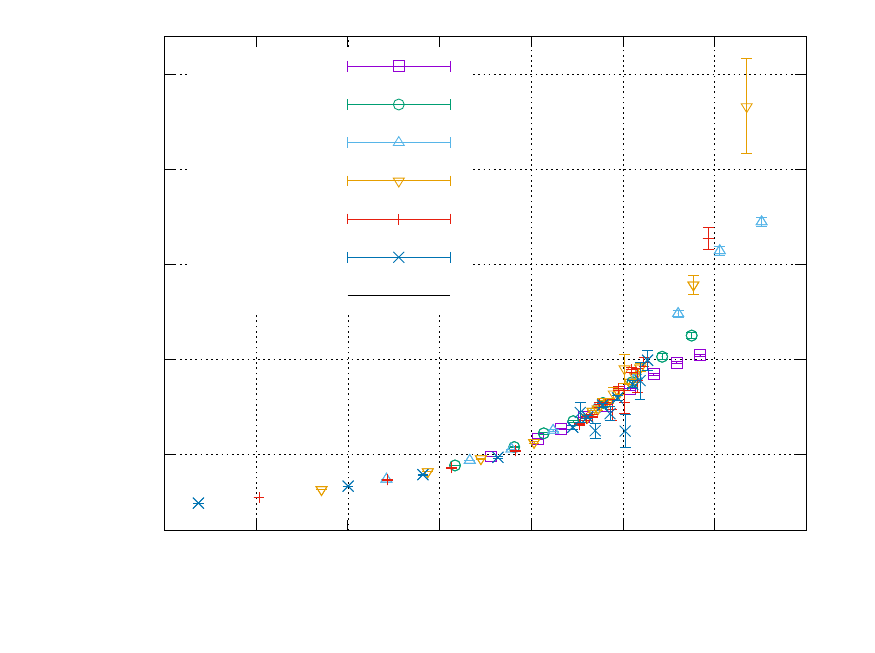}}%
    \gplfronttext
  \end{picture}%
\endgroup
\begingroup
  \inputencoding{latin1}%
  \makeatletter
  \providecommand\color[2][]{%
    \GenericError{(gnuplot) \space\space\space\@spaces}{%
      Package color not loaded in conjunction with
      terminal option `colourtext'%
    }{See the gnuplot documentation for explanation.%
    }{Either use 'blacktext' in gnuplot or load the package
      color.sty in LaTeX.}%
    \renewcommand\color[2][]{}%
  }%
  \providecommand\includegraphics[2][]{%
    \GenericError{(gnuplot) \space\space\space\@spaces}{%
      Package graphicx or graphics not loaded%
    }{See the gnuplot documentation for explanation.%
    }{The gnuplot epslatex terminal needs graphicx.sty or graphics.sty.}%
    \renewcommand\includegraphics[2][]{}%
  }%
  \providecommand\rotatebox[2]{#2}%
  \@ifundefined{ifGPcolor}{%
    \newif\ifGPcolor
    \GPcolortrue
  }{}%
  \@ifundefined{ifGPblacktext}{%
    \newif\ifGPblacktext
    \GPblacktexttrue
  }{}%
  \let\gplgaddtomacro\g@addto@macro
  \gdef\gplbacktext{}%
  \gdef\gplfronttext{}%
  \makeatother
  \ifGPblacktext
    \def\colorrgb#1{}%
    \def\colorgray#1{}%
  \else
    \ifGPcolor
      \def\colorrgb#1{\color[rgb]{#1}}%
      \def\colorgray#1{\color[gray]{#1}}%
      \expandafter\def\csname LTw\endcsname{\color{white}}%
      \expandafter\def\csname LTb\endcsname{\color{black}}%
      \expandafter\def\csname LTa\endcsname{\color{black}}%
      \expandafter\def\csname LT0\endcsname{\color[rgb]{1,0,0}}%
      \expandafter\def\csname LT1\endcsname{\color[rgb]{0,1,0}}%
      \expandafter\def\csname LT2\endcsname{\color[rgb]{0,0,1}}%
      \expandafter\def\csname LT3\endcsname{\color[rgb]{1,0,1}}%
      \expandafter\def\csname LT4\endcsname{\color[rgb]{0,1,1}}%
      \expandafter\def\csname LT5\endcsname{\color[rgb]{1,1,0}}%
      \expandafter\def\csname LT6\endcsname{\color[rgb]{0,0,0}}%
      \expandafter\def\csname LT7\endcsname{\color[rgb]{1,0.3,0}}%
      \expandafter\def\csname LT8\endcsname{\color[rgb]{0.5,0.5,0.5}}%
    \else
      \def\colorrgb#1{\color{black}}%
      \def\colorgray#1{\color[gray]{#1}}%
      \expandafter\def\csname LTw\endcsname{\color{white}}%
      \expandafter\def\csname LTb\endcsname{\color{black}}%
      \expandafter\def\csname LTa\endcsname{\color{black}}%
      \expandafter\def\csname LT0\endcsname{\color{black}}%
      \expandafter\def\csname LT1\endcsname{\color{black}}%
      \expandafter\def\csname LT2\endcsname{\color{black}}%
      \expandafter\def\csname LT3\endcsname{\color{black}}%
      \expandafter\def\csname LT4\endcsname{\color{black}}%
      \expandafter\def\csname LT5\endcsname{\color{black}}%
      \expandafter\def\csname LT6\endcsname{\color{black}}%
      \expandafter\def\csname LT7\endcsname{\color{black}}%
      \expandafter\def\csname LT8\endcsname{\color{black}}%
    \fi
  \fi
    \setlength{\unitlength}{0.0500bp}%
    \ifx\gptboxheight\undefined%
      \newlength{\gptboxheight}%
      \newlength{\gptboxwidth}%
      \newsavebox{\gptboxtext}%
    \fi%
    \setlength{\fboxrule}{0.5pt}%
    \setlength{\fboxsep}{1pt}%
\begin{picture}(5040.00,3772.00)%
    \gplgaddtomacro\gplbacktext{%
      \csname LTb\endcsname%
      \put(814,704){\makebox(0,0)[r]{\strut{}$0$}}%
      \csname LTb\endcsname%
      \put(814,1111){\makebox(0,0)[r]{\strut{}$0.1$}}%
      \csname LTb\endcsname%
      \put(814,1517){\makebox(0,0)[r]{\strut{}$0.2$}}%
      \csname LTb\endcsname%
      \put(814,1924){\makebox(0,0)[r]{\strut{}$0.3$}}%
      \csname LTb\endcsname%
      \put(814,2331){\makebox(0,0)[r]{\strut{}$0.4$}}%
      \csname LTb\endcsname%
      \put(814,2738){\makebox(0,0)[r]{\strut{}$0.5$}}%
      \csname LTb\endcsname%
      \put(814,3144){\makebox(0,0)[r]{\strut{}$0.6$}}%
      \csname LTb\endcsname%
      \put(814,3551){\makebox(0,0)[r]{\strut{}$0.7$}}%
      \csname LTb\endcsname%
      \put(3274,484){\makebox(0,0){\strut{}$U_c$}}%
      \csname LTb\endcsname%
      \put(3411,484){\makebox(0,0){\strut{}}}%
      \csname LTb\endcsname%
      \put(946,484){\makebox(0,0){\strut{}$1$}}%
      \csname LTb\endcsname%
      \put(1357,484){\makebox(0,0){\strut{}$1.5$}}%
      \csname LTb\endcsname%
      \put(1768,484){\makebox(0,0){\strut{}$2$}}%
      \csname LTb\endcsname%
      \put(2178,484){\makebox(0,0){\strut{}$2.5$}}%
      \csname LTb\endcsname%
      \put(2589,484){\makebox(0,0){\strut{}$3$}}%
      \csname LTb\endcsname%
      \put(3000,484){\makebox(0,0){\strut{}$3.5$}}%
      \csname LTb\endcsname%
      \put(3821,484){\makebox(0,0){\strut{}$4.5$}}%
      \csname LTb\endcsname%
      \put(4232,484){\makebox(0,0){\strut{}$5$}}%
      \csname LTb\endcsname%
      \put(4643,484){\makebox(0,0){\strut{}$5.5$}}%
    }%
    \gplgaddtomacro\gplfronttext{%
      \csname LTb\endcsname%
      \put(209,2127){\rotatebox{-270}{\makebox(0,0){\strut{}$m_s$}}}%
      \put(2794,154){\makebox(0,0){\strut{}$U$}}%
    }%
    \gplbacktext
    \put(0,0){\includegraphics{m_s_intensive-pdf}}%
    \gplfronttext
  \end{picture}%
\endgroup
 	}
	\caption{Left: Data collapse plot with the optimal parameters of $U_c$, $\nu$, and $\beta$ obtained from a simultaneous collapse fit to the gap $\Delta$ and the order parameter $m_s$.
	Note that the ``outliers'' are due to particularly small $\beta$ and are excluded from the analysis.
	For	$U< U_c \simeq \critU$ the order parameter vanishes.
	Right: The AFMI order parameter (staggered magnetization) $m_s$, with all quantities in units of $\kappa$, after the thermodynamic and continuum limit extrapolations.  
	We also show $m_s(U,\beta = \infty)$ as a solid black line with error band (calculated as in Ref.~\cite{semimetalmott}). The legend from the left plot applies to both.}
	\label{figure:structure factors}
\end{figure*}

\begin{figure*}[ht]
	\makebox[\linewidth]{
\begingroup
  \inputencoding{latin1}%
  \makeatletter
  \providecommand\color[2][]{%
    \GenericError{(gnuplot) \space\space\space\@spaces}{%
      Package color not loaded in conjunction with
      terminal option `colourtext'%
    }{See the gnuplot documentation for explanation.%
    }{Either use 'blacktext' in gnuplot or load the package
      color.sty in LaTeX.}%
    \renewcommand\color[2][]{}%
  }%
  \providecommand\includegraphics[2][]{%
    \GenericError{(gnuplot) \space\space\space\@spaces}{%
      Package graphicx or graphics not loaded%
    }{See the gnuplot documentation for explanation.%
    }{The gnuplot epslatex terminal needs graphicx.sty or graphics.sty.}%
    \renewcommand\includegraphics[2][]{}%
  }%
  \providecommand\rotatebox[2]{#2}%
  \@ifundefined{ifGPcolor}{%
    \newif\ifGPcolor
    \GPcolortrue
  }{}%
  \@ifundefined{ifGPblacktext}{%
    \newif\ifGPblacktext
    \GPblacktexttrue
  }{}%
  \let\gplgaddtomacro\g@addto@macro
  \gdef\gplbacktext{}%
  \gdef\gplfronttext{}%
  \makeatother
  \ifGPblacktext
    \def\colorrgb#1{}%
    \def\colorgray#1{}%
  \else
    \ifGPcolor
      \def\colorrgb#1{\color[rgb]{#1}}%
      \def\colorgray#1{\color[gray]{#1}}%
      \expandafter\def\csname LTw\endcsname{\color{white}}%
      \expandafter\def\csname LTb\endcsname{\color{black}}%
      \expandafter\def\csname LTa\endcsname{\color{black}}%
      \expandafter\def\csname LT0\endcsname{\color[rgb]{1,0,0}}%
      \expandafter\def\csname LT1\endcsname{\color[rgb]{0,1,0}}%
      \expandafter\def\csname LT2\endcsname{\color[rgb]{0,0,1}}%
      \expandafter\def\csname LT3\endcsname{\color[rgb]{1,0,1}}%
      \expandafter\def\csname LT4\endcsname{\color[rgb]{0,1,1}}%
      \expandafter\def\csname LT5\endcsname{\color[rgb]{1,1,0}}%
      \expandafter\def\csname LT6\endcsname{\color[rgb]{0,0,0}}%
      \expandafter\def\csname LT7\endcsname{\color[rgb]{1,0.3,0}}%
      \expandafter\def\csname LT8\endcsname{\color[rgb]{0.5,0.5,0.5}}%
    \else
      \def\colorrgb#1{\color{black}}%
      \def\colorgray#1{\color[gray]{#1}}%
      \expandafter\def\csname LTw\endcsname{\color{white}}%
      \expandafter\def\csname LTb\endcsname{\color{black}}%
      \expandafter\def\csname LTa\endcsname{\color{black}}%
      \expandafter\def\csname LT0\endcsname{\color{black}}%
      \expandafter\def\csname LT1\endcsname{\color{black}}%
      \expandafter\def\csname LT2\endcsname{\color{black}}%
      \expandafter\def\csname LT3\endcsname{\color{black}}%
      \expandafter\def\csname LT4\endcsname{\color{black}}%
      \expandafter\def\csname LT5\endcsname{\color{black}}%
      \expandafter\def\csname LT6\endcsname{\color{black}}%
      \expandafter\def\csname LT7\endcsname{\color{black}}%
      \expandafter\def\csname LT8\endcsname{\color{black}}%
    \fi
  \fi
    \setlength{\unitlength}{0.0500bp}%
    \ifx\gptboxheight\undefined%
      \newlength{\gptboxheight}%
      \newlength{\gptboxwidth}%
      \newsavebox{\gptboxtext}%
    \fi%
    \setlength{\fboxrule}{0.5pt}%
    \setlength{\fboxsep}{1pt}%
\begin{picture}(5040.00,3772.00)%
    \gplgaddtomacro\gplbacktext{%
      \csname LTb\endcsname%
      \put(876,918){\makebox(0,0)[r]{\strut{}$0$}}%
      \csname LTb\endcsname%
      \put(876,1273){\makebox(0,0)[r]{\strut{}$0.005$}}%
      \csname LTb\endcsname%
      \put(876,1629){\makebox(0,0)[r]{\strut{}$0.01$}}%
      \csname LTb\endcsname%
      \put(876,1985){\makebox(0,0)[r]{\strut{}$0.015$}}%
      \csname LTb\endcsname%
      \put(876,2341){\makebox(0,0)[r]{\strut{}$0.02$}}%
      \csname LTb\endcsname%
      \put(876,2697){\makebox(0,0)[r]{\strut{}$0.025$}}%
      \csname LTb\endcsname%
      \put(876,3053){\makebox(0,0)[r]{\strut{}$0.03$}}%
      \csname LTb\endcsname%
      \put(876,3409){\makebox(0,0)[r]{\strut{}$0.035$}}%
      \csname LTb\endcsname%
      \put(3297,484){\makebox(0,0){\strut{}$U_c$}}%
      \csname LTb\endcsname%
      \put(3431,484){\makebox(0,0){\strut{}}}%
      \csname LTb\endcsname%
      \put(1008,484){\makebox(0,0){\strut{}$1$}}%
      \csname LTb\endcsname%
      \put(1412,484){\makebox(0,0){\strut{}$1.5$}}%
      \csname LTb\endcsname%
      \put(1816,484){\makebox(0,0){\strut{}$2$}}%
      \csname LTb\endcsname%
      \put(2220,484){\makebox(0,0){\strut{}$2.5$}}%
      \csname LTb\endcsname%
      \put(2624,484){\makebox(0,0){\strut{}$3$}}%
      \csname LTb\endcsname%
      \put(3027,484){\makebox(0,0){\strut{}$3.5$}}%
      \csname LTb\endcsname%
      \put(3835,484){\makebox(0,0){\strut{}$4.5$}}%
      \csname LTb\endcsname%
      \put(4239,484){\makebox(0,0){\strut{}$5$}}%
      \csname LTb\endcsname%
      \put(4643,484){\makebox(0,0){\strut{}$5.5$}}%
    }%
    \gplgaddtomacro\gplfronttext{%
      \csname LTb\endcsname%
      \put(271,2127){\rotatebox{-270}{\makebox(0,0){\strut{}$s^{11}_{++}$}}}%
      \put(2825,154){\makebox(0,0){\strut{}$U$}}%
      \csname LTb\endcsname%
      \put(1932,3378){\makebox(0,0)[r]{\strut{}$\beta = 3$}}%
      \csname LTb\endcsname%
      \put(1932,3158){\makebox(0,0)[r]{\strut{}$\beta = 4$}}%
      \csname LTb\endcsname%
      \put(1932,2938){\makebox(0,0)[r]{\strut{}$\beta = 6$}}%
      \csname LTb\endcsname%
      \put(1932,2718){\makebox(0,0)[r]{\strut{}$\beta = 8$}}%
      \csname LTb\endcsname%
      \put(1932,2498){\makebox(0,0)[r]{\strut{}$\beta = 10$}}%
      \csname LTb\endcsname%
      \put(1932,2278){\makebox(0,0)[r]{\strut{}$\beta = 12$}}%
    }%
    \gplbacktext
    \put(0,0){\includegraphics{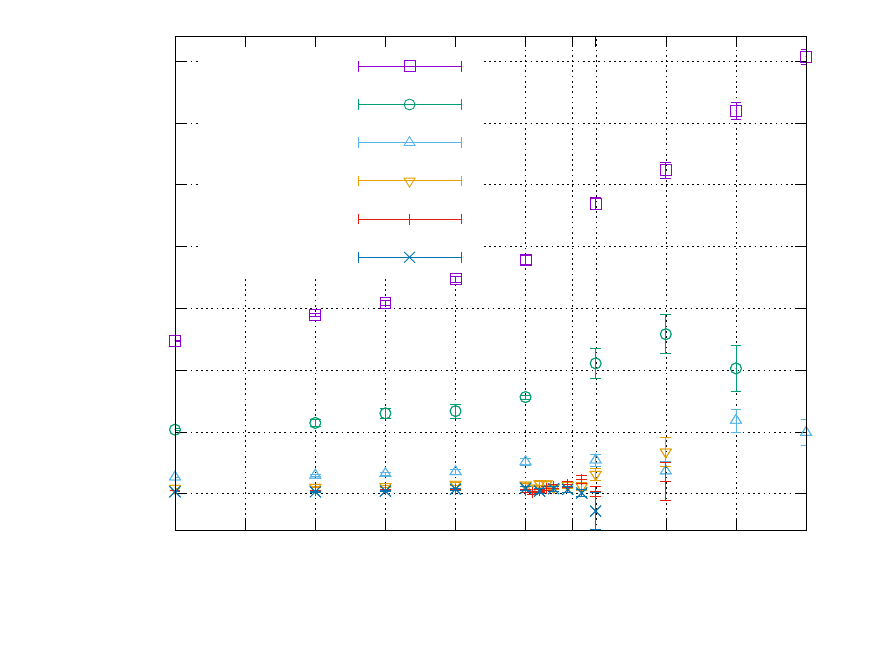}}%
    \gplfronttext
  \end{picture}%
\endgroup
\begingroup
  \inputencoding{latin1}%
  \makeatletter
  \providecommand\color[2][]{%
    \GenericError{(gnuplot) \space\space\space\@spaces}{%
      Package color not loaded in conjunction with
      terminal option `colourtext'%
    }{See the gnuplot documentation for explanation.%
    }{Either use 'blacktext' in gnuplot or load the package
      color.sty in LaTeX.}%
    \renewcommand\color[2][]{}%
  }%
  \providecommand\includegraphics[2][]{%
    \GenericError{(gnuplot) \space\space\space\@spaces}{%
      Package graphicx or graphics not loaded%
    }{See the gnuplot documentation for explanation.%
    }{The gnuplot epslatex terminal needs graphicx.sty or graphics.sty.}%
    \renewcommand\includegraphics[2][]{}%
  }%
  \providecommand\rotatebox[2]{#2}%
  \@ifundefined{ifGPcolor}{%
    \newif\ifGPcolor
    \GPcolortrue
  }{}%
  \@ifundefined{ifGPblacktext}{%
    \newif\ifGPblacktext
    \GPblacktexttrue
  }{}%
  \let\gplgaddtomacro\g@addto@macro
  \gdef\gplbacktext{}%
  \gdef\gplfronttext{}%
  \makeatother
  \ifGPblacktext
    \def\colorrgb#1{}%
    \def\colorgray#1{}%
  \else
    \ifGPcolor
      \def\colorrgb#1{\color[rgb]{#1}}%
      \def\colorgray#1{\color[gray]{#1}}%
      \expandafter\def\csname LTw\endcsname{\color{white}}%
      \expandafter\def\csname LTb\endcsname{\color{black}}%
      \expandafter\def\csname LTa\endcsname{\color{black}}%
      \expandafter\def\csname LT0\endcsname{\color[rgb]{1,0,0}}%
      \expandafter\def\csname LT1\endcsname{\color[rgb]{0,1,0}}%
      \expandafter\def\csname LT2\endcsname{\color[rgb]{0,0,1}}%
      \expandafter\def\csname LT3\endcsname{\color[rgb]{1,0,1}}%
      \expandafter\def\csname LT4\endcsname{\color[rgb]{0,1,1}}%
      \expandafter\def\csname LT5\endcsname{\color[rgb]{1,1,0}}%
      \expandafter\def\csname LT6\endcsname{\color[rgb]{0,0,0}}%
      \expandafter\def\csname LT7\endcsname{\color[rgb]{1,0.3,0}}%
      \expandafter\def\csname LT8\endcsname{\color[rgb]{0.5,0.5,0.5}}%
    \else
      \def\colorrgb#1{\color{black}}%
      \def\colorgray#1{\color[gray]{#1}}%
      \expandafter\def\csname LTw\endcsname{\color{white}}%
      \expandafter\def\csname LTb\endcsname{\color{black}}%
      \expandafter\def\csname LTa\endcsname{\color{black}}%
      \expandafter\def\csname LT0\endcsname{\color{black}}%
      \expandafter\def\csname LT1\endcsname{\color{black}}%
      \expandafter\def\csname LT2\endcsname{\color{black}}%
      \expandafter\def\csname LT3\endcsname{\color{black}}%
      \expandafter\def\csname LT4\endcsname{\color{black}}%
      \expandafter\def\csname LT5\endcsname{\color{black}}%
      \expandafter\def\csname LT6\endcsname{\color{black}}%
      \expandafter\def\csname LT7\endcsname{\color{black}}%
      \expandafter\def\csname LT8\endcsname{\color{black}}%
    \fi
  \fi
    \setlength{\unitlength}{0.0500bp}%
    \ifx\gptboxheight\undefined%
      \newlength{\gptboxheight}%
      \newlength{\gptboxwidth}%
      \newsavebox{\gptboxtext}%
    \fi%
    \setlength{\fboxrule}{0.5pt}%
    \setlength{\fboxsep}{1pt}%
\begin{picture}(5040.00,3772.00)%
    \gplgaddtomacro\gplbacktext{%
      \csname LTb\endcsname%
      \put(876,704){\makebox(0,0)[r]{\strut{}$-0.05$}}%
      \csname LTb\endcsname%
      \put(876,1416){\makebox(0,0)[r]{\strut{}$0$}}%
      \csname LTb\endcsname%
      \put(876,2128){\makebox(0,0)[r]{\strut{}$0.05$}}%
      \csname LTb\endcsname%
      \put(876,2839){\makebox(0,0)[r]{\strut{}$0.1$}}%
      \csname LTb\endcsname%
      \put(876,3551){\makebox(0,0)[r]{\strut{}$0.15$}}%
      \csname LTb\endcsname%
      \put(3297,484){\makebox(0,0){\strut{}$U_c$}}%
      \csname LTb\endcsname%
      \put(3431,484){\makebox(0,0){\strut{}}}%
      \csname LTb\endcsname%
      \put(1008,484){\makebox(0,0){\strut{}$1$}}%
      \csname LTb\endcsname%
      \put(1412,484){\makebox(0,0){\strut{}$1.5$}}%
      \csname LTb\endcsname%
      \put(1816,484){\makebox(0,0){\strut{}$2$}}%
      \csname LTb\endcsname%
      \put(2220,484){\makebox(0,0){\strut{}$2.5$}}%
      \csname LTb\endcsname%
      \put(2624,484){\makebox(0,0){\strut{}$3$}}%
      \csname LTb\endcsname%
      \put(3027,484){\makebox(0,0){\strut{}$3.5$}}%
      \csname LTb\endcsname%
      \put(3835,484){\makebox(0,0){\strut{}$4.5$}}%
      \csname LTb\endcsname%
      \put(4239,484){\makebox(0,0){\strut{}$5$}}%
      \csname LTb\endcsname%
      \put(4643,484){\makebox(0,0){\strut{}$5.5$}}%
    }%
    \gplgaddtomacro\gplfronttext{%
      \csname LTb\endcsname%
      \put(271,2127){\rotatebox{-270}{\makebox(0,0){\strut{}$q^2_{-}$}}}%
      \put(2825,154){\makebox(0,0){\strut{}$U$}}%
    }%
    \gplbacktext
    \put(0,0){\includegraphics{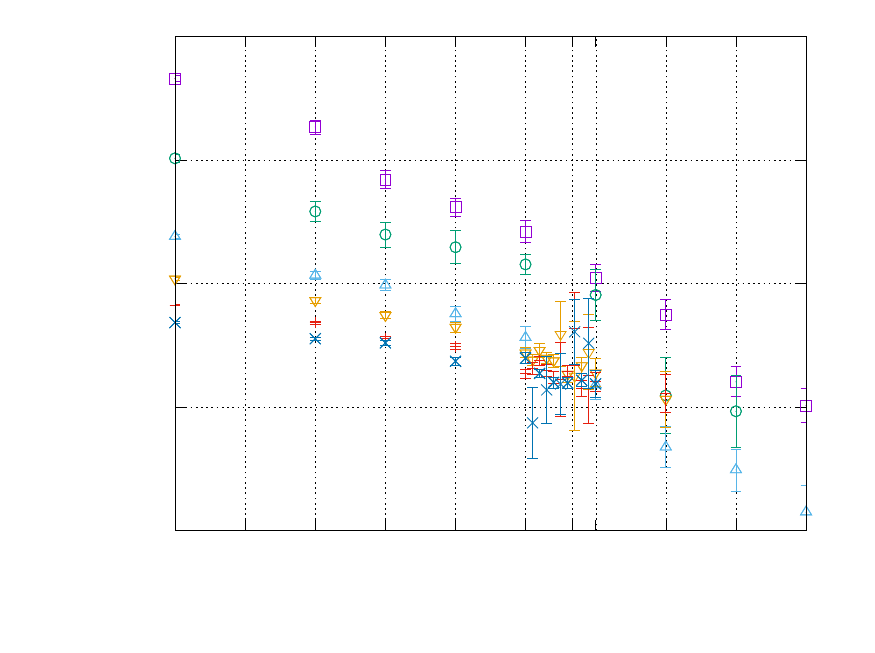}}%
    \gplfronttext
  \end{picture}%
\endgroup
 	}
	\caption{Ferromagnetic (left) and charge-density-wave (right) order parameters at different couplings, with all quantities in units of $\kappa$, after the thermodynamic and continuum limit extrapolations. The legend from the left plot applies to both.}
	\label{figure:s_f_and_cdw}
\end{figure*}

On the other hand, at very low temperatures, any finite non-zero value of $s^{ii}_{++}$, $s^{ii}_{--}$, or $\qm^2$ corresponds to an extensive ferromagnetic spin structure factor $\SF=S^{ii}_{++}/V$, 
antiferromagnetic spin structure factor $\SAF=S^{ii}_{--}/V$, or  staggered charge structure factor $S_\mathrm{CDW}=Q_-^2/V$, respectively, diverging linearly with the spatial volume.
The finite temperatures we use, however, provide a natural infrared cutoff for the correlation length, allowing different domains to cancel against one another.
Because the dynamical exponent $z=1$~\cite{PhysRevLett.97.146401} we can obtain intensive order parameters at any finite temperature by taking the thermodynamic limit of these extensive quantities and dividing them by $\beta^2$, rather than the spatial volume $V$. 
Finally, we use finite-size scaling in $\beta$ to remove the infrared regulator and determine zero-temperature properties.

We decompose the observables of interest into a set of operators with zero, one, or two fermion bilinears, each on a definite sublattice, and evaluate each separately.
This decomposition makes it easy to test the operators' individual behaviors in both non-interacting $U\to0$ and non-hopping $\kappa\to 0$ limits, which are analytically known.
We can also verify their expected scaling behaviors in Trotter error and volume.
We reconstruct the operators of physical interest by taking the appropriate linear combination.
We provide a detailed discussion and list of these operators in \autoref{tab:operators} of~\autoref{sect:15 ops} of the supplemental material.

\emph{Measurements and Extrapolation:} 
By inverting the fermion matrix on stochastic Fourier sources (see~\autoref{sect:stochastic sources}) we can compute fermion propagators and use Wick's theorem to contract the propagators into the relevant observables of interest (see \autoref{tab:observables} in~\autoref{sect:15 ops}).
Our structure factors measure two-point correlations between two bilinear operators at equal time; to fully exploit each configuration we leverage time-translation invariance and average measurements on many timeslices.
Finally, the average of all measurements across the configurations within an ensemble with fixed $U$, $\beta$, Trotter discretization $N_t$, and spatial extent $L$ gives an estimator for each observable.

We find that for both the anti- and ferromagnetic spin operators the $i$=$1$ and $3$ components for $\aaverage{S^i S^i}$ give different results configuration-by-configuration and ensemble-averaged\footnote{Note that $S^{11}_{--} = S^{22}_{--}$ at the operator level.}.
This is expected since our discretization breaks chiral symmetry which would have otherwise ensured their equality \cite{Buividovich:2016tgo}.  
However, the mixed differencing of our discretization mitigates this breaking, compared to a purely-forward or purely-backward differencing.
Furthermore, in the continuum limit this symmetry is restored.
Indeed, we observe that after continuum limit extrapolation, the different components agree up to uncertainties.
We therefore directly take
\begin{equation}
m_s^2 = \frac{\sum_i S^{ii}_{--}}{V(\kappa\beta)^2}= \frac{2S^{11}_{--}+S^{33}_{--}}{V(\kappa\beta)^2}, 
\end{equation}
as the staggered magnetization (and AFMI order parameter),
simplifying the subsequent analysis.

For each $U$ and $\beta$, we perform a simultaneous continuum- and infinite-spatial-volume extrapolation.
For numerical tractability, we extrapolate extensive quantities, for example $\SAF$, which is kept finite by the IR regulation provided by the finite temperature.
See \autoref{sect:extrap} of the supplementary material for an example of our extrapolations and a more detailed explanation of our extrapolation procedures.
As in Ref.~\cite{semimetalmott},  we use FSS relations to perform a simultaneous data-collapse of $m_s$ and the gap $\Delta$ and thus remove the dependence on $\beta$ in each of these quantities.
The collapse for $m_s$ is depicted in the left panel of \Figref{structure factors}.
Note that the data points below the transition, i.e.\@ $U-U_c < 0$, collapse to a single curve, whereas above the transition the points for different $\beta$ start to deviate because of the infrared cutoff imposed by our finite $\beta$ calculations.
This effect has regularly been observed before, see---for example---Fig.~7 in Ref.~\cite{White:1989zz} and Fig.~3 in Ref.~\cite{Buividovich:2018yar}.
For larger $\beta$ the collapse of the curve persists for a wider range above the transition.
Points with an obvious deviation from the global curve have been excluded from the fit so that a bias is avoided.
Since the collapse fit in $\Delta$, as done independently in Ref.~\cite{semimetalmott} yielding $U_c/\kappa=\critUold$ and $\nu=\critNuold$, requires only two free parameters $U_c$ and $\nu$, it is more stable and statistically more significant than the collapse fit of $m_s$ with the three free parameters $U_c$, $\nu$, and $\upbeta$. We therefore performed the simultaneous collapse fit of both $\Delta$ and $m_s$ giving the $\Delta$-fit a higher weight.
We find
\begin{align}
	\begin{split}
		\upbeta &= \critExp\,,\\
		\nu &= \critNu\,,\\
		\upbeta/\nu &=\critZeta\,,\\
		U_c/\kappa &= \critU\,.
	\end{split}\label{eq:final_results}
\end{align}
For the full correlation matrix and further details see \cref{sect:covariance}.
The critical coupling is essentially unchanged from our previous result~\cite{semimetalmott}.

We have also considered the inclusion of sub-leading corrections to the FSS analysis, along the lines of Ref.~\cite{Otsuka:2015iba}, where
an additional factor $1+c_L L^{-\omega}$ was introduced into the scaling relation for $m_s$.
While the FSS analysis in $L$ of Ref.~\cite{Otsuka:2015iba} found a clear signal for $c_L \neq 0$, an FSS analysis
of our MC data with an additional factor $1+c_\beta \beta^{-\omega^\prime}$ turned out consistent with $c_\beta = 0$. 
For any choice of $\omega^\prime$, we found that assuming $c_\beta \neq 0$ worsened the uncertainties of the
extracted parameters, though these remained consistent with the values quoted for $c_\beta = 0$. We thus conclude that our MC data does not support the scenario of sizable 
sub-leading corrections to the scaling in $\beta$, at least within the statistical accuracy and range of temperatures covered by our present set of ensembles. 
As the (expected) Lorentz-invariance should only emerge in the vicinity of the QCP, this apparent difference in the sub-leading
corrections may reflect the underlying structure of the Hubbard Hamiltonian.

The zero-temperature AFMI order parameter $m_s(\beta=\infty)$ is also given in the right panel of \Figref{structure factors}. This has been determined from MC data extrapolated
to infinite volume and the continuum limit, using the same strategy as for the zero-temperature single-particle gap in Ref.~\cite{semimetalmott}. We have extrapolated the 
$U/\kappa=4$ values to $\beta = \infty$ and used that result to constrain the overall scale of $m_s$ (given the critical exponents we have already fixed from FSS). Note that 
$m_s(\beta=\infty)$ so determined is expected to be strictly valid only in the vicinity of the QCP.

In \Figref{s_f_and_cdw}, we show our continuum and thermodynamic limit extrapolations of the FM ($s^{ii}_{++}$) and CDW ($q^2_{-}$) order parameters.    
While $q^2_-$ is a strictly semi-positive observable, we find that several individual values of $q^2_i$ are slightly negative (though with large error bars), 
in particular for large $U/\kappa$. This effect appears to be a consequence of residual extrapolation uncertainties, as our extrapolation does not enforce semi-positivity.
Still, we observe that both $s^{ii}_{++}$ and $q^2_{-}$ clearly vanish in the zero-temperature ($\beta\to\infty$) limit for all $U/\kappa$. This behavior is in contrast to that of
$m_s$, which remains non-zero for large $U/\kappa$, and confirms the AFM character of the QCP.

\emph{Discussion:} 
We have presented the first fully systematic, high-precision treatment of all operators that contribute to the AFM, FM, and CDW order parameters of the Hubbard Model on a honeycomb lattice, 
followed by an FSS analysis of the SM-AFMI transition in the inverse temperature $\beta$, culminating in our results~\eqref{eq:final_results}.
Within the accuracy of our MC data, we find that the QCP associated with the SM-AFMI transition coincides with the opening of the single-particle gap $\Delta$, which disfavors
the possibility of other exotic phenomena (such as an intermediate spin-liquid phase~\cite{Meng2010,Sorella:SciRep}) 
in the vicinity of the critical coupling where the AFMI order parameter $m_s$ appears. Also, the
vanishing of the FM and CDW order parameters throughout this transition strengthens the notion that the QCP is purely antiferromagnetic in character.

In addition to an unambiguous classification of the character of the QCP of the honeycomb Hubbard model, our results demonstrate the ability to perform high-precision 
calculations of strongly correlated electronic systems using lattice stochastic methods. 
A central component of our calculations is the Hasenbusch-accelerated HMC algorithm, 
as well as other state-of-the-art techniques originally developed for lattice QCD. This has allowed us to push our calculations to system sizes which are, to date, still the largest that have 
been performed, up to 102$\times$102 unit cells (or 20,808 lattice sites).

Our progress sets the stage for future high-precision calculations of additional observables of 
the Hubbard model and its extensions, as well as other Hamiltonian theories of strongly correlated electrons~\cite{PhysRevB.100.125116,PhysRevLett.122.077602,PhysRevLett.122.077601}. 
We anticipate the continued advancement of calculations with ever increasing system sizes, through the leveraging of additional state-of-the-art techniques from lattice QCD, 
such as multigrid solvers on GPU-accelerated architectures. We are actively pursuing research along these lines.

\emph{Acknowledgements:}
This work was funded, in part, through financial support from the Deutsche Forschungsgemeinschaft (DFG, German Research
Foundation) through the funds provided to the Sino-German Collaborative
Research Center TRR110 “Symmetries and the Emergence of Structure in QCD”
(DFG Project-ID 196253076 - TRR 110).
E.B. is supported by the U.S. Department of Energy under Contract No. DE-FG02-93ER-40762.
The authors gratefully acknowledge the computing time granted through JARA-HPC on the supercomputer JURECA~\cite{jureca} at Forschungszentrum J\"ulich.
We also gratefully acknowledge time on DEEP~\cite{DEEP}, an experimental modular supercomputer at the J\"ulich Supercomputing Centre.
The analysis was mostly done in \texttt{R}~\cite{R_language} using \texttt{hadron}~\cite{hadron}.
We are indebted to Bartosz Kostrzewa for helping us detect a compiler bug and for lending us his expertise on HPC hardware.

\begin{subappendices}
\section{Constructing spin structure factors from fundamental operators\label{sect:15 ops}}

For a fully systematic treatment of the different spin operators, we express these in terms of 15 operators $O_{0\dots14}$ of one or two bilinear operators.
Our basis of operators is complete for translationally-invariant equal-time observables (the generalization is straightforward) and we can express the observables of physical interest as linear combinations of the operators in our basis.

Our basis fixes one position on a sublattice.
We choose the operators with odd (even) index to act on $x\in A\ (B)$.
As we do not break the sublattice symmetry explicitly, we can always consider these pairs of operators simultaneously.
An overview of the $O_i$ based on the fermion matrix from~\cite{semimetalmott}
\begin{equation}
	\begin{split}
		M^{AA}_{(x,t)(y,t')} & = \delta_{xy}^{} \left(\delta_{t+1,t^\prime}^{} - \delta_{t,t^\prime}^{} \exp(-i\phidelta_{x,t}^{}) \right), \\
		M^{BB}_{(x,t)(y,t')} & = \delta_{xy}^{} \left(\delta_{t,t^\prime} - \delta_{t-1,t'} \exp(-i\phidelta_{x,t}^{}) \right), \\
		M^{AB}_{(x,t)(y,t')} = M^{BA}_{(x,t)(y,t^\prime)} & =- \kappadelta\,\delta_{\erwartung{x,y}} \delta_{t,t^\prime},
	\end{split}
\end{equation}
is given in Tab.~\ref{tab:operators}, and the various order parameters and structure factors are given in terms of the $O_i$ in Tab.~\ref{tab:observables}, along with the exact values of the physical observables in the $\kappa = 0$ and $U = 0$ limits and their leading error terms.
It should be noted that maximal ordering is not reached for $\beta < \infty$, thus the observables for $\kappa = 0$ are  
limited by the temporal correlation length (the largest inverse Matsubara frequency $\beta/\pi$).

\begin{table*}[ht]
\caption{Extensive operators $O_i$. Spin comes with a factor $1/2$, therefore the linear operators $O_{1,2}$ have to be divided by 2 and the quadratic ones $O_{3,\dots,14}$ by 4. 
Sums over repeated indices are implicit, so is the addition of the term with $p\leftrightarrow h$.
The two sites $x$ and $y$ are on the same lattice if $x\sim y$.
We define $\ti_x M\coloneqq\tr_S M^{-1}$ and $\rt_x M\coloneqq \real\tr_S M^{-1}$ where $S$ is the sublattice containing $x$. Analogously $\mathds{P}_x$ 
is the projector onto the sublattice containing $x$. The error estimation comes from an exact calculation of $\rt_xM$ at finite temperature and discretization 
in the non-interacting limit.
The exact values in the two limits $U=0$ and $\kappa=0$ are given for zero temperature.
We use the Bachmann–Landau notation to denote that an error in $\bigtheta x$ scales exactly as $x$, an error in $\ordnung{x}$ scales maximally as $x$, and an error in $\littleo{x}$ is asymptotically smaller than $x$.
\label{tab:operators}}
\makebox[\linewidth]{
	\begin{tabular}{c | c | cc | c | c | c | c}
		$O_i$ & Lattice & \multicolumn{2}{c|}{Operator/Implementation} & $U=0$ & $\kappa=0$ & Error & Diagram \\ \hline
		0 	& 			& \multicolumn{2}{c|}{$\mathds{1}$} & $V$ & $V$ & 0 & \feynmandiagram[horizontal=a to b]{a -- [fermion] b}; \\ 
			&&&&&&& \\
		\vspace{-.8cm}
		1,2 	& 			& \multicolumn{2}{c|}{$p_x^\pdagger p_x^\dagger$} & $V$ & $V$ & $\bigtheta{V\delta}$ & \\
			&			& \multicolumn{2}{c|}{$\rt_x M$} & & & &
			\feynmandiagram[horizontal=a to b]{a -- [fermion, half left] b -- [insertion=0.5, half left, edge label'=$x$] a}; \\
			&&&&&&& \\
		\vspace{-.8cm}
		3,4 	& $x\sim y$ 	& \multicolumn{2}{c|}{$p_x^\pdagger p_x^\dagger h_y^\pdagger h_y^\dagger$} & $V/2$ & $V$ & $\bigtheta{V\delta^2}$ & \\
			&			& \multicolumn{2}{c|}{$\frac{2}{V}\left|\ti_x M\right|^2$} & & & & 
			\feynmandiagram[horizontal=a to b]{a -- [fermion, half left] b -- [insertion=0.5, half left, edge label'=$x$] a}; 
			\feynmandiagram[horizontal=a to b]{a -- [anti fermion, half left] b -- [insertion=0.5, half left, edge label'=$y$] a}; \\
			&&&&&&& \\
		\vspace{-.8cm}
		5,6 	& $x\nsim y$ 	& \multicolumn{2}{c|}{$p_x^\pdagger p_x^\dagger h_y^\pdagger h_y^\dagger$} & $V/2$ & $0$ & $\ordnung{V\delta^2}$ & \\
			&			& \multicolumn{2}{c|}{$\frac{2}{V}\real\ti_x M\ti_y M^\dagger$} & & & &
			\feynmandiagram[horizontal=a to b]{a -- [fermion, half left] b -- [insertion=0.5, half left, edge label'=$x$] a}; 
			\feynmandiagram[horizontal=a to b]{a -- [anti fermion, half left] b -- [insertion=0.5, half left, edge label'=$y$] a}; \\
			&&&&&&& \\
		\vspace{-.9cm}
		7,8 	& $x\sim y$ 	& \multicolumn{2}{c|}{$p_x^\dagger h_x^\dagger h_y^\pdagger p_y^\pdagger$} & $1/2$ & $V$ & $\bigtheta{\delta}$ & \\ 
		      	&           		& \multicolumn{2}{c|}{$1-\frac{2}{V}(\rt_x M+\ti_x M\mathds{P}_yM^\dagger$)} & & & &  
			\feynmandiagram[horizontal=a to b]{a [dot, label=$x\ \ $] -- [fermion, half left] b [dot, label=$\ \ y$] -- [fermion, half left] a}; \\
			&&&&&&& \\
		\vspace{-.9cm}
		9,10 & $x\nsim y$ 	& \multicolumn{2}{c|}{$p_x^\dagger h_x^\dagger h_y^\pdagger p_y^\pdagger$} & $1/2$ & $V$ & $\ordnung{\delta}$ & \\ 
			&			& \multicolumn{2}{c|}{$\frac{1}{V}(\ti_xM\mathds{P}_yM^\dagger+\ti_yM\mathds{P}_xM^\dagger$)} & & & &
			\feynmandiagram[horizontal=a to b]{a [dot, label=$x\ \ $] -- [fermion, half left] b [dot, label=$\ \ y$] -- [fermion, half left] a}; \\
			&&&&&&& \\		
		\vspace{-.8cm}		
		11,12 	& $x\sim y$ 	& \multicolumn{2}{c|}{$p_x^\pdagger p_x^\dagger p_y^\pdagger p_y^\dagger$} & $(V+1)/2$ & $V$ & $\bigtheta{V\delta^2}$ & \\
				&			& \multicolumn{2}{c|}{$\frac{2}{V}(\rt_xM+\real\left(\ti_xM\right)^2-\ti_xM\mathds{P}_xM$)} & & & &
				\feynmandiagram[horizontal=a to b]{a -- [fermion, half left] b -- [insertion=0.5, half left, edge label'=$x$] a}; 
				\feynmandiagram[horizontal=a to b]{a -- [fermion, half left] b -- [insertion=0.5, half left, edge label'=$y$] a}; \\
			&&&&&&& \\
		\vspace{-.8cm}
		13,14 	& $x\nsim y$ 	& \multicolumn{2}{c|}{$p_x^\pdagger p_x^\dagger p_y^\pdagger p_y^\dagger$} & $(V-1)/2$ & $0$ & $\ordnung{V\delta^2}$ & \\
				&			& \multicolumn{2}{c|}{$\frac{2}{V}(\real\ti_xM\ti_yM-\real\ti_yM\mathds{P}_xM$)} & & & &
				\feynmandiagram[horizontal=a to b]{a -- [fermion, half left] b -- [insertion=0.5, half left, edge label'=$x$] a}; 
				\feynmandiagram[horizontal=a to b]{a -- [fermion, half left] b -- [insertion=0.5, half left, edge label'=$y$] a}; \\
	\end{tabular}}
\end{table*}

\begin{table*}[ht]
	\caption{Summary of physical observables (linear magnetizations and structure factors). The prime indicates that a given value can vary significantly due to finite temperature. 
	In particular $0^\prime$ only goes to zero at $\beta\rightarrow\infty$ and $V^\prime \approx \min\left(V,(\kappa\beta/\pi)^2\right)$.
	We use the Bachmann–Landau notation to denote that an error in $\bigtheta x$ scales exactly as $x$, an error in $\ordnung{x}$ scales maximally as $x$, and an error in $\littleo{x}$ is asymptotically smaller than $x$.
	\label{tab:observables}}
	\centering
	\begin{tabular}{r | c | c | c | r}
		Observable & Definition & $U=0$ & $\kappa=0$ & Error \\ \hline
		$\average{s^3_{-}}$ & $\frac 12\average{O_1-O_2}/V$ & $0$ & $0$ & $\littleo \delta$ \\
		$\average{s^3_{+}}$ & $1-\frac 12\average{O_1+O_2}/V$ & $0$ & $0$ & $\bigtheta \delta$ \\
		$S^{11}_{--}/V$ & $\frac 14\average{O_7+O_8+O_9+O_{10}}$ & $1/2$ & $V'$ & $\bigtheta \delta$ \\
		$S^{33}_{--}/V$ & $\frac 14\average{O_3+O_4-O_5-O_6+O_{11}+O_{12}-O_{13}-O_{14}}$ & $1/2$ & $V'$ & $\littleo{V\delta^2}$ \\
		$S^{11}_{++}/V$ & $\frac 14\average{O_7+O_8-O_9-O_{10}}$ & $0'$ & $0'$ & $\ordnung{\delta}$ \\
		$\average{S^{3}_{+}S^{3}_{+}}/V$ & $V-\average{O_1+O_2}+\frac 14\sum_{i\in\{3\dots 6,11\dots 14\}}\average{O_i}$ & $0'$ & $0'$ & $\bigtheta{V\delta^2}$ \\
		$S^{33}_{++}/V$ & $\average{S^{3}_{+}S^{3}_{+}}-V\average{s^3_{+}}^2$ & $0'$ & $0'$ & $\littleo{V\delta^2}$ \\
		$Q_{-}^2/V$ & $\frac14\average{-O_3-O_4+O_5+O_6+O_{11}+O_{12}-O_{13}-O_{14}}$ & $1/2$ & $0'$ & $\littleo{V\delta^2}$ \\
	\end{tabular}
\end{table*}

Let us consider the limits of no interaction ($U=0$) and no hopping ($\kappa=0$), for which the various observables can be exactly calculated for $T = 0$ ($\beta\to\infty$).
We start with the particle/hole number ($i = 1,2$ in Tab.~\ref{tab:operators}) which simply equals the number of unit cells $V$ at half-filling. Next, we consider the correlator 
between a particle and a hole ($i = 3-6$ in Tab.\ref{tab:operators}). For $U=0$, a particle does not see the holes. Therefore it is equally likely to encounter a hole 
on the same sublattice as a given particle, as it is to find a hole on the other sublattice. As there are $V$ holes in total, we find $V/2$ for each sublattice combination. 
For $\kappa = 0$, the lattice is frozen in a state of single up- or down-electrons per site with alternating sign. In terms of particles and holes, this translates to 
every $A$ site having a particle and a hole, and every $B$ site having neither (or vice-versa). Thus, a particle on a given site is perfectly correlated with all the holes on the same 
sublattice.

The coexistence of a particle and a hole as a pair in an exciton-like state ($i = 7-10$ in Tab.~\ref{tab:operators}) can be understood by a very similar consideration. 
At $U=0$ a particle is (due to charge conservation), only responsible for the existence of a single hole which can be on the same or the other sublattice yielding a correlator of $1/2$. 
$\kappa=0$ on the other hand leads to a maximal correlation between particles and holes as they can only coexist in pairs.

It remains to consider correlations between particles ($i = 11-14$ in Tab.~\ref{tab:operators}) which behave completely similarly to the particle-hole correlation but for the difference 
of self-correlation. This is not relevant in the case without hopping because it is maximally correlated anyway. In the non-interacting case however there are on 
average $(V+1)/2$ particles on the sublattice that contains the given particle and $(V-1)/2$ particles on the other sublattice.
\section{Extrapolation formulae\label{sect:extrap}}

The formulae used for the simultaneous thermodynamic and continuum limit extrapolation differ depending on the particular observable. There are however several rules applying to all of them. First of all the leading order error in the time discretisation is exactly linear $\bigtheta{\delta}$, though the coefficient is significantly reduced by our mixed differencing scheme. Next, in Ref.~\cite{semimetalmott} we showed that the spatial leading order error has to be $\bigtheta{L^{-3}}$ with the exception of observables prone to ensemble-wise positive (or negative) definite deviations. These observables cannot average out local errors over the Monte Carlo history and end up with an error in $\bigtheta{L^{-2}}$. Unfortunately the quadratic observables $S^{ii}_{--}$ and $m_s$ (\cref{eqn:extrapolation_s_af,eqn:extrapolation_m_s}) fall into this category.

From this starting point we derived the fit functions for all the magnetic observables by minor modifications. These modifications had to be made either because the leading order terms did not describe the data well enough (eq.~\eqref{eqn:extrapolation_s_1_f}), or because the in principle leading coefficient in $\delta$ (eq.~\eqref{eqn:extrapolation_s_3_s_3}) or $L$ (\cref{eqn:extrapolation_s_3_f,eqn:extrapolation_cdw}) could not be resolved numerically. A summary of the extrapolation formulae looks as follows:
\begin{align}
	\average{s^3_{+}} &= s^3_{+,0} + c^{s^3_{+}}_0\,\delta + c^{s^3_{+}}_1\,\delta \,L^{-3}\label{eqn:extrapolation_m_f}\\
	S_{AF} &= S_{AF,0} + c^{S_{AF}}_0\,\delta + c^{S_{AF}}_1\,L^{-2}\label{eqn:extrapolation_s_af}\\
	\kappa\beta m_s &= \kappa\beta m_{s,0} + c^{m_s}_0\,\delta + c^{m_s}_1 \,L^{-2}\label{eqn:extrapolation_m_s}\\
	S^{11}_{F} &= S^{11}_{F,0} + c^{S^{11}_{F}}_0\,\delta + c^{S^{11}_{F}}_1\,\delta^2 + c^{S^{11}_{F}}_2\,L^{-3} + c^{S^{11}_{F}}_3\,\delta \,L^{-3}\label{eqn:extrapolation_s_1_f}\\
	\average{S^{3}_{+}S^{3}_{+}}/V &= S^{33}_{F,0} + c^{S^{33}_{F}}_0\,\delta^2 \,L^2\label{eqn:extrapolation_s_3_s_3}\\
	S^{33}_{F} &= S^{33}_{F,0} + c^{S^{33}_{F}}_1\,\delta\label{eqn:extrapolation_s_3_f}\\
	S_\text{CDW} &= S_{\text{CDW},0} + c^{S_\text{CDW}}_0\,\delta\label{eqn:extrapolation_cdw}
\end{align}

There are two notable exceptions from the approach explained above. Firstly equation~\eqref{eqn:extrapolation_m_f} for $\average{s^3_{+}}$ can be derived analytically in the non-interacting limit. It turns out that the equation captures the leading order contributions very well even at finite coupling. $\average{s^3_{-}}$ on the other hand is always compatible with zero, even at finite $\delta$ and $L$, and is therefore not extrapolated at all.

\begin{figure}[ht]
	\includegraphics[width=0.45\columnwidth]{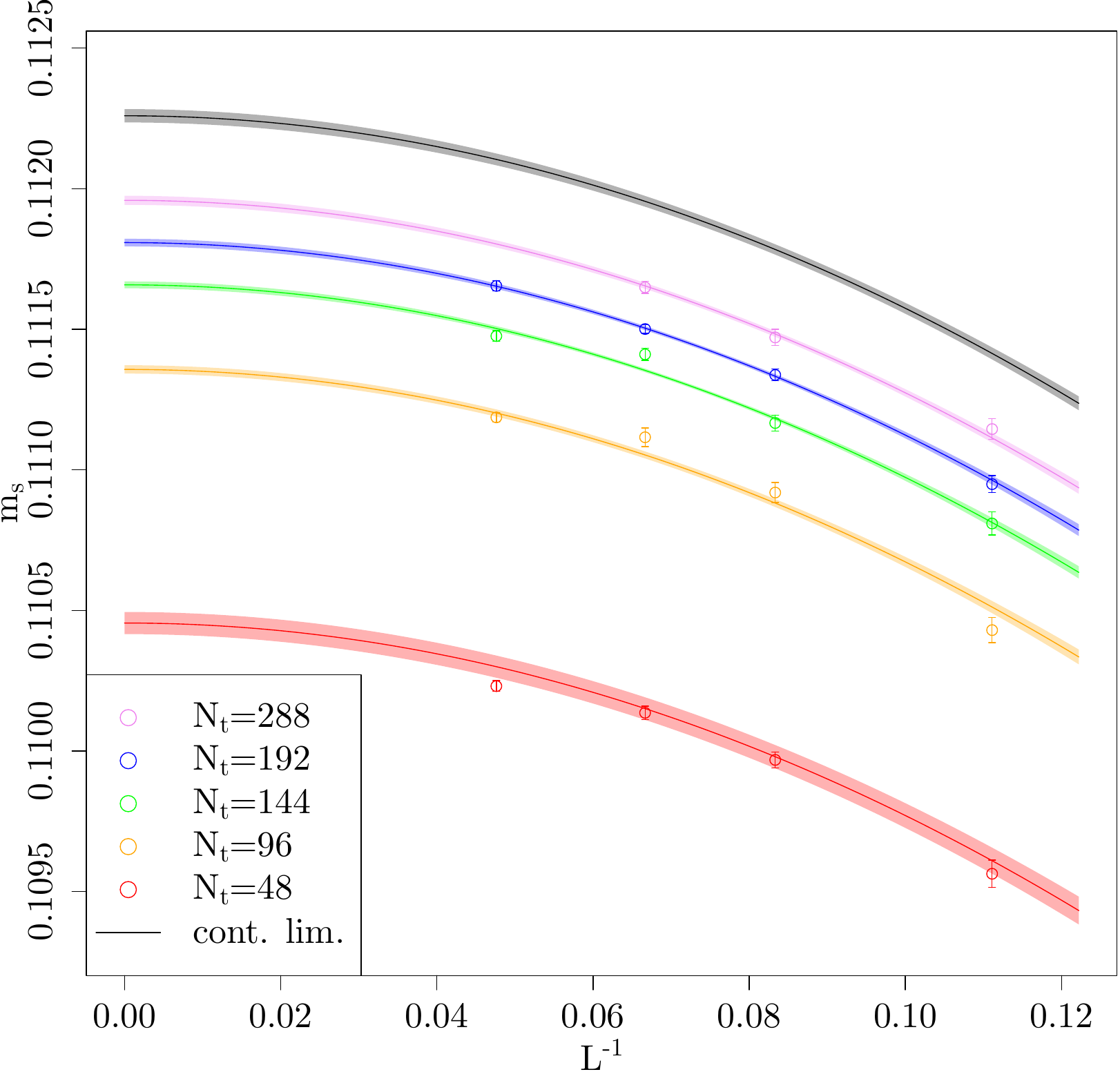}
	\includegraphics[width=0.45\columnwidth]{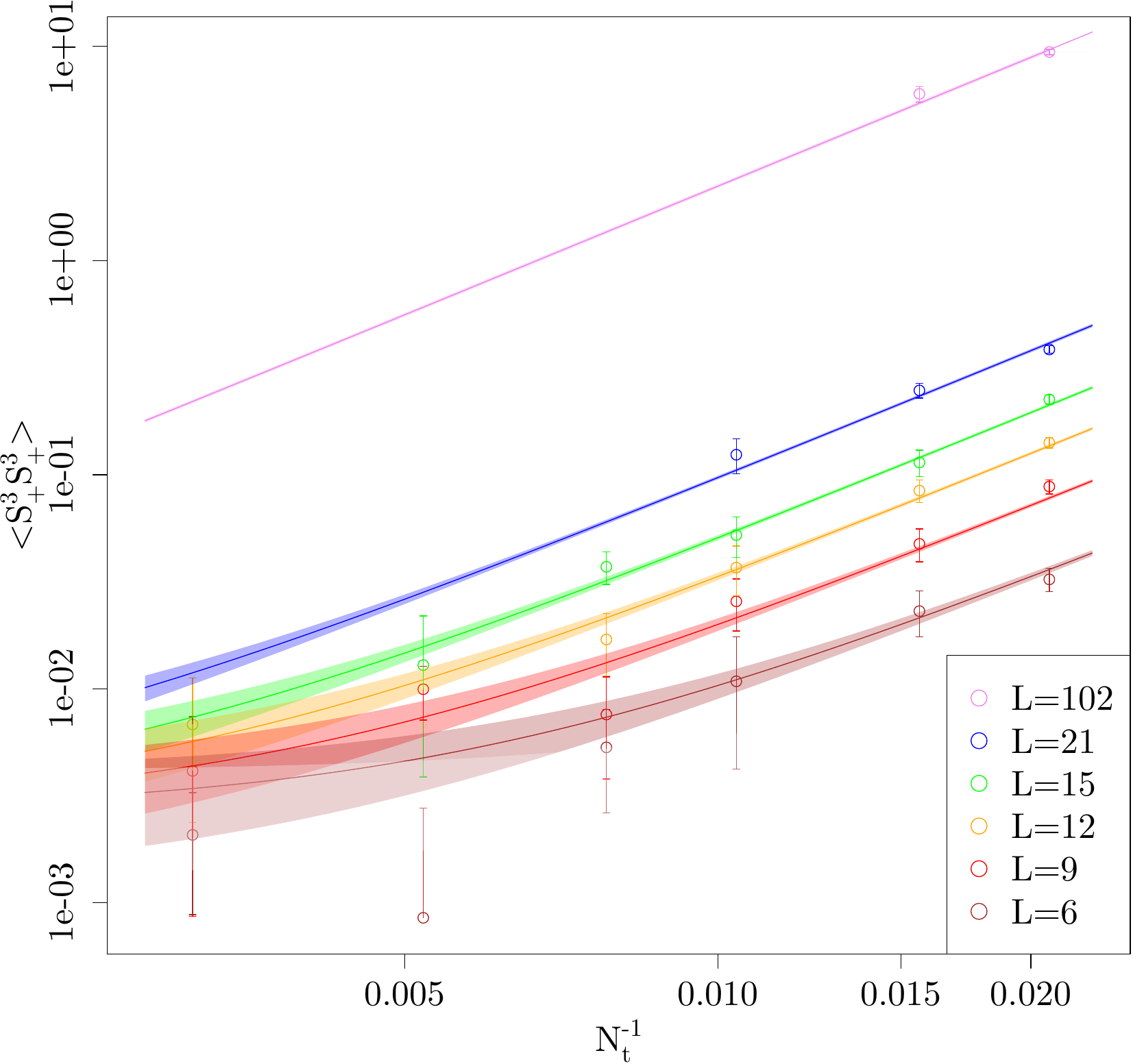}
	\caption{Simultaneous two-dimensional fit of $m_s$ using the extrapolation formula~\eqref{eqn:extrapolation_m_s} for $\beta=12$ and $U=\num{1}$ (left) and $\average{S^{3}_{+}S^{3}_{+}}$ using the extrapolation formula~\eqref{eqn:extrapolation_s_3_s_3} for $\beta=8$ and $U=\num{3.5}$ (right). All quantities are in units of $\kappa$. The fits are performed for $L\ge 9$ and $N_t/\beta\ge 6$.
	The fit on the left (right) has $\chi^2/\text{d.o.f.} \simeq \num{1.6}$ ($\chi^2/\text{d.o.f.} \simeq \num{1.1}$), corresponding to a p-value of $\simeq \num{0.08}$ ($\simeq \num{0.38}$).
	Note the double-logarithmic scale in the right panel.}
	\label{figure:extrapolation}
\end{figure}

We provide two examples for the simultaneous extrapolations in Figure~\ref{figure:extrapolation}. In the left panel we demonstrate the agreement of the data for the order parameter $m_s$ with the employed fit function~\eqref{eqn:extrapolation_m_s} even in the case of very small statistical errors.
Let us, with a look at the right panel, stress the importance of considering when to use the simple average $\average{\operator_1\operator_2}$ and when the connected version $\aaverage{\operator_1\operator_2}$. In the plot we clearly see the divergence of $\average{S^{3}_{+}S^{3}_{+}}$ in spatial volume as predicted in equation~\eqref{eqn:extrapolation_s_3_s_3}. This divergence foils any attempt of a statistically significant extrapolation, though the continuum limit would in principle yield the correct result. Should a thermodynamic limit extrapolation be attempted before the continuum limit extrapolation, one would even end up with infinite values. It is therefore of crucial importance in this case to use $S^{33}_{++}=\aaverage{S^{3}_{+}S^{3}_{+}}$ which is well behaved. On the other hand it makes sense to use $\average{\operator_1\operator_2}$ whenever $\average{\operator_1}\average{\operator_2}$ is known analytically to minimise noise. In particular $\average{S^{i}_{-}}$ and $\average{S^{1}_{+}}$ both vanish, which is why the subtraction need only be performed for $S^{33}_{++}$.
\section{Correlation matrix of the combined collapse fit\label{sect:covariance}}

The correlation matrix
\begin{align}
	\mathrm{corr}(U_c,\, \nu,\, \upbeta) &= \matr{S[round-mode=places,round-precision = 4]S[round-mode=places,round-precision = 4]S[round-mode=places,round-precision = 4]}{
		1.00000000 & -0.03947659 & -0.1197412\\
		-0.03947659 &  1.00000000 &  0.9944137\\
		-0.11974120 &  0.99441367 &  1.0000000
	}
\end{align}
clearly shows a strong correlation between $\nu$ and $\upbeta$, whereas $U_c$ is weakly correlated. Put differently, the ratio $\upbeta/\nu=\critZeta$ can be determined with extremely high precision.
We give enough digits to yield percent-level matching to our full numerical results when inverting the correlation matrix. 	%
\section{Stochastic Fourier Sources}\label{sect:stochastic sources}

We can estimate the trace of any matrix $D$ with dimension $d$ over a subset $I_0$ of the indices $I$ using the established method of noisy estimators~\cite{z2_noise,trace_estimation_overview} as follows. Sample a vector $\chi$ randomly with all the elements $\chi_i$ independent and identically distributed\footnote{We will later see that this is a sufficient but not a necessary condition.} (iid), such that
\begin{align}
	\erwartung{\chi}&=0\,,\\
	\erwartung{\chi_i^*\chi_j}&=\delta_{ij}\quad \text{if } i,j\in I_0\,,\label{eqn_orthonormal_noise}\\
	\chi_i&=0\quad \text{if } i\in I\setminus I_0\,.
\end{align}
Then the expectation value of $\chi^\dagger D\chi$, obtained by repeated calculation, yields the desired trace
\begin{equation}
	\tr_{I_0}D=\erwartung{\chi^\dagger D\chi}\,.
\end{equation}

\subsection{Comparison of different sources}
Ref.~\cite{trace_estimation_overview} gives an overview over the advantages and drawbacks of various distributions that can be employed for noisy trace estimation and also provides rigorous upper limits on the number of sources required to obtain a precision $\varepsilon$ with a probability of at least $1-\delta$. This upper bound in principle prefers Gaussian noise, but it is extremely loose and therefore not useful in practise. This is why we will not go into detail about these bounds. It suffices to know that such a bound exists and that the error on the trace estimation is proven to converge with $1/\sqrt n$ (as one would expect for a stochastic calculation) where $n$ is the number of noisy sources.

The approach presented in Ref.~\cite{z2_noise} gives a guideline how to identify efficient distributions that is far more useful. We present it here in a slightly modified form, but the conclusions are identical. Let us for now define $\chi^r$ to be the $r$-th noisy vector and $\erwartung{\cdot}$ denote the exact expectation value (where we usually use it for the MC-average). Then the two quantities
\begin{align}
	C_1(i) &\coloneqq \left|\erwartung{\frac1n\sum_r\left|\chi_i^r\right|^2}-1\right|\qquad \text{and}\\
	C_2(i,j) &\coloneqq \sqrt{\frac1n\erwartung{\left|\sum_r {\chi_i^r}^*\chi_j^r\right|^2}}\;,\;i\neq j
\end{align}
are good measures for the error. Here $C_1(i)$ quantifies misestimation of the diagonal entries of the matrix and $C_2(i,j)$ gives the error due to the off-diagonal elements being not suppressed perfectly.
Both $C_1(i)$ and $C_2(i,j)$ do not depend on the indices $i,j$ as long as the $\chi_i$ are identically (not necessarily independently) distributed for every $r$. We therefore drop the dependencies from now on. The expectation value of the total error for a trace estimation then reads
\begin{align}
	\sigma &= \frac{1}{\sqrt{n}}\sqrt{{C_1}^2\sum_{i\in I_0} \left|D_{ii}\right|^2+{C_2}^2\sum_{i,j\in I_0,i\neq j}\left|D_{ij}\right|^2}\,.
\end{align}
$C_1$ and $C_2$ are mostly independent from each other and we are going to minimize them individually in the following.

It is easy to see that $C_1$ can be eliminated completely by using noise on the complex unit circle, i.e.\ $\left|\chi_i^r\right|=1$ for all $i$ and $r$. This is besides other realised by (complex) Z2 noise which chooses with equal probability from $\lbrace \pm1\rbrace$ ($\lbrace\pm1,\pm\im\rbrace$).
Let us remark here that $C_1=0$ is a very important feature because noisy trace estimation is only efficient for diagonally dominated matrices. Z2 and similar noise calculate the trace of a diagonal matrix exactly with a single source whereas other distributions, e.g.\ Gaussian noise, would still require a high number of sources to get a decent approximation.

On the other hand $C_2$ is nearly independent of the distribution. In the case where all $\chi_i^r$ are iid, one finds
\begin{align}
	\erwartung{\eta^r} &= 0\,,\\
	\erwartung{\left|\eta^r\right|^2} &= \erwartung{\left({\chi_i^r}^*\chi_j^r\right)^*{\chi_i^r}^*\chi_j^r}
	= \erwartung{\chi_i^r{\chi_i^r}^*}\erwartung{{\chi_j^r}^*\chi_j^r}
	= 1
\end{align}
where we defined $\eta^r\coloneqq{\chi_i^r}^*\chi_j^r$ and dropped the indices $i\neq j$ as before. Thus $\eta^r$ is iid with zero expectation value and unit standard deviation. The central limit theorem now guarantees that for large $r$ the distribution approaches $\sum_r\eta^r \sim \mathcal{N}_{0,n}$. Moreover the variance is an additive quantity for independent variables which yields $C_2=1$ exactly.

This leads to the conclusion that any iid noise with numbers of unit modulus gives the same error and this error is the theoretically optimal one for iid noise.

In practice one is often not interested in the complete trace but for example in its real part only. Then real iid sources give the well~known result $C_2=1$. Remarkably however the same result also holds for complex noise where the deviation is split into the real and imaginary parts $C_2^R$ and $C_2^I$ respectively as ${C_2}^2={C_2^R}^2+{C_2^I}^2$. We observe that both projections can be minimised when $C_2^R=C_2^I=1/\sqrt{2}$. This is the case in complex Z2 noise which is the reason why it is usually preferred over real Z2 noise.

Note that neither of the possible error contributions $C_1$ and $C_2$ contains the condition $\erwartung{\chi}=0$ nor independence or identical distribution ($\erwartung{\eta^r}=0$ in contrast is required). We are therefore going to drop these superfluous requirements and see if we can find an even more efficient method for trace estimations. As $C_1$ cannot be optimised any further, we only have to consider $C_2$ which is the average of all variances and covariances of $\eta^r$ over $r$. We are now only considering noise with $\left|\chi_i^r\right|=1$, so $\left|\eta^r\right|=1$ as well and the variance of $\eta^r$ is guaranteed to be 1 as before. Therefore the only variable one can optimise is the covariance of $\eta^r$
\begin{align}
	\zeta^{rs} \coloneqq \mathrm{cov}\left({\eta^r}^*,\eta^s\right)
\end{align}
with which we can rewrite the off-diagonal error contribution
\begin{align}
	C_2 &= \sqrt{\frac1n\erwartung{\sum_{r,s}{\eta^r}^*\eta^s}}
	= \sqrt{\frac1n\erwartung{\sum_r\left|\eta^r\right|^2 + \sum_{r\neq s}{\eta^r}^*\eta^s}}
	= \sqrt{1+\frac1n\sum_{r\neq s}\zeta^{rs}}\,.
\end{align}
The symmetry in $r$ and $s$ suggests that $C_2$ is minimised when all $\zeta^{rs}$ are equal and maximally negative. This optimum is, quite intuitively, reached for all $\chi^r$ pairwise orthogonal. In this case the probability $p_r$ of redundancy one would have for iid noise is reduced to $p_r-1/(d-1)$ because one of the dimensions cannot be reached again. This difference already is the covariance we are looking for. Thus we obtain a lower bound for the error contribution
\begin{align}
	C_2 &\ge \sqrt{1+\frac1n\sum_{r\neq s}\frac{-1}{d-1}}
	= \sqrt{1-\frac{n(n-1)}{n(d-1)}}
	= \sqrt{1-\frac{n-1}{d-1}}
	= 1-\frac{n}{2d}+\ordnung{d^{-1}}+\ordnung{\left(\frac nd\right)^2}\,.
\end{align}
It is easy to convince oneself that this really is the optimum because the error vanishes for $n=d$ when one has sampled the full space. Therefore there is no deterministic nor stochastic algorithm that does not use any specific features of a given matrix and is more efficient than this method.

An efficient practical realisation would be to draw
\begin{align}
	\chi_i^0 &\sim \left\lbrace\eto{\im2\pi\frac kd} | k\in\lbrace0,\dots,d-1\rbrace\right\rbrace
\end{align}
iid and for all $r\ge1$ to sample $j_r\in\lbrace1,\dots,d-1\rbrace$ without repetition setting
\begin{align}
	\chi_i^r &= \eto{\im2\pi\frac {i}{d}j_r}\chi_i^0\,.
\end{align}

Last but not least we emphasise that for the trace estimation to be useful one has to reach the desired accuracy at some $n\ll d$. So the tiny gain from choosing all noise vectors pairwise orthogonal is negligible in practice.

\subsection{Second order observables}
The quartic operators $O_{3\dots14}$ cannot be expressed in terms of simple traces. Instead we have to generalise the noisy source estimation.
There are different possible approaches. In our simulations we chose to sample $\chi$ again as above, then define
\begin{align}
	\chi' &= M^{-1}\chi\,,\\
	\chi''&={M^{\dagger}}^{-1}\chi
\end{align}
and estimate the relevant quantities as
\begin{align}
	\sum_{x,y\in A}\left|\left(M^{-1}\right)_{(x,t),(y,t)}\right|^2 &= \erwartung{\sum_{x\in A}\left|\chi'_{x,t}\right|^2}\,,\\
	\sum_{x,y\in A}\left(\left(M^{-1}\right)_{(y,t),(x,t)}\left(M^{-1}\right)_{(x,t),(y,t)}\right) &= \erwartung{\sum_{x\in A}{\chi''_{x,t}}^*\chi'_{x,t}}\,.
\end{align}
Note that the right hand side term is not a simple scalar product $\chi'^\dagger\cdot \chi'$, respectively $\chi''^\dagger\cdot\chi'$. We have to include the projection here explicitly. This allows to compute $\sum_{x\in A,y\in B}$ nearly without additional computational effort, as only the projection sum has to be adjusted, not the linear solve.

\subsection{Average over time slices}
For the structure factors the averages over the time slices have to be included explicitly and cannot be absorbed in the traces. This means that the number of linear solves increases from the number of noisy sources $n$ to $N_t\cdot n$. As this is extremely expensive, we evaluate the average over all time slices by an average including only a fixed number of equidistant times, which is guaranteed to be equal in the infinite-statistics limit by translation invariance while saving computational cost without losing much accuracy, since neighbouring time slices on any given configuration are correlated.

Another approach would be to sample $\chi$ not from e.g.\@ $I_0=(A,t)$, but from $I$ always and include the projection later
\begin{equation}
	\tr_{I_0}D=\sum_{i\in I_0,j\in I}\erwartung{\chi_i^*D_{ij}\chi_j}\,.\label{eqn_trace_projected}
\end{equation}
Na\"ively one would expect that this saves a lot of linear solves, but it also introduces a lot of additional noise. The variance of the sum in equation~\eqref{eqn_trace_projected} increases linearly in the cardinality of $I$. In order to compensate this increase in statistical fluctuations we would have to increase the number of noisy sources by the same amount. Thus at equal statistical precision the number of linear solves stays the same.

\subsection{The time-shifted trace}
In general we do not only have to calculate traces, but also sums over off-diagonal matrix elements---correlators between operators at different times. This time shift can be implemented in the following way (again $\chi$ sampled as above):
\begin{align}
	\sum_t D_{t+\tau,t}&=\sum_{t,t'}D_{t',t}\delta_{t',t+\tau}
	=\sum_{t,t'}D_{t',t}\erwartung{\chi_{t'}^*\chi^{\vphantom{*}}_{t+\tau}}
	=\erwartung{\sum_t \left(D^\dagger\chi\right)^*_t\chi^{\vphantom{*}}_{t+\tau}}\,,
\end{align}
where $\tau$ is an arbitrary time shift.
 \end{subappendices}\clearpage{}%

	\chapter{Summary}
\section{The Ising model with HMC}

In this thesis we first (\Cref{chap:ising}) showed how to apply the hybrid Monte Carlo (HMC) algorithm to the Ising model, successfully applying an algorithm that uses only continuous state variables to a system with discrete degrees of freedom.
We find that the HMC algorithm generalises the Ising model very well to arbitrary geometries without much effort.
It has been presented here in its simplest form, and as such, we find that
 HMC is an extremely inefficient algorithm for the Ising model.
Although more flexible than the most efficient methods, such as cluster algorithms, its performance is wore than the standard Metropolis-Hastings algorithm.
The coefficient by which the Metropolis-Hastings algorithm surpasses the HMC decreases with dimension, so that HMC might be preferable in case of an extremely high number of nearest neighbours---in the case of less local coupling, for example.

Moreover, for physical systems that suffer from sign problems, one may hope to leverage complex Langevin, Lefschetz thimble, or other contour-optimizing methods (for a dramatically incomplete set of examples, consider, respectively, \Ref{Aarts:2010gr,Fodor:2015doa}, \Refs{Cristoforetti:2012su,Fujii:2013sra,Alexandru:2017oyw}, and \Refs{Alexandru:2017czx,Mori:2017pne,Kashiwa:2018vxr} and references therein).
The formulation in terms of continuous variables presented here is well-suited for these methods, while the methods that deal directly with the original discrete variables such as the Metropolis-Hastings, cluster, and worm algorithms, for example, are non-starters.
In that sense, our exact reformulation and HMC method can be seen as the first step towards solving otherwise-intractable problems.

Furthermore, the HMC algorithm could be optimised by more efficient integrators and different choices of the shift $C$, just to name the most obvious possibilities.
Many more methods have been developed to improve HMC performance and it is expected that some of them could also speed up the Ising model.

\section{The generalised eigenvalue and Prony methods}

Next, in \Cref{chap:prony} we clarified the relation among different methods
for the extraction of energy levels in lattice quantum chromodynamics (QCD) available in the
literature. We  proposed and tested a new combination of
generalised eigenvalue and Prony method (GEVM/PGEVM), which helps reduce excited state contaminations.

We first discussed the systematic effects in the Prony GEVM stemming
from states not resolved by the method. They decay exponentially fast
in time with $\exp(-\Delta E_{n, l} t_0)$ with $\Delta E_{n, l}
=E_{n} -E_l$ the difference of the first not resolved energy level
$E_{n}$ and the level of interest $E_l$. Using synthetic data we
have shown that this is indeed the leading correction.

Next we applied the method to a pion system and discussed its
ability to also determine backward propagating states, given high
enough statistical accuracy, see also
Ref.~\cite{Schiel:2015kwa}. Together with the results from the 
synthetic data we concluded that working at fixed
$\delta t$ is clearly advantageous compared to working at fixed $t_0$,
at least for data with little noise.

Finally, looking at lattice QCD examples for the $\eta$-meson and the
$\rho$-meson, we find that excited state contaminations can be reduced
significantly by using the combined GEVM/PGEVM. While it is not
clear whether also the statistical precision can be improved,
GEVM/PGEVM can significantly improve the confidence in the extraction
of energy levels, because plateaus start early enough in Euclidean
time. This is very much in line with the findings for the Prony method
in the version applied by the NPLQCD
collaboration~\cite{Beane:2009kya}.

The GEVM/PGEVM works particularly well, if in the first step
the GEVM removes as many intermediate states as possible and,
thus, the gap $\Delta E_{n, l}$ becomes as large as possible in the
PGEVM with moderately small $n$. The latter is important to
avoid numerical instabilities in the PGEVM.

\section{The quantum phase transition of the Hubbard model}

Our work in \Cref{chap:phase_transition,chap:afm_character} represents the first instance where the grand canonical Brower-Rebbi-Schaich (BRS) algorithm has been applied to the hexagonal Hubbard model 
(beyond mere proofs of principle), and we have found highly promising results. We emphasize that previously encountered issues related to the computational scaling and ergodicity 
of the HMC updates have been solved~\cite{Wynen:2018ryx}.
Our findings are split into two parts, broadly speaking \Cref{chap:phase_transition} focuses on electric and \Cref{chap:afm_character} on magnetic properties.

We calculated the single particle gap $\Delta$ as well as all operators that contribute to the antiferromagnetic (AFM), ferromagnetic (FM), and charge density wave (CDW) order parameters of the Hubbard Model on a honeycomb lattice.
Furthermore we provide a comprehensive analysis of the temporal continuum, thermodynamic and zero-temperature limits for all these quantities. The favorable scaling of the HMC
enabled us to simulate lattices with $L > 100$ and to
perform a highly systematic treatment of all three limits. The latter limit was taken by means of a finite-size scaling analysis, which determines the
critical coupling $U_c/\kappa=\critU$ as well as the critical exponents $\nu=\critNu$ and $\upbeta=\critExp$.

Depending on which symmetry is broken, the critical exponents of the hexagonal Hubbard model 
are expected to fall into one of the Gross-Neveu (GN) universality classes~\cite{PhysRevLett.97.146401}. The semimetal-antiferromagnetic Mott insulator (SM-AFMI) transition falls into the GN-Heisenberg 
$SU(2)$ universality class, as the staggered magnetisation $m_s$ is described by a vector with three components.

The GN-Heisenberg critical exponents have been studied by 
means of projection Monte Carlo (PMC) simulations of the hexagonal Hubbard model, by the $d = 4-\epsilon$ expansion around the upper 
critical dimension $d$, by large~$N$ calculations, and by functional renormalization group (FRG) methods.
In Table~\ref{tab:crit_values}, we give an up-to-date comparison with our results. Our value for $U_c/\kappa$ is in overall agreement with previous Monte Carlo (MC) simulations. 
For the critical exponents $\nu$ and $\upbeta$, the situation is less clear. Our results for $\nu$ (assuming $z = 1$ due to Lorentz invariance~\cite{PhysRevLett.97.146401}) agree 
best with the MC calculation (in the Blankenbecler-Sugar-Scalapino (BSS) formulation) of Ref.~\cite{Buividovich:2018yar}, followed by the FRG and large~$N$ calculations. 
On the other hand, our critical exponent $\nu$ is systematically larger than most PMC calculations and first-order $4-\epsilon$ expansion results. The agreement appears to be significantly 
improved when the $4-\epsilon$ expansion is taken to higher orders, although the discrepancy between expansions for $\nu$ and $1/\nu$ persists.
Finally, our critical exponent $\upbeta$ does not agree with any results previously derived in the literature. They have been clustering in two regions. The PMC methods and first order $4-\epsilon$ expansion yielding values between $0.7$ and $0.8$, the other methods predicting values larger than 1. Our result lies within this gap at approximately $0.9$ and our uncertainties do not overlap with any of the other results.

Thus, though we are confident to have pinned down the nature of the phase transition and to have performed a thorough analysis of the critical parameters, the values of $\nu$ and $\upbeta$ remain ambiguous. This is mostly due to a large spread of incompatible results existing in the literature prior to this work. With our values derived by an independent method we add a valuable confirmation for the critical coupling and some estimations of $\nu$ as well as a new but plausible result for $\upbeta$.

In addition to an unambiguous classification of the character of the quantum critical point (QCP) of the honeycomb Hubbard model, our results demonstrate the ability to perform high-precision 
calculations of strongly correlated electronic systems using lattice stochastic methods. 
A central component of our calculations is the Hasenbusch-accelerated HMC algorithm, 
as well as other state-of-the-art techniques originally developed for lattice QCD. This has allowed us to push our calculations to system sizes which are, to date, still the largest that have 
been performed, up to 102$\times$102 unit cells (or 20,808 lattice sites), which reaches a physically realistic size in the field of carbon-based nano-materials.
Thus, it may be plausible to put a 
particular experimental system (a nanotube, a graphene patch, or a topological insulator, for instance) into software, for a direct, first-principles 
Hubbard model calculation.

There are several future directions in which our present work can be developed that go beyond algorithmic improvements and simulations of yet larger lattices. For instance, while the AFMI phase may not be directly observable in graphene, 
we note that tentative empirical evidence for such a phase exists in carbon nanotubes~\cite{Deshpande02012009}, along with preliminary theoretical 
evidence from MC simulations presented in Ref.~\cite{Luu:2015gpl}. The MC calculation of the single-particle Mott gap in a (metallic) carbon nanotube is 
expected to be much easier, since the lattice dimension $L$ is determined by the physical nanotube radius used in the experiment (and by the number of unit 
cells in the longitudinal direction of the tube). As electron-electron interaction (or correlation) effects are expected to be more pronounced 
in the (1-dimensional) nanotubes, the treatment of flat graphene as the limiting case of an infinite-radius nanotube would be especially interesting.
Strong correlation effects could be even more pronounced in the (0-dimensional) carbon fullerenes (buckyballs), where we are also faced 
with a fermion sign problem due to the admixture of pentagons into the otherwise-bipartite honeycomb structure~\cite{leveragingML}.
This particular sign problem has the unusual property of vanishing as the system size becomes
large, as the number of pentagons in a buckyball is fixed by its Euler characteristic to be exactly~$12$, independent of the number of hexagons.

Our progress sets the stage for future high-precision calculations of additional observables of 
the Hubbard model and its extensions, as well as other Hamiltonian theories of strongly correlated electrons~\cite{PhysRevB.100.125116,PhysRevLett.122.077602,PhysRevLett.122.077601}. 
We anticipate the continued advancement of calculations with ever increasing system sizes, through the leveraging of additional state-of-the-art techniques from lattice QCD, 
such as multigrid solvers on GPU-accelerated architectures. We are actively pursuing research along these lines.

	\clearpage
	\pdfbookmark{Bibliography}{bibliography}
	\printbibliography[notkeyword=own]
	
	\cleardoublepage
\clearpage{}%
\pdfbookmark{Acknowledgements}{acknowledgements}
\chapter*{Acknowledgements}

This work would not have been possible without the support of many people. In particular the last year\footnote{Rejoice, dear reader from far future, shouldst thou not know the dark age whither I refer to.}
made the importance of human interactions and people to rely on painfully apparent to me. Therefore, in the first place I thank all those people that kept me from going insane (or helped me maintaining a constant level of insanity). Luckily, more people maintained my insanity than I can reasonably list by name here, starting with my family and friends, continuing with my colleagues, the members of student body, Physics Show as well as diverse choirs and ending with the very rare occasional stranger spicing up the monotony of everyday life. Let me for this reason name only Tillman Brehmer explicitly, the only person I ever brewed beer with.

I owe great thanks to my supervisors Carsten Urbach and Tom Luu. Not only did they provide me with thrilling research topics and help with any questions regarding these topics, they also gave me the opportunity to pursue independent and often exotic research.
With Tom my independence sometimes even went so far that it felt like I was his boss, not the other way around. Thank you, Tom, for making me feel so important.
My work with Carsten went far beyond research. Be it organising lectures, designing new exercises or even making music, I enjoyed our time together very much.

Furthermore I thank both groups I had the privilege to work with. It is a pity we could not spend (more) time together in person. In particular I thank Jan-Lukas Wynen and Marcel Rodekamp for many a productive session and my former office mates Matthias Fischer, Niko Schlage and Martin Ueding for the relaxed atmosphere and all our animated discussions.

During the last year I gained a precious collaborator in Manuel Schneider and practically a third supervisor in Karl Jansen. Not only my work profited from this acquaintance, it has also been an enrichment for me personally.

Thanks to Bernard Metsch for my promotion from tutor to lecturer assistant and for tolerating my thoughtless and often repetitive whistling.

I am deeply thankful to Micha Kajan, the guy who proof reads all the effusions of my keyboard.\clearpage{}%

\end{document}